\documentclass[11pt,a4paper]{article}

\usepackage[utf8]{inputenc}
\usepackage[T1]{fontenc}
\usepackage[english]{babel}
\usepackage{amsmath,amssymb}
\usepackage{geometry}
\usepackage[expansion=false]{microtype}
\usepackage{xcolor}
\usepackage{graphicx}
\usepackage[colorlinks=true,allcolors=blue]{hyperref}
\usepackage{enumitem}
\usepackage{booktabs}
\usepackage{tabularx}
\usepackage{longtable}
\usepackage{tikz}
\usetikzlibrary{arrows.meta,decorations.pathreplacing}

\newcommand{\Nsep}{\mathcal{N}_{\mathrm{sep}}}
\newcommand{\Npr}{N_{\mathrm{PR}}}

\title{\bfseries Toward a Thermodynamic Framework for Dissipative Solitons:\\
From Photonics to Turbulence and Bose--Einstein Condensate Analogies}

\author{Vladimir L.\ Kalashnikov$^{1,*}$ and Irina T.\ Sorokina$^{1,2}$ \\[0.3em]
\small $^{1}$Department of Physics, Norwegian University of Science and Technology,\\
\small 7491 Trondheim, Norway\\
\small $^{2}$ATLA lasers AS, Richard Birkelands vei 2B,\\
\small 7034 Trondheim, Norway\\
\small $^{*}$Correspondence: vladimir.kalashnikov@ntnu.no}

\begin{document}

\maketitle

\begin{abstract}
\noindent
Thermodynamic concepts are increasingly used in nonlinear photonics to describe Rayleigh--Jeans thermalization, optical wave turbulence, condensation, negative-temperature states, and statistical mode-locking phenomena. This raises a broader question: how far can thermodynamic reasoning be extended to localized structures maintained far from equilibrium by a continuous balance of gain, loss, dispersion, and nonlinearity? We address this question using strongly chirped dissipative solitons (DSs) of the complex cubic--quintic Ginzburg--Landau equation (CQGLE) as a model system. Their internal energy flows and separation of correlation scales provide a natural link between coherent solitary waves, semi-incoherent wave kinetics, driven open systems, and analogies with Bose--Einstein condensation. We review thermodynamic-like descriptions based on spectral entropy, internal energy, effective temperature, and condensation-like spectral restructuring, and relate them to dissipative-soliton resonance (DSR), stochastic mode-locking self-start, and redistribution between single- and multipulse attractors. Normal and anomalous group-delay dispersion (NGD and AGD) provide complementary realizations. In NGD, DSR is accompanied by spectral localization and increasing separation of two correlation scales, while stochastic dynamics show increasing accessibility of multipulse states. In AGD, the analytical spectrum has extended wings and dynamical robustness occupies only part of the existence domain. These results distinguish general features of nonequilibrium state selection from effects tied to a particular localization mechanism. We argue that thermodynamic-like observables are most useful as coarse-grained structural diagnostics rather than equilibrium state variables, and that DSs provide a photonic platform for exploring connections among nonequilibrium thermodynamics, wave turbulence, driven condensates, and statistical phase-transition concepts.
\end{abstract}


\vspace{1em}

\section{Introduction}

The most effective way to extract a large pulse energy directly from a
mode-locked oscillator is not to make the pulse more intense but to make it longer. In an all-normal-dispersion fiber laser \cite{Chong2008} or a
solid-state chirped-pulse oscillator \cite{Kalashnikov2005,Rudenkov2023}, the intracavity pulse is strongly chirped: dispersion and self-phase modulation stretch it in time, its peak power stays below the level at which nonlinearity destroys it, and the compression of the strongly chirped pulse is performed outside the cavity. Pulse stretching enables energy scaling while limiting the intracavity peak power. Within the reduced model used here, \emph{dissipative soliton resonance} (DSR) denotes the branch asymptotic in which the pulse energy can grow without bound through temporal stretching while the peak power and spectral width approach finite limits. This is a definition of an asymptotic solution family, not a statement that its arbitrarily large-energy members are dynamically accessible or stable \cite{Chang2008,Chang2009,GreluDSR2010}. Real oscillators do not deliver infinite pulse energy: they reach a finite maximum beyond which the single-pulse state becomes unstable. Multipulsing is one route to this limit. Other terminating mechanisms include continuous-wave breakthrough, breathing and period-doubling, $Q$-switched mode locking, noise-like operation, Raman-induced instabilities, dispersive-wave destabilization, and outright loss of mode locking. Which channel is reached first depends on the specific system and operating path. This paper is organized around one question: within the model adopted below, what fixes that maximum scalable energy?

The existence of a stationary DS is only the first level of the problem. A
solution may exist mathematically and may even be locally robust, while a noisy
laser preferentially reaches another attractor. This distinction is especially
important near the energy-scaling regime, where single- and multipulse states
can compete. We therefore distinguish throughout between \emph{existence},
\emph{dynamical robustness}, and \emph{noise-driven accessibility}. The last
of these is the natural point of contact with statistical and thermodynamic
language: it asks which of several available states is selected by the driven,
fluctuating dynamics, rather than which state minimizes a known equilibrium
potential.

A strongly chirped DS is a plausible object for such a description because the
chirp creates a pronounced internal spectral--temporal structure and
two well separated correlation scales. That structure is physically meaningful,
but it should not be confused with statistical incoherence. A deterministic
chirped pulse is still a coherent field realization. The statistical content
enters only when an ensemble is introduced through spontaneous-emission noise,
technical fluctuations, gain dynamics, or repeated self-start events. The
thermodynamic program pursued below is therefore modest but testable: the
spectral scale separation and the associated entropy-like indicators are asked
to organize, or predict, how the accessibility of single- and multipulse states
changes with energy. They are not assumed to be equilibrium state variables.

Thermodynamic language has already entered nonlinear optics with considerable success, along three largely independent routes. Weak-turbulence kinetics describes irreversible relaxation toward a Rayleigh--Jeans (RJ) spectrum, an $H$-theorem, and condensation of the wave energy into the lowest modes \cite{Picozzi2007,Picozzi2014,Connaughton2005,During2009}. The
equilibrium thermodynamics of highly multimoded systems supplies an extensive entropy, an equation of state, and RJ thermalization in a finite guided-mode set \cite{Wu2019,Pourbeyram2022,Baudin2020}. Negative-temperature RJ equilibrium states have subsequently been observed directly in multimode optical fiber \cite{Baudin2023}, while related negative-optical-temperature thermodynamic processes have been explored on other photonic platforms \cite{Mangini2022,Muniz2023}. The noise-driven statistical mechanics of mode locking supplies a Gibbs measure over cavity modes and treats pulse formation as a first-order phase transition
\cite{GordonFischer2002,Gat2004,Weill2010}. These three routes are set out in Section~3, because the framework developed here is best understood by contrast with them. What they share is decisive: in each, the modal basis or coarse-graining cutoff is a property of the setup---an ultraviolet cutoff imposed by discretization or by a finite band, the number of guided modes fixed by the fiber geometry, the coarse-graining correlation scale by the spectral filter---and that basis, as distinct from the occupations defined on it, does not respond to the energy stored in the field. It is the ``volume'' variable of those theories.

For a strongly chirped DS in NGD, resonance of the soliton
wavenumber with the linear dispersive branch and the requirement \(P\ge0\)
produce a finite cutoff \(\Delta\), while the spectral core has width \(\Xi\).
The resulting truncated Lorentzian spectrum has the RJ algebraic form and
admits the chemical-potential-like identification
\(-\mu_{\mathrm{shape}}=\Xi^{2}\)
\cite{Podivilov2005,Kalashnikov2009,Kalashnikov2024}. Its time-integrated field
autocorrelation consequently contains two scales: a short graining scale
\(l=\pi/\Delta\) set by the edge and a long collective scale
\(\Lambda=1/\Xi\) set by the core. In AGD there is no intrinsic hard cutoff;
Section~4.1 constructs the analogous short scale from an explicitly stated
spectral-dissipation/capture window. Their ratio
\begin{equation}
  r \;\equiv\; \frac{\Delta}{\Xi} ,
  \qquad
  \Nsep \;\sim\; \frac{\Lambda}{l},
  \label{eq:rdef}
\end{equation}
is the \emph{NGD spectral scale-separation index} and the prototype for the
operational AGD index defined later. Stated once, in the form that governs every later use: $r=\Delta/\Xi$ is a \emph{deterministic} scale parameter of the chirped solution, and interpreting it as an effective number of statistically occupied degrees of freedom requires an additional ensemble or coherent-mode calibration, which does not follow from the chirp. $r$ is used as the primary variable, and the relation between the two is written as $\sim$ rather than as an equality: the factor relating them is a correlation width convention, comparing the first zero of a sinc kernel with the decay length of an exponential. Different width-matching criteria give different numerical constants---the first zero of one profile against the half-width or the matched second moment of the other---so only the order-unity character of that factor is used in what follows, and no value is quoted for it. What is convention-free is the range and the ordering: $r\to0$ on the energy unscalable branch (Sec. 2.2), where the two spectral scales merge and the deterministic pulse becomes effectively a single-scale spectral structure; $r=1$ on the fidelity curve (Sec. 2.2); and $r\to\infty$ in the resonance limit (Sec. 5). Since the participation number $\Npr=(\sum_m\lambda_m)^2/\sum_m\lambda_m^2$ of an ensemble coherence operator satisfies $\Npr\ge1$ by construction, a useful calibration against it would be expected to approach the one-mode limit $\Npr\simeq1$ as $r\to0$, where the two spectral scales merge, and to be reproducibly non-decreasing over the tested range of $r$. Whether $\Npr$ remains unbounded as $r\to\infty$---the limit that corresponds to DSR---is not fixed by the deterministic scale separation and must be determined from the ensemble: finite detector bandwidth, gain correlations, noise statistics could, in principle, cause the participation number to saturate. The testable proposition is stated in Section~4.3: $\Npr$, computed from a defined ensemble, is a reproducible monotone function
$\Npr=f(r)$. What is established without that step is structural: $r$ occupies the place that the mode number occupies in conservative multimode thermodynamics. Still, it is generated by the solution rather than imposed on it, and it depends on the operating point
and on the energy.

Everything that follows is a consequence of that displacement, and the chain is
short enough to state here. First, in NGD the reduced adiabatic DSR limit
has a direct spectral representation: the core parameter \(\Xi\) tends to zero
at finite cutoff \(\Delta\), while the energy is accommodated by temporal
stretching. We call this \emph{condensation-like spectral narrowing} only
descriptively; it is not by itself evidence of a macroscopic statistical
occupation (Section~\ref{sec:bec}). AGD reaches DSR by a different
algebraic mechanism: the resonance locus must intersect the finite adiabatic
existence window, which requires sufficiently strong quintic phase saturation
\cite{Kalashnikov2026}. Thus the two dispersion regimes share energy-scaling
phenomenology but not an identical spectral-support mechanism. 

Second, on the
NGD branch this scale separation raises the shifted shape indicator \(H_s\),
because \(\Delta/\Xi\) grows. This is the specific sense in which the earlier
work contrasted the DS construction with the standard Bose--Einstein picture. However, this contrast is a statement about the shape functional, not about the total entropy in a fixed frequency bin, which also carries a falling scale term (Section~6.1). 

Third, along the specified isogain continuation and above a threshold energy, the internal-energy proxy passes through a maximum while the chosen entropy indicator continues to rise. The directional energy--entropy slope $\Theta_{s}=(\mathrm{d}U/\mathrm{d}s)/(\mathrm{d}H/\mathrm{d}s)$ therefore changes sign---formally the \emph{mirror image} of the textbook bounded-spectrum case, in which the entropy has an interior maximum and the temperature passes through infinity (or, depending on the convention for $\Theta_s$, through zero) (Section~6.3). This is a property of the continuation and of the adopted diagnostics, not an equilibrium absolute temperature. 

Fourth, at some energy, the additive entropy proxy of a multipulse complex carrying the same total energy at the same net gain may exceed that of its single-pulse counterpart, which we advance as a \emph{candidate} selection rule for fragmentation \cite{Renninger2010,Kalashnikov2025}. Because the pulses in such a complex share a single gain reservoir, the additivity underlying this comparison is an assumption that must be tested against transition statistics rather than taken for granted. On this reading, and in NGD, the ceiling on energy scaling need not be a loss of existence or of linear stability: it may be entropic in origin and arrive before the solution ceases to exist. That is a proposal, not a result---it presupposes the ensemble of Section~4.3 and the additivity of Section~6.4---and it does not transpose to AGD: along the continuation examined there, the adopted indicators show neither an entropy turnover nor a sign reversal, and, independently, the available finite-time noisy linearized scan displays a one-sided accessibility boundary inside the algebraic existence region, whose status as the asymptotic energy-limiting mechanism remains open (Sections~6.5 and 6.6). Neither claim excludes the other channels listed above. The model addresses one of them.

This article critically synthesizes that framework rather than surveying
DSs, for which comprehensive accounts already exist
\cite{Ankiewicz2008,Grelu2012}, or claiming a completed thermodynamics. Much of what follows was developed in three recent papers \cite{Kalashnikov2024,Kalashnikov2026,Kalashnikov2025}. The purpose here is to state that framework in one place, separate what it establishes from what it assumes, and specify the calculations and measurements that would decide the difference. Several things are attempted here that the primary papers do not. The adiabatic theories of normal and anomalous dispersion are brought onto a single master diagram, so that the two resonance mechanisms can be compared
rather than merely juxtaposed. The scale-separation index is confronted with the definition it would have to satisfy---the participation ratio of a
Karhunen--Lo\`eve decomposition of an ensemble coherence kernel
\cite{Wolf1982,StarikovWolf1982}---together with an explicit statement of the assumptions that definition requires and of the ones it is not yet known to satisfy \cite{Ponomarenko2004}. We distinguish the shifted spectral-shape indicator \(H_s\) used in the earlier work from the full fixed-bin differential entropy, retain the associated scale term explicitly, and examine which conclusions depend on that choice of functional. The correspondence with
driven-open condensates is put on a term-by-term basis, with the reservoir, noise, and geometry assumptions under which the reduction holds stated rather than suppressed, its sign convention fixed once, and the chirped DS located within the family of dissipative condensate models according to what each member conserves and what each minimizes
\cite{Pawlowski2017,WoutersCarusotto2007,He2015}. A density-of-states argument is used to separate the part of the divergence between DS spectral condensation and Bose--Einstein condensation that is kinematic from the part that is genuinely dissipative in origin.

The scope is correspondingly well-defined, and five limitations should be stated at the outset rather than discovered later. (i)~The analysis is confined to the $(1+1)$-dimensional CQGLE in its strongly chirped regime, where the group-delay dispersion and nonlinear phase terms dominate the corresponding dissipative corrections sufficiently for the adiabatic strong-chirp reduction to remain controlled. Outside that regime, the closed-form spectra are unavailable. (ii)~The statistical object is not yet secured. All spectra used below are those of a single deterministic solution, whose coherence kernel is of rank one. The identification of $\Nsep$ with a coherent-mode participation number presupposes an ensemble that is defined here (Section~4.3) but not yet constructed, and the eigenvalue calculation that would test the identification has not been performed. (iii)~The treatment is mean-field, and the modal weights that would enter any participation number are known to be closure-dependent \cite{Ponomarenko2004}. (iv)~The differential entropy of a continuous spectrum depends on the reference measure. We therefore fix a state-independent frequency bin, retain the scale term that this makes explicit, and report which conclusions survive its restoration and which do not (Section~6.1). (v)~Because the system is genuinely out of equilibrium, the entropy, temperature, and free energy defined below are \emph{thermodynamic-like indicators} evaluated along a named continuation path rather than state functions of an equilibrium theory. Namely, $\Theta$ is a directional slope whose numerical value is shown below to depend on the chosen continuation path, so it cannot in its present definition be interpreted as a path-independent state temperature; free-energy minimization does not select the stable state, and the entropy and internal-energy crossings acquire predictive content only once they are corroborated by direct dynamical simulation. Making the status of these quantities explicit, rather than assuming it, is one of the tasks of this paper. Table~\ref{tab:claimstatus} summarizes the status of these claims, and the same labels are used again in the conclusions.

{\small
\begin{longtable}{@{}p{0.34\textwidth} l p{0.42\textwidth}@{}}
\caption{Status of the principal claims. \textbf{E}~=~established result of the
adiabatic CQGLE analysis or of the cited literature; \textbf{D}~=~derived here
under the stated assumptions; \textbf{A}~=~formal analogy, conditional on a
mapping whose assumptions are listed where it is used; \textbf{C}~=~conjecture,
stated with the test that would decide it. The labels recur in
Section~8.}
\label{tab:claimstatus}\\
\toprule
Claim & Status & What would change the status \\
\midrule
\endfirsthead
\toprule
Claim & Status & What would change the status \\
\midrule
\endhead
Truncated-Lorentzian spectrum, cutoff $\Delta$, core $\Xi$ (Section~2.2)
 & E & Numerical validation outside the adiabatic window. \\
Two scales in the field autocorrelation, $l=\pi/\Delta$, $\Lambda=1/\Xi$
 (Section~4.1)
 & E &--- (direct consequence of the spectrum). \\
Strong chirp by itself establishes a statistical partial-coherence ensemble
 & C & A deterministic chirped field remains first-order coherent and has
   $\Npr=1$ for a single realization. Statistical partial coherence requires a
   specified ensemble or coarse graining and must be demonstrated through the
   coherence kernel (Section~4.3). \\
Under a specified stochastic CQGLE ensemble, $\Npr$ is a reproducible
 increasing function $f(r)$ of the deterministic scale ratio
 & C & Construct the ensemble, diagonalize $J(t_1,t_2)$ after removing timing,
   phase, frequency and energy jitter, and test monotonicity and reproducibility
   of $f$. Note that $f$ is a calibration and
   not an identity: $r$ spans $(0,\infty)$ while $\Npr\ge1$, so $f(r)\to1$ as
   $r\to0$ and the low-$r$ branch is the single-mode limit rather than a
   fractional mode count. \\
$-\mu_{\mathrm{shape}}=\Xi^{2}$ as a chemical potential
 & A & Identify the conserved or constrained quantity to which it is
   conjugate; fix the energy convention (Section~6.1). \\
The cutoff $\Delta$ is kinematic and does not require spectral filtering
   (Section~4.1)
 & E & The slaving relation \eqref{eq:slaving} contains no $\alpha$,
   and chirped solutions with the same $\Delta$ exist at $\alpha=0$
   \cite{Kalashnikov2024b}. \\
The Rayleigh--Jeans form of the spectrum, and every indicator built on it,
   requires spectral filtering (Section~4.1)
 & E & At $\alpha=0$ the profile is convex and not thermalized
   \cite{Kalashnikov2024b}, and the filter-free family is disjoint from the
   chirped branch, which requires $C>2/13$. Would change only if a filter-free
   solution with a Rayleigh--Jeans spectrum were exhibited. \\   
DSR as $\mu_{\mathrm{shape}}\to0^{-}$ at finite $\Delta$ (Section~5)
 & D & Algebraic consequence of the admissibility constraints. \\
Shape indicator $H_{s}(r)$ rises monotonically to $\ln4$ (Section~6.1)
 & D & A property of one selected functional; $H_{s}$ is not the unique
   Shannon entropy of the dimensionless spectral shape, which depends on
   whether $\Xi$ or $\Delta$ is used to nondimensionalize. \\
   $E_{\mathrm{AGD}}$ diverges logarithmically at the chirp-control line,
   Eq.~\eqref{eq:AGDlogdiv}
 & D & Coefficient verified numerically to $0.6\%$ over
   $\tilde\chi\in[1.2,3.0]$; an independent derivation of the
   $|\tilde\omega|^{-3}\to|\tilde\omega|^{-1}$ crossover would settle it. \\
$\Theta_{s}$ changes sign at the maximum of $U$, $r\simeq1.21$ (Section~6.3)
 & D & Numerically unchanged for the two entropy functionals tested along the
   isogain continuation, with both denominators verified nonzero there.
   Independence of the continuation path is \emph{ruled out}---Section~6.3
   obtains $r^{*}=1.44382$ on fixed-$C$ paths, $r\simeq1.21$ on isogains and
   $r\simeq1.01$--$1.09$ on branch-pairing curves, and no interior zero at all
   for $C\gtrsim1$. \\
``Negative absolute temperature''
 & C & Demonstrate path independence, a statistical measure and the
   conjugate variables; until then, negative energy--entropy slope. \\
Entropy-proxy crossing selects the multipulse state (Section~6.4)
 & C & Test additivity under shared gain saturation; compare with
   first-passage statistics in noisy simulation. \\
CQGLE $\leftrightarrow$ driven-dissipative GPE (Section~\ref{sec:dictionary})
 & A & Approximate, not term-for-term: division by $(i-\lambda_{\mathrm{P}})$
   mixes non-dissipative and dissipative coefficients, and a one-to-one reading
   requires $|\lambda_{\mathrm{P}}|\ll1$. Reservoir elimination, noise,
   trapping and geometry are not mapped. The sign of
   $\lambda_{\mathrm{P}}$ requires re-derivation. \\
Chirped DS and the KPZ phase as regimes of one field theory (Section~\ref{sec:drivenopen})
 & C & Requires the stochastic, long-wavelength reduction and a
   universality-class test on shot-resolved data. \\
The standard KPZ phase-hydrodynamic reduction is \emph{not established}
   for a finite-width localized DS, because the gapless
   diffusive phase mode $\omega\simeq-iDk^{2}$ from which it is built has no
   counterpart among the discrete zero modes of a localized DS (Section~\ref{sec:drivenopen})
 & C & Exhibit a genuinely gapless mode by long-wavelength
   expansion about the pulse---plausibly for a pulse train rather than a single
   pulse. The absence of such a mode in the localized solution is not an
   impossibility theorem: whether an emergent long-wavelength sector arises in a
   broad-plateau or DSR limit, where the flat top widens without bound, is an
   open spectral and stochastic question. \\
The finite-time $\Sigma$ boundary of Section~6.6 may reflect critical slowing of
   the background growth at the condensation threshold, rather than an instability of the
   pulse
 & C & Recompute the survival maps at $z=200$, $400$, $800$ and
   test whether the $50\%$ contour collapses onto $\sigma z=\mathrm{const}$. \\
Numerical values $E^{*}=6.0$--$6.3$ for the sign
   reversal and $E^{*}=10.9$--$17.3$ for the entropy-proxy crossing under the
   explicit pairing reconstruction used here
 & D & These values are reproducible for the equations and pairing rule
   stated in this manuscript. They are lower than the approximate values quoted
   in Ref.~\cite{Kalashnikov2025}; until the original numerical continuation and
   normalization are reproduced line by line, the shift should be treated as a
   mapping/reproducibility issue rather than as a conceptual contradiction. \\
 Along the continuation examined, anomalous dispersion shows no entropic
   turnover. Independently,
   a finite-time noisy linearized scan shows a one-sided accessibility boundary
   inside the algebraic existence region. Its sign asymmetry is not
   explained by the leading even near-resonance scale-separation factor alone;
   whether it is the asymptotic energy-limiting
   mechanism is open (Section~6.6)
 & D & Read from the quantum-noise stability maps of
   Ref.~\cite{Kalashnikov2026}; these maps are based on a linearized calculation run to $z=200$ with
   $N_{\mathrm{shot}}=32$. A longer, fully nonlinear scan extended above
   $\tilde E=10$, and a scan symmetric in $\delta C$, would change it. \\
\bottomrule
\end{longtable}
}

The paper is organized as follows. Section~2 sets out the model and the
adiabatic theory, derives the spectra in both dispersion regimes, and
constructs the master diagram on which the rest of the argument is conducted. Section~3 reviews the three traditions of optical thermodynamics and isolates the assumption they share. Section~4 applies these ideas to the chirped DS and states the underlying assumptions precisely: it establishes the two spectral scales, shows why a Schr\"odinger soliton has only one, defines the scale-separation index, and states what would have to be computed before that index could be called a count. Section~5 reformulates DSR in these terms and separates the
intrinsic from the conditional resonance. Section~6 defines the 
thermodynamic-like indicators, fixes the conventions they depend on, and
examines the limits of energy scaling---growth of the shape indicator, the
turnover of the differential entropy, the sign reversal of the energy--entropy slope, and DS fragmentation. Section~7 assesses the analogies with wave turbulence, with the noise-driven theory of mode locking, and with driven-open condensates, stating what each explains and where each fails. Section~8 collects the conclusions, the open questions, and the outlook.

\section{The Chirped Dissipative Soliton and Its Adiabatic Theory}

Everything that follows is built on a closed-form asymptotic solution for the strongly chirped DS, so we present that solution first. The order is deliberate. Every descriptor introduced later---the two spectral scales, the scale-separation index, the entropy-like functionals---is computed from a \emph{particular} deterministic solution, and none can be formulated until that solution is available in the spectral domain. That is a statement about where the descriptors come from, not a claim that they are statistical. Statistical statements require an ensemble, which is supplied nowhere in this paper and is specified as future work in Section~4.3. This section supplies the solution: the model and the routes available for solving it (Section~2.1), the adiabatic construction itself (Section~2.2), and the parametric space that it generates
(Section~2.3).

We fix three conventions here and use them throughout.

\emph{Fourier transform:}
$\tilde a(\omega)=\int a(t)\,e^{-i\omega t}\,\mathrm{d}t$,
$a(t)=(2\pi)^{-1}\!\int\tilde a(\omega)\,e^{i\omega t}\,\mathrm{d}\omega$.
The spectral power is $p(\omega)=|\tilde a(\omega)|^{2}$, the energy
$E=\int|a|^{2}\,\mathrm{d}t=(2\pi)^{-1}\!\int p\,\mathrm{d}\omega$, and linear
waves are written as $a\propto e^{i\omega t-ikz}$. \emph{Dimensions:} $z$ is a
propagation length (for a cavity map, a round-trip number, so that all
$L^{-1}$ below are per round trip) and $t$ a retarded time, so that
Table~\ref{tab:units} applies. \emph{Provenance:} equations reproduced from the
primary papers, corrected here, or newly derived here are marked in the text at
the point of use, and the numerical values are collected with their status in
Table~\ref{tab:claimstatus}.

\begin{table}[t]
\caption{Symbols, roles and dimensions for \eqref{eq:cqgle}. $L$~=~propagation
length (or round trip), $T$~=~time, $P$~=~power. All composite quantities used
later---$C$, $\Sigma$, $E^{*}$, $\omega^{*}$, $r$---are dimensionless.}
\label{tab:units}
\centering\small
\begin{tabular}{@{}llll@{}}
\toprule
Symbol & Role & Dimension & Notes \\
\midrule
$z$, $t$ & propagation, retarded time & $L$, $T$ & \\
$a$, $P=|a|^{2}$ & field, power & $P^{1/2}$, $P$ & \\
$\sigma$ & saturated net loss & $L^{-1}$ & $\sigma\approx\vartheta(E/E_{\mathrm{cw}}-1)$ \\
$\alpha$ & spectral filtering (inverse squared bandwidth) & $T^{2}L^{-1}$ & \\
$\beta$ & group-delay dispersion & $T^{2}L^{-1}$ & sign fixes the regime \\
$\gamma$ & self-phase modulation & $P^{-1}L^{-1}$ & \\
$\kappa$ & self-amplitude modulation & $P^{-1}L^{-1}$ & \\
$\zeta$ & SAM saturation & $P^{-1}$ & \\
$\chi$ & quintic (saturable) SPM & $P^{-2}L^{-1}$ & \\
$\Delta$, $\Xi$ & spectral support, spectral core & $T^{-1}$ & outputs, not parameters \\
$E$ & pulse energy & $PT$ & $E^{*}=E\kappa\sqrt{\zeta/(\beta\gamma)}$ \\
\bottomrule
\end{tabular}
\end{table}

\subsection{The cubic--quintic Ginzburg--Landau equation, and the routes to
its solution}

The master equation of a mode-locked laser and, under spatio-temporal duality,
of a broad class of driven-open nonlinear media is the $(1+1)$-dimensional
complex cubic--quintic Ginzburg--Landau equation (CQGLE)
\cite{Kalashnikov2026,vanSaarloos1992,Moores1993}
\begin{equation}
  \frac{\partial a}{\partial z}
  = \Big[-\sigma + (\alpha + i\beta)\frac{\partial^{2}}{\partial t^{2}}
         - i\big(\gamma-\chi P\big)P + \kappa\big(1-\zeta P\big)P \Big] a(z,t) ,
  \qquad P(z,t) = |a(z,t)|^{2} ,
  \label{eq:cqgle}
\end{equation}
Here \(z\) is the evolution coordinate---propagation distance in a distributed
model, or the round-trip index after cavity normalization---and \(t\) is the
local retarded time. Four coefficients are dissipative---the saturated net loss
$\sigma$, the squared inverse spectral-filter bandwidth $\alpha$, the
self-amplitude modulation (SAM) $\kappa$ and its saturation $\zeta$---and three
are not: the group-delay dispersion (GDD) $\beta$, positive for NGD and negative for AGD, the self-phase modulation (SPM)
$\gamma$, and its quintic correction $\chi$ ($\chi>0$ for SPM saturation,
$\chi<0$ for self-enhancement) \cite{Kalashnikov2026}. For the derivation below,
we set $\chi=0$. A finite $\chi$ deforms the branches and resonance interval
perturbatively without changing the NGD construction qualitatively, but
it becomes essential for the strongly chirped AGD energy-scaling branch
(Section~5.2). The interplay of these coefficients is what distinguishes a DS from a
conservative nonlinear Schr\"odinger soliton: setting $\sigma=\alpha=\kappa=0$
removes gain, loss, filtering and self-amplitude modulation and leaves the
conservative cubic--quintic nonlinear Schr\"odinger equation (NLS) (with $\zeta$ then irrelevant), and setting
$\chi=0$ as well gives the cubic NLS.
Equation~\eqref{eq:cqgle} is not integrable, and four principal routes to its solution have been pursued.

\emph{(i) Exact solitary-wave solutions.} With $\zeta=0$, \eqref{eq:cqgle}
reduces to Haus's master equation, which admits the exact chirped profile
$a(t)=\sqrt{P_{0}}\,\mathrm{sech}(t/T_{\mathrm{DS}})^{1+i\psi}$, and the
cubic--quintic case admits its own closed forms of arbitrary amplitude
\cite{Haus2000,Akhmediev1995,Renninger2008,Akhmediev2000}. These are exact but isolated: they
exist only where the coefficients satisfy an algebraic constraint, so they
define a lower-dimensional constraint manifold in the coefficient
space and, at a given
operating point, fix the energy rather than leaving it free. Precisely the continuum that makes
a DS thermodynamically interesting is invisible to them.
The limiting case is instructive. Two reductions of the cubic GL equation
tighten the constraint until no free coefficient survives at all---bandwidth-limited
gain with nonlinear dispersion, and broadband gain with linear dispersion and
nonlinear losses---and each then possesses a single exact solitary pulse of the
strongly chirped form $\mathrm{sech}(\cdot)^{1+i\psi}$ with the \emph{same}
numerical chirp $\psi=\sqrt2$ \cite{Malomed1997}. In such parameter-free models
the separation of the bound pulses discussed in Section~6.4 is likewise a pure
number rather than a tunable quantity. That is the opposite pole from the object
studied here: the entire content of the present framework lies in the
two-parameter continuum $(C,\Sigma)$ that these reductions collapse.

\emph{(ii) Reduced-variable methods.} Projecting \eqref{eq:cqgle} onto a
low-dimensional ansatz---amplitude, width, chirp, phase---by a variational or
moment procedure yields ordinary differential equations in $z$ whose fixed
points are the solitons and whose bifurcations describe multipulsing
\cite{Bale2008}. The variational version of this reduction shows which parts of the conservative formalism survive once the dynamics becomes dissipative. A
Lagrangian density can still be written for the non-dissipative part of
\eqref{eq:cqgle}, with gain, loss, filtering and SAM collected into a source on
the right-hand side of the Euler--Lagrange equations. The parameter conjugate to
the overall phase then returns not a conservation law but the energy-balance
relation, whose vanishing \emph{is} the condition of stationarity---the
dissipative surrogate for the Noether argument that ties phase-shift symmetry to
conservation of energy in the Schr\"odinger case \cite{Ankiewicz2007}. The
reduction is exact in the limited sense that, for a given trial function, it
reproduces the method of moments term by term, and it extends to pulsating
solitons, which appear as limit cycles rather than fixed points of the reduced
flow \cite{Ankiewicz2007}. Its limitation is the ansatz, and the cubic--quintic
case is where that limitation bites. The chirped trial form
$a(t)\exp[\mathrm{i}d\ln a(t)]$ spans the complete set of bright solutions of the
\emph{cubic} Ginzburg-Landau equation. Applied to the cubic--quintic equation, the same construction
recovers only a small subclass of its solutions \cite{Ankiewicz2007}. The method is therefore powerful for stability questions. However, it assumes the pulse shape, whereas for a
strongly chirped DS the shape is part of the solution rather than an input: the truncated
Lorentzian derived below would have to be guessed in advance.

\emph{(iii) Perturbation theory about the Schr\"odinger soliton}, treating gain,
loss and filtering as small corrections \cite{Haus1991,Kapitula1998}. This is
sound when the dissipative terms are weak \emph{and} the GDD is anomalous. In
NGD, there is no soliton to perturb, and the strongly chirped DS is
precisely a structure with no conservative limit on the branch of interest.

\emph{(iv) Direct numerical propagation}, which is decisive but local: every run
returns one point of a multidimensional parameter space, and the space of
\eqref{eq:cqgle} has seven dimensions before the noise is added. It is by this
route that the stability domains of cubic--quintic pulses were first mapped, and
found to be finite rather than coextensive with the existence domains
\cite{SotoCrespo1997}---a distinction that recurs in Section~6.

The adiabatic theory used here is a fifth route, and it exploits as a small
parameter the very feature that defeats (i)--(iii): the chirp.

\subsection{The adiabatic approximation: from the time domain to the
spectrum}

Two physical conditions define the regime. The non-dissipative terms dominate,
$\gamma\gg\kappa$ and $\beta\gg\alpha$ with $\beta>0$ (NGD); and the
resulting chirp $\psi\approx 3/(\alpha/\beta+\kappa/\gamma)$ is large,
$\psi\gg1$ \cite{Kalashnikov2005,Podivilov2005}. (Here and in Section~2,
$\psi$ denotes the chirp; in Section~3.1 the same letter is the field of a
nonlinear Schr\"odinger equation, following the sources of each.) A large chirp
means a strongly \emph{inhomogeneous local phase}: the instantaneous frequency
sweeps across the pulse, so different temporal slices of the DS oscillate at
different frequencies and are, in that sense, distinguishable. The DS acquires a
nontrivial internal structure, and the analysis below is an account of that
structure.

Substituting the soliton ansatz
\begin{equation}
  a(z,t) = \sqrt{P(t)}\;e^{\,i\varphi(t)-iqz} ,
  \qquad
  \Omega(t) \equiv \frac{d\varphi}{dt} ,
  \label{eq:ansatz}
\end{equation}
into \eqref{eq:cqgle} and separating real and imaginary parts gives two coupled
equations for $P$ and $\Omega$ \cite{Podivilov2005,Kalashnikov2024}. The
adiabatic assumption is that the envelope varies slowly compared with the phase,
$d^{2}\sqrt{P}/dt^{2}\to0$ with $\alpha\ll\beta$. The first equation then
collapses to an algebraic relation
\begin{equation}
  \gamma P(t) = q - \beta\,\Omega(t)^{2} .
  \label{eq:slaving}
\end{equation}
This is the pivot of the whole construction. The power is \emph{slaved} to the
instantaneous frequency: time enters only through $\Omega$, the map
$t\mapsto\Omega$ is monotonic, and the internal structure of the pulse can be
read entirely in the frequency variable. The adiabatic method is, in this
precise sense, a transfer of the problem from the time domain to the spectral
one, and it is legitimate exactly because the chirp is large.

Two consequences follow immediately. First, since $P\ge0$, relation
\eqref{eq:slaving} \emph{bounds the spectrum}:
\begin{equation}
  \Omega^{2} \le \Delta^{2} \equiv q/\beta ,
  \label{eq:cutoff}
\end{equation}
which is the same condition as resonance with the linear dispersive waves
$k=\beta\omega^{2}$ at $k=q$ \cite{Podivilov2005,Kalashnikov2025}. The cutoff
$\Delta$ is not imposed. It is produced by the requirement that the power stay
non-negative. Second, the remaining equation for $\Omega(t)$ carries a singular
prefactor. Excluding the non-physical singularity---the regularization procedure
of Refs.~\cite{Podivilov2005,Kalashnikov2024}---leaves
\begin{equation}
  \frac{d\Omega}{dt}
  = \frac{\beta\kappa\zeta}{3\gamma^{2}}\,
    \big(\Delta^{2}-\Omega^{2}\big)\big(\Xi^{2}+\Omega^{2}\big) ,
  \qquad
  \Xi^{2} = \frac{\gamma}{\zeta\beta}\Big(1+C-\tfrac{5}{3}\zeta P_{0}\Big) ,
  \label{eq:freqODE}
\end{equation}
and, as a by-product, a \emph{quadratic} condition on the peak power with two
roots,
\begin{equation}
  \zeta P_{0}^{\pm}
  = \frac{3}{4}\left[\,1-\frac{C}{2}
    \pm \sqrt{\Big(1-\frac{C}{2}\Big)^{2}-\Sigma}\;\right] ,
  \qquad
  C \equiv \frac{\alpha\gamma}{\beta\kappa} ,
  \quad
  \Sigma \equiv \frac{4\zeta\sigma}{\kappa} ,
  \label{eq:branches}
\end{equation}
with $\Delta^{2}=\gamma P_{0}/\beta$. The control parameter $C$ measures
dissipative against non-dissipative effects and the ``soliton condition''
$C\simeq1$ expresses their balance \cite{Kalashnikov2024}---a
\emph{dissipative-balance} condition, which the master diagram will place at the
upper end of the DSR interval where the fidelity curve meets $\Sigma=0$
(Section~2.3), and which should not be confused with the exact-potential
condition \eqref{eq:gibbscubic}, $C=-1$, discussed in Section~4. $\Sigma$ is
the normalized saturated net loss, and it is the variable through which the
cavity is tuned. Everything else in the theory is a function of these two.

Integrating \eqref{eq:freqODE} gives $t(\Omega)$ in closed form and a temporal
width $T_{\mathrm{DS}}=3\gamma^{2}/[\beta\zeta\kappa\Delta(\Delta^{2}+\Xi^{2})]$
(written $T$ in the sources; renamed here to free $T$ for the temperature).
Because the power is slaved by \eqref{eq:slaving}, the field is known once
$\Omega(t)$ is, and its Fourier transform can be evaluated by the method of
stationary phase---again justified by $\psi\gg1$ \cite{Bender1999}. The result
is the spectral power
\begin{equation}
  p(\omega) = \frac{6\pi\gamma}{\zeta\kappa\,\big(\Xi^{2}+\omega^{2}\big)}
  \quad\text{for } |\omega|\le\Delta,
  \qquad p(\omega)=0 \ \text{otherwise} ,
  \label{eq:DSspectrum}
\end{equation}
carrying a spectral phase that the same construction defines only implicitly.
Writing $t_{\omega}$ for the stationary point, $\Omega(t_{\omega})=\omega$,
\begin{equation}
  \Phi(\omega) = \varphi(t_{\omega}) - \omega\,t_{\omega}
                 + \frac{\pi}{4}\,\mathrm{sgn}\big[\dot\Omega(t_{\omega})\big] ,
  \qquad
  \Phi'(\omega) = -\,t_{\omega} ,
  \qquad
  \Phi''(\omega) = -\frac{1}{\dot\Omega(t_{\omega})} ,
  \label{eq:specphase}
\end{equation}
so that \eqref{eq:freqODE} fixes the curvature in closed form,
\begin{equation}
  \Phi''(\omega)
  = -\,\frac{3\gamma^{2}}
           {\beta\kappa\zeta\,\big(\Delta^{2}-\omega^{2}\big)
                              \big(\Xi^{2}+\omega^{2}\big)} ,
  \label{eq:specphase2}
\end{equation}
while $\Phi$ itself follows from $t_{\omega}$, i.e.\ from integrating
\eqref{eq:freqODE}, and has no elementary closed form\footnote{A rational expression
for $\Phi$ appears in earlier statements of this construction, our own included.
It reproduces neither $\Phi'=-t_{\omega}$ nor \eqref{eq:specphase2} and is
withdrawn. Nothing downstream depends on it: the spectral \emph{power}
\eqref{eq:DSspectrum}, which is what every indicator below is built from,
follows from $p=2\pi P(t_{\omega})/|\dot\Omega(t_{\omega})|$ and is unaffected.}.
With the Fourier
convention $E=\int|a(t)|^{2}\,\mathrm{d}t=(2\pi)^{-1}\!\int p(\omega)\,
\mathrm{d}\omega$ implied by \eqref{eq:DSspectrum}, direct integration gives the
energy
\begin{equation}
  E = \frac{1}{2\pi}\int_{-\Delta}^{\Delta} p(\omega)\,\mathrm{d}\omega
    = \frac{6\gamma}{\zeta\kappa\,\Xi}\,\tan^{-1}\!\Big(\frac{\Delta}{\Xi}\Big) .
  \label{eq:DSenergy}
\end{equation}
The factor $\Xi^{-1}$ is not cosmetic: it carries the dimensions, and it governs
the divergence structure at the resonance, where $\Xi\to0$ at fixed $\Delta$.

\begin{figure}[t]
\centering
\begin{tikzpicture}[x=1cm,y=1cm,
  ax/.style={draw=black,line width=0.5pt,-{Latex[length=1.5mm]}}]
\draw[ax] (-0.15,0) -- (4.55,0);
\draw[ax] (0,-1.15) -- (0,2.55);
\node[font=\scriptsize,anchor=north] at (2.1,-0.2) {local time $t$};
\fill[black!12] (0.000,0.000) (0.016,0.116) (0.033,0.229) (0.050,0.339) (0.069,0.445) (0.088,0.549) (0.108,0.649) (0.129,0.746) (0.151,0.840) (0.174,0.931) (0.198,1.018) (0.224,1.103) (0.251,1.184) (0.279,1.263) (0.309,1.338) (0.341,1.410) (0.375,1.479) (0.411,1.544) (0.450,1.607) (0.491,1.666) (0.535,1.723) (0.581,1.776) (0.632,1.826) (0.686,1.873) (0.744,1.917) (0.807,1.957) (0.875,1.995) (0.949,2.029) (1.029,2.060) (1.115,2.088) (1.208,2.113) (1.308,2.135) (1.415,2.154) (1.529,2.169) (1.650,2.181) (1.777,2.191) (1.908,2.197) (2.041,2.200) (2.176,2.200) (2.309,2.196) (2.439,2.190) (2.565,2.180) (2.685,2.167) (2.799,2.151) (2.905,2.132) (3.004,2.110) (3.096,2.085) (3.182,2.056) (3.261,2.025) (3.333,1.990) (3.401,1.952) (3.463,1.911) (3.521,1.867) (3.575,1.820) (3.625,1.769) (3.671,1.716) (3.715,1.659) (3.755,1.599) (3.793,1.536) (3.829,1.470) (3.863,1.401) (3.894,1.329) (3.924,1.253) (3.953,1.174) (3.980,1.093) (4.005,1.008) (4.029,0.920) (4.052,0.828) (4.074,0.734) (4.095,0.636) (4.115,0.536) (4.134,0.432) (4.152,0.325) (4.169,0.215) (4.186,0.102) -- (4.2,0) -- (0,0) -- cycle;
\draw[line width=0.9pt] plot coordinates {(0.000,0.000) (0.016,0.116) (0.033,0.229) (0.050,0.339) (0.069,0.445) (0.088,0.549) (0.108,0.649) (0.129,0.746) (0.151,0.840) (0.174,0.931) (0.198,1.018) (0.224,1.103) (0.251,1.184) (0.279,1.263) (0.309,1.338) (0.341,1.410) (0.375,1.479) (0.411,1.544) (0.450,1.607) (0.491,1.666) (0.535,1.723) (0.581,1.776) (0.632,1.826) (0.686,1.873) (0.744,1.917) (0.807,1.957) (0.875,1.995) (0.949,2.029) (1.029,2.060) (1.115,2.088) (1.208,2.113) (1.308,2.135) (1.415,2.154) (1.529,2.169) (1.650,2.181) (1.777,2.191) (1.908,2.197) (2.041,2.200) (2.176,2.200) (2.309,2.196) (2.439,2.190) (2.565,2.180) (2.685,2.167) (2.799,2.151) (2.905,2.132) (3.004,2.110) (3.096,2.085) (3.182,2.056) (3.261,2.025) (3.333,1.990) (3.401,1.952) (3.463,1.911) (3.521,1.867) (3.575,1.820) (3.625,1.769) (3.671,1.716) (3.715,1.659) (3.755,1.599) (3.793,1.536) (3.829,1.470) (3.863,1.401) (3.894,1.329) (3.924,1.253) (3.953,1.174) (3.980,1.093) (4.005,1.008) (4.029,0.920) (4.052,0.828) (4.074,0.734) (4.095,0.636) (4.115,0.536) (4.134,0.432) (4.152,0.325) (4.169,0.215) (4.186,0.102)};
\draw[line width=0.9pt,densely dashed] plot coordinates {(0.000,0.000) (0.016,0.029) (0.033,0.059) (0.050,0.088) (0.069,0.118) (0.088,0.147) (0.108,0.176) (0.129,0.206) (0.151,0.235) (0.174,0.264) (0.198,0.294) (0.224,0.323) (0.251,0.353) (0.279,0.382) (0.309,0.411) (0.341,0.441) (0.375,0.470) (0.411,0.500) (0.450,0.529) (0.491,0.558) (0.535,0.588) (0.581,0.617) (0.632,0.646) (0.686,0.676) (0.744,0.705) (0.807,0.735) (0.875,0.764) (0.949,0.793) (1.029,0.823) (1.115,0.852) (1.208,0.881) (1.308,0.911) (1.415,0.940) (1.529,0.970) (1.650,0.999) (1.777,1.028) (1.908,1.058) (2.041,1.087) (2.176,1.117) (2.309,1.146) (2.439,1.175) (2.565,1.205) (2.685,1.234) (2.799,1.263) (2.905,1.293) (3.004,1.322) (3.096,1.352) (3.182,1.381) (3.261,1.410) (3.333,1.440) (3.401,1.469) (3.463,1.498) (3.521,1.528) (3.575,1.557) (3.625,1.587) (3.671,1.616) (3.715,1.645) (3.755,1.675) (3.793,1.704) (3.829,1.734) (3.863,1.763) (3.894,1.792) (3.924,1.822) (3.953,1.851) (3.980,1.880) (4.005,1.910) (4.029,1.939) (4.052,1.969) (4.074,1.998) (4.095,2.027) (4.115,2.057) (4.134,2.086) (4.152,2.115) (4.169,2.145) (4.186,2.174)};
\node[font=\scriptsize,anchor=south east] at (2.6,2.2) {$P(t)$};
\node[font=\scriptsize,anchor=north east] at (2.6,1) {$\Omega(t)$};
\draw[dotted,line width=0.5pt] (0,2.2) -- (4.35,2.2);
\draw[dotted,line width=0.5pt] (0,0) -- (4.35,0);
\node[font=\scriptsize,anchor=west] at (4.38,2.2) {$+\Delta$};
\node[font=\scriptsize,anchor=west] at (4.38,0.0) {$-\Delta$};
\node[font=\footnotesize\bfseries,anchor=north] at (2.1,-1.75) {(a) time domain};
\draw[-{Latex[length=2mm]},line width=0.7pt] (4.95,1.1) -- (5.85,1.1);
\node[font=\scriptsize,anchor=south,align=center] at (5,1.1)
  {$\gamma P=q-\beta\Omega^{2}$};
\node[font=\scriptsize,anchor=north,align=center] at (5.40,1.0)
  {slaving};
\draw[ax] (5.85,0) -- (10.55,0);
\draw[ax] (8.1,0) -- (8.1,2.55);
\node[font=\scriptsize,anchor=north] at (8.1,-0.2) {detuning $\omega$};
\fill[black!12] (6.000,0.182) (6.056,0.191) (6.112,0.201) (6.168,0.212) (6.224,0.223) (6.280,0.236) (6.337,0.249) (6.393,0.264) (6.449,0.280) (6.505,0.297) (6.561,0.316) (6.617,0.336) (6.673,0.359) (6.729,0.384) (6.785,0.411) (6.841,0.441) (6.897,0.474) (6.954,0.510) (7.010,0.551) (7.066,0.595) (7.122,0.645) (7.178,0.700) (7.234,0.761) (7.290,0.829) (7.346,0.905) (7.402,0.988) (7.458,1.080) (7.515,1.180) (7.571,1.290) (7.627,1.406) (7.683,1.529) (7.739,1.656) (7.795,1.782) (7.851,1.903) (7.907,2.012) (7.963,2.101) (8.019,2.165) (8.075,2.197) (8.132,2.194) (8.188,2.158) (8.244,2.091) (8.300,1.999) (8.356,1.888) (8.412,1.767) (8.468,1.640) (8.524,1.514) (8.580,1.391) (8.636,1.275) (8.692,1.167) (8.749,1.068) (8.805,0.977) (8.861,0.895) (8.917,0.821) (8.973,0.753) (9.029,0.693) (9.085,0.639) (9.141,0.590) (9.197,0.545) (9.253,0.506) (9.310,0.469) (9.366,0.437) (9.422,0.407) (9.478,0.380) (9.534,0.356) (9.590,0.334) (9.646,0.313) (9.702,0.295) (9.758,0.277) (9.814,0.262) (9.870,0.247) (9.927,0.234) (9.983,0.222) (10.039,0.210) (10.095,0.200) (10.151,0.190) -- (10.2,0) -- (6.0,0) -- cycle;
\draw[line width=0.9pt] plot coordinates {(6.000,0.182) (6.056,0.191) (6.112,0.201) (6.168,0.212) (6.224,0.223) (6.280,0.236) (6.337,0.249) (6.393,0.264) (6.449,0.280) (6.505,0.297) (6.561,0.316) (6.617,0.336) (6.673,0.359) (6.729,0.384) (6.785,0.411) (6.841,0.441) (6.897,0.474) (6.954,0.510) (7.010,0.551) (7.066,0.595) (7.122,0.645) (7.178,0.700) (7.234,0.761) (7.290,0.829) (7.346,0.905) (7.402,0.988) (7.458,1.080) (7.515,1.180) (7.571,1.290) (7.627,1.406) (7.683,1.529) (7.739,1.656) (7.795,1.782) (7.851,1.903) (7.907,2.012) (7.963,2.101) (8.019,2.165) (8.075,2.197) (8.132,2.194) (8.188,2.158) (8.244,2.091) (8.300,1.999) (8.356,1.888) (8.412,1.767) (8.468,1.640) (8.524,1.514) (8.580,1.391) (8.636,1.275) (8.692,1.167) (8.749,1.068) (8.805,0.977) (8.861,0.895) (8.917,0.821) (8.973,0.753) (9.029,0.693) (9.085,0.639) (9.141,0.590) (9.197,0.545) (9.253,0.506) (9.310,0.469) (9.366,0.437) (9.422,0.407) (9.478,0.380) (9.534,0.356) (9.590,0.334) (9.646,0.313) (9.702,0.295) (9.758,0.277) (9.814,0.262) (9.870,0.247) (9.927,0.234) (9.983,0.222) (10.039,0.210) (10.095,0.200) (10.151,0.190)};
\draw[dashed,line width=0.6pt] (6.0,0) -- (6.0,2.35);
\draw[dashed,line width=0.6pt] (10.2,0) -- (10.2,2.35);
\node[font=\scriptsize,anchor=south] at (6.0,2.33) {$-\Delta$};
\node[font=\scriptsize,anchor=south] at (10.2,2.33) {$+\Delta$};
\draw[{Latex[length=1.2mm]}-{Latex[length=1.2mm]},line width=0.5pt]
   (7.47,1.1) -- (8.73,1.1);
\node[font=\scriptsize,anchor=south] at (8.25,1.1) {$\Xi$};
\node[font=\scriptsize,anchor=west] at (10.2,1.6) {$p(\omega)$};
\node[font=\footnotesize\bfseries,anchor=north] at (8.1,-1.75)
  {(b) spectral domain};
\node[font=\scriptsize,anchor=north,align=center] at (8.1,-0.55)
  {truncated Lorentzian, Eq.~\eqref{eq:DSspectrum}};
\end{tikzpicture}
\caption{What the adiabatic approximation does. (\textbf{a})~Under a strong
chirp, the instantaneous frequency $\Omega(t)$ (dashed) sweeps monotonically
across the pulse, so each temporal slice carries its own frequency: this is the
inhomogeneous local phase that gives the DS an internal structure. The envelope
$P(t)$ (shaded) is not independent of it but slaved to it by
\eqref{eq:slaving}, and vanishes exactly where $\Omega$ reaches $\pm\Delta$---
the cutoff is a consequence of $P\ge0$, not an assumption.
(\textbf{b})~Because the map $t\mapsto\Omega$ is one-to-one, the whole problem
can be carried into the frequency variable, where stationary phase yields the
truncated Lorentzian \eqref{eq:DSspectrum}: a core of half-width $\Xi$ inside a
spectral window with a half-width $\Delta$. Curves are computed from \eqref{eq:freqODE} and
\eqref{eq:DSspectrum} for $\Xi/\Delta=0.3$; the vertical scales of $P$ and
$\Omega$ in (a) are separate.}
\label{fig:adiabatic}
\end{figure}
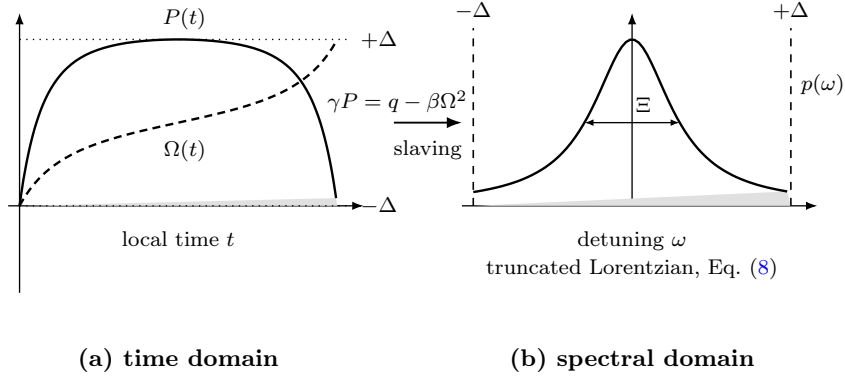

Equation~\eqref{eq:DSspectrum} is a \emph{truncated Lorentzian}: a Lorentzian of
half-width $\Xi$ cut off at $\pm\Delta$ (Figure~\ref{fig:adiabatic}). It is
reproduced from Refs.~\cite{Podivilov2005,Kalashnikov2024}, and its status
should be stated with it. Stationary phase is a leading-order \emph{interior}
asymptotic, and \eqref{eq:specphase2} says exactly where it fails: since
$\dot\Omega\to0$ as $|\omega|\to\Delta$, the curvature $\Phi''$ diverges, the
stationary point becomes degenerate, and the ordinary approximation is
nonuniform in a neighborhood of the edge. The sharp truncation at $\pm\Delta$
is therefore the leading-order boundary of the adiabatic support, not a
demonstrated discontinuity of the physical spectrum. What the canonical form of
the edge layer is---whether the degeneracy is of the kind that yields an
Airy-type uniform approximation or of some other order---requires the local
normal form to be derived, which we have not done, and we make no claim about
it. The uniform machinery for coalescing saddles
\cite{Bender1999,ChesterFriedmanUrsell1957} is where such a derivation would
start. One piece of evidence already points the same way: the same adiabatic construction applied to the filter-free equation, $\alpha=0$, returns
a profile that vanishes continuously at $\pm\Delta$ rather than jumping
\cite{Kalashnikov2024b}, so the discontinuity is tied to the filtered case and not to the truncation mechanism, which is common to both (Section~4.1).

Three
consequences are carried forward: quantities dominated by the interior---the
core width, the energy, the shape indicator of Section~6.1---are safe at leading
order. Quantities sensitive to the edge---the fine structure of the
autocorrelation, and any entropy computed with weight near $|\omega|=\Delta$---
should be checked by excluding a boundary layer $|\omega|>(1-\epsilon)\Delta$
and demonstrating convergence as $\epsilon\to0$. A direct comparison with
Fourier transforms of numerically propagated CQGLE solutions near the edge
remains to be made. Two independent scales therefore
characterize the internal structure, and they appear together in the
time-integrated field autocorrelation
$R(\tau)=\int a(t)a^{*}(t+\tau)\,\mathrm{d}t$---the Wiener--Khinchin partner of
the spectrum for a single deterministic realization, and not, without an
ensemble average, a first-order coherence function in the statistical sense
(Section~4.1)---which takes the form of an exponential envelope carrying a sinc fine structure
\cite{Kalashnikov2025}
\begin{equation}
  R(\tau) \;\propto\; \int_{-\infty}^{\infty}
     e^{-\Xi|t|}\,\mathrm{sinc}\big(\Delta(\tau-t)\big)\,dt ,
  \label{eq:autocorr}
\end{equation}
whence a short ``graining'' time $l=\pi/\Delta$ set by the spectral edge and a
long ``confining'' time $\Lambda=1/\Xi$ set by the spectral core. Their
coincidence, $\Xi=\Delta$, is the point at which the spectral phase is closest
to quadratic, and \eqref{eq:specphase2} makes this precise: writing
$x=\omega^{2}$, the curvature $\Phi''\propto-[(\Delta^{2}-x)(\Xi^{2}+x)]^{-1}$
varies with $x$ at the rate $\Delta^{2}-\Xi^{2}-2x$, which vanishes at $x=0$
exactly when $\Xi=\Delta$. There, and only there, is $\Phi''$ stationary at
zero detuning---a chirp as nearly frequency-independent, and
hence optimal external compressibility of DS is provided \cite{Kalashnikov2024}.

In AGD with saturable quintic SPM, the same program goes through with
one modification: the spectrum is not truncated, but decays algebraically, and
the role of the cutoff is taken over by an effective spectral-dissipation window
\cite{Kalashnikov2026}. The two-scale structure survives.

The point to retain is that $\Delta$ and $\Xi$ are \emph{outputs} of the theory.
By \eqref{eq:branches} and \eqref{eq:freqODE} they are functions of the
operating point and, through \eqref{eq:DSenergy}, of the pulse energy---not
parameters of the oscillator. Section~4 is largely an unpacking of that one fact.

\subsection{The master diagram}

\begin{figure}[t]
\centering
\begin{tikzpicture}[x=1cm,y=1cm,
  ax/.style={draw=black,line width=0.5pt},
  iso/.style={draw=black!55,line width=0.7pt},
  isod/.style={draw=black!55,line width=0.7pt,dashed}]
\fill[black!10] (4.097,2.700) (4.184,2.647) (4.270,2.597) (4.357,2.548) (4.443,2.502) (4.530,2.458) (4.616,2.415) (4.703,2.375) (4.789,2.337) (4.876,2.301) (4.962,2.266) (5.049,2.234) (5.135,2.203) (5.222,2.174) (5.308,2.147) (5.395,2.121) (5.481,2.097) (5.568,2.075) (5.654,2.054) (5.741,2.034) (5.827,2.015) (5.914,1.998) (6.000,1.982) (6.087,1.967) (6.173,1.954) (6.260,1.941) (6.346,1.929) (6.433,1.918) (6.519,1.908) (6.605,1.899) (6.692,1.890) (6.778,1.882) (6.865,1.875) (6.951,1.868) (7.038,1.862) (7.124,1.856) (7.211,1.851) (7.297,1.846) (7.384,1.842) (7.470,1.838) (7.557,1.835) (7.643,1.831) (7.730,1.828) (7.816,1.826) (7.903,1.823) (7.989,1.821) (8.076,1.819) (8.162,1.817) (8.249,1.816) (8.335,1.814) (8.422,1.813) (8.508,1.812) (8.595,1.810) (8.681,1.809) (8.768,1.808) (8.854,1.808) (8.941,1.807) (9.027,1.806) (9.114,1.806) (9.200,1.805) (9.200,0.422) (9.114,0.423) (9.027,0.424) (8.941,0.425) (8.854,0.426) (8.768,0.427) (8.681,0.428) (8.595,0.429) (8.508,0.431) (8.422,0.432) (8.335,0.434) (8.249,0.436) (8.162,0.438) (8.076,0.441) (7.989,0.444) (7.903,0.446) (7.816,0.450) (7.730,0.453) (7.643,0.457) (7.557,0.461) (7.470,0.466) (7.384,0.471) (7.297,0.477) (7.211,0.483) (7.124,0.489) (7.038,0.497) (6.951,0.504) (6.865,0.513) (6.778,0.522) (6.692,0.532) (6.605,0.543) (6.519,0.555) (6.433,0.567) (6.346,0.581) (6.260,0.596) (6.173,0.612) (6.087,0.629) (6.000,0.647) (5.914,0.667) (5.827,0.688) (5.741,0.710) (5.654,0.734) (5.568,0.760) (5.481,0.787) (5.395,0.816) (5.308,0.847) (5.222,0.880) (5.135,0.914) (5.049,0.951) (4.962,0.990) (4.876,1.031) (4.789,1.075) (4.703,1.120) (4.616,1.168) (4.530,1.219) (4.443,1.272) (4.357,1.327) (4.270,1.386) (4.184,1.446) (4.097,1.510) -- cycle;
\draw[ax,-{Latex[length=1.6mm]}] (0,0) -- (9.7,0);
\draw[ax,-{Latex[length=1.6mm]}] (0,0) -- (0,5.9);
\node[font=\scriptsize,anchor=north] at (4.8,-0.55) {normalized energy $E^{*}$};
\node[font=\scriptsize,rotate=90,anchor=south] at (-0.75,2.7) {control parameter $C$};
\draw[ax] (0.357,0) -- (0.357,-0.10);
\node[font=\scriptsize,anchor=north] at (0.357,-0.14) {0.5};
\draw[ax] (1.443,0) -- (1.443,-0.10);
\node[font=\scriptsize,anchor=north] at (1.443,-0.14) {1};
\draw[ax] (2.529,0) -- (2.529,-0.10);
\node[font=\scriptsize,anchor=north] at (2.529,-0.14) {2};
\draw[ax] (3.965,0) -- (3.965,-0.10);
\node[font=\scriptsize,anchor=north] at (3.965,-0.14) {5};
\draw[ax] (5.051,0) -- (5.051,-0.10);
\node[font=\scriptsize,anchor=north] at (5.051,-0.14) {10};
\draw[ax] (6.137,0) -- (6.137,-0.10);
\node[font=\scriptsize,anchor=north] at (6.137,-0.14) {20};
\draw[ax] (7.573,0) -- (7.573,-0.10);
\node[font=\scriptsize,anchor=north] at (7.573,-0.14) {50};
\draw[ax] (8.659,0) -- (8.659,-0.10);
\node[font=\scriptsize,anchor=north] at (8.659,-0.14) {100};
\draw[ax] (0,0.000) -- (-0.10,0.000);
\node[font=\scriptsize,anchor=east] at (-0.14,0.000) {0};
\draw[ax] (0,1.350) -- (-0.10,1.350);
\node[font=\scriptsize,anchor=east] at (-0.14,1.350) {0.5};
\draw[ax] (0,2.700) -- (-0.10,2.700);
\node[font=\scriptsize,anchor=east] at (-0.14,2.700) {1};
\draw[ax] (0,4.050) -- (-0.10,4.050);
\node[font=\scriptsize,anchor=east] at (-0.14,4.050) {1.5};
\draw[ax] (0,5.400) -- (-0.10,5.400);
\node[font=\scriptsize,anchor=east] at (-0.14,5.400) {2};
\draw[iso] plot coordinates {(9.194,1.689) (7.693,1.713) (7.155,1.737) (6.811,1.761) (6.554,1.784) (6.346,1.808) (6.172,1.832) (6.020,1.856) (5.885,1.880) (5.764,1.903) (5.653,1.927) (5.551,1.951) (5.456,1.975) (5.368,1.999) (5.284,2.022) (5.205,2.046) (5.131,2.070) (5.059,2.094) (4.991,2.117) (4.926,2.141) (4.863,2.165) (4.802,2.189) (4.743,2.213) (4.687,2.236) (4.632,2.260) (4.579,2.284) (4.527,2.308) (4.476,2.332) (4.427,2.355) (4.379,2.379) (4.332,2.403) (4.286,2.427) (4.241,2.450) (4.196,2.474) (4.153,2.498) (4.110,2.522) (4.068,2.546) (4.027,2.569) (3.986,2.593) (3.946,2.617) (3.907,2.641) (3.868,2.665) (3.829,2.688) (3.791,2.712) (3.753,2.736) (3.716,2.760) (3.679,2.783) (3.643,2.807) (3.606,2.831) (3.570,2.855) (3.534,2.879) (3.499,2.902) (3.463,2.926) (3.428,2.950) (3.393,2.974) (3.358,2.998) (3.323,3.021) (3.289,3.045) (3.254,3.069) (3.219,3.093) (3.184,3.117) (3.150,3.140) (3.115,3.164) (3.080,3.188) (3.045,3.212) (3.009,3.235) (2.974,3.259) (2.938,3.283) (2.902,3.307) (2.865,3.331) (2.828,3.354) (2.791,3.378) (2.753,3.402) (2.714,3.426) (2.674,3.450) (2.633,3.473) (2.591,3.497) (2.547,3.521) (2.501,3.545) (2.452,3.568) (2.401,3.592) (2.344,3.616) (2.280,3.640) (2.202,3.664) (2.080,3.687)};
\draw[isod] plot coordinates {(1.758,0.054) (1.738,0.102) (1.718,0.149) (1.699,0.197) (1.681,0.244) (1.663,0.292) (1.646,0.339) (1.629,0.387) (1.612,0.435) (1.596,0.482) (1.581,0.530) (1.566,0.577) (1.551,0.625) (1.537,0.672) (1.523,0.720) (1.510,0.768) (1.497,0.815) (1.485,0.863) (1.473,0.910) (1.461,0.958) (1.450,1.005) (1.439,1.053) (1.429,1.101) (1.419,1.148) (1.409,1.196) (1.400,1.243) (1.391,1.291) (1.383,1.338) (1.375,1.386) (1.367,1.434) (1.360,1.481) (1.353,1.529) (1.346,1.576) (1.340,1.624) (1.334,1.671) (1.329,1.719) (1.324,1.767) (1.320,1.814) (1.315,1.862) (1.312,1.909) (1.308,1.957) (1.306,2.004) (1.303,2.052) (1.301,2.100) (1.300,2.147) (1.299,2.195) (1.298,2.242) (1.298,2.290) (1.298,2.337) (1.299,2.385) (1.301,2.433) (1.303,2.480) (1.305,2.528) (1.308,2.575) (1.312,2.623) (1.317,2.671) (1.322,2.718) (1.328,2.766) (1.335,2.813) (1.343,2.861) (1.351,2.908) (1.361,2.956) (1.371,3.004) (1.383,3.051) (1.396,3.099) (1.411,3.146) (1.427,3.194) (1.445,3.241) (1.465,3.289) (1.488,3.337) (1.513,3.384) (1.543,3.432) (1.577,3.479) (1.618,3.527) (1.668,3.574) (1.735,3.622) (1.841,3.670)};
\draw[iso] plot coordinates {(8.790,1.440) (7.959,1.451) (7.534,1.463) (7.243,1.475) (7.019,1.487) (6.835,1.499) (6.679,1.511) (6.544,1.523) (6.423,1.535) (6.313,1.547) (6.214,1.558) (6.122,1.570) (6.036,1.582) (5.956,1.594) (5.881,1.606) (5.810,1.618) (5.743,1.630) (5.679,1.642) (5.618,1.654) (5.559,1.666) (5.503,1.677) (5.448,1.689) (5.396,1.701) (5.346,1.713) (5.297,1.725) (5.249,1.737) (5.203,1.749) (5.158,1.761) (5.115,1.773) (5.072,1.784) (5.030,1.796) (4.990,1.808) (4.950,1.820) (4.911,1.832) (4.873,1.844) (4.835,1.856) (4.799,1.868) (4.762,1.880) (4.727,1.892) (4.692,1.903) (4.657,1.915) (4.624,1.927) (4.590,1.939) (4.557,1.951) (4.524,1.963) (4.492,1.975) (4.460,1.987) (4.429,1.999) (4.397,2.010) (4.366,2.022) (4.336,2.034) (4.305,2.046) (4.275,2.058) (4.245,2.070) (4.215,2.082) (4.185,2.094) (4.156,2.106) (4.127,2.117) (4.097,2.129) (4.068,2.141) (4.039,2.153) (4.010,2.165) (3.981,2.177) (3.951,2.189) (3.922,2.201) (3.893,2.213) (3.863,2.225) (3.834,2.236) (3.804,2.248) (3.774,2.260) (3.744,2.272) (3.713,2.284) (3.682,2.296) (3.651,2.308) (3.618,2.320) (3.586,2.332) (3.552,2.343) (3.517,2.355) (3.480,2.367) (3.442,2.379) (3.402,2.391) (3.357,2.403) (3.308,2.415) (3.249,2.427) (3.163,2.439)};
\draw[isod] plot coordinates {(2.864,0.054) (2.851,0.084) (2.839,0.113) (2.826,0.143) (2.814,0.173) (2.803,0.203) (2.791,0.232) (2.780,0.262) (2.769,0.292) (2.758,0.322) (2.748,0.351) (2.738,0.381) (2.728,0.411) (2.718,0.441) (2.709,0.470) (2.699,0.500) (2.690,0.530) (2.682,0.559) (2.673,0.589) (2.665,0.619) (2.657,0.649) (2.649,0.678) (2.642,0.708) (2.634,0.738) (2.627,0.768) (2.620,0.797) (2.614,0.827) (2.607,0.857) (2.601,0.887) (2.595,0.916) (2.590,0.946) (2.584,0.976) (2.579,1.005) (2.574,1.035) (2.569,1.065) (2.565,1.095) (2.561,1.124) (2.557,1.154) (2.553,1.184) (2.550,1.214) (2.546,1.243) (2.543,1.273) (2.541,1.303) (2.538,1.333) (2.536,1.362) (2.535,1.392) (2.533,1.422) (2.532,1.451) (2.531,1.481) (2.530,1.511) (2.530,1.541) (2.530,1.570) (2.531,1.600) (2.532,1.630) (2.533,1.660) (2.535,1.689) (2.537,1.719) (2.539,1.749) (2.542,1.779) (2.546,1.808) (2.550,1.838) (2.554,1.868) (2.560,1.897) (2.565,1.927) (2.572,1.957) (2.579,1.987) (2.587,2.016) (2.597,2.046) (2.607,2.076) (2.618,2.106) (2.630,2.135) (2.644,2.165) (2.660,2.195) (2.678,2.225) (2.698,2.254) (2.721,2.284) (2.749,2.314) (2.782,2.343) (2.823,2.373) (2.879,2.403) (2.978,2.433)};
\draw[iso] plot coordinates {(8.344,0.982) (7.893,0.988) (7.590,0.994) (7.360,1.000) (7.173,1.005) (7.015,1.011) (6.877,1.017) (6.754,1.023) (6.643,1.029) (6.542,1.035) (6.448,1.041) (6.360,1.047) (6.279,1.053) (6.201,1.059) (6.128,1.065) (6.058,1.071) (5.991,1.077) (5.927,1.083) (5.865,1.089) (5.805,1.095) (5.746,1.101) (5.690,1.107) (5.634,1.112) (5.580,1.118) (5.527,1.124) (5.475,1.130) (5.423,1.136) (5.372,1.142) (5.322,1.148) (5.271,1.154) (5.220,1.160) (5.169,1.166) (5.118,1.172) (5.066,1.178) (5.012,1.184) (4.956,1.190) (4.897,1.196) (4.833,1.202) (4.760,1.208) (4.666,1.214)};
\draw[isod] plot coordinates {(3.962,0.054) (3.957,0.066) (3.952,0.078) (3.948,0.090) (3.943,0.102) (3.939,0.113) (3.935,0.125) (3.931,0.137) (3.927,0.149) (3.923,0.161) (3.919,0.173) (3.915,0.185) (3.911,0.197) (3.908,0.209) (3.904,0.221) (3.901,0.232) (3.897,0.244) (3.894,0.256) (3.891,0.268) (3.888,0.280) (3.885,0.292) (3.882,0.304) (3.879,0.316) (3.876,0.328) (3.874,0.339) (3.871,0.351) (3.869,0.363) (3.866,0.375) (3.864,0.387) (3.862,0.399) (3.860,0.411) (3.858,0.423) (3.856,0.435) (3.854,0.446) (3.852,0.458) (3.851,0.470) (3.849,0.482) (3.848,0.494) (3.847,0.506) (3.846,0.518) (3.845,0.530) (3.844,0.542) (3.843,0.554) (3.842,0.565) (3.842,0.577) (3.841,0.589) (3.841,0.601) (3.841,0.613) (3.841,0.625) (3.841,0.637) (3.841,0.649) (3.841,0.661) (3.842,0.672) (3.843,0.684) (3.844,0.696) (3.845,0.708) (3.846,0.720) (3.847,0.732) (3.849,0.744) (3.850,0.756) (3.852,0.768) (3.854,0.779) (3.857,0.791) (3.859,0.803) (3.862,0.815) (3.865,0.827) (3.868,0.839) (3.871,0.851) (3.875,0.863) (3.879,0.875) (3.883,0.887) (3.888,0.898) (3.893,0.910) (3.898,0.922) (3.904,0.934) (3.910,0.946) (3.916,0.958) (3.923,0.970) (3.930,0.982) (3.938,0.994) (3.946,1.005) (3.955,1.017) (3.965,1.029) (3.975,1.041) (3.987,1.053) (3.999,1.065) (4.012,1.077) (4.026,1.089) (4.042,1.101) (4.059,1.112) (4.078,1.124) (4.100,1.136) (4.124,1.148) (4.152,1.160) (4.185,1.172) (4.224,1.184) (4.276,1.196) (4.352,1.208)};
\draw[line width=1.0pt] plot coordinates {(8.797,1.808) (8.460,1.812) (8.218,1.816) (8.029,1.820) (7.873,1.824) (7.740,1.828) (7.624,1.832) (7.521,1.836) (7.428,1.840) (7.343,1.844) (7.265,1.848) (7.194,1.852) (7.127,1.856) (7.064,1.860) (7.005,1.864) (6.950,1.868) (6.897,1.872) (6.847,1.876) (6.799,1.880) (6.754,1.884) (6.710,1.888) (6.668,1.892) (6.628,1.896) (6.589,1.900) (6.552,1.904) (6.515,1.908) (6.480,1.912) (6.447,1.916) (6.414,1.920) (6.382,1.924) (6.351,1.928) (6.321,1.932) (6.291,1.936) (6.263,1.940) (6.235,1.944) (6.208,1.948) (6.181,1.952) (6.155,1.956) (6.130,1.960) (6.105,1.964) (6.081,1.968) (6.057,1.972) (6.033,1.976) (6.010,1.980) (5.988,1.984) (5.966,1.988) (5.944,1.992) (5.923,1.996) (5.902,2.000) (5.882,2.004) (5.861,2.008) (5.841,2.012) (5.822,2.016) (5.803,2.020) (5.784,2.024) (5.765,2.028) (5.747,2.032) (5.729,2.036) (5.711,2.040) (5.693,2.044) (5.676,2.048) (5.659,2.052) (5.642,2.056) (5.625,2.060) (5.608,2.064) (5.592,2.068) (5.576,2.072) (5.560,2.076) (5.544,2.080) (5.529,2.084) (5.514,2.088) (5.498,2.092) (5.483,2.096) (5.469,2.100) (5.454,2.104) (5.440,2.108) (5.425,2.112) (5.411,2.116) (5.397,2.120) (5.383,2.124) (5.369,2.128) (5.356,2.132) (5.342,2.136) (5.329,2.140) (5.316,2.144) (5.302,2.148) (5.289,2.152) (5.277,2.156) (5.264,2.160) (5.251,2.164) (5.239,2.168) (5.226,2.172) (5.214,2.176) (5.202,2.180) (5.190,2.184) (5.177,2.188) (5.166,2.192) (5.154,2.196) (5.142,2.200) (5.130,2.204) (5.119,2.208) (5.107,2.212) (5.096,2.216) (5.085,2.220) (5.074,2.224) (5.062,2.228) (5.051,2.232) (5.040,2.236) (5.030,2.240) (5.019,2.244) (5.008,2.249) (4.997,2.253) (4.987,2.257) (4.976,2.261) (4.966,2.265) (4.956,2.269) (4.945,2.273) (4.935,2.277) (4.925,2.281) (4.915,2.285) (4.905,2.289) (4.895,2.293) (4.885,2.297) (4.875,2.301) (4.865,2.305) (4.856,2.309) (4.846,2.313) (4.836,2.317) (4.827,2.321) (4.817,2.325) (4.808,2.329) (4.798,2.333) (4.789,2.337) (4.780,2.341) (4.771,2.345) (4.761,2.349) (4.752,2.353) (4.743,2.357) (4.734,2.361) (4.725,2.365) (4.716,2.369) (4.707,2.373) (4.699,2.377) (4.690,2.381) (4.681,2.385) (4.672,2.389) (4.664,2.393) (4.655,2.397) (4.647,2.401) (4.638,2.405) (4.630,2.409) (4.621,2.413) (4.613,2.417) (4.604,2.421) (4.596,2.425) (4.588,2.429) (4.579,2.433) (4.571,2.437) (4.563,2.441) (4.555,2.445) (4.547,2.449) (4.539,2.453) (4.531,2.457) (4.523,2.461) (4.515,2.465) (4.507,2.469) (4.499,2.473) (4.491,2.477) (4.483,2.481) (4.475,2.485) (4.468,2.489) (4.460,2.493) (4.452,2.497) (4.445,2.501) (4.437,2.505) (4.429,2.509) (4.422,2.513) (4.414,2.517) (4.407,2.521) (4.399,2.525) (4.392,2.529) (4.384,2.533) (4.377,2.537) (4.370,2.541) (4.362,2.545) (4.355,2.549) (4.348,2.553) (4.340,2.557) (4.333,2.561) (4.326,2.565) (4.319,2.569) (4.312,2.573) (4.305,2.577) (4.298,2.581) (4.290,2.585) (4.283,2.589) (4.276,2.593) (4.269,2.597) (4.262,2.601) (4.255,2.605) (4.248,2.609) (4.242,2.613) (4.235,2.617) (4.228,2.621) (4.221,2.625) (4.214,2.629) (4.207,2.633) (4.201,2.637) (4.194,2.641) (4.187,2.645) (4.180,2.649) (4.174,2.653) (4.167,2.657) (4.160,2.661) (4.154,2.665) (4.147,2.669) (4.141,2.673) (4.134,2.677) (4.128,2.681) (4.121,2.685) (4.114,2.689) (4.108,2.693) (4.102,2.697) (4.095,2.701) (4.089,2.705) (4.082,2.709) (4.076,2.713) (4.069,2.717) (4.063,2.721) (4.057,2.725) (4.050,2.729) (4.044,2.733) (4.038,2.737) (4.032,2.741) (4.025,2.745) (4.019,2.749) (4.013,2.753) (4.007,2.757) (4.001,2.761) (3.994,2.765) (3.988,2.769) (3.982,2.773) (3.976,2.777) (3.970,2.781) (3.964,2.785) (3.958,2.789) (3.952,2.793) (3.946,2.797) (3.940,2.801) (3.934,2.805) (3.928,2.809) (3.922,2.813) (3.916,2.817) (3.910,2.821) (3.904,2.825) (3.898,2.829) (3.892,2.833) (3.886,2.837) (3.880,2.841) (3.874,2.845) (3.868,2.849) (3.862,2.853) (3.857,2.857) (3.851,2.861) (3.845,2.865) (3.839,2.869) (3.833,2.873) (3.828,2.877) (3.822,2.881) (3.816,2.885) (3.810,2.889) (3.805,2.893) (3.799,2.897) (3.793,2.901) (3.788,2.905) (3.782,2.909) (3.776,2.913) (3.771,2.917) (3.765,2.921) (3.759,2.925) (3.754,2.929) (3.748,2.933) (3.743,2.937) (3.737,2.941) (3.731,2.945) (3.726,2.949) (3.720,2.953) (3.715,2.957) (3.709,2.961) (3.704,2.965) (3.698,2.969) (3.693,2.973) (3.687,2.977) (3.682,2.981) (3.676,2.985) (3.671,2.989) (3.666,2.993) (3.660,2.997) (3.655,3.001) (3.649,3.005) (3.644,3.009) (3.638,3.013) (3.633,3.017) (3.628,3.021) (3.622,3.025) (3.617,3.029) (3.612,3.033) (3.606,3.037) (3.601,3.041) (3.596,3.045) (3.590,3.049) (3.585,3.053) (3.580,3.057) (3.574,3.061) (3.569,3.065) (3.564,3.069) (3.559,3.073) (3.553,3.077) (3.548,3.081) (3.543,3.085) (3.538,3.089) (3.533,3.093) (3.527,3.097) (3.522,3.101) (3.517,3.105) (3.512,3.109) (3.507,3.113) (3.501,3.117) (3.496,3.121) (3.491,3.125) (3.486,3.129) (3.481,3.133) (3.476,3.137) (3.471,3.141) (3.465,3.145) (3.460,3.150) (3.455,3.154) (3.450,3.158) (3.445,3.162) (3.440,3.166) (3.435,3.170) (3.430,3.174) (3.425,3.178) (3.420,3.182) (3.415,3.186) (3.410,3.190) (3.405,3.194) (3.400,3.198) (3.395,3.202) (3.390,3.206) (3.385,3.210) (3.380,3.214) (3.375,3.218) (3.370,3.222) (3.365,3.226) (3.360,3.230) (3.355,3.234) (3.350,3.238) (3.345,3.242) (3.340,3.246) (3.335,3.250) (3.330,3.254) (3.325,3.258) (3.320,3.262) (3.315,3.266) (3.310,3.270) (3.305,3.274) (3.300,3.278) (3.295,3.282) (3.291,3.286) (3.286,3.290) (3.281,3.294) (3.276,3.298) (3.271,3.302) (3.266,3.306) (3.261,3.310) (3.256,3.314) (3.252,3.318) (3.247,3.322) (3.242,3.326) (3.237,3.330) (3.232,3.334) (3.227,3.338) (3.222,3.342) (3.218,3.346) (3.213,3.350) (3.208,3.354) (3.203,3.358) (3.198,3.362) (3.194,3.366) (3.189,3.370) (3.184,3.374) (3.179,3.378) (3.174,3.382) (3.170,3.386) (3.165,3.390) (3.160,3.394) (3.155,3.398) (3.151,3.402) (3.146,3.406) (3.141,3.410) (3.136,3.414) (3.132,3.418) (3.127,3.422) (3.122,3.426) (3.117,3.430) (3.113,3.434) (3.108,3.438) (3.103,3.442) (3.098,3.446) (3.094,3.450) (3.089,3.454) (3.084,3.458) (3.080,3.462) (3.075,3.466) (3.070,3.470) (3.065,3.474) (3.061,3.478) (3.056,3.482) (3.051,3.486) (3.047,3.490) (3.042,3.494) (3.037,3.498) (3.033,3.502) (3.028,3.506) (3.023,3.510) (3.019,3.514) (3.014,3.518) (3.009,3.522) (3.005,3.526) (3.000,3.530) (2.995,3.534) (2.991,3.538) (2.986,3.542) (2.981,3.546) (2.977,3.550) (2.972,3.554) (2.967,3.558) (2.963,3.562) (2.958,3.566) (2.953,3.570) (2.949,3.574) (2.944,3.578) (2.939,3.582) (2.935,3.586) (2.930,3.590) (2.926,3.594) (2.921,3.598) (2.916,3.602) (2.912,3.606) (2.907,3.610) (2.902,3.614) (2.898,3.618) (2.893,3.622) (2.889,3.626) (2.884,3.630) (2.879,3.634) (2.875,3.638) (2.870,3.642) (2.866,3.646) (2.861,3.650) (2.856,3.654) (2.852,3.658) (2.847,3.662) (2.843,3.666) (2.838,3.670) (2.833,3.674) (2.829,3.678) (2.824,3.682) (2.820,3.686) (2.815,3.690) (2.810,3.694) (2.806,3.698) (2.801,3.702) (2.797,3.706) (2.792,3.710) (2.787,3.714) (2.783,3.718) (2.778,3.722) (2.774,3.726) (2.769,3.730) (2.764,3.734) (2.760,3.738) (2.755,3.742) (2.751,3.746) (2.746,3.750) (2.742,3.754) (2.737,3.758) (2.732,3.762) (2.728,3.766) (2.723,3.770) (2.719,3.774) (2.714,3.778) (2.709,3.782) (2.705,3.786) (2.700,3.790) (2.696,3.794) (2.691,3.798) (2.687,3.802) (2.682,3.806) (2.677,3.810) (2.673,3.814) (2.668,3.818) (2.664,3.822) (2.659,3.826) (2.654,3.830) (2.650,3.834) (2.645,3.838) (2.641,3.842) (2.636,3.846) (2.631,3.850) (2.627,3.854) (2.622,3.858) (2.618,3.862) (2.613,3.866) (2.608,3.870) (2.604,3.874) (2.599,3.878) (2.595,3.882) (2.590,3.886) (2.585,3.890) (2.581,3.894) (2.576,3.898) (2.572,3.902) (2.567,3.906) (2.562,3.910) (2.558,3.914) (2.553,3.918) (2.549,3.922) (2.544,3.926) (2.539,3.930) (2.535,3.934) (2.530,3.938) (2.526,3.942) (2.521,3.946) (2.516,3.950) (2.512,3.954) (2.507,3.958) (2.502,3.962) (2.498,3.966) (2.493,3.970) (2.488,3.974) (2.484,3.978) (2.479,3.982) (2.475,3.986) (2.470,3.990) (2.465,3.994) (2.461,3.998) (2.456,4.002) (2.451,4.006) (2.447,4.010) (2.442,4.014) (2.437,4.018) (2.433,4.022) (2.428,4.026) (2.423,4.030) (2.419,4.034) (2.414,4.038) (2.409,4.042) (2.405,4.046) (2.400,4.050) (2.395,4.055) (2.391,4.059) (2.386,4.063) (2.381,4.067) (2.376,4.071) (2.372,4.075) (2.367,4.079) (2.362,4.083) (2.358,4.087) (2.353,4.091) (2.348,4.095) (2.343,4.099) (2.339,4.103) (2.334,4.107) (2.329,4.111) (2.324,4.115) (2.320,4.119) (2.315,4.123) (2.310,4.127) (2.305,4.131) (2.301,4.135) (2.296,4.139) (2.291,4.143) (2.286,4.147) (2.282,4.151) (2.277,4.155) (2.272,4.159) (2.267,4.163) (2.262,4.167) (2.258,4.171) (2.253,4.175) (2.248,4.179) (2.243,4.183) (2.238,4.187) (2.233,4.191) (2.229,4.195) (2.224,4.199) (2.219,4.203) (2.214,4.207) (2.209,4.211) (2.204,4.215) (2.199,4.219) (2.195,4.223) (2.190,4.227) (2.185,4.231) (2.180,4.235) (2.175,4.239) (2.170,4.243) (2.165,4.247) (2.160,4.251) (2.155,4.255) (2.150,4.259) (2.146,4.263) (2.141,4.267) (2.136,4.271) (2.131,4.275) (2.126,4.279) (2.121,4.283) (2.116,4.287) (2.111,4.291) (2.106,4.295) (2.101,4.299) (2.096,4.303) (2.091,4.307) (2.086,4.311) (2.081,4.315) (2.076,4.319) (2.071,4.323) (2.066,4.327) (2.061,4.331) (2.055,4.335) (2.050,4.339) (2.045,4.343) (2.040,4.347) (2.035,4.351) (2.030,4.355) (2.025,4.359) (2.020,4.363) (2.015,4.367) (2.010,4.371) (2.004,4.375) (1.999,4.379) (1.994,4.383) (1.989,4.387) (1.984,4.391) (1.978,4.395) (1.973,4.399) (1.968,4.403) (1.963,4.407) (1.958,4.411) (1.952,4.415) (1.947,4.419) (1.942,4.423) (1.937,4.427) (1.931,4.431) (1.926,4.435) (1.921,4.439) (1.915,4.443) (1.910,4.447) (1.905,4.451) (1.899,4.455) (1.894,4.459) (1.889,4.463) (1.883,4.467) (1.878,4.471) (1.872,4.475) (1.867,4.479) (1.861,4.483) (1.856,4.487) (1.851,4.491) (1.845,4.495) (1.840,4.499) (1.834,4.503) (1.829,4.507) (1.823,4.511) (1.818,4.515) (1.812,4.519) (1.806,4.523) (1.801,4.527) (1.795,4.531) (1.790,4.535) (1.784,4.539) (1.778,4.543) (1.773,4.547) (1.767,4.551) (1.761,4.555) (1.756,4.559) (1.750,4.563) (1.744,4.567) (1.739,4.571) (1.733,4.575) (1.727,4.579) (1.721,4.583) (1.715,4.587) (1.710,4.591) (1.704,4.595) (1.698,4.599) (1.692,4.603) (1.686,4.607) (1.680,4.611) (1.674,4.615) (1.668,4.619) (1.662,4.623) (1.656,4.627) (1.650,4.631) (1.644,4.635) (1.638,4.639) (1.632,4.643) (1.626,4.647) (1.620,4.651) (1.614,4.655) (1.608,4.659) (1.602,4.663) (1.596,4.667) (1.589,4.671) (1.583,4.675) (1.577,4.679) (1.571,4.683) (1.564,4.687) (1.558,4.691) (1.552,4.695) (1.545,4.699) (1.539,4.703) (1.533,4.707) (1.526,4.711) (1.520,4.715) (1.513,4.719) (1.507,4.723) (1.500,4.727) (1.494,4.731) (1.487,4.735) (1.481,4.739) (1.474,4.743) (1.467,4.747) (1.461,4.751) (1.454,4.755) (1.447,4.759) (1.441,4.763) (1.434,4.767) (1.427,4.771) (1.420,4.775) (1.413,4.779) (1.406,4.783) (1.399,4.787) (1.392,4.791) (1.385,4.795) (1.378,4.799) (1.371,4.803) (1.364,4.807) (1.357,4.811) (1.350,4.815) (1.343,4.819) (1.335,4.823) (1.328,4.827) (1.321,4.831) (1.314,4.835) (1.306,4.839) (1.299,4.843) (1.291,4.847) (1.284,4.851) (1.276,4.855) (1.269,4.859) (1.261,4.863) (1.253,4.867) (1.246,4.871) (1.238,4.875) (1.230,4.879) (1.222,4.883) (1.214,4.887) (1.207,4.891) (1.199,4.895) (1.191,4.899) (1.182,4.903) (1.174,4.907) (1.166,4.911) (1.158,4.915) (1.150,4.919) (1.141,4.923) (1.133,4.927) (1.125,4.931) (1.116,4.935) (1.108,4.939) (1.099,4.943) (1.090,4.947) (1.082,4.951) (1.073,4.956) (1.064,4.960) (1.055,4.964) (1.046,4.968) (1.037,4.972) (1.028,4.976) (1.019,4.980) (1.010,4.984) (1.000,4.988) (0.991,4.992) (0.982,4.996) (0.972,5.000) (0.962,5.004) (0.953,5.008) (0.943,5.012) (0.933,5.016) (0.923,5.020) (0.913,5.024) (0.903,5.028) (0.893,5.032) (0.883,5.036) (0.872,5.040) (0.862,5.044) (0.851,5.048) (0.841,5.052) (0.830,5.056) (0.819,5.060) (0.808,5.064) (0.797,5.068) (0.786,5.072) (0.775,5.076) (0.763,5.080) (0.752,5.084) (0.740,5.088) (0.728,5.092) (0.716,5.096) (0.704,5.100) (0.692,5.104) (0.680,5.108) (0.667,5.112) (0.655,5.116) (0.642,5.120) (0.629,5.124) (0.616,5.128) (0.603,5.132) (0.589,5.136) (0.575,5.140) (0.562,5.144) (0.548,5.148) (0.533,5.152) (0.519,5.156) (0.504,5.160) (0.490,5.164) (0.475,5.168) (0.459,5.172) (0.444,5.176) (0.428,5.180) (0.412,5.184) (0.396,5.188) (0.379,5.192) (0.362,5.196) (0.345,5.200) (0.328,5.204) (0.310,5.208) (0.292,5.212) (0.273,5.216) (0.255,5.220) (0.235,5.224) (0.216,5.228) (0.196,5.232) (0.175,5.236) (0.154,5.240)};
\draw[line width=0.9pt,densely dotted] plot coordinates {(9.332,0.421) (8.769,0.427) (8.432,0.432) (8.188,0.438) (7.997,0.443) (7.838,0.449) (7.703,0.454) (7.585,0.460) (7.480,0.465) (7.385,0.471) (7.298,0.477) (7.219,0.482) (7.145,0.488) (7.076,0.493) (7.012,0.499) (6.951,0.504) (6.894,0.510) (6.840,0.515) (6.788,0.521) (6.739,0.526) (6.692,0.532) (6.647,0.538) (6.603,0.543) (6.562,0.549) (6.521,0.554) (6.483,0.560) (6.445,0.565) (6.409,0.571) (6.374,0.576) (6.339,0.582) (6.306,0.587) (6.274,0.593) (6.243,0.598) (6.212,0.604) (6.182,0.610) (6.153,0.615) (6.125,0.621) (6.097,0.626) (6.070,0.632) (6.043,0.637) (6.018,0.643) (5.992,0.648) (5.967,0.654) (5.943,0.659) (5.919,0.665) (5.895,0.671) (5.872,0.676) (5.850,0.682) (5.827,0.687) (5.805,0.693) (5.784,0.698) (5.763,0.704) (5.742,0.709) (5.721,0.715) (5.701,0.720) (5.681,0.726) (5.662,0.732) (5.642,0.737) (5.623,0.743) (5.605,0.748) (5.586,0.754) (5.568,0.759) (5.550,0.765) (5.532,0.770) (5.514,0.776) (5.497,0.781) (5.480,0.787) (5.463,0.793) (5.446,0.798) (5.430,0.804) (5.413,0.809) (5.397,0.815) (5.381,0.820) (5.366,0.826) (5.350,0.831) (5.335,0.837) (5.319,0.842) (5.304,0.848) (5.289,0.854) (5.274,0.859) (5.260,0.865) (5.245,0.870) (5.231,0.876) (5.217,0.881) (5.203,0.887) (5.189,0.892) (5.175,0.898) (5.161,0.903) (5.147,0.909) (5.134,0.915) (5.121,0.920) (5.107,0.926) (5.094,0.931) (5.081,0.937) (5.068,0.942) (5.056,0.948) (5.043,0.953) (5.030,0.959) (5.018,0.964) (5.006,0.970) (4.993,0.976) (4.981,0.981) (4.969,0.987) (4.957,0.992) (4.945,0.998) (4.933,1.003) (4.922,1.009) (4.910,1.014) (4.898,1.020) (4.887,1.025) (4.875,1.031) (4.864,1.037) (4.853,1.042) (4.842,1.048) (4.831,1.053) (4.820,1.059) (4.809,1.064) (4.798,1.070) (4.787,1.075) (4.776,1.081) (4.766,1.086) (4.755,1.092) (4.745,1.097) (4.734,1.103) (4.724,1.109) (4.713,1.114) (4.703,1.120) (4.693,1.125) (4.683,1.131) (4.673,1.136) (4.663,1.142) (4.653,1.147) (4.643,1.153) (4.633,1.158) (4.623,1.164) (4.613,1.170) (4.604,1.175) (4.594,1.181) (4.584,1.186) (4.575,1.192) (4.565,1.197) (4.556,1.203) (4.547,1.208) (4.537,1.214) (4.528,1.219) (4.519,1.225) (4.510,1.231) (4.500,1.236) (4.491,1.242) (4.482,1.247) (4.473,1.253) (4.464,1.258) (4.455,1.264) (4.447,1.269) (4.438,1.275) (4.429,1.280) (4.420,1.286) (4.411,1.292) (4.403,1.297) (4.394,1.303) (4.386,1.308) (4.377,1.314) (4.369,1.319) (4.360,1.325) (4.352,1.330) (4.343,1.336) (4.335,1.341) (4.327,1.347) (4.318,1.353) (4.310,1.358) (4.302,1.364) (4.294,1.369) (4.285,1.375) (4.277,1.380) (4.269,1.386) (4.261,1.391) (4.253,1.397) (4.245,1.402) (4.237,1.408) (4.229,1.414) (4.221,1.419) (4.214,1.425) (4.206,1.430) (4.198,1.436) (4.190,1.441) (4.182,1.447) (4.175,1.452) (4.167,1.458) (4.159,1.463) (4.152,1.469) (4.144,1.475) (4.137,1.480) (4.129,1.486) (4.122,1.491) (4.114,1.497) (4.107,1.502) (4.099,1.508) (4.092,1.513) (4.084,1.519) (4.077,1.524) (4.070,1.530) (4.062,1.536) (4.055,1.541) (4.048,1.547) (4.041,1.552) (4.034,1.558) (4.026,1.563) (4.019,1.569) (4.012,1.574) (4.005,1.580) (3.998,1.585) (3.991,1.591) (3.984,1.596) (3.977,1.602) (3.970,1.608) (3.963,1.613) (3.956,1.619) (3.949,1.624) (3.942,1.630) (3.935,1.635) (3.928,1.641) (3.921,1.646) (3.915,1.652) (3.908,1.657) (3.901,1.663) (3.894,1.669) (3.888,1.674) (3.881,1.680) (3.874,1.685) (3.867,1.691) (3.861,1.696) (3.854,1.702) (3.847,1.707) (3.841,1.713) (3.834,1.718) (3.828,1.724) (3.821,1.730) (3.815,1.735) (3.808,1.741) (3.802,1.746) (3.795,1.752) (3.789,1.757) (3.782,1.763) (3.776,1.768) (3.769,1.774) (3.763,1.779) (3.757,1.785) (3.750,1.791) (3.744,1.796) (3.738,1.802) (3.731,1.807) (3.725,1.813) (3.719,1.818) (3.712,1.824) (3.706,1.829) (3.700,1.835) (3.694,1.840) (3.688,1.846) (3.681,1.852) (3.675,1.857) (3.669,1.863) (3.663,1.868) (3.657,1.874) (3.651,1.879) (3.644,1.885) (3.638,1.890) (3.632,1.896) (3.626,1.901) (3.620,1.907) (3.614,1.913) (3.608,1.918) (3.602,1.924) (3.596,1.929) (3.590,1.935) (3.584,1.940) (3.578,1.946) (3.572,1.951) (3.566,1.957) (3.560,1.962) (3.555,1.968) (3.549,1.974) (3.543,1.979) (3.537,1.985) (3.531,1.990) (3.525,1.996) (3.519,2.001) (3.514,2.007) (3.508,2.012) (3.502,2.018) (3.496,2.023) (3.490,2.029) (3.485,2.035) (3.479,2.040) (3.473,2.046) (3.468,2.051) (3.462,2.057) (3.456,2.062) (3.450,2.068) (3.445,2.073) (3.439,2.079) (3.433,2.084) (3.428,2.090) (3.422,2.096) (3.416,2.101) (3.411,2.107) (3.405,2.112) (3.400,2.118) (3.394,2.123) (3.389,2.129) (3.383,2.134) (3.377,2.140) (3.372,2.145) (3.366,2.151) (3.361,2.156) (3.355,2.162) (3.350,2.168) (3.344,2.173) (3.339,2.179) (3.333,2.184) (3.328,2.190) (3.322,2.195) (3.317,2.201) (3.311,2.206) (3.306,2.212) (3.301,2.217) (3.295,2.223) (3.290,2.229) (3.284,2.234) (3.279,2.240) (3.274,2.245) (3.268,2.251) (3.263,2.256) (3.257,2.262) (3.252,2.267) (3.247,2.273) (3.241,2.278) (3.236,2.284) (3.231,2.290) (3.225,2.295) (3.220,2.301) (3.215,2.306) (3.210,2.312) (3.204,2.317) (3.199,2.323) (3.194,2.328) (3.189,2.334) (3.183,2.339) (3.178,2.345) (3.173,2.351) (3.168,2.356) (3.162,2.362) (3.157,2.367) (3.152,2.373) (3.147,2.378) (3.142,2.384) (3.136,2.389) (3.131,2.395) (3.126,2.400) (3.121,2.406) (3.116,2.412) (3.110,2.417) (3.105,2.423) (3.100,2.428) (3.095,2.434) (3.090,2.439) (3.085,2.445) (3.080,2.450) (3.075,2.456) (3.069,2.461) (3.064,2.467) (3.059,2.473) (3.054,2.478) (3.049,2.484) (3.044,2.489) (3.039,2.495) (3.034,2.500) (3.029,2.506) (3.024,2.511) (3.019,2.517) (3.014,2.522) (3.009,2.528) (3.004,2.534) (2.999,2.539) (2.994,2.545) (2.989,2.550) (2.984,2.556) (2.979,2.561) (2.974,2.567) (2.969,2.572) (2.964,2.578) (2.959,2.583) (2.954,2.589) (2.949,2.595) (2.944,2.600) (2.939,2.606) (2.934,2.611) (2.929,2.617) (2.924,2.622) (2.919,2.628) (2.914,2.633) (2.909,2.639) (2.904,2.644) (2.899,2.650) (2.894,2.655) (2.889,2.661) (2.884,2.667) (2.879,2.672) (2.874,2.678) (2.870,2.683) (2.865,2.689) (2.860,2.694) (2.855,2.700) (2.850,2.705) (2.845,2.711) (2.840,2.716) (2.835,2.722) (2.830,2.728) (2.826,2.733) (2.821,2.739) (2.816,2.744) (2.811,2.750) (2.806,2.755) (2.801,2.761) (2.796,2.766) (2.792,2.772) (2.787,2.777) (2.782,2.783) (2.777,2.789) (2.772,2.794) (2.767,2.800) (2.763,2.805) (2.758,2.811) (2.753,2.816) (2.748,2.822) (2.743,2.827) (2.738,2.833) (2.734,2.838) (2.729,2.844) (2.724,2.850) (2.719,2.855) (2.714,2.861) (2.710,2.866) (2.705,2.872) (2.700,2.877) (2.695,2.883) (2.690,2.888) (2.686,2.894) (2.681,2.899) (2.676,2.905) (2.671,2.911) (2.667,2.916) (2.662,2.922) (2.657,2.927) (2.652,2.933) (2.647,2.938) (2.643,2.944) (2.638,2.949) (2.633,2.955) (2.628,2.960) (2.624,2.966) (2.619,2.972) (2.614,2.977) (2.609,2.983) (2.605,2.988) (2.600,2.994) (2.595,2.999) (2.590,3.005) (2.586,3.010) (2.581,3.016) (2.576,3.021) (2.571,3.027) (2.567,3.033) (2.562,3.038) (2.557,3.044) (2.552,3.049) (2.548,3.055) (2.543,3.060) (2.538,3.066) (2.534,3.071) (2.529,3.077) (2.524,3.082) (2.519,3.088) (2.515,3.094) (2.510,3.099) (2.505,3.105) (2.500,3.110) (2.496,3.116) (2.491,3.121) (2.486,3.127) (2.481,3.132) (2.477,3.138) (2.472,3.143) (2.467,3.149) (2.463,3.154) (2.458,3.160) (2.453,3.166) (2.448,3.171) (2.444,3.177) (2.439,3.182) (2.434,3.188) (2.430,3.193) (2.425,3.199) (2.420,3.204) (2.415,3.210) (2.411,3.215) (2.406,3.221) (2.401,3.227) (2.396,3.232) (2.392,3.238) (2.387,3.243) (2.382,3.249) (2.378,3.254) (2.373,3.260) (2.368,3.265) (2.363,3.271) (2.359,3.276) (2.354,3.282) (2.349,3.288) (2.345,3.293) (2.340,3.299) (2.335,3.304) (2.330,3.310) (2.326,3.315) (2.321,3.321) (2.316,3.326) (2.311,3.332) (2.307,3.337) (2.302,3.343) (2.297,3.349) (2.292,3.354) (2.288,3.360) (2.283,3.365) (2.278,3.371) (2.273,3.376) (2.269,3.382) (2.264,3.387) (2.259,3.393) (2.254,3.398) (2.250,3.404) (2.245,3.410) (2.240,3.415) (2.235,3.421) (2.231,3.426) (2.226,3.432) (2.221,3.437) (2.216,3.443) (2.212,3.448) (2.207,3.454) (2.202,3.459) (2.197,3.465) (2.193,3.471) (2.188,3.476) (2.183,3.482) (2.178,3.487) (2.173,3.493) (2.169,3.498) (2.164,3.504) (2.159,3.509) (2.154,3.515) (2.149,3.520) (2.145,3.526) (2.140,3.532) (2.135,3.537) (2.130,3.543) (2.125,3.548) (2.121,3.554) (2.116,3.559) (2.111,3.565) (2.106,3.570) (2.101,3.576) (2.097,3.581) (2.092,3.587) (2.087,3.593) (2.082,3.598) (2.077,3.604) (2.072,3.609) (2.067,3.615) (2.063,3.620) (2.058,3.626) (2.053,3.631) (2.048,3.637) (2.043,3.642) (2.038,3.648) (2.033,3.653) (2.028,3.659) (2.024,3.665) (2.019,3.670) (2.014,3.676) (2.009,3.681) (2.004,3.687) (1.999,3.692) (1.994,3.698) (1.989,3.703) (1.984,3.709) (1.979,3.714) (1.974,3.720) (1.970,3.726) (1.965,3.731) (1.960,3.737) (1.955,3.742) (1.950,3.748) (1.945,3.753) (1.940,3.759) (1.935,3.764) (1.930,3.770) (1.925,3.775) (1.920,3.781) (1.915,3.787) (1.910,3.792) (1.905,3.798) (1.900,3.803) (1.895,3.809) (1.890,3.814) (1.885,3.820) (1.880,3.825) (1.875,3.831) (1.870,3.836) (1.865,3.842) (1.859,3.848) (1.854,3.853) (1.849,3.859) (1.844,3.864) (1.839,3.870) (1.834,3.875) (1.829,3.881) (1.824,3.886) (1.819,3.892) (1.814,3.897) (1.808,3.903) (1.803,3.909) (1.798,3.914) (1.793,3.920) (1.788,3.925) (1.783,3.931) (1.777,3.936) (1.772,3.942) (1.767,3.947) (1.762,3.953) (1.757,3.958) (1.751,3.964) (1.746,3.970) (1.741,3.975) (1.736,3.981) (1.730,3.986) (1.725,3.992) (1.720,3.997) (1.715,4.003) (1.709,4.008) (1.704,4.014) (1.699,4.019) (1.693,4.025) (1.688,4.031) (1.683,4.036) (1.677,4.042) (1.672,4.047) (1.667,4.053) (1.661,4.058) (1.656,4.064) (1.650,4.069) (1.645,4.075) (1.640,4.080) (1.634,4.086) (1.629,4.092) (1.623,4.097) (1.618,4.103) (1.612,4.108) (1.607,4.114) (1.601,4.119) (1.596,4.125) (1.590,4.130) (1.585,4.136) (1.579,4.141) (1.574,4.147) (1.568,4.152) (1.563,4.158) (1.557,4.164) (1.551,4.169) (1.546,4.175) (1.540,4.180) (1.534,4.186) (1.529,4.191) (1.523,4.197) (1.517,4.202) (1.512,4.208) (1.506,4.213) (1.500,4.219) (1.494,4.225) (1.489,4.230) (1.483,4.236) (1.477,4.241) (1.471,4.247) (1.465,4.252) (1.460,4.258) (1.454,4.263) (1.448,4.269) (1.442,4.274) (1.436,4.280) (1.430,4.286) (1.424,4.291) (1.418,4.297) (1.412,4.302) (1.406,4.308) (1.400,4.313) (1.394,4.319) (1.388,4.324) (1.382,4.330) (1.376,4.335) (1.370,4.341) (1.364,4.347) (1.358,4.352) (1.352,4.358) (1.345,4.363) (1.339,4.369) (1.333,4.374) (1.327,4.380) (1.320,4.385) (1.314,4.391) (1.308,4.396) (1.302,4.402) (1.295,4.408) (1.289,4.413) (1.283,4.419) (1.276,4.424) (1.270,4.430) (1.263,4.435) (1.257,4.441) (1.250,4.446) (1.244,4.452) (1.237,4.457) (1.231,4.463) (1.224,4.469) (1.217,4.474) (1.211,4.480) (1.204,4.485) (1.197,4.491) (1.191,4.496) (1.184,4.502) (1.177,4.507) (1.170,4.513) (1.163,4.518) (1.157,4.524) (1.150,4.530) (1.143,4.535) (1.136,4.541) (1.129,4.546) (1.122,4.552) (1.115,4.557) (1.108,4.563) (1.101,4.568) (1.093,4.574) (1.086,4.579) (1.079,4.585) (1.072,4.591) (1.064,4.596) (1.057,4.602) (1.050,4.607) (1.042,4.613) (1.035,4.618) (1.028,4.624) (1.020,4.629) (1.013,4.635) (1.005,4.640) (0.997,4.646) (0.990,4.651) (0.982,4.657) (0.974,4.663) (0.966,4.668) (0.959,4.674) (0.951,4.679) (0.943,4.685) (0.935,4.690) (0.927,4.696) (0.919,4.701) (0.911,4.707) (0.903,4.712) (0.894,4.718) (0.886,4.724) (0.878,4.729) (0.870,4.735) (0.861,4.740) (0.853,4.746) (0.844,4.751) (0.836,4.757) (0.827,4.762) (0.818,4.768) (0.810,4.773) (0.801,4.779) (0.792,4.785) (0.783,4.790) (0.774,4.796) (0.765,4.801) (0.756,4.807) (0.747,4.812) (0.738,4.818) (0.728,4.823) (0.719,4.829) (0.710,4.834) (0.700,4.840) (0.690,4.846) (0.681,4.851) (0.671,4.857) (0.661,4.862) (0.651,4.868) (0.641,4.873) (0.631,4.879) (0.621,4.884) (0.611,4.890) (0.601,4.895) (0.590,4.901) (0.580,4.907) (0.569,4.912) (0.558,4.918) (0.547,4.923) (0.537,4.929) (0.525,4.934) (0.514,4.940) (0.503,4.945) (0.492,4.951) (0.480,4.956) (0.469,4.962) (0.457,4.968) (0.445,4.973) (0.433,4.979) (0.421,4.984) (0.409,4.990) (0.396,4.995) (0.384,5.001) (0.371,5.006) (0.358,5.012) (0.345,5.017) (0.332,5.023) (0.319,5.029) (0.305,5.034) (0.292,5.040) (0.278,5.045) (0.264,5.051) (0.249,5.056) (0.235,5.062) (0.220,5.067) (0.205,5.073) (0.190,5.078) (0.175,5.084) (0.159,5.090)};
\draw[line width=1.0pt,dashdotted] plot coordinates {(4.322,1.350) (4.321,1.353) (4.321,1.357) (4.320,1.360) (4.319,1.364) (4.319,1.367) (4.318,1.370) (4.317,1.374) (4.317,1.377) (4.316,1.380) (4.316,1.384) (4.315,1.387) (4.314,1.391) (4.314,1.394) (4.313,1.397) (4.312,1.401) (4.312,1.404) (4.311,1.408) (4.310,1.411) (4.310,1.414) (4.309,1.418) (4.308,1.421) (4.308,1.424) (4.307,1.428) (4.306,1.431) (4.306,1.435) (4.305,1.438) (4.305,1.441) (4.304,1.445) (4.303,1.448) (4.303,1.452) (4.302,1.455) (4.301,1.458) (4.301,1.462) (4.300,1.465) (4.299,1.468) (4.299,1.472) (4.298,1.475) (4.298,1.479) (4.297,1.482) (4.296,1.485) (4.296,1.489) (4.295,1.492) (4.294,1.495) (4.294,1.499) (4.293,1.502) (4.292,1.506) (4.292,1.509) (4.291,1.512) (4.291,1.516) (4.290,1.519) (4.289,1.523) (4.289,1.526) (4.288,1.529) (4.287,1.533) (4.287,1.536) (4.286,1.539) (4.286,1.543) (4.285,1.546) (4.284,1.550) (4.284,1.553) (4.283,1.556) (4.282,1.560) (4.282,1.563) (4.281,1.567) (4.281,1.570) (4.280,1.573) (4.279,1.577) (4.279,1.580) (4.278,1.583) (4.278,1.587) (4.277,1.590) (4.276,1.594) (4.276,1.597) (4.275,1.600) (4.274,1.604) (4.274,1.607) (4.273,1.611) (4.273,1.614) (4.272,1.617) (4.271,1.621) (4.271,1.624) (4.270,1.627) (4.270,1.631) (4.269,1.634) (4.268,1.638) (4.268,1.641) (4.267,1.644) (4.266,1.648) (4.266,1.651) (4.265,1.655) (4.265,1.658) (4.264,1.661) (4.263,1.665) (4.263,1.668) (4.262,1.671) (4.262,1.675) (4.261,1.678) (4.260,1.682) (4.260,1.685) (4.259,1.688) (4.259,1.692) (4.258,1.695) (4.257,1.698) (4.257,1.702) (4.256,1.705) (4.256,1.709) (4.255,1.712) (4.254,1.715) (4.254,1.719) (4.253,1.722) (4.253,1.726) (4.252,1.729) (4.251,1.732) (4.251,1.736) (4.250,1.739) (4.250,1.742) (4.249,1.746) (4.248,1.749) (4.248,1.753) (4.247,1.756) (4.247,1.759) (4.246,1.763) (4.245,1.766) (4.245,1.770) (4.244,1.773) (4.244,1.776) (4.243,1.780) (4.242,1.783) (4.242,1.786) (4.241,1.790) (4.241,1.793) (4.240,1.797) (4.240,1.800) (4.239,1.803) (4.238,1.807) (4.238,1.810) (4.237,1.814) (4.237,1.817) (4.236,1.820) (4.235,1.824) (4.235,1.827) (4.234,1.830) (4.234,1.834) (4.233,1.837) (4.232,1.841) (4.232,1.844) (4.231,1.847) (4.231,1.851) (4.230,1.854) (4.230,1.858) (4.229,1.861) (4.228,1.864) (4.228,1.868) (4.227,1.871) (4.227,1.874) (4.226,1.878) (4.226,1.881) (4.225,1.885) (4.224,1.888) (4.224,1.891) (4.223,1.895) (4.223,1.898) (4.222,1.902) (4.221,1.905) (4.221,1.908) (4.220,1.912) (4.220,1.915) (4.219,1.918) (4.219,1.922) (4.218,1.925) (4.217,1.929) (4.217,1.932) (4.216,1.935) (4.216,1.939) (4.215,1.942) (4.215,1.945) (4.214,1.949) (4.213,1.952) (4.213,1.956) (4.212,1.959) (4.212,1.962) (4.211,1.966) (4.211,1.969) (4.210,1.973) (4.209,1.976) (4.209,1.979) (4.208,1.983) (4.208,1.986) (4.207,1.989) (4.207,1.993) (4.206,1.996) (4.206,2.000) (4.205,2.003) (4.204,2.006) (4.204,2.010) (4.203,2.013) (4.203,2.017) (4.202,2.020) (4.202,2.023) (4.201,2.027) (4.200,2.030) (4.200,2.033) (4.199,2.037) (4.199,2.040) (4.198,2.044) (4.198,2.047) (4.197,2.050) (4.197,2.054) (4.196,2.057) (4.195,2.061) (4.195,2.064) (4.194,2.067) (4.194,2.071) (4.193,2.074) (4.193,2.077) (4.192,2.081) (4.192,2.084) (4.191,2.088) (4.190,2.091) (4.190,2.094) (4.189,2.098) (4.189,2.101) (4.188,2.105) (4.188,2.108) (4.187,2.111) (4.187,2.115) (4.186,2.118) (4.185,2.121) (4.185,2.125) (4.184,2.128) (4.184,2.132) (4.183,2.135) (4.183,2.138) (4.182,2.142) (4.182,2.145) (4.181,2.148) (4.181,2.152) (4.180,2.155) (4.179,2.159) (4.179,2.162) (4.178,2.165) (4.178,2.169) (4.177,2.172) (4.177,2.176) (4.176,2.179) (4.176,2.182) (4.175,2.186) (4.175,2.189) (4.174,2.192) (4.173,2.196) (4.173,2.199) (4.172,2.203) (4.172,2.206) (4.171,2.209) (4.171,2.213) (4.170,2.216) (4.170,2.220) (4.169,2.223) (4.169,2.226) (4.168,2.230) (4.168,2.233) (4.167,2.236) (4.166,2.240) (4.166,2.243) (4.165,2.247) (4.165,2.250) (4.164,2.253) (4.164,2.257) (4.163,2.260) (4.163,2.264) (4.162,2.267) (4.162,2.270) (4.161,2.274) (4.161,2.277) (4.160,2.280) (4.160,2.284) (4.159,2.287) (4.158,2.291) (4.158,2.294) (4.157,2.297) (4.157,2.301) (4.156,2.304) (4.156,2.308) (4.155,2.311) (4.155,2.314) (4.154,2.318) (4.154,2.321) (4.153,2.324) (4.153,2.328) (4.152,2.331) (4.152,2.335) (4.151,2.338) (4.151,2.341) (4.150,2.345) (4.149,2.348) (4.149,2.352) (4.148,2.355) (4.148,2.358) (4.147,2.362) (4.147,2.365) (4.146,2.368) (4.146,2.372) (4.145,2.375) (4.145,2.379) (4.144,2.382) (4.144,2.385) (4.143,2.389) (4.143,2.392) (4.142,2.395) (4.142,2.399) (4.141,2.402) (4.141,2.406) (4.140,2.409) (4.140,2.412) (4.139,2.416) (4.139,2.419) (4.138,2.423) (4.138,2.426) (4.137,2.429) (4.136,2.433) (4.136,2.436) (4.135,2.439) (4.135,2.443) (4.134,2.446) (4.134,2.450) (4.133,2.453) (4.133,2.456) (4.132,2.460) (4.132,2.463) (4.131,2.467) (4.131,2.470) (4.130,2.473) (4.130,2.477) (4.129,2.480) (4.129,2.483) (4.128,2.487) (4.128,2.490) (4.127,2.494) (4.127,2.497) (4.126,2.500) (4.126,2.504) (4.125,2.507) (4.125,2.511) (4.124,2.514) (4.124,2.517) (4.123,2.521) (4.123,2.524) (4.122,2.527) (4.122,2.531) (4.121,2.534) (4.121,2.538) (4.120,2.541) (4.120,2.544) (4.119,2.548) (4.119,2.551) (4.118,2.555) (4.118,2.558) (4.117,2.561) (4.117,2.565) (4.116,2.568) (4.116,2.571) (4.115,2.575) (4.115,2.578) (4.114,2.582) (4.114,2.585) (4.113,2.588) (4.113,2.592) (4.112,2.595) (4.112,2.598) (4.111,2.602) (4.111,2.605) (4.110,2.609) (4.110,2.612) (4.109,2.615) (4.109,2.619) (4.108,2.622) (4.108,2.626) (4.107,2.629) (4.107,2.632) (4.106,2.636) (4.106,2.639) (4.105,2.642) (4.105,2.646) (4.104,2.649) (4.104,2.653) (4.103,2.656) (4.103,2.659) (4.102,2.663) (4.102,2.666) (4.101,2.670) (4.101,2.673) (4.100,2.676) (4.100,2.680) (4.099,2.683) (4.099,2.686) (4.098,2.690) (4.098,2.693) (4.097,2.697) (4.097,2.700)};
\draw[line width=0.7pt,dashdotted,draw=black!45] plot coordinates {(4.523,0.432) (4.523,0.435) (4.522,0.438) (4.521,0.441) (4.520,0.444) (4.520,0.447) (4.519,0.450) (4.518,0.453) (4.517,0.457) (4.516,0.460) (4.516,0.463) (4.515,0.466) (4.514,0.469) (4.513,0.472) (4.513,0.475) (4.512,0.478) (4.511,0.481) (4.510,0.484) (4.510,0.487) (4.509,0.490) (4.508,0.493) (4.507,0.496) (4.507,0.500) (4.506,0.503) (4.505,0.506) (4.504,0.509) (4.504,0.512) (4.503,0.515) (4.502,0.518) (4.501,0.521) (4.501,0.524) (4.500,0.527) (4.499,0.530) (4.498,0.533) (4.498,0.536) (4.497,0.539) (4.496,0.543) (4.495,0.546) (4.495,0.549) (4.494,0.552) (4.493,0.555) (4.493,0.558) (4.492,0.561) (4.491,0.564) (4.490,0.567) (4.490,0.570) (4.489,0.573) (4.488,0.576) (4.487,0.579) (4.487,0.582) (4.486,0.586) (4.485,0.589) (4.484,0.592) (4.484,0.595) (4.483,0.598) (4.482,0.601) (4.482,0.604) (4.481,0.607) (4.480,0.610) (4.479,0.613) (4.479,0.616) (4.478,0.619) (4.477,0.622) (4.476,0.625) (4.476,0.628) (4.475,0.632) (4.474,0.635) (4.474,0.638) (4.473,0.641) (4.472,0.644) (4.471,0.647) (4.471,0.650) (4.470,0.653) (4.469,0.656) (4.469,0.659) (4.468,0.662) (4.467,0.665) (4.466,0.668) (4.466,0.671) (4.465,0.675) (4.464,0.678) (4.464,0.681) (4.463,0.684) (4.462,0.687) (4.461,0.690) (4.461,0.693) (4.460,0.696) (4.459,0.699) (4.459,0.702) (4.458,0.705) (4.457,0.708) (4.456,0.711) (4.456,0.714) (4.455,0.718) (4.454,0.721) (4.454,0.724) (4.453,0.727) (4.452,0.730) (4.452,0.733) (4.451,0.736) (4.450,0.739) (4.449,0.742) (4.449,0.745) (4.448,0.748) (4.447,0.751) (4.447,0.754) (4.446,0.757) (4.445,0.761) (4.445,0.764) (4.444,0.767) (4.443,0.770) (4.442,0.773) (4.442,0.776) (4.441,0.779) (4.440,0.782) (4.440,0.785) (4.439,0.788) (4.438,0.791) (4.438,0.794) (4.437,0.797) (4.436,0.800) (4.436,0.803) (4.435,0.807) (4.434,0.810) (4.434,0.813) (4.433,0.816) (4.432,0.819) (4.431,0.822) (4.431,0.825) (4.430,0.828) (4.429,0.831) (4.429,0.834) (4.428,0.837) (4.427,0.840) (4.427,0.843) (4.426,0.846) (4.425,0.850) (4.425,0.853) (4.424,0.856) (4.423,0.859) (4.423,0.862) (4.422,0.865) (4.421,0.868) (4.421,0.871) (4.420,0.874) (4.419,0.877) (4.419,0.880) (4.418,0.883) (4.417,0.886) (4.417,0.889) (4.416,0.893) (4.415,0.896) (4.415,0.899) (4.414,0.902) (4.413,0.905) (4.413,0.908) (4.412,0.911) (4.411,0.914) (4.411,0.917) (4.410,0.920) (4.409,0.923) (4.409,0.926) (4.408,0.929) (4.407,0.932) (4.407,0.936) (4.406,0.939) (4.405,0.942) (4.405,0.945) (4.404,0.948) (4.403,0.951) (4.403,0.954) (4.402,0.957) (4.401,0.960) (4.401,0.963) (4.400,0.966) (4.399,0.969) (4.399,0.972) (4.398,0.975) (4.397,0.979) (4.397,0.982) (4.396,0.985) (4.395,0.988) (4.395,0.991) (4.394,0.994) (4.393,0.997) (4.393,1.000) (4.392,1.003) (4.392,1.006) (4.391,1.009) (4.390,1.012) (4.390,1.015) (4.389,1.018) (4.388,1.021) (4.388,1.025) (4.387,1.028) (4.386,1.031) (4.386,1.034) (4.385,1.037) (4.384,1.040) (4.384,1.043) (4.383,1.046) (4.382,1.049) (4.382,1.052) (4.381,1.055) (4.381,1.058) (4.380,1.061) (4.379,1.064) (4.379,1.068) (4.378,1.071) (4.377,1.074) (4.377,1.077) (4.376,1.080) (4.375,1.083) (4.375,1.086) (4.374,1.089) (4.374,1.092) (4.373,1.095) (4.372,1.098) (4.372,1.101) (4.371,1.104) (4.370,1.107) (4.370,1.111) (4.369,1.114) (4.369,1.117) (4.368,1.120) (4.367,1.123) (4.367,1.126) (4.366,1.129) (4.365,1.132) (4.365,1.135) (4.364,1.138) (4.363,1.141) (4.363,1.144) (4.362,1.147) (4.362,1.150) (4.361,1.154) (4.360,1.157) (4.360,1.160) (4.359,1.163) (4.358,1.166) (4.358,1.169) (4.357,1.172) (4.357,1.175) (4.356,1.178) (4.355,1.181) (4.355,1.184) (4.354,1.187) (4.354,1.190) (4.353,1.193) (4.352,1.196) (4.352,1.200) (4.351,1.203) (4.350,1.206) (4.350,1.209) (4.349,1.212) (4.349,1.215) (4.348,1.218) (4.347,1.221) (4.347,1.224) (4.346,1.227) (4.346,1.230) (4.345,1.233) (4.344,1.236) (4.344,1.239) (4.343,1.243) (4.342,1.246) (4.342,1.249) (4.341,1.252) (4.341,1.255) (4.340,1.258) (4.339,1.261) (4.339,1.264) (4.338,1.267) (4.338,1.270) (4.337,1.273) (4.336,1.276) (4.336,1.279) (4.335,1.282) (4.335,1.286) (4.334,1.289) (4.333,1.292) (4.333,1.295) (4.332,1.298) (4.332,1.301) (4.331,1.304) (4.330,1.307) (4.330,1.310) (4.329,1.313) (4.329,1.316) (4.328,1.319) (4.327,1.322) (4.327,1.325) (4.326,1.329) (4.326,1.332) (4.325,1.335) (4.324,1.338) (4.324,1.341) (4.323,1.344) (4.323,1.347) (4.322,1.350)};
\fill (4.097,2.700) circle (1.6pt);
\fill (4.322,1.350) circle (1.6pt);
\draw[line width=0.5pt,dashed] (0.15,1.800) -- (9.35,1.800);
\node[font=\scriptsize,anchor=west] at (7.80,2) {$C=2/3$};
\node[font=\scriptsize,anchor=west] at (4.217,2.800) {$C=1$};
\node[font=\scriptsize,anchor=north east] at (4.190,1.6) {$C=\tfrac12$};

\node[font=\scriptsize,anchor=west] at (0.9,5.15) {$\Sigma=0$: no DS above};
\node[font=\scriptsize,anchor=west] at (2.05,3.75) {$P_{0}^{+}$ (scalable)};
\node[font=\scriptsize,anchor=west] at (0.35,1.05) {$P_{0}^{-}$ (unscalable)};
\node[font=\scriptsize,anchor=west] at (6.05,1.35) {\textbf{DSR}};
\node[font=\scriptsize,anchor=south west,align=left] at (5.05,2.20) {fidelity $\Xi=\Delta$};
\draw[line width=0.4pt,-{Latex[length=1mm]}] (5.02,2.26) -- (4.30,2.00);
\node[font=\scriptsize,anchor=west] at (4.5,0.5) {branch divider};
\node[font=\scriptsize,anchor=east] at (3,2.3) {isogains $\Sigma=$ const};
\end{tikzpicture}
\caption{The master diagram, computed from Eqs.~\eqref{eq:branches},
\eqref{eq:DSenergy} and \eqref{eq:dimensionless}. Grey curves are isogains
$\Sigma=\mathrm{const}$ (solid: the scalable branch $P_{0}^{+}$; dashed: the
unscalable branch $P_{0}^{-}$), shown for $\Sigma=0.1$, $0.3$ and $0.6$. The
solid black curve is the vacuum-stability threshold $\Sigma=0$, above which no
DS exists; the dotted curve is the branch divider, on which the discriminant of
\eqref{eq:branches} vanishes, and the two roots merge; the dash-dotted curve is
the fidelity curve $\Xi=\Delta$, where the two correlation scales coincide, and
external compressibility is optimal. The region bounded by these three curves is the region of DSR: there $\Xi<\Delta$, and
$E^{*}\to\infty$ as $\Xi\to0$. Each curve is drawn only on the branch
and over the interval on which it is defined. The vacuum-stability curve exists
for $2/3<C<2$: $\Xi^{*2}=\tfrac94C-\tfrac32$ vanishes at $C=2/3$, where the
energy diverges (dashed horizontal asymptote), and $\Delta^{*2}=\tfrac32(1-C/2)$
vanishes at $C=2$, where the curve terminates. The fidelity curve
$\Sigma_{\mathrm{fid}}=\tfrac34(1-C^{2})$ lies on the scalable root only for
$\tfrac12\le C\le1$ (heavy dash-dot); it meets $\Sigma=0$ at $C=1$,
$E^{*}=5.44$, and touches the branch divider tangentially at $C=1/2$,
$E^{*}=6.28$ (dots). Below $C=1/2$ it continues on the \emph{unscalable} root
(light dash-dot) and no longer bounds the DSR region, whose upper boundary is
taken over there by the branch divider, as \eqref{eq:DSRband} records; this is
why the shading continues below $C=1/2$ while the heavy dash-dot does not. The
value $C=2$ quoted in
earlier statements of this construction, our own included, is the terminus of
the $\Sigma=0$ curve and not a point of the fidelity curve, whose endpoint is
$C=1$, in agreement with \eqref{eq:DSRinterval}.}
\label{fig:masterdiagram}
\end{figure}
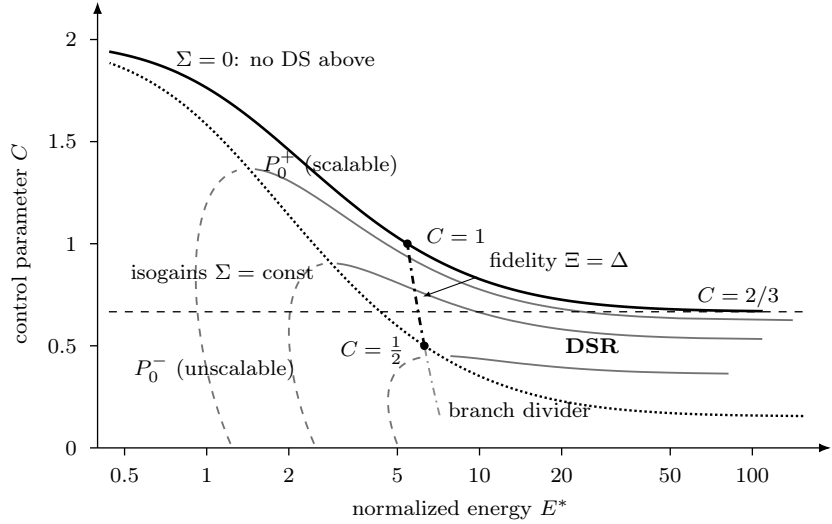

Because $\Delta$ and $\Xi$ depend on the coefficients of \eqref{eq:cqgle} only
through $C$ and $\Sigma$, so does the energy \eqref{eq:DSenergy}. Introducing
the dimensionless variables $E^{*}=E\kappa\sqrt{\zeta/(\beta\gamma)}$ and
$\omega^{*}=\omega\sqrt{\zeta\beta/\gamma}$, together with the spectral rescaling
$p^{*}=(\kappa/\beta)\,p$ reduces the
spectrum to
\begin{equation}
  p^{*}(\omega^{*}) = \frac{6\pi}{\Xi^{*2}+\omega^{*2}}
  \quad\text{for } |\omega^{*}|\le\Delta^{*} ,
  \qquad
  \Delta^{*2}=\zeta P_{0} ,
  \qquad
  \Xi^{*2}=1+C-\tfrac{5}{3}\Delta^{*2} ,
  \label{eq:dimensionless}
\end{equation}
so that the six-parameter problem collapses onto a plane. Plotting $C$ against
$E^{*}$ at fixed $\Sigma$ generates a manifold of \emph{isogains}, and the
resulting two-dimensional \textbf{master diagram} (Figure~\ref{fig:masterdiagram})
is the reduced parametric space of the DS at fixed $\Sigma$ and fixed model
assumptions
\cite{Rudenkov2023,Kalashnikov2024,Kalashnikov2025}. The link to the
laboratory runs through $\Sigma$: near threshold
$\sigma\approx\vartheta(E/E_{\mathrm{cw}}-1)$, with $E_{\mathrm{cw}}$ the
continuous-wave energy, so an isogain is a curve of constant average pump power
\cite{Kalashnikov2006}.

The diagram is partitioned by three curves, and each of them will reappear in a
thermodynamic guise.

\begin{itemize}[leftmargin=*,itemsep=0.25em]
  \item The \textbf{vacuum-stability threshold} $\Sigma=0$ separates the region
        where a DS can exist from the region where the continuous wave is
        unstable, and no solution of the present adiabatic chirped branch
        survives.\footnote{The last clause is a
        statement about the adiabatic chirped family, and should not be read as
        a statement about localized structures in general. In cubic GL models
        whose trivial solution is unstable, the exact pulse persists for a long
        time before the background instability destroys it, and does so longer
        in a wide computational window than in a narrow one
        \cite{Malomed1997}. In a two-component GL system with the gain confined
        to one core, robust localized pulses have been found sitting
        \emph{on top of} an unstable background---a small-amplitude standing
        wave just above threshold, and a chaotic one further above---which
        neither the instability of that background nor its subsequent
        chaotization destroys \cite{SakaguchiMalomed2000}. Those models are
        cubic, and the second is two-component, so no quantitative transfer is
        intended. What transfers is a caution about wording: $\Sigma=0$ is where
        the vacuum loses stability, not a demonstration that nothing localized
        is observable beyond it. The distinction is not idle here, because the
        resonance limit $\Sigma\to0^{+}$ approaches this boundary from
        within.}
  \item The \textbf{branch-dividing curve} separates the two roots of
        \eqref{eq:branches}. The upper root $P_{0}^{+}$ is
        \emph{energy-scalable}: at fixed $C$ its energy diverges
        asymptotically. The lower root $P_{0}^{-}$ is \emph{unscalable} and
        possesses a conservative-soliton limit \cite{Kalashnikov2024,Kalashnikov2025}.
        The existence of two branches carrying the same $\Sigma$ is what will
        later permit a single scalable pulse and a complex of several
        unscalable ones to be compared at equal energy.
  \item The \textbf{fidelity curve} $\Xi=\Delta$, on which the two correlation
        scales coincide, and external compressibility is locally
        favourable in the sense established below---the spectral-phase curvature
        \eqref{eq:specphase2} is stationary at zero detuning there. That is a
        local statement within the present asymptotic description, and is not a
        proof of a global optimum of the externally compressed pulse
        duration. It also marks the lower
        energy bound of the region in which $\Xi<\Delta$---the region of
        DSR. Combining \eqref{eq:branches} with
        \eqref{eq:dimensionless} puts that curve at
        $\Sigma=\tfrac{3}{4}(1-C^{2})$, so the resonance occupies a finite
        interval of $C$: at $\Sigma=0$ that
        interval runs from $C=2/3$, where $\Xi$ vanishes, and the energy diverges, to $C=1$, where the fidelity curve meets the
        vacuum-stability curve, in agreement with \eqref{eq:DSRinterval}. On the scalable root, the fidelity curve is defined only for $\tfrac12\le C\le1$, touching the branch divider tangentially at the
        lower end \cite{Kalashnikov2025}.
\end{itemize}

The strategy of reading a two-parameter plane as a superposition of
regimes has a direct precedent in the mode-locking literature. Haus outlined continuous-wave (CW) passive mode locking by examining small-signal saturable-absorber loading in relation to small-signal gain. He divided the analysis into three regimes: the existence of steady-state single-pulse solutions, stability against relaxation oscillations, and self-starting operation. For effective operation, the system must operate within the overlap of these three regimes, which can be achieved by gradually increasing the pump while remaining in the stable region \cite{Haus1976}. Two features of that construction survive the
change of coordinates. The first is the explicit warning that the existence of
steady-state mode-locked solutions does not guarantee that they will be
realized, which is the existence-versus-accessibility distinction that
Section~6.6 is forced to make for the anomalous branch. The second is that the
upper boundary of the single-pulse region is precisely where multiple-pulse
solutions appear, so that the multipulse transition enters as a boundary of the
diagram rather than as a separate phenomenon---the same placement it receives in
Section~6.4. The differences are equally worth stating: Haus's axes are
externally set control parameters, whereas the abscissa here is the pulse
energy, a property of the solution. His regimes are stability regions of a fast- or slow-absorber theory, whereas the curves here are existence and shape boundaries of the adiabatic family. The master diagram is not a version of Haus's diagram. What it inherits is the method.

The master diagram organizes what follows. It is an
\emph{existence and continuation diagram}: its curves are boundaries of
existence and of admissibility, not phase boundaries in the thermodynamic sense, since no phases, order parameters, or coexistence conditions have been defined for this system. Once the indicators of Section~6 are attached to its coordinates, it also carries their level sets, which we call indicator contours rather than an entropy or temperature landscape.

The device of projecting a soliton family onto a plane and reading its
properties off the geometry has a conservative ancestor, and the comparison fixes both what the master diagram inherits and what it cannot. For solitary waves of a nonintegrable but conservative NLS equation with a local nonlinearity, the family is customarily plotted on the plane of its two invariants, Hamiltonian $H$ against energy $E$. The resulting curve carries a complete stability theory: its slope returns the propagation constant, $dH/dE=-q$; branches with $H''(E)<0$ are stable and those with $H''(E)>0$ are not; stability can change only at cusps; and where two branches coexist at the same $E$, the one of lower $H$ is the stable one \cite{Akhmediev1999HQ}. The construction is moreover profile-free: $H$ and $E$ follow from single-peakedness and localization alone, without solving for the soliton \cite{Akhmediev1999HQ}. That is the exact inverse of the situation here, where the profile is the object to be found, and the invariants do not exist. The master diagram keeps the strategy---project, then read existence, branch structure, and the location of
transitions off the geometry---but it cannot keep the theorem. Its abscissa is a control parameter, and its ordinate is an energy at fixed net gain, not a pair of conserved quantities. There is no $H$ to be concave, and the branch pair of \eqref{eq:branches} is accordingly not ordered by any minimum principle. No theorem of comparable status replaces the conservative concavity
criterion. Section~6 instead explores entropy- and temperature-like diagnostics
as candidate organizers of branch selection; any predictive use of them must be
corroborated by direct dynamics.

The same comparison says something about the branch that does have a
conservative limit. For the conservative cubic--quintic nonlinearity, the entire $H(E)$ family is concave down, so every member of it is stable, and for one sign of the quintic coefficient the family terminates at a finite maximum energy $E_{\max}\propto|\chi|^{-1/2}$ \cite{Akhmediev1999HQ}. The conservative limit of the unscalable root is therefore not merely unscalable but bounded, and stable throughout---which is what its role played in Section~6.4, as the branch into which a scalable pulse can discharge.

What the adiabatic theory delivers, however, is a solution manifold, not yet thermodynamics. The master diagram charts stationary states. By itself, it says nothing about entropy, temperature, or how many degrees of freedom a DS possesses. Supplying that requires a statistical vocabulary, and nonlinear optics has built one---three of them, in fact, along largely independent routes. Section~3 sets out these three traditions and isolates the assumptions they share. Section~4 applies them to the object just described and identifies where those assumptions must be modified.

\section{Three Traditions of Optical Thermodynamics, and What They Omit}

Over the past two decades, thermodynamic and statistical-mechanical language has
migrated from its traditional home in many-body physics into nonlinear optics
along three largely independent routes: a kinetic one, an equilibrium one, and a
noise-driven one. They are set out here in some detail, because the
DS framework developed below reproduces several of their
ingredients, modifies others, and is most clearly understood through the contrast.
Each route is described in the notation of its own sources; where a symbol has
already been used in a different sense, the collision is flagged in place.

\subsection{The kinetic route: wave turbulence and wave condensation}

Weak-turbulence theory applied to nonlinear optical waves shows that an
incoherent field governed by a nonlinear Schr\"odinger-type equation
\begin{equation}
  i\partial_z\psi = -\alpha\nabla^{2}\psi + \beta\,\partial_{tt}\psi
                    + g|\psi|^{2}\psi ,
  \label{eq:nls}
\end{equation}
\noindent relaxes towards a Rayleigh--Jeans (RJ) equilibrium spectrum, and that this relaxation can be accompanied by condensation of the wave energy into the lowest modes, by the formation of incoherent and semicoherent solitons, and by long-range order emerging out of a fluctuating background \cite{Picozzi2007,Picozzi2014}. The scope of that statement should be made explicit before it is used. The Hamiltonian dynamics of \eqref{eq:nls} is itself time reversible. Irreversible relaxation, and the $H$-theorem that certifies it, arise only after the statistical assumptions underlying a wave-kinetic closure---weak nonlinearity, random phases, scale separation, and the continuum and resonance approximations---have been introduced. This matters beyond bookkeeping: it is the first instance of the pattern that organizes the whole paper, in which thermodynamics enters through a \emph{statistical level of description} rather than by rewriting a deterministic wave equation. Two notational points are important here: $\psi$ denotes the field, not the chirp of Section~2, while $\alpha$ and $\beta$ denote the transverse dispersion and the GDD rather than the filter bandwidth and GDD of \eqref{eq:cqgle}. Each route below keeps the notation of its own sources. Equation~\eqref{eq:nls} conserves the power $N=\int|\psi|^{2}\,d\mathbf{r}\,dt$ and the total energy $H=H_{l}+H_{nl}$; the kinetic closure retains only the linear (dispersive) part $H_{l}$, the nonlinear contribution being higher order in the small parameter $\varepsilon=H_{nl}/H_{l}$. The available closure depends on the field statistics. For \emph{inhomogeneous} (quasi-homogeneous) statistics, it is reached at first order in $\varepsilon$ and yields a Vlasov-like equation, in which the incoherent field behaves as an ensemble of independent quasiparticles moving in their own self-consistent
mean-field potential $V_{\mathrm{eff}}=2g\widetilde{N}(z,\mathbf{r})$, with effective dispersion $\tilde\omega=\alpha k^{2}+V_{\mathrm{eff}}$---the structure that underlies incoherent solitons \cite{Picozzi2007,Picozzi2009}. Because it is reversible, it cannot describe relaxation. The mean-field step carries a validity condition of its own, independent of the smallness of $\varepsilon$. It presumes that the medium responds to the \emph{average} intensity, which
requires a response slow compared with the phase-fluctuation correlation time and, less obviously, negligible intensity fluctuations of the source. When those fluctuations are large, one has
$\langle n_{nl}(I)\rangle\neq n_{nl}(\langle I\rangle)$. The self-induced
potential changes shape, and with it the modal content of the resulting
incoherent soliton \cite{Ponomarenko2004}. We return to this in
Section~4.3, because it bears on how confidently the microstate count is
defined. For \emph{homogeneous} statistics, the closure requires second-order perturbation theory and produces a Boltzmann-like collision integral, and the resulting kinetic equation is irreversible,
$dS/dz\ge 0$, with the nonequilibrium entropy
$S(z)=V\!\int\!\ln n_{\mathbf{k},\omega}(z)\,d\mathbf{k}\,d\omega$
\cite{Picozzi2007}. This entropy is a \emph{logarithmic} functional of the occupancies rather than a Gibbs--Shannon entropy; we return to this distinction in Section~3.2.

Maximizing $S$ subject to conservation of $N$ and $H_{l}$ gives the RJ spectrum
\begin{equation}
  n^{\mathrm{eq}}_{\mathbf{k},\omega}
  = \frac{T}{K(\mathbf{k},\omega)-\mu} ,
  \qquad
  K(\mathbf{k},\omega)=\alpha k^{2}-\beta\omega^{2} ,
  \label{eq:RJ}
\end{equation}
in which the chemical potential sets the associated
coherence scales. The scaling follows from the transform of \eqref{eq:RJ}: a
spectrum $\propto(\alpha k^{2}+|\mu|)^{-1}$ has the real-space correlation
$\exp(-|x|\sqrt{|\mu|/\alpha}\,)$, so that
\begin{equation}
  \lambda_{c}\simeq\sqrt{\frac{\alpha}{|\mu|}} ,
  \qquad
  \tau_{c}\simeq\sqrt{\frac{|\beta|}{|\mu|}} ,
  \label{eq:RJscales}
\end{equation}
with the square roots demanded by dimensions
\cite{Picozzi2007}.\footnote{The linear forms $\alpha/|\mu|$, $|\beta|/|\mu|$ appear in
some summaries of this result, including our own, and continue to be
proposed. They are dimensionally
inconsistent for a quadratic denominator and are corrected here. The DS scales
of Section~4.1 are unaffected, being read off the transform directly:
$\Lambda=1/\Xi=|\mu|^{-1/2}$ is already the square-root form.}
Four properties of this equilibrium recur throughout the review.

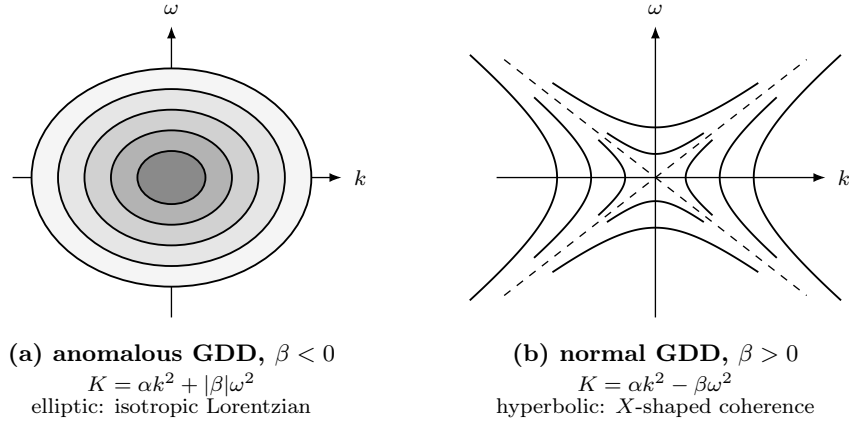
\begin{figure}[t]
\centering
\begin{tikzpicture}[x=1cm,y=1cm,
  ax/.style={draw=black,line width=0.5pt,-{Latex[length=1.6mm]}},
  iso/.style={draw=black,line width=0.7pt}]

\begin{scope}
  \draw[ax] (-2.1,0) -- (2.25,0); \node[font=\scriptsize,anchor=west] at (2.28,0) {$k$};
  \draw[ax] (0,-1.85) -- (0,2.0); \node[font=\scriptsize,anchor=south] at (0,2.03) {$\omega$};
  \foreach \s/\g in {1.85/4,1.50/10,1.15/18,0.80/30,0.45/46}
     {\draw[iso,fill=black!\g] (0,0) ellipse ({\s} and {0.78*\s});}
  \node[font=\footnotesize\bfseries,anchor=north] at (0,-2.05)
     {(a)\ anomalous GDD, $\beta<0$};
  \node[font=\scriptsize,anchor=north,align=center] at (0,-2.45)
     {$K=\alpha k^{2}+|\beta|\omega^{2}$\\[-1pt]elliptic: isotropic Lorentzian};
\end{scope}

\begin{scope}[shift={(6.4,0)}]
  \draw[ax] (-2.1,0) -- (2.25,0); \node[font=\scriptsize,anchor=west] at (2.28,0) {$k$};
  \draw[ax] (0,-1.85) -- (0,2.0); \node[font=\scriptsize,anchor=south] at (0,2.03) {$\omega$};
  \draw[dashed,line width=0.5pt] (-2.0,-1.56) -- (2.0,1.56);
  \draw[dashed,line width=0.5pt] (-2.0,1.56) -- (2.0,-1.56);
  \foreach \a in {0.40,0.85,1.30}{
    \draw[iso] plot[domain=-1.25:1.25,samples=60]
        ({\a*cosh(\x)},{0.78*\a*sinh(\x)});
    \draw[iso] plot[domain=-1.25:1.25,samples=60]
        ({-\a*cosh(\x)},{0.78*\a*sinh(\x)});}
  \foreach \b in {0.40,0.85}{
    \draw[iso] plot[domain=-1.25:1.25,samples=60]
        ({\b*sinh(\x)},{0.78*\b*cosh(\x)});
    \draw[iso] plot[domain=-1.25:1.25,samples=60]
        ({\b*sinh(\x)},{-0.78*\b*cosh(\x)});}
  \node[font=\footnotesize\bfseries,anchor=north] at (0,-2.05)
     {(b)\ normal GDD, $\beta>0$};
  \node[font=\scriptsize,anchor=north,align=center] at (0,-2.45)
     {$K=\alpha k^{2}-\beta\omega^{2}$\\[-1pt]hyperbolic: $X$-shaped coherence};
\end{scope}

\end{tikzpicture}
\caption{The sign of dispersion changes the conservative equilibrium spectrum
qualitatively. Curves are level sets of the linear dispersion
$K(\mathbf{k},\omega)$ entering the Rayleigh--Jeans distribution
\eqref{eq:RJ}; occupancy is largest where $K-\mu$ is smallest, i.e.\ on the
innermost level set. (\textbf{a})~For anomalous GDD the level sets are ellipses,
and the equilibrium spectrum is an isotropic Lorentzian, shaded here by
occupancy. (\textbf{b})~For normal GDD they are hyperbolas with the asymptotes
$\omega=\pm\sqrt{\alpha/\beta}\,k$ (dashed), and coherence is skewed along
space--time trajectories rather than being separately spatial or temporal. The
normal/anomalous asymmetry that organizes the argument is thus already present in
conservative kinetics \cite{Picozzi2007}.}
\label{fig:elliptic-hyperbolic}
\end{figure}

\begin{enumerate}[leftmargin=*,itemsep=0.25em]
  \item \textbf{The sign of dispersion changes the equilibrium spectrum
        qualitatively.} For AGD ($\beta<0$), the surfaces
        $K=\mathrm{const}$ are elliptic and $n^{\mathrm{eq}}$ is the familiar
        isotropic Lorentzian. For NGD ($\beta>0$), they are hyperbolic,
        and the equilibrium spectrum acquires an $X$-shaped spatiotemporal
        structure---coherence skewed along space--time trajectories rather than
        separately spatial or temporal \cite{Picozzi2007}. The normal/anomalous
        asymmetry that organizes much of the argument is thus already present in
        conservative kinetics (Figure~\ref{fig:elliptic-hyperbolic})
        \cite{Picozzi2007}.
  \item \textbf{The RJ distribution is only a \emph{formal} solution.} Both $N$
        and $H_{l}$ diverge when \eqref{eq:RJ} is integrated over all
        $\mathbf{k},\omega$, so an ultraviolet cutoff $k_{c},\omega_{c}$ must be
        supplied from outside: by the spatial discretization of a simulation, by
        viscosity or diffusion at the microscopic scale of a medium
        \cite{Connaughton2005}, by the finite bandwidth of a guided-wave or laser
        configuration \cite{Picozzi2007}, or---in the laboratory realization
        described at the end of this subsection---by the Debye screening length
        of a photorefractive crystal \cite{Sun2012}. Whatever its provenance,
        the thermodynamics that follows (condensate fraction, critical energy,
        critical temperature) is expressed \emph{in terms of it}.
  \item \textbf{Condensation is the limit $\mu\to0^{-}$ at finite $T$,} in which
        $\lambda_{c}$ and $\tau_{c}$ diverge. The condensate fraction then
        follows BEC-like laws, $N_{0}/N = 1-E/E_{\mathrm{tr}}$ with a critical
        energy $E_{\mathrm{tr}}/N=k_{c}^{2}/3$ per unit power, or equivalently
        $N_{0}/N=1-T/T_{\mathrm{tr}}$ with
        $T_{\mathrm{tr}}=3E_{\mathrm{tr}}/(4\pi Vk_{c}^{3})$
        \cite{Picozzi2007,Connaughton2005}. The continuum form is quantitatively
        inadequate: agreement with simulation requires the \emph{discrete} mode
        sum,
        $N_{0}/N = 1-(E/N)\sum'_{\mathbf{k}}(k_{x}^{2}+k_{y}^{2}+k_{z}^{2})^{-1}
        /\sum'_{\mathbf{k}}1$, so the microstate count enters as an actual
        enumeration of modes and not merely as a regularization
        \cite{Connaughton2005}. In two dimensions, the infrared divergence drives
        $T_{c}\to0$ in the thermodynamic limit, yet condensation is
        re-established at any finite system size, with
        $T_{c}=\rho A/\sum'_{\mathbf{k}}(\alpha k^{2})^{-1}$
        \cite{Picozzi2007,Connaughton2005,ZakharovNazarenko2005}.
  \item \textbf{Retaining the interaction energy makes the transition first
        order.} The estimates above use only the linear part of the energy.
        Adapting the Bogoliubov expansion of a weakly interacting Bose gas to
        the classical wave problem replaces the free dispersion
        $\omega(k)=k^{2}$ by $\omega_{B}(k)=\sqrt{k^{4}+2\rho_{0}k^{2}}$, with
        $\rho_{0}=n_{0}/V$ the condensed density and equilibrium occupancies
        $\varphi^{\mathrm{eq}}_{k}=T/\omega_{B}(k)$. The resulting closed
        relation between $\langle H\rangle$ and $n_{0}$ agrees with direct
        simulation without adjustable parameters and shows the condensation
        transition to be \emph{subcritical}, i.e.\ of first order
        \cite{Connaughton2005,During2009}. 
\end{enumerate}

The most instructive result of this tradition, however, is its \emph{entropic}
reading of coherent-structure formation. Since $H=H_{l}+H_{nl}$ is conserved
while relaxation increases the fluctuation content measured by $H_{l}$, the
field can reach its most disordered state only by minimizing $H_{nl}$---that is,
by generating a plane-wave condensate (defocusing) or a soliton (focusing)
immersed in a sea of small-scale fluctuations which store the information needed
for reversibility \cite{Picozzi2007,Connaughton2005,Rumpf2001,Dyachenko1992}. An
increase of ``disorder'' therefore \emph{requires} the appearance of order. The
same equilibrium distribution reproduces the fundamental relation
\begin{equation}
  T\,dS = dH_{l} - \mu\,dN ,
  \label{eq:TdS}
\end{equation}
and, once linear momentum is included, an analog of thermodynamic pressure
\cite{Picozzi2007}. It also permits a clean second-law thought experiment: if the
transverse area of a multimode waveguide is suddenly increased at some $z_{0}$,
the field relaxes to a new equilibrium of higher entropy---the optical
counterpart of removing a piston \cite{Picozzi2007}. We shall meet a version of
this in which the ``volume'' changes by itself.

The program also has a direct experimental realization. Kinetic condensation
of classical waves has been observed by propagating a random-phase beam prepared
with spatial light modulators through a self-defocusing photorefractive crystal
and imaging the output in both real and momentum space \cite{Sun2012}. The
measurements recover the apparatus quantitatively: the chemical potential
approaches $\mu_{c}=k_{0}^{2}/2k_{L}$ as the energy per particle is lowered, the
condensate fraction grows accordingly, and the uncondensed modes settle onto the
algebraic $k^{-2}$ tail of the RJ distribution---with a measured exponent
$-2.02\pm0.06$---which is the equipartition of energy among the excited modes.
Formal reversibility was demonstrated as well: phase-conjugating the recorded
output and recycling it through the crystal reverses the flow of condensation
and recovers the initial thermal cloud. Two details of that experiment matter
here. Condensation occurs in a \emph{bounded} two-dimensional system, so power
accumulates not at $k=0$ but at the lowest available wavenumber $k_{0}\neq0$
fixed by the finite beam area---the geometry, once again, supplying the mode
set. And the authors are explicit that their thermalization contrasts with the
wave dynamics of optical mode locking and of self-focusing turbulence, which
they classify as inherently dissipative and driven far from equilibrium: in the
focusing case, modulation instability dominates, the dynamics is governed by the
potential rather than the kinetic energy, the inverse cascade is unidirectional,
and ``condensation'' appears as coherent soliton structures \cite{Sun2012}. The
demarcation is thus drawn from the other side as well, and it is precisely the
territory beyond it---focusing, soliton-shaped, dissipative, driven---that this
review occupies.

\subsection{The equilibrium route: coherent structures at fixed invariants,
and multimode thermodynamics}

A complementary formulation asks not how the field relaxes but which macrostate
is statistically preferred at fixed invariants. For a nonintegrable,
collapse-free NLS equation on a bounded interval, the canonical Gibbs measure
$\propto\exp[-\beta(H-\lambda N)]$ is not normalizable in the focusing case,
because $H-\lambda N$ is unbounded below along amplitude dilations of a ground
state. Even the conditioned measures that \emph{are} normalizable place
infinite kinetic energy in a typical realization: each mode carries a finite share---the optical form of the Jeans ultraviolet catastrophe
\cite{Jordan2000,Lebowitz1988}. Jordan, Turkington and Zirbel resolved this by
constructing a mean-field maximum-entropy ensemble on an $n$-mode spectral
truncation, with the inverse temperature rescaled as
$\beta^{(n)}=n/(H_{0}-H^{*}_{n})$ so that the mean energy remains finite as
$n\to\infty$ \cite{Jordan2000}. The outcome is a sharp version of the ``coherent
structure plus radiation'' picture. The mean field is deterministic and solves
the ground-state equation, minimizing $H$ at fixed $N=N_{0}$. The residual
energy resides in independent Gaussian fluctuations with
\begin{equation}
  \mathrm{Var}(u_{k}) = \mathrm{Var}(v_{k})
  = \frac{H_{0}-H^{*}_{n}}{n\lambda_{k}} ,
  \label{eq:JTZvar}
\end{equation}
so that the power spectral density behaves as $1/\lambda_{k}\sim k^{-2}$ away
from the coherent structure while the \emph{kinetic} energy per mode is
equipartitioned. The entropy of the ensemble is, up to constants depending only
on $n$,
\begin{equation}
  S = \mathrm{const}(n)
      + n\ln\!\left[\frac{H_{0}-H_{n}(\langle\psi\rangle)}{n}\right] ,
  \label{eq:JTZentropy}
\end{equation}
the logarithm of the kinetic energy stored in infinitesimally fine-scale
fluctuations, so that entropy maximization and energy minimization of the
coherent structure are literally the same variational problem \cite{Jordan2000}.
The ensembles concentrate on the microcanonical manifold $H_{n}=H_{0}$,
$N_{n}=N_{0}$ as $n\to\infty$, which makes the mean-field approximation
asymptotically exact. One structural feature will be needed in Section~6: the
coherent structure and the fluctuation bath are cleanly separated, the mean
field carrying all the potential energy and no entropy, the bath all the entropy
and no coherence.

For a nonlinear multimode optical structure supporting a \emph{finite} number
$M$ of bound states, the same maximum-entropy logic acquires a complete
thermodynamic apparatus. A conserved optical power $\mathcal{P}$ and a conserved
(linear-dominated) internal energy $U$, together with a weakly nonintegrable
nonlinear coupling, make the system ergodic within its invariant manifolds.
Maximizing the number of accessible microstates then yields the RJ occupancies
$|c_i|^2 = -T/(\varepsilon_i+\mu)$, an extensive entropy
$S = \sum_i \ln |c_i|^2$, and an equation of state
\begin{equation}
  U - \mu\,\mathcal{P} = M T ,
  \label{eq:eos-conservative}
\end{equation}
the optical counterpart of $pV = N k_B T$ \cite{Wu2019}. The framework is
remarkably complete: it supplies an Euler relation, an ``optical pressure''
conjugate to the mode number, isentropic invariants,
negative-temperature states in which power flows towards the \emph{highest}
modes, and even Carnot-like all-optical cycles \cite{Wu2019,Muniz2023}.

Of the three routes, this is the one whose predictions have been tested in the
greatest detail, and the manner of the testing matters here as much as its
outcome. Mode-resolved measurements in graded-index multimode fibers--- off-axis
digital holography followed by a spectrally resolved modal decomposition, which
returns the whole occupancy distribution and not merely the fundamental-mode
fraction---show the field relaxing irreversibly onto the RJ law, with $T$ and
$\mu$ \emph{predicted} from the launch conditions through
\eqref{eq:eos-conservative} rather than fitted to the thermalized distribution
($T=0.40$ against $0.42\,\mathrm{mm}^{-1}$ measured), and with power
equipartitioned among the modes of each degenerate group \cite{Pourbeyram2022}.
The same work realizes two distinct ensembles: a single launch condition, for
which $\mathcal{P}$ and $H$ are separately invariant, is microcanonical, whereas
an ensemble of statistically equivalent speckle inputs, for which $H$ fluctuates
about a fixed expectation, is canonical-like \cite{Pourbeyram2022}. Independent
holographic mode-decomposition experiments confirm that $T$ and $\mu$ are fixed
by the laser-- fiber coupling condition alone---being, remarkably, unchanged as
the input pulse duration is varied from $174\,\mathrm{fs}$ to
$435\,\mathrm{ps}$---while the Hamiltonian and the mode parity stay constant as
the input power is raised \cite{Mangini2022}.

Two aspects of this experimental literature are used repeatedly below, and a
third is a caveat that has to be settled before the DS entropy can be defined at
all.

First, the \textbf{cutoff is a precondition, not a convenience}. The accessible
mode set terminates at a maximum group index fixed by the fiber cutoff, and a
finite bandwidth is exactly what prevents the entropy from diverging at low
temperature---the guided-wave form of the ultraviolet catastrophe of
Section~3.1 \cite{Mangini2022}. Its regularizing role is the same as that of
$k_{c}$ in the kinetic route and of $n$ in the mean-field ensemble. What
distinguishes the dissipative problem, as Section~4 argues, is only where the
cutoff comes from.

Second, the entropy has been measured by a decomposition that transposes
directly to a single pulse. Writing $S = M\ln\mathcal{P} + \sum_{i}\ln|f_{i}|^{2}$
with $|f_{i}|^{2}$ the \emph{normalized} occupancies, the second term---the
\emph{configuration entropy} $\tilde{S}$---responds only to the reshaping of the
distribution, not to trivial extensive growth. Fiber-cutback experiments
follow $\tilde{S}$ along the propagation coordinate and find it rising and then
leveling off as equilibrium is approached, which is what the theory requires:
at fixed energy per unit power the equilibrium configuration entropy is
independent of $\mathcal{P}$, while the full entropy continues to grow through
the $M\ln\mathcal{P}$ term \cite{Mangini2024}. Section~6 takes this over as a
measurement protocol for the DS.

Third, and less comfortably, \textbf{the entropy functional is not uniform across this literature}. Multimode thermodynamics uses the logarithmic Boltzmann form \( S = \sum_{i} \ln |c_{i}|^{2} \)—the discrete counterpart of the kinetic entropy \( \int \ln n \) discussed in Section 3.1, and the functional that RJ actually maximizes \cite{Wu2019,Mangini2024}. In contrast, analyses that need to discuss equipartition among non-degenerate groups revert to the Gibbs–Shannon form \( -\sum_{i} p_{i} \ln p_{i} \) \cite{Zitelli2024,Jaynes1965}. These two forms are not equivalent, and the choice between them is significant, not merely academic. The Boltzmann functional diverges whenever occupancy vanishes, which in the experiments necessitated discarding all measured mode fractions below a certain threshold \cite{Mangini2024}. For a \emph{truncated} Lorentzian—where occupancy vanishes at the spectral edge by design—the functional would diverge precisely at the boundary that encapsulates the physics discussed in Section~4. Therefore, we utilize the Gibbs–Shannon functional throughout this work. We state the cost of this choice clearly: we give up the property that the RJ form maximizes the functional we adopt. This trade-off is acceptable here only because the DS spectrum is not derived by maximizing anything; rather, it is provided by the deterministic adiabatic solution presented in Section 2. Thus, the functional measures disorder for a given distribution rather than serving as a variational principle. The implications of this choice are discussed in Section~6.

Within the same conservative setting, finally, the RJ law is not the last word.
Mode-resolved data spanning the linear, quasi-soliton and soliton regimes are
better fitted by a weighted Bose--Einstein distribution, of which RJ is the
classical limit $|\mu'+\epsilon_{i}|\ll|Tn_{0}|$ \cite{Zitelli2024}. More
pertinently here, strong random mode coupling suppresses \emph{global}
condensation into the fundamental mode while leaving intact steady states in
which power condenses \emph{locally} into a higher-order group---``glassy''
states, intermediate between the disordered low-energy state and the fully
condensed one \cite{Zitelli2024}. Here is a conservative system in which the
global equilibrium attractor fails to select the observed configuration, and in
which the configuration selected is a set of sub-condensates rather than one.
Both features recur, for quite different reasons, in the multipulse physics of
Section~6.

\subsection{The noise-driven route: mode locking as a first-order phase
transition}

The third tradition starts from the laser rather than from a conservative wave
equation, and is on that ground the closest existing relative of the present
work. In the master-equation description of a passively mode-locked laser
\cite{Haus2000}, the envelope evolves under saturable gain, fast saturable
absorption and spectral filtering, driven by the unavoidable spontaneous-emission
noise. When the deterministic part of the evolution is a gradient flow,
$G=-\delta\mathsf{H}/\delta\psi^{*}$, additive white Gaussian noise makes the
stationary distribution an \emph{exact} Gibbs measure
$\rho\propto\exp(-\mathsf{H}[\psi]/LT)$, in which the intracavity noise power
$T$ plays the role of temperature and
\begin{equation}
  \mathsf{H}[\psi]
  = \int_{0}^{L}\!\Big(-\tfrac{\gamma_{s}}{2}|\psi|^{4}
      + \gamma_{g}|\psi'|^{2}\Big)dx
    + L\,U(\mathcal{P})
  \label{eq:GFhamiltonian}
\end{equation}
combines the destabilizing quartic self-interaction supplied by the absorber,
the stiffness supplied by the filter, and a gain-saturation term $U$ that acts
as a chemical potential and tames the otherwise unbounded quartic term
\cite{GordonFischer2002,Gat2004}. Gordon and Fischer showed that pulse formation
is then a \emph{first-order phase transition}---a spontaneous ordering of mode
phases, the disordered phase being multimode continuous-wave operation---and
Gat, Gordon and Fischer solved the corresponding coarse-grained model exactly
\cite{GordonFischer2002,Gat2004}. Four of their results bear on what follows. In
quoting them, we write $\mathcal{N}$ for the number of degrees of freedom and $m$
for the order parameter, denoted $N$ and $M$ in the original, to avoid collision
with the power of Section~3.1 and the mode number of Section~3.2.

\begin{itemize}[leftmargin=*,itemsep=0.25em]
  \item \textbf{The coarse-graining scale is set by the filter.} Spectral
        filtering introduces a length over which the envelope is smooth, so the
        field is represented by $\mathcal{N}$ complex degrees of freedom,
        $\mathcal{N}$ being the ratio of the cavity length to the filter-limited
        pulse width and ranging from $\sim10^{2}$ to $\sim10^{9}$ in practice
        \cite{Gat2004}. This is the closest existing analog of the microstate
        count constructed in Section~4---but it is fixed by the cavity, not by
        the pulse.
  \item \textbf{The thermodynamics collapses onto a single dimensionless group,}
        $\gamma=\gamma_{s}\mathcal{P}^{2}/T$. The free energy per degree of
        freedom is $f(\gamma,y)=-\big[\tfrac{\gamma}{2}y^{2}+\ln(1-y)\big]$ with
        $y=m^{2}$; a second minimum appears at $\gamma=4$, and the two minima
        exchange stability at $\gamma^{*}\simeq4.91$, giving a genuine
        first-order transition with coexistence, metastability, hysteresis,
        superheating and supercooling \cite{GordonFischer2002,Gat2004}. Since
        $m^{2}$ is the power residing in a single degree of freedom, ordering
        means locking into \emph{one} pulse.
  \item \textbf{Ensembles are equivalent.} The canonical (fixed-power)
        and grand-canonical (variable-power) descriptions give the same
        thermodynamics, so the transition occurs at the same
        $\gamma^{*}$ whatever the form of the gain-saturation function.
        The mean power then follows from
        $1+\mathcal{P}u'(\mathcal{P})-\gamma\bar{y}(\gamma)^{2}=0$,
        and the susceptibility $\chi=\mathcal{P}'(\gamma)$ is strictly
        positive in the mode-locked state \cite{Gat2004}.
  \item \textbf{Finite $\mathcal{N}$ softens the transition.} A uniform
        asymptotic expansion in $1/\mathcal{N}$ shows that the sharp
        thermodynamic-limit discontinuity becomes a measurable crossover of width
        $|\gamma-\gamma^{*}|\sim1/\mathcal{N}$, within which the metastable
        branch contributes appreciably \cite{Gat2004}.
\end{itemize}

The NGD results of Ref.~\cite{Kalashnikov2025} provide a direct physical bridge to this statistical mode-locking picture. Their pulse-number statistics, built from 150 stochastic realizations, show that increasing the control energy moves probability weight from the single-pulse state toward multipulse states. Even more importantly, the self-start calculation shows that the mean formation time of the single DS is nonmonotonic, while at sufficiently high energy the two-pulse state can form faster than the single-pulse state. Thus, the relevant change is not simply the disappearance of a stationary solution. It is a change in which attractor is most readily reached from noise.

This is precisely the conceptual level on which thermodynamic language is useful here. The NGD calculation does not prove that the laser undergoes an equilibrium Gibbs transition, nor does it prove that entropy causes the breakup. It does show that energy scaling is accompanied by a redistribution of stochastic accessibility among competing macrostates. Any proposed entropy or temperature-like diagnostic should ultimately be judged by whether it organizes that redistribution.

Two of these have direct counterparts in the adiabatic theory of Section~2. The
collapse of the whole thermodynamics onto the single group $\gamma$ is mirrored
by the reduction of the DS parametric space to $(C,\Sigma)$. The difference is
that those groups are built from deterministic cavity parameters, and the
analogue of $\gamma$ is a \emph{pair} of coordinates rather than one. And the
isogain foliation of the master diagram plays the part corresponds the
gain-saturation function $U(\mathcal{P})$: it selects the operating point
without altering the underlying thermodynamics, in the same sense in which the
fixed-power and variable-power ensembles are equivalent \cite{Gat2004}.

\subsection{What the three traditions share}

\begin{table}[t]
\caption{The three traditions of Section~3, and the DS
treated here. The fourth column is the one on which the argument
turns: in every established framework the microstate count is a property of the
apparatus, whereas for a chirped DS it is a function of state.}
\label{tab:traditions}
\centering
\footnotesize
\begin{tabularx}{\textwidth}{@{}p{2.35cm} p{2.9cm} p{2.75cm} X p{2.6cm}@{}}
\toprule
\textbf{Framework} & \textbf{Dynamics; invariants} &
\textbf{Stationary law} & \textbf{Microstate count and its origin} &
\textbf{What selects the state} \\
\midrule
Wave-turbulence kinetics
  \cite{Picozzi2007,Picozzi2014}
& conservative NLS; $N$, $H_{l}$
& RJ, $n^{\mathrm{eq}}=T/(K-\mu)$
& $k_{c},\omega_{c}$--- discretization, microscopic damping, or waveguide
  bandwidth; \emph{fixed}
& $H$-theorem; $S=\int\!\ln n$ maximal \\[0.35em]
Wave condensation in a bounded system
  \cite{Connaughton2005,During2009,Sun2012}
& conservative; $N$, $H_{l}$; observed in a photorefractive crystal
& RJ plus condensate at $k_{0}$
& $k_{c}$ from the Debye length, $k_{0}$ from the finite beam area;
  \emph{fixed by geometry}
& as above; first order once $H_{nl}$ is restored \\[0.35em]
Mean-field NLS ensemble
  \cite{Jordan2000,Lebowitz1988}
& conservative, nonintegrable; $N_{0}$, $H_{0}$
& coherent structure $+$ Gaussian bath, $\mathrm{Var}\propto1/\lambda_{k}$
& $n$--- spectral truncation, with $\beta^{(n)}$ rescaled as $n\to\infty$;
  \emph{fixed}
& maximum entropy $\equiv$ minimum $H$ at fixed $N$ \\[0.35em]
Multimode optical thermodynamics
  \cite{Wu2019,Pourbeyram2022,Mangini2022,Mangini2024}
& conservative, weakly nonintegrable; $\mathcal{P}$, $U$
& RJ, $|c_i|^{2}=-T/(\varepsilon_i+\mu)$; $U-\mu\mathcal{P}=MT$
& $M$--- guided modes below cutoff, set by core diameter, NA and wavelength;
  \emph{fixed}
& maximum number of microstates; $S=\sum_i\ln|c_i|^{2}$ \\[0.35em]
Mode locking as a phase transition
  \cite{GordonFischer2002,Gat2004}
& dissipative but gradient flow; $T$ imported as noise power
& exact Gibbs measure $\rho\propto e^{-\mathsf{H}/LT}$
& $\mathcal{N}$--- cavity length over filter-limited pulse width;
  \emph{fixed by the cavity}
& exchange of free-energy minima at $\gamma^{*}$ \\[0.35em]
\textbf{Chirped DS} (this work)
& genuinely dissipative, no potential; nothing conserved, $E$ free
& NGD: truncated Lorentzian with
  $-\mu_{\mathrm{shape}}=\Xi^{2}$; AGD: two-horn core with algebraic wings
& NGD: $\Nsep\sim\Delta/\Xi$ generated by the pulse; AGD: an operational
  scale ratio requires an explicit dissipation/capture window
& no variational criterion; entropy- and temperature-like crossings are
  candidate diagnostics whose branch-selection role remains to be tested dynamically \\
\bottomrule
\end{tabularx}
\end{table}

Three structural assumptions run through all of them. Table~\ref{tab:traditions}
sets the three routes side by side, together with the DS for
comparison.

\textbf{(i) Conservative dynamics, or dissipative dynamics admitting a potential representation}. The routes of Sections~3.1 and 3.2 assume that gain and loss are
absent or perturbative and that the Hamiltonian is dominated by its linear part.
The route of Section~3.3 \emph{is} genuinely dissipative, but it purchases its
Gibbs measure at a price: the evolution must derive from a potential
$\mathsf{H}$,
and a Gibbs construction is available only under additional potential and noise
conditions. Generic dispersive--Kerr mode locking need not satisfy those
conditions and can instead possess a non-Gibbsian stationary state
\cite{Gat2004,Katz2006}. For the cubic--quintic equation used here, the
deterministic coefficient-collinearity conditions are derived explicitly in
Section~4 and require \emph{both} $C=-1$ and $\tilde\chi=1$. In addition,
the temperature is supplied from outside, as the spontaneous-emission noise
power, rather than emerging from the deterministic dynamics \cite{Gat2004}.

\textbf{(ii) Degrees of freedom given in advance.} Every construction above
needs a microstate count, and in every case that count is imposed: the
ultraviolet cutoff $k_{c}$ regularizing the RJ integrals \cite{Picozzi2007}, the
lowest attainable wavenumber $k_{0}$ set by a finite beam area \cite{Sun2012},
the spectral truncation $n$ of the mean-field ensemble \cite{Jordan2000}, the
number $M$ of guided modes---fixed by core diameter, numerical aperture and
wavelength, and required to be finite if the entropy is to exist at all
\cite{Wu2019,Mangini2022}---and the ratio $\mathcal{N}$ of cavity length to
filter-limited pulse width \cite{Gat2004}. This number is the ``volume''
variable of the theory. It is a property of the apparatus, and it does not
respond to the energy stored in the field.

\textbf{(iii) An extrema principle selects the stationary state.} Detailed
balance holds, or an effective potential exists, so that the stationary state is
a genuine equilibrium: the RJ spectrum annihilates the collision integral, the
coherent structure minimizes $H$ at fixed $N$, and mode locking is selected by
the crossing of two minima of a free energy. Stability is thereby decided
without solving the dynamics.

\begin{figure}[t]
\centering
\begin{tikzpicture}[x=1cm,y=1cm,
  bar/.style={draw=black,fill=black!22,line width=0.35pt},
  box/.style={draw=black,line width=0.35pt,fill=black!8},
  ax/.style={draw=black,line width=0.5pt,-{Latex[length=1.6mm]}}]

\node[anchor=west,font=\footnotesize\bfseries] at (-0.4,4.55)
   {(a)\ \ Imposed mode set: $M$ fixed by the waveguide};

\begin{scope}[shift={(0,2.4)}]
  \draw[ax] (0,0) -- (4.1,0);
  \draw[ax] (0,0) -- (0,1.55);
  \node[font=\scriptsize,anchor=north] at (2.0,-0.12) {mode index $i$};
  \node[font=\scriptsize,rotate=90,anchor=south] at (-0.35,0.75) {$|c_i|^2$};
  \foreach \i/\h in {1/0.22,2/0.50,3/0.78,4/0.70,5/0.50,6/0.34,7/0.20,8/0.13}
     {\draw[bar] ({0.30+0.42*(\i-1)},0) rectangle ({0.30+0.42*(\i-1)+0.28},\h);}
  \draw[dashed,line width=0.6pt] (3.72,-0.12) -- (3.72,1.45);
  \node[font=\scriptsize,anchor=south east,align=right] at (3.68,1.05)
     {cutoff\\[-1pt]$M$};
  \node[font=\scriptsize,anchor=north] at (1.75,1.75) {low power};
\end{scope}

\begin{scope}[shift={(5.6,2.4)}]
  \draw[ax] (0,0) -- (4.1,0);
  \draw[ax] (0,0) -- (0,1.55);
  \node[font=\scriptsize,anchor=north] at (2.0,-0.12) {mode index $i$};
  \foreach \i/\h in {1/0.11,2/0.13,3/0.16,4/0.19,5/0.25,6/0.34,7/0.56,8/1.30}
     {\draw[bar] ({0.30+0.42*(\i-1)},0) rectangle ({0.30+0.42*(\i-1)+0.28},\h);}
  \draw[dashed,line width=0.6pt] (3.72,-0.12) -- (3.72,1.45);
  \node[font=\scriptsize,anchor=south east,align=right] at (4.2,1.05)
     {cutoff\\[-1pt]$M$};
  \node[font=\scriptsize,anchor=north] at (2.0,1.75) {high power (RJ)};
\end{scope}

\node[font=\scriptsize,align=center] at (4.72,3) {$=$\\[-1pt]\emph{same} count};

\node[anchor=west,font=\footnotesize\bfseries] at (-0.4,0.55)
   {(b)\ \ Self-generated scale separation: $\Nsep\sim\Delta/\Xi$ grows with energy};

\begin{scope}[shift={(0,-2.0)}]
  \draw[ax] (0.05,0) -- (4.1,0);
  \draw[ax] (2.05,0) -- (2.05,1.55);
  \node[font=\scriptsize,anchor=north] at (2.05,-0.12) {detuning $\omega$};
  \node[font=\scriptsize,rotate=90,anchor=south] at (-0.30,0.75) {$n(\omega)$};
  \draw[line width=0.9pt,fill=black!12]
     plot[domain=-1.55:1.55,samples=90] ({2.05+\x},{1.30*0.3844/(\x*\x+0.3844)})
     -- (3.60,0) -- (0.50,0) -- cycle;
  \draw[dashed,line width=0.6pt] (3.60,0) -- (3.60,1.45);
  \draw[dashed,line width=0.6pt] (0.50,0) -- (0.50,1.45);
  \node[font=\scriptsize,anchor=south] at (3.60,1.42) {$\Delta$};
  \node[font=\scriptsize,anchor=south] at (0.50,1.42) {$-\Delta$};
  \draw[{Latex[length=1.2mm]}-{Latex[length=1.2mm]},line width=0.5pt]
     (2.05-0.62,0.65) -- (2.05+0.62,0.65);
  \node[font=\scriptsize,anchor=south] at (2.05,0.66) {$\Xi$};
  \node[font=\scriptsize,anchor=north] at (2.05,2) {low energy};
\end{scope}

\begin{scope}[shift={(5.6,-2.0)}]
  \draw[ax] (0.05,0) -- (4.1,0);
  \draw[ax] (2.05,0) -- (2.05,1.55);
  \node[font=\scriptsize,anchor=north] at (2.05,-0.12) {detuning $\omega$};
  \draw[line width=0.9pt,fill=black!12]
     plot[domain=-1.72:1.72,samples=200] ({2.05+\x},{1.35*0.0240/(\x*\x+0.0240)})
     -- (3.77,0) -- (0.33,0) -- cycle;
  \draw[dashed,line width=0.6pt] (3.77,0) -- (3.77,1.45);
  \draw[dashed,line width=0.6pt] (0.33,0) -- (0.33,1.45);
  \node[font=\scriptsize,anchor=south] at (3.77,1.42) {$\Delta$};
  \node[font=\scriptsize,anchor=south] at (0.33,1.42) {$-\Delta$};
  \draw[{Latex[length=1.2mm]}-{Latex[length=1.2mm]},line width=0.5pt]
     (2.05-0.155,0.95) -- (2.05+0.155,0.95);
  \node[font=\scriptsize,anchor=west] at (2.28,0.98) {$\Xi\!\to\!0$};
  \node[font=\scriptsize,anchor=north] at (2.05,2) {high energy (DSR)};
\end{scope}

\node[font=\scriptsize,align=center] at (4.72,-1.25) {$<$\\[-1pt]separation \emph{grows}};

\end{tikzpicture}
\caption{Fixed external mode number versus self-generated spectral scale separation.
(\textbf{a}) In a multimode waveguide, the eigenvalue ladder is a property of the
geometry: raising the power reshuffles the occupancies $|c_i|^2$ towards the
Rayleigh--Jeans law, but the number $M$ of available states---and hence the
``volume'' variable of the equation of state \eqref{eq:eos-conservative}---is the
same before and after.
(\textbf{b}) For a strongly chirped DS the spectrum supplies its
own support. The spectral edge $\Delta$ saturates while the Lorentzian width
$\Xi$ collapses, so the two correlation scales decouple and the scale-separation index
$\Nsep\sim\Delta/\Xi$ of Eq.~\eqref{eq:Neff} increases with the pulse
energy. In the deterministic solution this is a scale-separation index and
not a literal count of independent microstates; a counting interpretation would
require the ensemble-coherence calibration of Section~4.3. Bars and curves are schematic.}
\label{fig:microstates}
\end{figure}
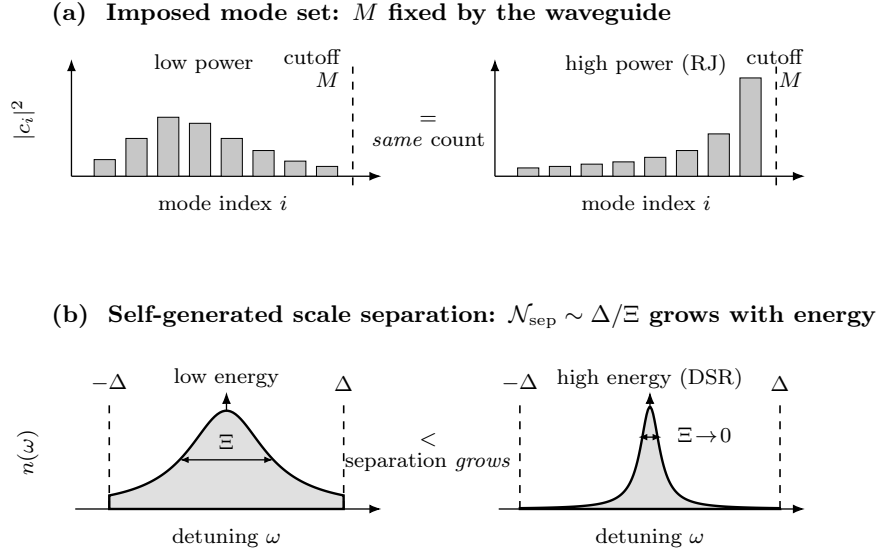

None of these three assumptions holds for the objects considered here.
It is the second that is decisive, and Figure~\ref{fig:microstates} states the
contrast in the form in which the rest of the introduction develops it.

\section{Dissipative Solitons as Thermodynamic Objects}

The solution of Section~2 and the vocabulary of Section~3 can now be brought
together. One negative point should be recorded before they are: the CQGLE with
both dispersion and nonlinearity retained is \emph{not} a gradient
flow at a generic operating point, so the exact Gibbs measure of
Section~3.3 is unavailable, and the statistical description has to be built on
the deterministic solution itself rather than on an invariant measure
\cite{Gat2004}. That is why the adiabatic theory had to come first.
The qualification is worth making precise, because the exception is smaller than
it first appears, and locating it exactly turns a worry into a result.
Writing \eqref{eq:cqgle} as
$\partial_{z}a=-\sigma a+A\,\partial_{t}^{2}a+B|a|^{2}a+Q|a|^{4}a$ with
$A=\alpha+i\beta$, $B=\kappa-i\gamma$ and $Q=-\kappa\zeta+i\chi$, a common
real potential requires all three coefficients to be collinear in the complex
plane, $A=e^{i\theta}g_{1}$, $B=-e^{i\theta}g_{2}$, $Q=-e^{i\theta}g_{3}$ with
$g_{i}$ real. Two independent conditions follow. The cubic one,
\begin{equation}
  \operatorname{Im}\frac{B}{A}=0
  \;\Longleftrightarrow\;
  \alpha\gamma+\beta\kappa=0
  \;\Longleftrightarrow\;
  C=-1 ,
  \label{eq:gibbscubic}
\end{equation}
is the one usually quoted, and by itself it does define a line. But the quintic
coefficient must be collinear too,
\begin{equation}
  \operatorname{Im}\frac{Q}{B}=0
  \;\Longleftrightarrow\;
  \kappa(\chi-\gamma\zeta)=0
  \;\Longleftrightarrow\;
  \tilde\chi=1 ,
  \label{eq:gibbsquintic}
\end{equation}
so, for the cubic--quintic equation, the exact-potential locus is not a line but
the intersection $C=-1$, $\tilde\chi=1$, of codimension two.

\paragraph{Necessary versus sufficient conditions.}
The coefficient-collinearity conditions
\eqref{eq:gibbscubic}--\eqref{eq:gibbsquintic} identify the locus on which the
deterministic derivative and nonlinear terms can share a common phase, and hence
admit a common real-potential representation. The accurate name for what
they define is therefore a \emph{coefficient-collinearity locus}, on which the
dispersive/filtering and the conservative/dissipative nonlinear sectors become
gradient compatible; ``exact-potential point'' is retained below only as the
shorthand for it. Two further requirements separate that locus from a genuine
potential structure, and they should be stated in order rather than conflated.
Establishing a globally defined potential for the \emph{complete deterministic
flow} additionally requires the remaining linear terms---the saturated net gain
and the rotating-frame wavenumber---to be incorporated consistently, since
collinearity of the derivative and nonlinear coefficients alone leaves them
outside the gradient structure. They should not, in turn, be read
as a sufficient condition for the \emph{full stochastic laser model} to possess
a Gibbs invariant measure. Such a measure additionally requires the noise
statistics, the gain constraint and the remaining linear terms to satisfy the
assumptions of the corresponding statistical mode-locking construction. Away
from those assumptions, dispersive Kerr mode locking is generically
non-Gibbsian \cite{Katz2006}. What is established below is therefore a statement
about a gradient structure of the deterministic flow, not about an invariant
measure of the noisy dynamics; ``exact-potential point'' is used throughout in
that narrower sense.

Where does that point sit? Exactly in the corner. The anomalous resonance locus
\eqref{eq:AGDlocus} is $C=-1/\tilde\chi$, which passes through $C=-1$ precisely
at $\tilde\chi=1$; and the existence threshold evaluated there is
$\Sigma_{\mathrm{th}}(-1,1)=0$, so the admissible interval
$0<\Sigma<\Sigma_{\mathrm{th}}$ is empty. The exact-potential point is therefore
the single point at which the resonance locus, the quintic threshold
\eqref{eq:AGDcriterion} and the vacuum-stability limit $\Sigma\to0^{+}$ all
meet---the corner of the anomalous diagram at which the energy diverges and the
strongly chirped solution simultaneously ceases to exist. It is approached, never
occupied. Two conclusions follow, and they point in opposite directions. The
reassuring one is that no operating point used anywhere in this paper is an
equilibrium point in disguise: at any finite loss $\tilde\chi\ge\tilde\chi_{\min}(\Sigma)>1$,
so \eqref{eq:gibbsquintic} is violated throughout the anomalous window, and the
statement that a variational criterion is unavailable stands without exception.
The suggestive one is that the equilibrium point is not somewhere irrelevant: it
is the accumulation point of exactly the limit the paper cares about.
This geometrical coincidence should not, however, be interpreted as an
approach to thermodynamic equilibrium. An equilibrium interpretation would
additionally require the stochastic forcing, the gain constraint and the
dissipative terms to approach a detailed-balance or fluctuation--dissipation
structure in the same limit, and no such joint limiting procedure is
established here. Accordingly, whether any of the spectral indicators of
Section~6 acquires an equilibrium meaning near this corner remains a separate
question rather than a consequence of coefficient collinearity.

What that solution supplies is an \emph{internal deterministic structure}. The strong chirp makes the DS phase-inhomogeneous: energy is continuously redistributed inside the pulse---injected near the spectral center, where the gain exceeds the spectral loss, and drained at the spectral wings, where the chirp has carried the instantaneous frequency far from the filter maximum. The pulse therefore acquires a resolved frequency--time structure with two characteristic scales. It does not thereby become a statistical mixture, and the distinction is the pivot of this section. The three subsections below trace the consequence: in NGD, the
spectral cutoff and spectral core generate two autocorrelation scales, while, in
AGD, an analogous separation appears only after the physically explicit
dissipation/capture scale is supplied (Section~4.1). A Schr\"odinger soliton generates only one (Section~4.2). The ratio of the correlation scales for the chirped DS is proposed as a deterministic analog of the scale-counting parameter that may correlate with an ensemble quasiparticle or participation count of
incoherent-soliton kinetic theory in a deterministic setting (Section~4.3).

\subsection{Two correlation scales, and where the cutoff comes from}

Read the normalized DS spectrum as a distribution over frequency and ask what it
implies for the field autocorrelation
$R(\tau)=\int a(t)a^{*}(t+\tau)\,\mathrm{d}t
=\pi^{-1}\!\int_{0}^{\infty}\!S(\omega)\cos(\omega\tau)\,\mathrm{d}\omega$ and
its normalized form $\hat g(\tau)=R(\tau)/R(0)$. Two cautions about names are
needed before the calculation, because both are load-bearing later. First, for a
single deterministic pulse $R$ is a \emph{time-integrated autocorrelation}, not
a first-order coherence function: the full mutual-coherence kernel of a
nonstationary field is the two-time object $J(t_{1},t_{2})$ of Section~4.3,
which for one realization equals $a(t_{1})a^{*}(t_{2})$ and depends on both
times separately, not on their difference. Second, the decay of $R$ over a
finite delay reflects finite bandwidth and nothing more. A chirped deterministic
field has $|\hat g|$ of exactly the same form as a partially coherent field with
the same spectrum, and the two are distinguished only by an ensemble
\cite{Glauber1963a,Glauber1963b}. 

One corollary is that, for a stationary field with Gaussian statistics, the intensity autocorrelation follows from the field correlation by the Siegert relation $g^{(2)}(\tau)=1+|g^{(1)}(\tau)|^{2}$, which is a standard Gaussian-moment identity and holds only under those two hypotheses. Neither hypothesis is satisfied by a single chirped DS, which is neither stationary in $t$ nor an ensemble at all. An intensity autocorrelation measured on it therefore does not return $|\hat g|^{2}$, and the familiar route from intensity data back to coherence is unavailable. This is the same obstruction that reappears in Section~6.5 as the insufficiency of dispersive Fourier transformation for the two-time kernel \eqref{eq:kernel}, arriving there from the spectral side and here from the temporal one.

What follows therefore establishes two
\emph{scales}, which is a statement about the spectrum, and defers to
Section~4.3 the separate question of what, if anything, they count. The answer
differs in form between the two dispersion regimes and agrees in substance, and
the difference is instructive.

\emph{Normal GDD.} The spectrum \eqref{eq:DSspectrum} is a Lorentzian of
half-width $\Xi$ confined within $|\omega|\le\Delta$. Its cosine
transform is accordingly the transform of the Lorentzian \emph{convolved} with
the transform of the truncation window,
\begin{equation}
  R(\tau) \;\propto\;
  \int_{-\infty}^{\infty} e^{-\Xi|t|}\,
  \operatorname{sinc}\!\big(\Delta(\tau-t)\big)\,dt ,
  \label{eq:autocorr2}
\end{equation}
which is not an oscillator with a damping rate but a \emph{convolution}: the Lorentzian core transforms to an exponential of width $\Lambda=1/\Xi$, the hard window transforms to a sinc kernel of width $l=\pi/\Delta$, and the product of the two spectral factors becomes the convolution of the two temporal ones
\cite{Kalashnikov2025}. The two scales are not two features of one curve but the
two factors of a convolution---an envelope and a resolution.

The decomposition is not an artifact of the closed form. Independent simulations
of flat-topped DSR pulses---in a model with saturable gain and a
linear--quadratic intensity dependence of the nonlinear loss, and therefore with
none of the adiabatic apparatus used here---resolve the same two components in
the spectrum and assign them to different parts of the pulse. The steep leading
and trailing edges, across which the intensity gradient and hence the
instantaneous-frequency excursion are large, generate a broad rectangular
pedestal. The flat central region, across which the frequency barely varies,
generates a much narrower bell-shaped feature at zero detuning $\omega=0$
\cite{Komarov2013}. Read through the slaving relation of Section~2.2, this is
precisely the statement that the edges supply the cutoff $\Delta$ and the core
supplies $\Xi$. The two scales are therefore separately visible in the spectrum
itself, before any autocorrelation is taken.

\emph{Anomalous GDD.} Here, the mechanism that bounds the spectrum in NGD
is absent. With $\beta<0$ the cutoff parameter changes sign, the non-negativity
of the power no longer truncates anything, and the stationary-phase envelope
extends to arbitrarily large detuning, decaying algebraically as
$S_{\mathrm{AGD}}\sim|\omega|^{-3}$ with a two-horn core and a central
depression \cite{Kalashnikov2026}. The energy is finite regardless, so nothing
forces a cutoff in the solution. An effective one is nevertheless present in any
physical system, and it is dissipative rather than kinematic---but it must be
defined operationally, because the spectral filter that produces it has no sharp edge.
Over an evolution interval $L$ the filter of \eqref{eq:cqgle} attenuates as
$\exp(-\alpha L\omega^{2})$, so the bandwidth at any fixed attenuation threshold
scales as
\begin{equation}
  \omega_{\mathrm{f}} \sim (\alpha L)^{-1/2} ,
  \label{eq:filterBW}
\end{equation}
with the prefactor set by the threshold and not by $\alpha$ alone; $\alpha^{-1/2}$
is the special case $L=1$ in the normalization of Table~\ref{tab:units}. We
therefore define the \emph{occupied window} $\omega_{\mathrm{cap}}$ as a
centered cumulative-energy radius of the spectrum---the smallest half-width $\Omega\ge0$ such that $[-\Omega,\Omega]$ contains a fixed fraction
$1-\eta$ of the spectral energy, with $\eta$ stated wherever a number is
quoted, and \emph{not} an ordinary one-sided quantile, which for an even
spectrum would vanish at $f=1/2$ (Eq.~\eqref{eq:centeredquantile})---rather than by a formula in $\alpha$, subject always to
\begin{equation}
  \omega_{\mathrm{cap}}(E) \;\le\; \min\{\omega_{\mathrm{f}},
  \omega_{\mathrm{det}}\} ,
  \label{eq:capbound}
\end{equation}
since only $\omega_{\mathrm{cap}}$ may respond to the state. With that convention, we take
$S^{(\mathrm{cap})}=S_{|\omega|\le\omega_{\mathrm{cap}}}$. 

With that
convention, and by the convolution theorem, the windowed autocorrelation is
again a convolution with a sinc kernel, of width
\begin{equation}
  \ell \sim \frac{\pi}{\omega_{\mathrm{cap}}} ,
  \qquad\text{against}\qquad
  \varrho \sim \frac{1}{\omega_{\mathrm{core}}} .
  \label{eq:scalesAGD}
\end{equation}
The long scale is set by the width $\omega_{\mathrm{core}}$ of the central
core, defined operationally by a cumulative-energy fraction in the same way
\cite{Kalashnikov2026}. The short scale $\ell$ is a first-zero width and the
long scale $\varrho$ is a decay length; the factor $\pi$ between the two
conventions is carried explicitly here and is the reason $\Nsep$ and $r$ differ
by $\pi$ in \eqref{eq:rdef}. The hierarchy $\ell\ll\varrho$ survives the loss of the
truncation.

The wings are not passive in this construction, and they provide a second, sharper statement about the short scale. Split the windowed transform at a matching frequency $\omega_{0}$ beyond which the algebraic form
\eqref{eq:AGDtail} is accurate. The tail then contributes \begin{gather}
R_{\mathrm{wing}}(\tau) \simeq
\frac{K}{\pi}\int_{\omega_{0}}^{\omega_{\mathrm{cap}}}
\frac{\cos\omega\tau}{\omega^{3}}\,\mathrm{d}\omega
= R_{\mathrm{wing}}(0)
- \frac{K}{2\pi}\,\tau^{2}\,
\ln\frac{\omega_{\mathrm{cap}}}{\omega_{0}}
+ O\!\left[K\tau^{4}\big(\omega_{\mathrm{cap}}^{2}-\omega_{0}^{2}\big)\right] ,\\ \nonumber
\qquad
K \propto \frac{1}{\big(1+C\tilde\chi\big)^{2}\sqrt{\tilde\chi}} ,
\label{eq:AGDwing}
\end{gather}
for $|\tau|\,\omega_{\mathrm{cap}}\ll1$, as follows by expanding $\cos(\omega\tau)$ inside the $|\omega|^{-3}$ tail integrand. Two things follow, and the third is a warning. First, the curvature of the correlation peak--- an alternative and, for a noisy measurement, a more accessible measure of the graining scale than the first zero $\ell$--- depends on the occupied window only \emph{logarithmically}, so the two measures of the same scale respond to $\omega_{\mathrm{cap}}$ at quite different rates and should not be quoted interchangeably. Second, the prefactor in \eqref{eq:AGDwing} is the same
$(1+C\tilde\chi)^{-2}$ that produces the logarithmic energy divergence
\eqref{eq:AGDlogdiv}: the widening of the two-scale separation toward the anomalous resonance and the divergence itself are carried by the same term of the tail. This point matters because the two observations cannot be used to corroborate each other---they are two manifestations of the same asymptotic fact, not independent evidence.

\begin{figure}[t]
\centering
\begin{tikzpicture}[x=1cm,y=1cm,
  ax/.style={draw=black,line width=0.5pt,-{Latex[length=1.5mm]}},
  cv/.style={draw=black,line width=0.9pt}]

\begin{scope}
  \draw[ax] (-0.15,0) -- (4.35,0);
  \draw[ax] (2.0,0) -- (2.0,2.35);
  \fill[black!12] (0.000,0.092) (0.040,0.096) (0.080,0.100) (0.120,0.104) (0.160,0.108) (0.201,0.113) (0.241,0.118) (0.281,0.123) (0.321,0.128) (0.361,0.134) (0.401,0.141) (0.441,0.148) (0.481,0.155) (0.521,0.163) (0.561,0.171) (0.602,0.180) (0.642,0.190) (0.682,0.200) (0.722,0.212) (0.762,0.224) (0.802,0.238) (0.842,0.252) (0.882,0.268) (0.922,0.286) (0.962,0.305) (1.003,0.326) (1.043,0.349) (1.083,0.374) (1.123,0.402) (1.163,0.433) (1.203,0.467) (1.243,0.505) (1.283,0.547) (1.323,0.594) (1.363,0.647) (1.404,0.705) (1.444,0.770) (1.484,0.841) (1.524,0.921) (1.564,1.009) (1.604,1.105) (1.644,1.209) (1.684,1.320) (1.724,1.436) (1.764,1.554) (1.805,1.670) (1.845,1.778) (1.885,1.872) (1.925,1.943) (1.965,1.987) (2.005,2.000) (2.045,1.979) (2.085,1.928) (2.125,1.850) (2.165,1.752) (2.206,1.642) (2.246,1.525) (2.286,1.407) (2.326,1.292) (2.366,1.182) (2.406,1.080) (2.446,0.986) (2.486,0.900) (2.526,0.823) (2.566,0.753) (2.607,0.690) (2.647,0.633) (2.687,0.582) (2.727,0.536) (2.767,0.495) (2.807,0.458) (2.847,0.425) (2.887,0.395) (2.927,0.368) (2.967,0.343) (3.008,0.320) (3.048,0.300) (3.088,0.281) (3.128,0.264) (3.168,0.249) (3.208,0.234) (3.248,0.221) (3.288,0.209) (3.328,0.198) (3.368,0.187) (3.409,0.178) (3.449,0.169) (3.489,0.161) (3.529,0.153) (3.569,0.146) (3.609,0.139) (3.649,0.133) (3.689,0.127) (3.729,0.122) (3.769,0.116) (3.810,0.112) (3.850,0.107) (3.890,0.103) (3.930,0.099) (3.970,0.095) -- (4.0,0) -- (0,0) -- cycle;
  \draw[cv] plot coordinates {(0.000,0.092) (0.040,0.096) (0.080,0.100) (0.120,0.104) (0.160,0.108) (0.201,0.113) (0.241,0.118) (0.281,0.123) (0.321,0.128) (0.361,0.134) (0.401,0.141) (0.441,0.148) (0.481,0.155) (0.521,0.163) (0.561,0.171) (0.602,0.180) (0.642,0.190) (0.682,0.200) (0.722,0.212) (0.762,0.224) (0.802,0.238) (0.842,0.252) (0.882,0.268) (0.922,0.286) (0.962,0.305) (1.003,0.326) (1.043,0.349) (1.083,0.374) (1.123,0.402) (1.163,0.433) (1.203,0.467) (1.243,0.505) (1.283,0.547) (1.323,0.594) (1.363,0.647) (1.404,0.705) (1.444,0.770) (1.484,0.841) (1.524,0.921) (1.564,1.009) (1.604,1.105) (1.644,1.209) (1.684,1.320) (1.724,1.436) (1.764,1.554) (1.805,1.670) (1.845,1.778) (1.885,1.872) (1.925,1.943) (1.965,1.987) (2.005,2.000) (2.045,1.979) (2.085,1.928) (2.125,1.850) (2.165,1.752) (2.206,1.642) (2.246,1.525) (2.286,1.407) (2.326,1.292) (2.366,1.182) (2.406,1.080) (2.446,0.986) (2.486,0.900) (2.526,0.823) (2.566,0.753) (2.607,0.690) (2.647,0.633) (2.687,0.582) (2.727,0.536) (2.767,0.495) (2.807,0.458) (2.847,0.425) (2.887,0.395) (2.927,0.368) (2.967,0.343) (3.008,0.320) (3.048,0.300) (3.088,0.281) (3.128,0.264) (3.168,0.249) (3.208,0.234) (3.248,0.221) (3.288,0.209) (3.328,0.198) (3.368,0.187) (3.409,0.178) (3.449,0.169) (3.489,0.161) (3.529,0.153) (3.569,0.146) (3.609,0.139) (3.649,0.133) (3.689,0.127) (3.729,0.122) (3.769,0.116) (3.810,0.112) (3.850,0.107) (3.890,0.103) (3.930,0.099) (3.970,0.095)};
  \draw[dashed,line width=0.6pt] (0,0) -- (0,2.25);
  \draw[dashed,line width=0.6pt] (4.0,0) -- (4.0,2.25);
  \node[font=\scriptsize,anchor=south] at (0,2.22) {$-\Delta$};
  \node[font=\scriptsize,anchor=south] at (4.0,2.22) {$+\Delta$};
  \draw[{Latex[length=1.2mm]}-{Latex[length=1.2mm]},line width=0.5pt] (1.56,1.0) -- (2.44,1.0);
  \node[font=\scriptsize,anchor=south] at (2.15,1.0) {$\Xi$};
  \node[font=\scriptsize,anchor=north] at (2.0,-0.12) {$\omega$};
  \node[font=\footnotesize\bfseries,anchor=north] at (2.0,-0.5) {(a) normal GDD};
  \node[font=\scriptsize,anchor=north,align=center] at (2.0,-0.85)
    {cutoff \emph{intrinsic}: $P\ge0$ truncates at $\Delta$};
\end{scope}

\begin{scope}[shift={(5.8,0)}]
  \draw[ax] (-0.15,0) -- (4.35,0);
  \draw[ax] (2.0,0) -- (2.0,2.35);
  \fill[black!12] (0.000,0.009) (0.033,0.009) (0.067,0.010) (0.100,0.010) (0.134,0.011) (0.167,0.011) (0.200,0.012) (0.234,0.013) (0.267,0.013) (0.301,0.014) (0.334,0.015) (0.367,0.016) (0.401,0.017) (0.434,0.018) (0.467,0.019) (0.501,0.021) (0.534,0.022) (0.568,0.023) (0.601,0.025) (0.634,0.027) (0.668,0.029) (0.701,0.031) (0.735,0.034) (0.768,0.036) (0.801,0.039) (0.835,0.043) (0.868,0.046) (0.902,0.050) (0.935,0.055) (0.968,0.060) (1.002,0.066) (1.035,0.073) (1.068,0.080) (1.102,0.089) (1.135,0.099) (1.169,0.110) (1.202,0.123) (1.235,0.138) (1.269,0.155) (1.302,0.176) (1.336,0.200) (1.369,0.228) (1.402,0.262) (1.436,0.303) (1.469,0.351) (1.503,0.410) (1.536,0.481) (1.569,0.567) (1.603,0.672) (1.636,0.800) (1.669,0.953) (1.703,1.135) (1.736,1.343) (1.770,1.570) (1.803,1.790) (1.836,1.955) (1.870,1.994) (1.903,1.840) (1.937,1.502) (1.970,1.140) (2.003,1.004) (2.037,1.202) (2.070,1.579) (2.104,1.888) (2.137,2.000) (2.170,1.929) (2.204,1.748) (2.237,1.524) (2.270,1.300) (2.304,1.096) (2.337,0.920) (2.371,0.772) (2.404,0.650) (2.437,0.549) (2.471,0.466) (2.504,0.397) (2.538,0.341) (2.571,0.294) (2.604,0.255) (2.638,0.222) (2.671,0.195) (2.705,0.171) (2.738,0.152) (2.771,0.135) (2.805,0.120) (2.838,0.107) (2.871,0.096) (2.905,0.087) (2.938,0.079) (2.972,0.071) (3.005,0.065) (3.038,0.059) (3.072,0.054) (3.105,0.049) (3.139,0.045) (3.172,0.042) (3.205,0.039) (3.239,0.036) (3.272,0.033) (3.306,0.031) (3.339,0.029) (3.372,0.027) (3.406,0.025) (3.439,0.023) (3.472,0.022) (3.506,0.020) (3.539,0.019) (3.573,0.018) (3.606,0.017) (3.639,0.016) (3.673,0.015) (3.706,0.014) (3.740,0.013) (3.773,0.013) (3.806,0.012) (3.840,0.011) (3.873,0.011) (3.907,0.010) (3.940,0.010) (3.973,0.009) -- (4.0,0) -- (0,0) -- cycle;
  \draw[cv] plot coordinates {(0.000,0.009) (0.033,0.009) (0.067,0.010) (0.100,0.010) (0.134,0.011) (0.167,0.011) (0.200,0.012) (0.234,0.013) (0.267,0.013) (0.301,0.014) (0.334,0.015) (0.367,0.016) (0.401,0.017) (0.434,0.018) (0.467,0.019) (0.501,0.021) (0.534,0.022) (0.568,0.023) (0.601,0.025) (0.634,0.027) (0.668,0.029) (0.701,0.031) (0.735,0.034) (0.768,0.036) (0.801,0.039) (0.835,0.043) (0.868,0.046) (0.902,0.050) (0.935,0.055) (0.968,0.060) (1.002,0.066) (1.035,0.073) (1.068,0.080) (1.102,0.089) (1.135,0.099) (1.169,0.110) (1.202,0.123) (1.235,0.138) (1.269,0.155) (1.302,0.176) (1.336,0.200) (1.369,0.228) (1.402,0.262) (1.436,0.303) (1.469,0.351) (1.503,0.410) (1.536,0.481) (1.569,0.567) (1.603,0.672) (1.636,0.800) (1.669,0.953) (1.703,1.135) (1.736,1.343) (1.770,1.570) (1.803,1.790) (1.836,1.955) (1.870,1.994) (1.903,1.840) (1.937,1.502) (1.970,1.140) (2.003,1.004) (2.037,1.202) (2.070,1.579) (2.104,1.888) (2.137,2.000) (2.170,1.929) (2.204,1.748) (2.237,1.524) (2.270,1.300) (2.304,1.096) (2.337,0.920) (2.371,0.772) (2.404,0.650) (2.437,0.549) (2.471,0.466) (2.504,0.397) (2.538,0.341) (2.571,0.294) (2.604,0.255) (2.638,0.222) (2.671,0.195) (2.705,0.171) (2.738,0.152) (2.771,0.135) (2.805,0.120) (2.838,0.107) (2.871,0.096) (2.905,0.087) (2.938,0.079) (2.972,0.071) (3.005,0.065) (3.038,0.059) (3.072,0.054) (3.105,0.049) (3.139,0.045) (3.172,0.042) (3.205,0.039) (3.239,0.036) (3.272,0.033) (3.306,0.031) (3.339,0.029) (3.372,0.027) (3.406,0.025) (3.439,0.023) (3.472,0.022) (3.506,0.020) (3.539,0.019) (3.573,0.018) (3.606,0.017) (3.639,0.016) (3.673,0.015) (3.706,0.014) (3.740,0.013) (3.773,0.013) (3.806,0.012) (3.840,0.011) (3.873,0.011) (3.907,0.010) (3.940,0.010) (3.973,0.009)};
  \draw[{Latex[length=1.2mm]}-{Latex[length=1.2mm]},line width=0.5pt] (1.33,1.35) -- (2.67,1.35);
  \node[font=\scriptsize,anchor=south] at (2.7,1.35) {$\omega_{\mathrm{core}}$};
  \draw[-{Latex[length=1.6mm]},line width=0.5pt] (4.05,0.45) -- (4.55,0.45);
  \node[font=\scriptsize,anchor=west,align=left] at (4.05,0.85)
    {$\sim|\omega|^{-3}$\\[-2pt]to $\omega_{\mathrm{cap}}$};
  \node[font=\scriptsize,anchor=north] at (2.0,-0.12) {$\omega$};
  \node[font=\footnotesize\bfseries,anchor=north] at (2.0,-0.5) {(b) anomalous GDD};
  \node[font=\scriptsize,anchor=north,align=center] at (2.0,-0.85)
    {cutoff \emph{extrinsic}: the filter sets $\omega_{\mathrm{cap}}$};
\end{scope}

\begin{scope}[shift={(0.35,-4.6)}]
  \draw[ax] (-0.15,0) -- (9.8,0);
  \draw[ax] (0,-0.75) -- (0,2.25);
  \draw[cv] plot coordinates {(0.000,1.900) (0.031,1.886) (0.063,1.846) (0.094,1.780) (0.125,1.690) (0.157,1.579) (0.188,1.448) (0.220,1.302) (0.251,1.144) (0.282,0.977) (0.314,0.807) (0.345,0.636) (0.376,0.468) (0.408,0.308) (0.439,0.159) (0.471,0.023) (0.502,-0.096) (0.533,-0.198) (0.565,-0.281) (0.596,-0.343) (0.627,-0.385) (0.659,-0.408) (0.690,-0.412) (0.721,-0.399) (0.753,-0.370) (0.784,-0.329) (0.816,-0.278) (0.847,-0.218) (0.878,-0.154) (0.910,-0.088) (0.941,-0.023) (0.972,0.039) (1.004,0.095) (1.035,0.144) (1.067,0.184) (1.098,0.215) (1.129,0.234) (1.161,0.243) (1.192,0.242) (1.223,0.230) (1.255,0.209) (1.286,0.181) (1.317,0.147) (1.349,0.108) (1.380,0.066) (1.412,0.023) (1.443,-0.018) (1.474,-0.058) (1.506,-0.093) (1.537,-0.123) (1.568,-0.146) (1.600,-0.163) (1.631,-0.172) (1.663,-0.173) (1.694,-0.167) (1.725,-0.155) (1.757,-0.136) (1.788,-0.113) (1.819,-0.085) (1.851,-0.055) (1.882,-0.023) (1.913,0.008) (1.945,0.038) (1.976,0.066) (2.008,0.090) (2.039,0.109) (2.070,0.123) (2.102,0.132) (2.133,0.135) (2.164,0.132) (2.196,0.123) (2.227,0.110) (2.259,0.093) (2.290,0.071) (2.321,0.048) (2.353,0.023) (2.384,-0.002) (2.415,-0.026) (2.447,-0.049) (2.478,-0.069) (2.509,-0.086) (2.541,-0.098) (2.572,-0.107) (2.604,-0.110) (2.635,-0.109) (2.666,-0.103) (2.698,-0.093) (2.729,-0.079) (2.760,-0.063) (2.792,-0.044) (2.823,-0.023) (2.855,-0.002) (2.886,0.018) (2.917,0.038) (2.949,0.055) (2.980,0.070) (3.011,0.081) (3.043,0.089) (3.074,0.093) (3.105,0.093) (3.137,0.089) (3.168,0.081) (3.200,0.070) (3.231,0.056) (3.262,0.040) (3.294,0.023) (3.325,0.005) (3.356,-0.013) (3.388,-0.029) (3.419,-0.045) (3.451,-0.058) (3.482,-0.068) (3.513,-0.076) (3.545,-0.080) (3.576,-0.081) (3.607,-0.078) (3.639,-0.072) (3.670,-0.063) (3.701,-0.051) (3.733,-0.038) (3.764,-0.023) (3.796,-0.007) (3.827,0.008) (3.858,0.023) (3.890,0.037) (3.921,0.049) (3.952,0.059) (3.984,0.066) (4.015,0.070) (4.046,0.071) (4.078,0.069) (4.109,0.065) (4.141,0.057) (4.172,0.047) (4.203,0.036) (4.235,0.023) (4.266,0.009) (4.297,-0.005) (4.329,-0.018) (4.360,-0.031) (4.392,-0.042) (4.423,-0.051) (4.454,-0.058) (4.486,-0.062) (4.517,-0.064) (4.548,-0.063) (4.580,-0.059) (4.611,-0.053) (4.642,-0.044) (4.674,-0.034) (4.705,-0.023) (4.737,-0.011) (4.768,0.002) (4.799,0.014) (4.831,0.026) (4.862,0.036) (4.893,0.045) (4.925,0.051) (4.956,0.056) (4.988,0.058) (5.019,0.057) (5.050,0.054) (5.082,0.049) (5.113,0.042) (5.144,0.033) (5.176,0.023) (5.207,0.012) (5.238,0.000) (5.270,-0.011) (5.301,-0.022) (5.333,-0.031) (5.364,-0.039) (5.395,-0.046) (5.427,-0.050) (5.458,-0.052) (5.489,-0.052) (5.521,-0.050) (5.552,-0.046) (5.584,-0.039) (5.615,-0.032) (5.646,-0.023) (5.678,-0.013) (5.709,-0.002) (5.740,0.008) (5.772,0.018) (5.803,0.027) (5.834,0.035) (5.866,0.041) (5.897,0.045) (5.929,0.048) (5.960,0.048) (5.991,0.047) (6.023,0.043) (6.054,0.038) (6.085,0.031) (6.117,0.022) (6.148,0.013) (6.180,0.004) (6.211,-0.006) (6.242,-0.015) (6.274,-0.024) (6.305,-0.031) (6.336,-0.037) (6.368,-0.041) (6.399,-0.044) (6.430,-0.045) (6.462,-0.044) (6.493,-0.041) (6.525,-0.036) (6.556,-0.030) (6.587,-0.022) (6.619,-0.014) (6.650,-0.005) (6.681,0.004) (6.713,0.012) (6.744,0.020) (6.776,0.028) (6.807,0.034) (6.838,0.038) (6.870,0.041) (6.901,0.042) (6.932,0.041) (6.964,0.038) (6.995,0.034) (7.026,0.029) (7.058,0.022) (7.089,0.015) (7.121,0.006) (7.152,-0.002) (7.183,-0.010) (7.215,-0.018) (7.246,-0.025) (7.277,-0.030) (7.309,-0.035) (7.340,-0.038) (7.372,-0.039) (7.403,-0.039) (7.434,-0.037) (7.466,-0.033) (7.497,-0.028) (7.528,-0.022) (7.560,-0.015) (7.591,-0.007) (7.622,0.000) (7.654,0.008) (7.685,0.015) (7.717,0.022) (7.748,0.028) (7.779,0.032) (7.811,0.035) (7.842,0.037) (7.873,0.036) (7.905,0.035) (7.936,0.032) (7.968,0.027) (7.999,0.022) (8.030,0.015) (8.062,0.008) (8.093,0.001) (8.124,-0.006) (8.156,-0.013) (8.187,-0.020) (8.218,-0.025) (8.250,-0.030) (8.281,-0.033) (8.313,-0.034) (8.344,-0.034) (8.375,-0.033) (8.407,-0.031) (8.438,-0.027) (8.469,-0.022) (8.501,-0.016) (8.532,-0.009) (8.564,-0.002) (8.595,0.005) (8.626,0.011) (8.658,0.018) (8.689,0.023) (8.720,0.027) (8.752,0.030) (8.783,0.032) (8.814,0.033) (8.846,0.032) (8.877,0.029) (8.909,0.026) (8.940,0.021) (8.971,0.016) (9.003,0.010) (9.034,0.003) (9.065,-0.003) (9.097,-0.010) (9.128,-0.016) (9.160,-0.021) (9.191,-0.025) (9.222,-0.028) (9.254,-0.030) (9.285,-0.031) (9.316,-0.030) (9.348,-0.028) (9.379,-0.025)};
  \draw[dashed,line width=0.9pt] plot coordinates {(0.000,1.900) (0.105,1.849) (0.209,1.799) (0.314,1.750) (0.418,1.702) (0.523,1.656) (0.627,1.612) (0.732,1.568) (0.836,1.526) (0.941,1.484) (1.046,1.444) (1.150,1.405) (1.255,1.367) (1.359,1.330) (1.464,1.294) (1.568,1.259) (1.673,1.225) (1.778,1.192) (1.882,1.159) (1.987,1.128) (2.091,1.098) (2.196,1.068) (2.300,1.039) (2.405,1.011) (2.509,0.983) (2.614,0.957) (2.719,0.931) (2.823,0.906) (2.928,0.881) (3.032,0.857) (3.137,0.834) (3.241,0.812) (3.346,0.790) (3.451,0.768) (3.555,0.747) (3.660,0.727) (3.764,0.708) (3.869,0.688) (3.973,0.670) (4.078,0.652) (4.182,0.634) (4.287,0.617) (4.392,0.600) (4.496,0.584) (4.601,0.568) (4.705,0.553) (4.810,0.538) (4.914,0.523) (5.019,0.509) (5.123,0.495) (5.228,0.482) (5.333,0.469) (5.437,0.456) (5.542,0.444) (5.646,0.432) (5.751,0.420) (5.855,0.409) (5.960,0.398) (6.065,0.387) (6.169,0.376) (6.274,0.366) (6.378,0.356) (6.483,0.347) (6.587,0.337) (6.692,0.328) (6.796,0.319) (6.901,0.311) (7.006,0.302) (7.110,0.294) (7.215,0.286) (7.319,0.278) (7.424,0.271) (7.528,0.263) (7.633,0.256) (7.737,0.249) (7.842,0.243) (7.947,0.236) (8.051,0.230) (8.156,0.223) (8.260,0.217) (8.365,0.212) (8.469,0.206) (8.574,0.200) (8.679,0.195) (8.783,0.190) (8.888,0.184) (8.992,0.179) (9.097,0.175) (9.201,0.170) (9.306,0.165)};
  \draw[line width=0.4pt] (0.000,-0.10) -- (0.000,0.10);
  \draw[line width=0.4pt] (0.476,-0.10) -- (0.476,0.10);
  \draw[line width=0.4pt] (0.953,-0.10) -- (0.953,0.10);
  \draw[line width=0.4pt] (1.429,-0.10) -- (1.429,0.10);
  \draw[line width=0.4pt] (1.905,-0.10) -- (1.905,0.10);
  \draw[line width=0.4pt] (2.382,-0.10) -- (2.382,0.10);
  \draw[line width=0.4pt] (2.858,-0.10) -- (2.858,0.10);
  \draw[line width=0.4pt] (3.334,-0.10) -- (3.334,0.10);
  \draw[line width=0.4pt] (3.810,-0.10) -- (3.810,0.10);
  \draw[decorate,decoration={brace,amplitude=3pt,mirror},line width=0.4pt]
     (0,-0.30) -- (3.810,-0.30);
  \node[font=\scriptsize,anchor=north] at (1.905,-0.42)
     {$\Nsep\sim\Lambda/l$ graining cells};
  \draw[{Latex[length=1.2mm]}-{Latex[length=1.2mm]},line width=0.5pt]
     (0,1.55) -- (0.476,1.55);
  \node[font=\scriptsize,anchor=west] at (0.2,1.225) {$l=\pi/\Delta$};
  \node[font=\scriptsize,anchor=south west] at (3.810,0.62)
     {$\Lambda=1/\Xi$};
  \node[font=\scriptsize,anchor=north] at (5.0,-1.05) {delay $\tau$};
  \node[font=\footnotesize\bfseries,anchor=north] at (5.0,-1.45)
     {(c) autocorrelation: a resolution and an envelope scale};
\end{scope}
\end{tikzpicture}
\caption{The two correlation scales, and the two ways of obtaining them.
(\textbf{a})~In normal GDD the spectrum is a Lorentzian of half-width $\Xi$
truncated at $\pm\Delta$, the truncation being forced by $P\ge0$ through
\eqref{eq:slaving}: the ultraviolet cutoff is a property of the solution.
(\textbf{b})~In anomalous GDD the same construction gives a two-horn core with
algebraic $|\omega|^{-3}$ wings and \emph{no} truncation; the boundary of the
occupied spectrum is instead supplied from outside, by the spectral filter,
and defined here as an energy quantile $\omega_{\mathrm{cap}}$ of the windowed
envelope \eqref{eq:filterBW} \cite{Kalashnikov2026}.
(\textbf{c})~Either way, the normalized time-integrated field
autocorrelation is a convolution
\eqref{eq:autocorr2} of a broad envelope (dashed), whose width is the reciprocal
of the spectral core, with a narrow sinc kernel (solid), whose width is the
reciprocal of the cutoff. The kernel is the finest correlation interval
resolved by the occupied spectrum---one ``graining cell''---and the number of
such cells spanned by the envelope is the scale-separation index
\eqref{eq:Neff}, drawn here for eight graining cells. Whether that number also counts
statistically independent degrees of freedom is the open question of
Section~4.3. A Schr\"odinger soliton has no cutoff, hence no
kernel distinct from its envelope, hence $\Nsep\sim1$
(Section~4.2).}
\label{fig:twoscales}
\end{figure}
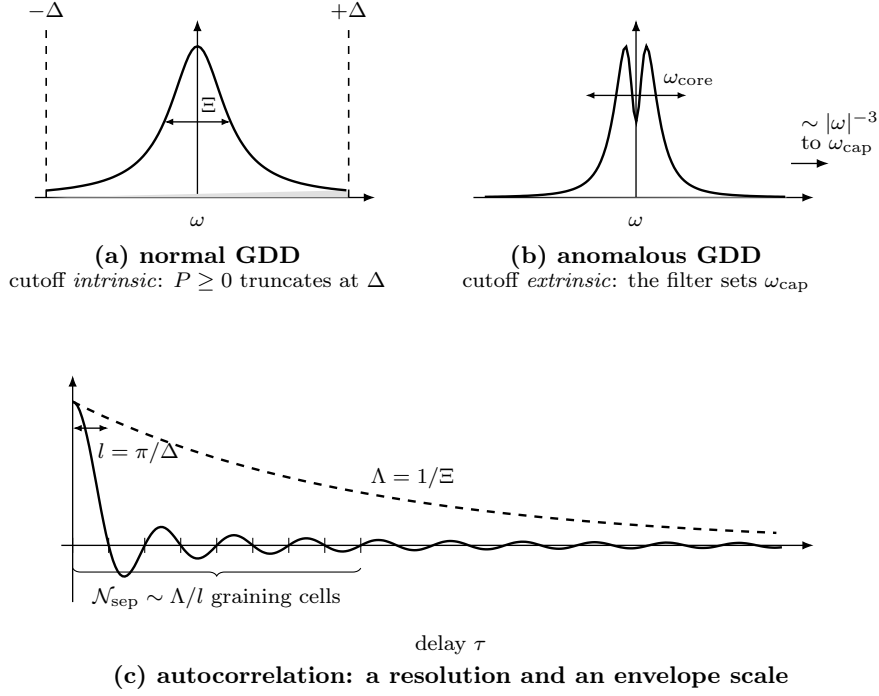

The two regimes therefore reach the same two-scale structure by different
routes, and Figure~\ref{fig:twoscales} sets them side by side. \emph{In both regimes the short scale is set by the high-frequency boundary of
the occupied spectrum and the long scale by the spectral core. What differs is
the provenance of that boundary, and how sharply it is defined}---intrinsic and kinematic in NGD, where $P\ge0$ terminates the
spectrum at $\Delta$; extrinsic and dissipative in AGD, where the
spectral filter supplies $\omega_{\mathrm{cap}}$. Spectral dissipation shapes both mechanisms, though it is not, as we previously wrote, indispensable to DS formation: the filter-free case treated in the next paragraph shows that chirped solutions survive its removal. What does not survive is their thermalized form, so the second mechanism is no less intrinsic to the physics for being extrinsic to the solution. 

This is more than a bookkeeping remark. It is exactly the situation in driven-open condensates,
where first-order correlation functions are routinely regularized by an
ultraviolet cutoff introduced by hand to control short-time behavior
\cite{deLeeuw2014}. The two constructions should
be kept apart, because the ultraviolet cutoffs that appear in the review
literature on driven-open photon fluids \cite{Carusotto2013} are of a different
kind: a lattice spacing renormalizing a contact interaction, and a cell volume
$V$ that must satisfy $\Gamma\gg g/V$ before a truncated-Wigner simulation
reproduces the quantum dynamics. Those regularize the \emph{interaction} and
the \emph{noise}, not $g^{(1)}$ itself, and only Ref.~\cite{deLeeuw2014}
carries the first-order-coherence statement made here. The anomalous-dispersion DS reproduces that
construction with the cutoff supplied by a physical filter (i.e., ``by hand'') rather than by
regularization, which is one reason the condensate analogy of Section~7 is
closer there than in normal dispersion.

\emph{NGD without spectral dissipation.} The cleanest test of that conjecture is to remove the filter altogether, which
Ref.~\cite{Kalashnikov2024b} does: the CQGLE at $\alpha=0$ in NGD retains chirped solutions---a spike on a constant background, a tabletop, a truncated spike---and the adiabatic construction can be repeated on them. Three things follow, and they separate what the cutoff owes to the filter from what it does not. First, the cutoff owes it nothing. The slaving relation \eqref{eq:slaving} contains no $\alpha$, so $P\ge0$ truncates the spectrum at the same $\Delta=\sqrt{q/\beta}$ whether a filter is present or not: the normal-dispersion cutoff is kinematic in the strict sense, and this is the independent check of that claim. Second, the \emph{shape} owes it everything. At $\alpha=0$ the profile is not a truncated Lorentzian but a convex one that vanishes at $\pm\Delta$ instead of jumping, and Ref.~\cite{Kalashnikov2024b} states plainly that it is not of Rayleigh--Jeans form and is therefore not thermalized. Since $-\mu_{\mathrm{shape}}=\Xi^{2}$, $H_{s}$, $U$ and $\Theta_{s}$ are all read off the Rayleigh--Jeans denominator, every indicator
of Section~6 is a property of the \emph{filtered} problem and not of the chirp alone. That is a limitation of scope, and we state it as one. Third, the two families are disjoint rather than continuously connected. Removing the filter sends $C=\alpha\gamma/\beta\kappa\to0$, whereas $\Xi^{*2}>0$ requires
$\sqrt{(1-C/2)^{2}-\Sigma}<\tfrac{13}{10}C-\tfrac15$ and hence $C>2/13$: at
$C\to0$ one finds $\Xi^{*2}=-1.50$, $-1.44$, $-1.13$ and $-0.65$ for
$\Sigma=0$, $0.1$, $0.5$ and $0.9$, all inadmissible. The filter-free solution is thus not the $\alpha\to0$ limit of the chirped branch, but a separate family, and the master diagram does not reach it---$2/13$ being, consistently, the same endpoint that Ref.~\cite{Kalashnikov2024b} quotes for the DSR interval. Two consequences are carried forward: the continuous vanishing at $\pm\Delta$ supports the reading of Section~2.2 that the sharp truncation is a leading-order artifact of stationary phase rather than a physical discontinuity. The runaway of the direct cascade in the filter-free case, discussed in Section~\ref{sec:turbulence}, identifies the filter as the ultraviolet sink of the turbulence analogy.

AGD also supplies a second, sharper observable. Along near-resonant
paths the windowed envelopes are approximately self-similar,
$S^{(\mathrm{cap})}(\omega;E)\approx\xi(E)f(\omega)$ with $f$ nearly
energy-independent, so that
\begin{equation}
  R(\tau;E) \approx \xi(E)\,R_{f}(\tau) ,
  \qquad
  \hat g(\tau;E) \approx \hat g_{f}(\tau) :
  \label{eq:selfsimilar}
\end{equation}
raising the energy rescales the \emph{amplitude} of the autocorrelation while
leaving its \emph{shape} almost invariant \cite{Kalashnikov2026}. Correlation
strength and correlation shape separate, and only the former carries the energy.
A shot-to-shot measurement can test this directly---and, because such a
measurement supplies exactly the ensemble average that Section~4.3 requires, the
same data---if recorded with phase, see Section~6.5---would convert $\hat g$
into a genuine $g^{(1)}$ and calibrate $f(r)$ at the same time. We return to it in Section~6.

\subsection{What the Schr\"odinger soliton lacks}

The contrast with the conservative case is not a matter of degree. Delete the
dissipative terms from \eqref{eq:cqgle} and the surviving object is the
Schr\"odinger soliton $a(t)=\sqrt{P_{0}}\,\mathrm{sech}(t/T_{\mathrm{DS}})$,
whose phase is uniform across the pulse. Three properties follow, and each of
them is a property the DS does \emph{not} have.

First, the chirp vanishes. With $\Omega(t)\equiv0$ every temporal slice
oscillates at the same frequency. There is no map $t\mapsto\Omega$, no slaving
relation \eqref{eq:slaving}, and therefore nothing in the \emph{time--frequency}
structure to distinguish one slice from another. This is not to say the pulse
lacks structure---a transform-limited $\mathrm{sech}$ has definite amplitude and
spectral profiles---but that it lacks the particular internal separation between
a spectral core and a spectral edge on which everything below depends.

Second, the spectrum has no edge. It is $\mathrm{sech}$-shaped, decaying
exponentially with a single width $\sim1/T_{\mathrm{DS}}$, so there is no scale
in it other than the one already carried by the pulse duration. The cosine
transform of a one-scale spectrum is a one-scale correlation function: $l$ and
$\Lambda$ collapse onto each other, the convolution \eqref{eq:autocorr2} becomes
trivial, and the ratio that will define $\Nsep$ is of order unity. A
conservative soliton is a single coherent degree of freedom, not an ensemble of
many.

Third — and here a common shorthand needs correcting —the equation coefficients do not fix the energy of a conservative soliton. The cubic NLS family, in a schematic normalization,
\begin{equation}
  \psi(t,z)=\eta\,\mathrm{sech}\!\left[\eta\,(t-t_{0}-vz)\right]
  \exp\!\left\{i\left[\tfrac{v}{2}(t-t_{0})
  +\left(\eta^{2}-\tfrac{v^{2}}{4}\right)z+\phi_{0}\right]\right\} ,
  \label{eq:nlsfamily}
\end{equation}
with the exact numerical factors depending on the NLS normalization adopted, is
a continuum parameterized by amplitude $\eta$, velocity $v$, position $t_{0}$
and phase $\phi_{0}$
\cite{ZakharovShabat1972,Hasegawa1973}, and ensembles of such solitons possess a statistical mechanics---soliton-gas kinetics, with its own equation of state and thermodynamic limit \cite{ElKamchatnov2005,Suret2024}. Dissipation there is
not merely destructive, a soliton condensate having been observed to
\emph{emerge} under dissipation in a nonlinear electrical transmission line, by a rearrangement that existing hydrodynamic theory does not capture \cite{Fache2025}. What the conservative family lacks is not variability but \emph{selection}, and the statement needs one qualification to be correct. Phase, temporal position and velocity are generated by continuous symmetries of the conservative equation---global $U(1)$, translation and the Galilean boost---so the members they connect are images of one another. The amplitude $\eta$ is not of that kind: at fixed equation normalization it labels physically distinct members of the family, with distinct widths and energies. What is true of the whole family is the weaker and sufficient statement that nothing internal to the conservative dynamics prefers one member over another; and
a thermodynamics of the single soliton would have no dissipative flux to
describe. The DS family is selected by a gain--loss balance, which is what makes its members distinguishable states of one driven system rather than images of one another. Together with the absence of a second scale, this is what leaves the single conservative soliton with nothing for the present construction to act on---not an absence of a continuum, and not an absence of statistical mechanics in the conservative world generally.

It is worth being precise about which ingredient does the work, because
``dissipative'' alone is not the answer. The CQGLE in anomalous dispersion also supports a \emph{weakly} chirped, soliton-like branch outside the adiabatic existence window, whose profile is of Pereira--Stenflo type and whose spectrum fills in the central dip and develops oscillatory wings \cite{Kalashnikov2026,PereiraStenflo1977}. That branch is dissipative and is nonetheless closer to the conservative case in the respect that matters here: it is not strongly chirped, so it lacks the frequency-resolved internal structure. The strong chirp, not the dissipation as such, is what generates the structure. The origin of the cutoff, by contrast, is branch dependent, and the manuscript's own results require the distinction. On the strongly chirped normal-dispersion branch, the finite spectral support already follows from the existence and positivity conditions of the asymptotic solution---the slaving relation \eqref{eq:slaving} contains no $\alpha$, and chirped solutions with the same $\Delta$ persist as the filter is taken to its limiting value---so there the edge is kinematic. In anomalous dispersion, the spectrum is not truncated, and the operational edge is tied much more directly to dissipative regularization and to the capture rule that defines it. Dissipation is what makes the strong chirp possible in both cases; what it supplies in addition, and where, is not the same on the two branches.

\subsection{Quasiparticles, the scale-separation index, and what would make it
a count}
\label{sec:quasiparticles}

\textit{Conditional nature of this subsection}. All thermodynamic statements
below presuppose that the scale-separation index $r=\Delta/\Xi$ can be
calibrated against a genuine participation number $\Npr$ via
Eq.~\eqref{eq:calibration}. We have not yet performed that calibrationed.
Without it, $r$ is a spectral shape parameter, not a statistical degree of
freedom, and the quantities defined here are structural indicators whose
thermodynamic interpretation is conjectural.

The structure just described invites an established interpretation, and the
purpose of this subsection is to state the invitation precisely enough to see
what it presupposes. A quasi-homogeneous \emph{incoherent} field obeying a
nonlinear Schr\"odinger-type equation can, at leading kinetic order, be described
by a Vlasov-like equation in which statistical quasiparticles move in a
self-consistent potential \(V_{\mathrm{eff}}=2g\widetilde N\). This structure
underlies kinetic descriptions of incoherent solitons
\cite{Picozzi2007,Picozzi2014,Picozzi2009}. The strongly chirped DS provides a
formal two-scale analogue: a short correlation scale and a longer confinement
scale \cite{Kalashnikov2026,Kalashnikov2025}. The deterministic CQGLE solution
does not by itself establish a Vlasov quasiparticle ensemble: the kinetic
derivation starts from a fluctuating field with specified statistics and invokes
a statistical closure, neither of which follows from the deterministic
stationary pulse. Accordingly, we retain the phrase
\emph{semi-incoherent soliton} only when referring to the established kinetic
analogy in the cited literature; for the deterministic solution itself we use
the more precise phrase \emph{two-scale strongly chirped DS}.
In the geometric-optics description of
highly multimoded incoherent solitons, the beam is a distribution of rays whose
density is a function of the ray invariant, and the self-consistency
requirement---that the beam be guided uniformly by the waveguide its own
intensity induces---fixes not the intensity profile but the \emph{incoherence}:
the angular spread of diffuse irradiation needed at each intensity in the beam
cross section, given by a universal expression in the nonlinearity
\cite{Snyder1998}. Two consequences inverse the situation here. A stationary
incoherent soliton may carry an arbitrary intensity profile, with the
nonlinearity setting the required angular distribution instead, whereas the
profile of a coherent soliton---and of the chirped DS---is dictated by the
equation. And the multimoded regime is entered through a size condition, the
beam radius exceeding the coherent-soliton radius by a large factor. The chirped
DS meets neither: its profile is determined, and it carries no free angular or
modal distribution left to be fixed by self-consistency. The reference system is
in any case spatial and conservative, so nothing transfers quantitatively. The
point of the comparison is only that an incoherent soliton has a genuine
internal distribution to be solved for, and the deterministic DS has a ratio of
two scales. Their ratio,
\begin{equation}
  \Nsep \;\sim\; \frac{\text{collective scale}}{\text{graining scale}}
 \;\sim\;
  \begin{cases}
    \Lambda/l = \Delta/\pi\Xi, & \text{NGD},\\[2pt]
    \varrho/\ell = \omega_{\mathrm{cap}}/\pi\omega_{\mathrm{core}},
      & \text{AGD},
  \end{cases}
  \label{eq:Neff}
\end{equation}
relates them. Where the conservative equilibrium theory possesses a single
coherence time, $\tau_{c}\simeq\sqrt{|\beta|/|\mu|}$ by \eqref{eq:RJscales}, the DS possesses two, and it is their \emph{ratio} rather than either one separately that every indicator below is a function of.

Two features of \eqref{eq:Neff} should be recorded before it is used. The relation is written as $\sim$ and not as an equality. The factor $\pi$ compares the first zero of a sinc kernel with the decay length of an exponential, and matched full widths at half maximum, or matched second moments. No numerical value of $\Nsep$ carries physical content, and we work with $r=\Delta/\Xi$ throughout, quoting $\Nsep$ only where the graining picture is being invoked.

The ordering carries content, and it is convention-free. On the unscalable branch $r\ll1$: the core is far wider than the spectral window, the spectrum is effectively rectangular over $|\omega|\le\Delta$, the convolution \eqref{eq:autocorr2} collapses onto the sinc kernel alone, and the DS is a single-scale spectral structure in the same sense as the Schr\"odinger soliton of Section~4.2---a statement about the number of spectral scales, not about coherence, which is rank-one on either branch. That is consistent with that branch being the one that possesses a conservative-soliton limit (Section~2.3). On the fidelity curve $r=1$, the two scales coincide, which is why that curve is a threshold and not merely an indicator contour. Above it the scales separate, and $r\to\infty$ in the resonance limit, where $\mu_{\mathrm{shape}}\to0^{-}$ at fixed $\Delta$. The three regimes---one scale, threshold, two separated scales---are what the index is for. Whether that ordering is also an ordering by a count of statistical degrees of freedom is what has to be earned, and the next paragraphs state what earning it requires.

What a genuine count would require is fixed by the theory of partially coherent
fields, and it is worth writing down in full, because the gap between it and
\eqref{eq:Neff} is the principal open problem of this paper. The object that
carries modal information is the two-time mutual-coherence kernel
\begin{equation}
  J(t_{1},t_{2}) = \big\langle a_{n}(t_{1})\,a_{n}^{*}(t_{2})\big\rangle_{n} ,
  \label{eq:kernel}
\end{equation}
in which $\langle\cdot\rangle_{n}$ is an average over an \emph{explicitly
specified} ensemble $\{a_{n}\}$. For a Hermitian, positive-semidefinite $J$, the
Mercer--Karhunen--Lo\`eve theorem supplies an orthonormal set and a set of
occupancies \cite{Wolf1982,Christodoulides2001}
\begin{equation}
  J(t_{1},t_{2}) = \sum_{m}\lambda_{m}\,\varphi_{m}(t_{1})\varphi_{m}^{*}(t_{2}) ,
  \qquad
  \int\! J(t,t')\,\varphi_{m}(t')\,\mathrm{d}t'
  = \lambda_{m}\varphi_{m}(t) ,
  \label{eq:KL}
\end{equation}
and the effective number of degrees of freedom is the participation number
\begin{equation}
  \Npr = \frac{\big(\sum_{m}\lambda_{m}\big)^{2}}{\sum_{m}\lambda_{m}^{2}} ,
  \label{eq:NPR}
\end{equation}
which is the quantity for which Starikov and Wolf established the connection to
the coherence properties of a source \cite{StarikovWolf1982,Starikov1982}. Three things
follow, and they should be kept separate.

\emph{(i) A deterministic pulse has $\Npr=1$.} If the ensemble contains one
realization, \eqref{eq:kernel} reduces to $J=a(t_{1})a^{*}(t_{2})$, a rank-one
kernel with a single nonzero eigenvalue, whatever the chirp and whatever the
bandwidth. The adiabatic solution of Section~2 is such a realization. No
manipulation of its spectrum can produce a nontrivial modal spectrum, and the
finite width of the autocorrelation \eqref{eq:autocorr2} does not indicate
otherwise: a coherent field may have arbitrary temporal dependence and arbitrary
spectrum \cite{Glauber1963a,Glauber1963b}. This is the sharpest correction we
make to earlier statements of the framework, our own included, in which the
chirp was said to make the pulse lose internal coherence.
That statement is exact, and it is also the reason a resolution has to
be named. The rank-one result holds at unlimited resolution, but the
classification of a signal as deterministic or stochastic is not itself
resolution-independent. Analyses built on the $(\epsilon,\tau)$ entropy and the
finite-size Lyapunov exponent show that one record can be legitimately called
deterministic on one range of scales and stochastic on another, and that a
deterministic system with enough degrees of freedom becomes practically
indistinguishable from a stochastic one once the attainable resolution or
embedding dimension falls short. The recommendation is to classify behavior at
a stated scale rather than to answer the question of the system's ``true''
character \cite{Cencini2000}. That is the right frame here, and it cuts in both
directions. It licenses treating the DS statistically at resolutions coarser
than the graining scale that the spectrum itself supplies, with no contradiction
of $\Npr=1$ at unlimited resolution---coarse-graining is the entire content of
the graining scale. It equally forbids the converse move: no coarse-grained
entropy can show that the pulse \emph{is} a statistical object, because the
classification belongs to the description at a scale and not to the solution.
What Eq.~\eqref{eq:calibration} proposes to test is accordingly not whether the
DS is really an ensemble, but whether a named ensemble, examined at a named
resolution, reproduces the deterministic scale ratio.

\emph{(ii) The ensemble has to be named, and cleaned.} Four candidates are
physically available in a mode-locked oscillator, and they are not equivalent:
quantum noise (spontaneous emission entering each round trip); technical and
shot-to-shot fluctuations; slow gain fluctuations, whose correlation time
exceeds the round trip; and the unresolved fast degrees of freedom eliminated by
the adiabatic approximation itself. Only the first three are accessible to
experiment. In each case $J$ is to be accumulated over realizations and
diagonalized, and the accumulation is not innocent: global phase drift, timing
jitter, carrier-frequency drift, pulse-energy fluctuation and position wandering
are all \emph{extrinsic} degrees of freedom that inflate the eigenvalue spectrum
and can by themselves manufacture several coherent modes from a single
deterministic pulse. A specification of the ensemble must therefore include the
alignment protocol---recentring in time\footnote{A shot-dependent global phase
$e^{i\varphi_{j}}$ cancels identically in $J(t_{1},t_{2})=\langle
a(t_{1})a^{*}(t_{2})\rangle$ and need not be removed; conditioning also
\emph{raises} apparent coherence, so raw and conditioned kernels should both be
reported},
compensation of carrier drift, and either normalization or explicit modeling of
energy fluctuations---and must report how $\Npr$ varies as each is switched off.
Position wandering is not a hypothetical entry on that list. In a GL system
whose background has become chaotic, the centroid of a surviving pulse executes
a random walk that is diffusive to good accuracy, $\langle(\Delta X)^{2}\rangle
\propto\Delta t$, in a mechanism its authors liken to the Gordon--Haus effect
\cite{SakaguchiMalomed2000}. A diffusive centroid has a variance that grows
without bound over the observation window, so an unaligned kernel would be
dominated by it and $\Npr$ would report the length of the record rather than any
property of the pulse. Recentring is accordingly not a cosmetic step but the difference between a convergent and a divergent statistic.
Nothing in the present paper performs that calculation.

\emph{(iii) The relation to be tested is a calibration, not an identity.} Since $r$ ranges over $(0,\infty)$ across the diagram while $\Npr\ge1$ by construction, no proportionality can relate them. A useful calibration should approach the one-mode limit $\Npr\simeq1$ when the two deterministic scales merge, and be reproducibly nondecreasing over the tested range of $r$. Whether $\Npr$ remains unbounded as $r\to\infty$ is \emph{not} fixed by the deterministic scale separation and must be determined from the ensemble: finite detector bandwidth, gain correlations, noise statistics and the alignment protocol may in principle cause the participation number to saturate, and a monotone bounded calibration such as $f(r)=1+r/(1+r)$ is not excluded apriori.
The possible conjecture is accordingly
\begin{equation}
  \Npr \;=\; f(r) , \qquad f \ \text{to be determined from the specified ensemble}.
  \label{eq:calibration}
\end{equation}

\noindent Equation~\eqref{eq:calibration} is best read as a possible bridge between two descriptions, not as a prerequisite for using the two-scale picture at all. The deterministic ratio $r$ is already a well-defined structural coordinate of the chirped DS. What remains open is whether repeated noisy realizations assign to that structure an effective statistical multiplicity. If such a calibration exists, it would strengthen the quasiparticle language; if not, the spectral-scale separation and the dynamical results below remain valid, while the literal ``degree-count'' interpretation should be dropped.

The existing stochastic evidence should therefore be separated by the physical question it answers. In NGD, Ref.~\cite{Kalashnikov2025} follows noise-driven self-start and final pulse-number selection. It shows that accessibility shifts toward multipulse states as energy increases. In AGD,
Ref.~\cite{Kalashnikov2026} asks whether a prepared strongly chirped pulse remains close to the analytical branch under noise, and finds a finite robust region inside the existence domain. These are complementary observations. They should not be reduced to a comparison of ensemble sizes: one probes \emph{selection from noise}, the other \emph{survival of a prepared state}. Together they motivate the same conceptual hierarchy---existence, robustness, and accessibility---without requiring that $r$ already be interpreted as a literal number of microstates.

Thus, the spectrum itself can be read statistically, subject to
the same qualification. The truncated Lorentzian of half-width $\Xi$ has exactly the algebraic form of a Rayleigh--Jeans distribution with a negative offset \eqref{eq:RJ}, with the cutoff no longer imported but generated by the pulse
\cite{Podivilov2005,Kalashnikov2024,Kalashnikov2025}. We write
$\mu_{\mathrm{shape}}$ for that offset and record its status plainly: in
equilibrium statistical mechanics, a chemical potential is conjugate to a
conserved or statistically constrained wave action, whereas the CQGLE does not conserve the pulse norm---gain and loss fix it dynamically---so
$-\mu_{\mathrm{shape}}=\Xi^{2}$ occupies the algebraic position of a chemical potential without yet being one. Two symbols are therefore kept apart in what follows. $\mu_{\mathrm{shape}}$, with $-\mu_{\mathrm{shape}}=\Xi^{2}$, is the offset of the spectral denominator---a shape parameter, fixed once the spectrum is written and free of any energy convention. A chemical-potential-like $\mu_{\mathrm{eff}}$ exists only once a quasiparticle energy is chosen: writing
the RJ law as
$n(\omega)=\Theta_{\mathrm{RJ}}/[\varepsilon(\omega)-\mu_{\mathrm{eff}}]$ gives
$\mu_{\mathrm{eff}}=-\beta\Xi^{2}$ for $\varepsilon=\beta\omega^{2}$ and
$\mu_{\mathrm{eff}}=-\Xi^{2}/2$ for $\varepsilon=\omega^{2}/2$, and the internal
energy \eqref{eq:DSinternal} uses the latter. Statements about ``the chemical potential vanishing at resonance'' refer to either symbol, since both vanish with $\Xi$. Statements about numerical values refer to $\mu_{\mathrm{eff}}$ and require the convention to be quoted with them. The same reading applies in anomalous GDD, with the windowed spectral envelope in place of the truncation
\cite{Kalashnikov2026}.

This is the point at which the present framework departs from all three
traditions described in Section~3. There, the microstate count was supplied externally---$k_{c}$ by discretization or by hand, $n$ by spectral truncation, $M$ by the waveguide geometry \cite{Wu2019,Mangini2022}, $\mathcal{N}$ by the
cavity length together with the filter bandwidth. In the experiments of
Section~3.2, this is literal: the same fiber, at the same wavelength, offers
the same $M$ whatever the launched power, and the thermodynamic parameters $T$ and $\mu$ move over a fixed ladder of eigenvalues \cite{Mangini2022}. For a DS, the analogous quantity is \emph{self-generated}: the pulse creates its own spectral support, its own graining scale, and hence its own scale-separation index, all of which depend on the operating point and on the pulse energy
(Figure~\ref{fig:microstates}b). Energy scaling is therefore accompanied by a change in the available spectral scale separation rather than merely by a redistribution of power within a fixed, externally imposed bandwidth.

The question of whether self-generated scale separation can be translated into a state-dependent effective mode count is at the core of the ensemble-calibration question presented in Eq.~\eqref{eq:calibration}. If a reproducible relationship between $\Nsep$ (the scale separation) and the participation number $\Npr$ can be established, then the multimode ``volume'' analogy described in Eq.~\eqref{eq:eos-conservative} would become natural. In this scenario, an effective number of degrees of freedom would be determined by the state itself, rather than being solely dictated by the apparatus. As a result, the piston in this conservative thought experiment would move autonomously. Until this calibration is achieved, we will regard ``state-dependent volume'' merely as a heuristic analogy, rather than as a recognized thermodynamic extensive variable.

\section{Dissipative Soliton Resonance, Revisited}

Dissipative soliton resonance (DSR) was introduced as the observation that, in certain regions of CQGLE parameter space, the energy of a stable single-pulse solution diverges as the parameters approach a hypersurface, with the pulse broadening into a flat-topped, linearly chirped structure whose peak power saturates at the continuous-wave value and whose spectral width does not
collapse \cite{Chang2008,Chang2009,GreluDSR2010}. Systematic numerical exploration established that the resonance locus is continuous across the dispersion-free point and occupies substantial intervals in both dispersion domains, and that the sign and magnitude of the quintic nonlinearity control where it lies \cite{Chang2008,GreluDSR2010}. The adiabatic theory adds that the locus can be written down, and that this form reveals that the two dispersion regimes are different things.

Throughout this subsection, we use the operational definition: DSR denotes a stationary solution branch along which, under continuation of the specified control parameter with the remaining parameters fixed, the pulse energy can grow without bound---here in the vacuum-stability limit $\Sigma\to0^{+}$---while the peak power and the relevant spectral scale remain bounded and the solution remains admissible
\cite{Kalashnikov2024,Kalashnikov2026}.

\subsection{Normal dispersion: an intrinsic property of the chirped branch}

Here everything follows from the two algebraic relations of Section~2.3. With
$\chi=0$, the energy \eqref{eq:DSenergy} in dimensionless form is
\begin{equation}
  E^{*} = \frac{6}{\Xi^{*}}\arctan\!\Big(\frac{\Delta^{*}}{\Xi^{*}}\Big) ,
  \label{eq:Estar}
\end{equation}
so a divergence requires $\Xi^{*}\to0$ at finite $\Delta^{*}$: the Lorentzian
core collapses while the cutoff saturates. By \eqref{eq:dimensionless}, that is
the condition $\Delta^{*2}=\tfrac{3}{5}(1+C)$, and combining it with the
scalable root of \eqref{eq:branches} gives the resonance locus in the closed form
\begin{equation}
  \Sigma_{\Xi}(C) = \frac{36}{25}\Big(\frac{2}{3}-C\Big)\big(1+C\big) .
  \label{eq:DSRlocus}
\end{equation}
The same construction places the fidelity curve $\Xi=\Delta$, which is the lower
energy boundary of the resonant region, at
\begin{equation}
  \Sigma_{\mathrm{fid}}(C) = \frac{3}{4}\big(1-C^{2}\big) ,
  \label{eq:fidlocus}
\end{equation}
and the region in which $\Xi<\Delta$---the DSR region proper---is the band
between them:
\begin{equation}
  \max\big\{0,\;\Sigma_{\Xi}(C)\big\} \;<\; \Sigma \;<\;
  \Sigma_{\mathrm{fid}}(C) ,
  \qquad \tfrac{1}{2}\le C<1 .
  \label{eq:DSRband}
\end{equation}
\footnote{For $C<\tfrac12$ the upper bound is taken over by the branch-dividing curve
$\Sigma=(1-C/2)^{2}$, which the fidelity curve touches tangentially at
$C=\tfrac12$. The band survives there but is progressively squeezed.} In the
vacuum-stability limit, the statement collapses to a single interval,
\begin{equation}
  \Sigma\to0^{+}: \qquad
  \frac{2}{3} < C < 1 ,
  \label{eq:DSRinterval}
\end{equation}
with the energy diverging as $C\to\tfrac{2}{3}^{+}$ and the fidelity curve
reached at $C=1$. All of \eqref{eq:DSRlocus}--\eqref{eq:DSRinterval} are
elementary consequences of \eqref{eq:branches} and
\eqref{eq:dimensionless}. The endpoint $C=2/3$ reproduces the value obtained
numerically in the reduced model \cite{Kalashnikov2026,Kalashnikov2025}. They are
drawn in Figure~\ref{fig:masterdiagram}.

The physical content is the one anticipated in Section~4: the chemical
potential $-\mu_{\mathrm{shape}}=\Xi^{2}$ tends to zero at finite $\Delta$, so the spectrum condenses into a ``finger'' at zero detuning, the peak power becomes bounded from above, and the energy is harvested purely by temporal stretching
\cite{Kalashnikov2009,Kalashnikov2024,Kalashnikov2025}. The limit $\mu\to0^{-}$ is formally the wave-condensation limit of Section~3.1, in which the correlation length and time diverge \cite{Picozzi2007,Connaughton2005}. The
essential difference is that there the cutoff $k_{c}$ is a fixed external
parameter, whereas here $\Delta$ saturates dynamically, so that the \emph{ratio}
of the two scales---and with it the index $\Nsep\sim r$
of \eqref{eq:Neff}---diverges rather than merely the coherence length. Three experimentally recognizable signatures accompany the crossing: saturation of the spectral broadening, appearance and growth of a Lorentzian spike at the spectrum center, and reversal of the energy-scaling mechanism from pulse shortening to
asymptotic stretching. All three have been observed in a Kerr-lens mode-locked
Cr$^{2+}$:ZnS chirped-pulse oscillator \cite{Rudenkov2023,Kalashnikov2025}.

Two of the three also appear in numerical work that reaches the resonance by an entirely different route. Raising the pump in a mode-locked laser with saturable gain and cubic--quintic nonlinear loss first raises the peak power and then stops doing so: the peak saturates at the level fixed by the ratio of the two loss nonlinearities, the pulse turns rectangular and thereafter lengthens monotonically, and a narrow bell-shaped peak emerges at the center of the otherwise rectangular spectrum and grows until it dominates the profile \cite{Komarov2013}. That model carries no quintic SPM whatever, which is the state of affairs to be expected if normal-dispersion DSR is intrinsic to the chirped branch and already present in the cubic limit. The parametric trend reported there points the same way: weaker Kerr nonlinearity and larger normal GDD are found to favor the resonance \cite{Komarov2013}, and both move the operating point in the direction of decreasing $C=\alpha\gamma/\beta\kappa$, towards the endpoint $C\to2/3^{+}$ at which \eqref{eq:Estar} diverges. The gain is saturated dynamically by the intracavity energy rather than held at a prescribed $\Sigma$---but the direction is the one the master diagram predicts.

The asymptotic shape reached in this limit also admits a description in a
different language, and it is worth recording because it connects the resonance to a separate branch of Ginzburg--Landau theory. As $E^{*}\to\infty$, the DS tends to a plateau of fixed height. The peak power is bounded above and joined to the vacuum by two steep edges. Asymptotically, that is a pair of back-to-back \emph{fronts} separated by a flat interior, and fronts, together with the pulses that can be assembled from them, are among the elementary solutions of generalized Ginzburg--Landau equations \cite{vanSaarloos1992}. The reading is consistent with everything above. The interior, across which the instantaneous
frequency barely moves, is the spectral core: it supplies $\Xi$, and it is what collapses at resonance. The fronts carry the entire frequency excursion: they supply the cutoff $\Delta$ and hence the graining scale $l=\pi/\Delta$, and they are unchanged by the stretching. This is why $\Delta$ saturates while $\Xi\to0$---and why the broad pedestal of \cite{Komarov2013} does not narrow as the pulse lengthens, even though the central spike grows. Closed-form kink solutions are available for generalized Ginzburg-Landau equations carrying higher-order terms via bilinear methods \cite{Liu2017}, and they explicitly exhibit the front structure. They are subject, however, to the limitation of Section~2.1 in a stronger form: the constraints accompanying them fix several coefficients of the equation in terms of the parameters of the solution, rather than merely
restricting them. Such solutions illustrate the asymptotic geometry: they cannot chart it across the parameter space of an oscillator.

A finite quintic SPM does not disturb any of this. The resonance interval is
deformed only perturbatively,
$C_{\mathrm{DSR},\pm}(\tilde\chi)=C_{\mathrm{DSR},\pm}(0)+\delta C_{\pm}\tilde\chi+\mathcal{O}(\tilde\chi^{2})$,
with saturable SPM ($\tilde\chi>0$) displacing the divergence towards larger $C$ and self-enhancing SPM ($\tilde\chi<0$) towards smaller $C$
\cite{Kalashnikov2026}. Normal-dispersion DSR is thus an \emph{intrinsic}
property of the chirped branch: it is already present in the cubic limit, and
the quintic term only moves it.

\subsection{Anomalous dispersion: a conditional resonance}

In anomalous GDD, the argument cannot even begin the same way, because the object it would begin from does not exist. As established in Section~4.1, the spectrum is not truncated: there is no $\Delta$ to saturate and no Lorentzian core to collapse. The divergence has to come from somewhere else, and it comes from the far wings. The algebraic tail of the stationary-phase envelope carries a prefactor controlled by the combination $1+C\tilde\chi$
\begin{equation}
  S_{\mathrm{AGD}}(\omega) \;\sim\;
  \frac{1}{\big(1+C\tilde\chi\big)^{2}\sqrt{\tilde\chi}\,|\omega|^{3}} ,
  \qquad |\omega|\to\infty ,
  \label{eq:AGDtail}
\end{equation}
which is integrable in general but becomes non-uniform as that combination
vanishes. This identifies the resonance not with a condensation but with a
``chirp--control'' line \cite{Kalashnikov2026}:
\begin{equation}
  1 + C\tilde\chi = 0
  \qquad\Longrightarrow\qquad
  C^{(\mathrm{AGD})}_{\mathrm{DSR}}(\tilde\chi) = -\frac{1}{\tilde\chi} .
  \label{eq:AGDlocus}
\end{equation}
Equation~\eqref{eq:AGDlocus} is necessary but not sufficient, and this is the decisive structural difference. The line \eqref{eq:AGDlocus} produces an actual divergence only if it lies \emph{inside} the adiabatic existence window of the strongly chirped solution, which in anomalous dispersion is the finite domain
\begin{equation}
  C<0,\quad \tilde\chi>0,\quad D_{a}>0,\qquad
  0<\Sigma<\min\big\{\Sigma_{\mathrm{th}},\;\Sigma_{Q}\big\} ,
  \label{eq:AGDwindow}
\end{equation}
\begin{equation}
\Sigma_{\mathrm{th}}(C,\tilde\chi)
   = \frac{3\big(\tilde\chi(C+4)-3\big)}{4\,\tilde\chi^{2}} ,
  \qquad
  \Sigma_{Q}(C,\tilde\chi) = \frac{(C-2)^{2}}{4\,(1-C\tilde\chi)},
  \label{eq:AGDthresholds}
\end{equation}
with $D_{a}=-\tilde\Delta^{2}_{-}>0$ corresponding to the anomalous-dispersion
\cite{Kalashnikov2026}\footnote{\emph{A note on the normalization of $\Sigma$.} Equations
\eqref{eq:AGDwindow}--\eqref{eq:AGDthresholds}, the bound $\Sigma_{\max}$ and
the softened threshold $\tilde\chi_{\min}$ below are taken from
Ref.~\cite{Kalashnikov2026} but \emph{rescaled} to the normalization of this
paper. That reference sets $\Sigma=\zeta\sigma/\kappa$, whereas
\eqref{eq:branches} uses $\Sigma=4\zeta\sigma/\kappa$. The two differ by exactly
a factor of four, so each threshold quoted there has been multiplied by four
here. A reader comparing the two papers will therefore find
$\Sigma_{\mathrm{th}}$, $\Sigma_{Q}$ and $\Sigma_{\max}$ larger by that factor
in the present notation, and $\tilde\chi_{\min}\simeq1+\Sigma/3$ in place of
$1+4\Sigma/3$. The rescaling may be checked at $\tilde\chi\to0$, where
$\Sigma_{Q}\to(C-2)^{2}/4=(1-C/2)^{2}$, which is precisely the branch divider of
\eqref{eq:branches}: both express the reality of the same discriminant. A single
normalization is now used throughout the paper, including Section~6.5 and
Figure~\ref{fig:agdentropy}. The criterion $\tilde\chi>1$ of
\eqref{eq:AGDcriterion} is a $\Sigma\to0$ statement and is unaffected by the
choice, as are all the qualitative conclusions of Sections~5.2 and 6.5.}. 

Two consequences follow immediately. A strongly chirped
anomalous-dispersion DS exists at all only above a quintic threshold,
$\tilde\chi>3/(C+4)$, and the admissible loss is bounded by
$\Sigma_{\max}(C)=(C+4)^{2}/16$, attained at $\tilde\chi=6/(C+4)$. And, in the vacuum-stability limit, the window shrinks to $C\in\big(3/\tilde\chi-4,\,0\big)$, so that requiring $-1/\tilde\chi$ to fall inside it gives the resonance criterion in a single line:
\begin{equation}
  -\frac{1}{\tilde\chi} > \frac{3}{\tilde\chi}-4
  \qquad\Longleftrightarrow\qquad
  \tilde\chi > 1 ,
  \label{eq:AGDcriterion}
\end{equation}
softened at finite loss to
the exact root of $\Sigma\tilde\chi^{2}-3\tilde\chi+3=0$ that joins
$\tilde\chi=1$ as $\Sigma\to0$,
\begin{equation}
  \tilde\chi_{\min}(\Sigma)
  = \frac{3-\sqrt{9-12\Sigma}}{2\Sigma}
  = 1+\frac{\Sigma}{3}+\frac{2\Sigma^{2}}{9}+O(\Sigma^{3}) ,
  \label{eq:AGDchimin}
\end{equation}
obtained by evaluating $\Sigma_{\mathrm{th}}$ of \eqref{eq:AGDthresholds} on the
resonance locus itself, $\Sigma_{\mathrm{th}}(-1/\tilde\chi,\tilde\chi)
=3(\tilde\chi-1)/\tilde\chi^{2}$
\cite{Kalashnikov2026}. 

Anomalous-dispersion DSR therefore requires a sufficiently strong saturable quintic SPM and is simply absent in the cubic limit. This reconciles the numerical finding that anomalous-dispersion DSR
depends on a strong quintic term \cite{Chang2009,GreluDSR2010} with the experimental reports of DSR-like rectangular pulses in anomalous-dispersion fiber lasers \cite{Duan2012,Gene2023,Huang2015}. It identifies which parameter paths in a real oscillator can and cannot lead to energy scaling: only those on which enough nonlinear phase is accumulated to push $\tilde\chi$ past $\tilde\chi_{\min}(\Sigma)$, which exceeds unity at any finite loss.

\subsection{Why these are not the same phenomenon}

It is tempting to read the two cases as one resonance seen from two sides of the dispersion-free point, and the adiabatic theory shows that they are not. The contrast is sharpest when stated in the vocabulary of Section~4.

\emph{The order of the divergence.} The two divergences are not even of the same strength, which is worth recording because it is easily hidden by the word ``resonance''. In normal dispersion, $E^{*}=(6/\Xi^{*})\arctan r$ diverges as $1/\Xi$. In anomalous dispersion, the leading $\tilde\omega^{2}$ term of $G$ carries the factor $1+C\tilde\chi$, so on the chirp-control line the tail \eqref{eq:AGDtail} softens from $|\tilde\omega|^{-3}$ to $|\tilde\omega|^{-1}$ and the energy diverges only logarithmically,
\begin{equation}
E_{\mathrm{AGD}} \;\simeq\;
\frac{\sqrt{\tilde\chi}}{\pi\,(\tilde\chi-1)^{2}}\,
\ln\frac{1}{\big|1+C\tilde\chi\big|} ,
\label{eq:AGDlogdiv}
\end{equation}
\noindent a coefficient we have verified numerically to $0.6\%$ at
$\tilde\chi=1.2$, $1.5$, $2.5$ and $3.0$, by fitting $E_{\mathrm{AGD}}$
against $\ln|1+C\tilde\chi|^{-1}$ along the approach to the chirp-control
line.

The divergence \eqref{eq:AGDlogdiv} is an algebraic statement about the
stationary solution, and it does not by itself imply that the divergent path is
dynamically accessible. The finite-time quantum-noise maps of Section~6.6 show
that the accessible region is bounded in detuning and that the divergence is not
reached within the simulated interval. Whether it is approached asymptotically,
or cut off by an instability before it is reached, is open. Its blow-up as  $\tilde\chi\to1^{+}$ is the same statement as the threshold \eqref{eq:AGDcriterion}: the resonance becomes available exactly where the coefficient ceases to be finite. Anomalous-dispersion energy scaling is therefore not merely conditional on the geometry of Section~5.2 but intrinsically slower, which is a second reason not to read the two regimes as one phenomenon.

\emph{Where the divergence lives.} Normal-dispersion DSR is an infrared phenomenon: $\Xi\to0$, the spectral core collapses onto $\omega=0$, and the energy accumulates in the condensate-like finger while the support $\Delta$ stands still. Anomalous-dispersion DSR is an ultraviolet phenomenon: the core
keeps its shape---indeed the normalized envelope is nearly energy-independent, Eq.~\eqref{eq:selfsimilar}---and the divergence is carried by the amplitude of the $|\omega|^{-3}$ wings through \eqref{eq:AGDtail}.

\emph{How the scale separation grows.} Both regimes increase $\Nsep$, and they do so from opposite ends of the spectrum. In normal dispersion, the graining scale $l=\pi/\Delta$ is pinned, and the collective scale $\Lambda=1/\Xi$ diverges. In anomalous dispersion, the core scale $\varrho$ is nearly fixed by the self-similar envelope, while the growing wing weight pushes an increasing fraction of the energy towards the edge of the window $\omega_{\mathrm{cap}}$, so that the effective graining scale
$\ell\sim\pi/\omega_{\mathrm{cap}}$ shrinks \cite{Kalashnikov2026}. This
holds along the energy-scaling path at fixed $C$, on which the envelope is
self-similar. It does \emph{not} hold along the approach to the resonance in
$C$, where the quantile $\omega_{\mathrm{core}}$ moves outward with the wing
weight even though the envelope shape does not change. Section~6.5 separates the
two senses of ``core'' and the two continuations. Here a distinction of symbols must be kept, because three different bandwidths have been conflated in discussions of this point. The spectral filter bandwidth $\omega_{\mathrm{f}}=\alpha^{-1/2}$ is a fixed property of the oscillator. The detector or spectrometer window $\omega_{\mathrm{det}}$ is a fixed property of the apparatus. And $\omega_{\mathrm{cap}}$ is the effective spectral-dissipation
window of Section~4.1. Only the last is state-dependent. Everything below refers to $\omega_{\mathrm{cap}}$, and the sensitivity of the results to $\omega_{\mathrm{f}}$ and $\omega_{\mathrm{det}}$---which is a sensitivity of $\Nsep$ to the operational definition of the core fraction \cite{Kalashnikov2026}---remains to be mapped. The two-scale separation widens either way---which is why the entropic argument of
Section~6 applies to both---but the mechanism is not shared.

\emph{Intrinsic versus conditional.} Normal-dispersion DSR follows from the admissibility constraints alone, Eqs.~\eqref{eq:DSRlocus}--\eqref{eq:DSRinterval}, and survives the cubic limit. Anomalous-dispersion DSR is a geometric coincidence between two independently specified objects---a resonance line and
an existence window---and holds only above the threshold \eqref{eq:AGDcriterion}.

\emph{Existence versus accessibility.} Finally, in anomalous dispersion, even admissibility is not the end of the matter. A linearized quantum-noise analysis shows that the analytically admissible branch is not uniformly stable: robust single-pulse operation occupies a finite region whose boundary lies to one side of the exact resonance line, so that the numerically accessible part of the divergent path is bounded in detuning before it is bounded in energy. Section~6.6 sets out that calculation and reads the maps.

None of this asymmetry is peculiar to the dissipative problem. Already in
conservative kinetics, the sign of $\beta$ decides whether the equilibrium
spectrum is an isotropic Lorentzian or a hyperbolic, $X$-shaped structure
(Figure~\ref{fig:elliptic-hyperbolic}) \cite{Picozzi2007}. What the DS adds is that the same sign decides whether the spectrum has an edge at all, and hence whether energy scaling proceeds by condensing a core or by inflating a tail.

\section{Where Energy Scaling Stops: Shape Indicators, Directional
Energy--Entropy Slopes, and Fragmentation}

DSR promises unbounded energy, but real oscillators deliver a finite maximum. The thermodynamic reading locates a candidate obstruction, and does so in a way that the pulse-shape analysis alone cannot. It rests on the conditional result of Section~4: \emph{if} the scale-separation index of a strongly chirped DS counts the internal degrees of freedom of a defined ensemble, then an entropy, an internal energy, and a temperature-like slope are definable for a single pulse. The \emph{interpretation} of everything below rests on that one supposition,
and we ask the reader to carry it through the section as a supposition. The
algebra does not: $H_{s}$, $U$ and the locations of their extrema are
well-defined functionals of the deterministic spectrum \eqref{eq:DSspectrum}
whatever the supposition turns out to be worth, and they would remain the
correct statements about that spectrum even if the ensemble test of Section~4.3
failed. What the supposition buys is the right to call them entropy, internal
energy and temperature. The first subsection makes the dependence explicit and, in doing so, settles a question of definition that earlier treatments left open.

\subsection{Thermodynamic-like spectral indicators of a chirped DS}

Take the truncated Lorentzian \eqref{eq:DSspectrum} not as a spectrum but as an
unnormalized density over frequency detuning, and normalize it on its own support
\cite{Kalashnikov2024,Kalashnikov2025}:
\begin{equation}
  \tilde{p}(\omega) = \frac{p(\omega)}{Z}
   = \frac{\Xi}{2\big(\Xi^{2}+\omega^{2}\big)\arctan(\Delta/\Xi)} ,
  \qquad
  Z = \int_{-\Delta}^{\Delta}\! p\,\mathrm{d}\omega
    = \frac{12\pi\gamma}{\zeta\kappa\,\Xi}\arctan\!\Big(\frac{\Delta}{\Xi}\Big) .
  \label{eq:DSpdf}
\end{equation}
The prefactor $\gamma/\zeta\kappa$ cancels in $\tilde p$ and was omitted from $Z$ in earlier statements of this construction. It is restored here, since $Z$ is otherwise dimensionally inconsistent with \eqref{eq:DSspectrum}. In the dimensionless variables of \eqref{eq:dimensionless}, the same quantity reads
$Z^{*}=(12\pi/\Xi^{*})\arctan(\Delta^{*}/\Xi^{*})$, and all numerical values
below are computed in those variables. The effective chemical potential is read off from the Rayleigh--Jeans form directly, and the entropy-like functional is the Gibbs--Shannon one of Section~3.2 evaluated on \eqref{eq:DSpdf}:
\begin{gather}
  -\mu_{\mathrm{shape}} = \Xi^{2} ,
  \qquad
  -\mu_{\mathrm{eff}} = \tfrac{1}{2}\Xi^{2}
  \quad\big[\varepsilon(\omega)=\omega^{2}/2\big] ,
  \label{eq:DSmu}\\[2pt]
  h \;=\; -\!\int \tilde p\,\ln\!\big(\tilde p\,\delta\omega\big)\,\mathrm{d}\omega
    \;=\; \underbrace{\ln\!\frac{2\,\Xi\arctan(\Delta/\Xi)}{\delta\omega}}
          _{\textstyle h_{\mathrm{scale}}}
      \;+\; \underbrace{\frac{1}{\arctan(\Delta/\Xi)}
          \int_{0}^{\Delta/\Xi}\frac{\ln(1+x^{2})}{1+x^{2}}\,\mathrm{d}x}
          _{\textstyle H_{s}} ,
  \label{eq:DSentropy}\\[2pt]
  U = \frac{1}{2}\int_{-\Delta}^{\Delta}\tilde{p}(\omega)\,\omega^{2}\,d\omega
    = \frac{\Xi}{2}\left[\frac{\Delta}{\arctan(\Delta/\Xi)}-\Xi\right] ,
  \label{eq:DSinternal}\\[2pt]
  \Theta_{s}[H] = \frac{\mathrm{d}U/\mathrm{d}s}{\mathrm{d}H/\mathrm{d}s} ,
  \qquad
  \mathcal{F}_{s}[H] = U - \Theta_{s}[H]\,H .
  \label{eq:DStemp}
\end{gather}

Four definitional choices are made explicit here. The earlier papers
\cite{Kalashnikov2024,Kalashnikov2025} intentionally used the shifted quantity
\(H_s\) as a positive spectral-shape measure; the present review goes one step
further by separating that shape measure from a fixed-bin differential entropy
and by tracking which derivative-based conclusions depend on this choice.

\emph{The reference measure.} A differential entropy is not invariant: changing the frequency unit shifts $h$ by an additive term, and if that shift is state-dependent, it changes derivatives, extrema, and crossings. Equation \eqref{eq:DSentropy} therefore refers the density to a bin $\delta\omega$ that
is \emph{fixed by the apparatus}---in practice the resolution of the
spectrometer, which is a bin width; the cavity filter bandwidth is a
transmission scale and not a bin, and the two should not be interchanged---and never to a scale carried by the state. A relative entropy may be used instead, but only under a condition that must be stated, because in general
\begin{equation}
  D_{\mathrm{KL}}(\tilde p\,\|\,q)
   = -h_{\mathrm{diff}}(\tilde p) - \int \tilde p\,\ln q\,\mathrm{d}\omega ,
  \label{eq:KLid}
\end{equation}
whose second term depends on the state unless $q$ is constant over a support
common to all states compared. Fixing a bounded detector window $[-\Omega_{\mathrm{det}},\Omega_{\mathrm{det}}]$ and taking $q$ uniform on it makes that term a constant, and only then does $D_{\mathrm{KL}}$ differ from \eqref{eq:DSentropy} by a constant and share its derivatives
\cite{Shannon1948,KullbackLeibler1951}. All comparisons below use one fixed $\delta\omega$ and one fixed window. We do not invoke a general equivalence \cite{Jaynes1965,Hnizdo2010}.

\emph{The split is a choice, not a canonical decomposition.}
The full fixed-bin differential entropy contains a term that depends on
\(\Xi\) and \(\Delta\), and is therefore not a state-independent constant. In
Appendix~D of Ref.~\cite{Kalashnikov2025}, the authors explicitly introduced a
\emph{shifted} entropy \(H_s\) to remove the \(\Xi\to0\) divergence and obtain a
positive shape-oriented quantity. That operation is legitimate as a definition
of a spectral-shape proxy; it should not, however, be confused with the
differential entropy relative to a fixed experimental bin. In the latter
convention \(h\to\ln(4\pi\Xi/\delta\omega)\to-\infty\) as the spectrum contracts
onto zero detuning. We therefore keep both quantities and name them
differently, rather than treating one as a correction of the other. Nondimensionalizing
with the core, $x=\omega/\Xi$, gives
\begin{equation}
  h = \ln\frac{\Xi}{\delta\omega} + h_{x}(r) ,
  \qquad
  h_{x}(r) = \ln\big(2\arctan r\big) + H_{s}(r) ,
  \label{eq:splitx}
\end{equation}
whereas nondimensionalizing with the support, $y=\omega/\Delta$, gives
\begin{equation}
  h = \ln\frac{\Delta}{\delta\omega} + h_{y}(r) ,
  \qquad
  h_{y}(r) = \ln\!\Big(\frac{2\arctan r}{r}\Big) + H_{s}(r) .
  \label{eq:splity}
\end{equation}
The shape entropy of the dimensionless spectrum is denoted as \( h_{x} \) or \( h_{y} \), depending on the chosen unit. The term \( H_{s} \) refers to the component that is common to both measurements, while the factor \( \arctan r \), which differentiates them, is inherently shape-dependent. Thus, \( H_{s} \) is best described as a ``scale-invariant spectral-shape indicator'' derived from the earlier construction, rather than as the unique Shannon entropy associated with a dimensionless spectral shape. 

It remains a valid and useful monotone functional of \( r \), and it is the quantity that is plotted and compared throughout this analysis. However, statements like ``the entropy of a DS'' can be ambiguous unless the specific functional and units are defined; normalizing the spectrum does not eliminate this ambiguity. 

The relationship between \( H_{s} \) and the sum-of-logarithms configuration entropy, \( \tilde S = \sum_{i} \ln |f_{i}|^{2} \), from multimode experiments is one of \textit{selected analogy}. Both are constructed from normalized occupancy and both omit an extensive scale term, but they are not identical: as noted in Section 3.2, the two functionals exhibit differences.

\begin{table}[ht]
\caption{The three entropy-like functionals used in this work. Status
labels as in Table~\ref{tab:claimstatus}. The two turnover values are distinct
quantities, not a discrepancy: $h$ attains its maximum at
$r\simeq1.449$ (Eq.~\eqref{eq:hmax}), whereas the maximum of $U$---and hence the
zero of $\Theta_{s}$---lies at $r\simeq1.21$ (Eq.~\eqref{eq:Umax}), both on
isogains.}
\label{tab:entropyfunctionals}
\centering
\begin{tabular}{@{}llll@{}}
\toprule
Functional & Definition & Behaviour along an isogain & Status \\
\midrule
$H_{s}$ & scale-invariant shape term, Eq.~\eqref{eq:splitx}
        & monotone, saturating at $\ln4$ & D (selected indicator) \\
$h$     & $-\int\tilde p\,\ln(\tilde p\,\delta\omega)\,\mathrm{d}\omega$
        & maximum at $r\simeq1.449$      & D (full differential entropy) \\
$\tilde S$ & $\sum_{i}\ln|f_{i}|^{2}$
        & extensive growth               & A (multimode analogy) \\
\bottomrule
\end{tabular}
\end{table}

The notation $\Theta_{s}[H]$ is used deliberately. Unlike an equilibrium
temperature, this quantity depends both on the selected entropy functional $H$
and on the path followed through the stationary solution manifold.
Reparameterizing the \emph{same} path leaves the ratio unchanged, whereas
changing the path generally does not. It is therefore a directional
energy--entropy slope rather than a state temperature, and the same caution
applies to $\mathcal{F}_{s}[H]$. Where the three functionals of Section~6.1 must
be distinguished, we write $\Theta_{s}^{(H_{s})}$, $\Theta_{s}^{(h)}$ and
$\Theta_{s}^{(\ln r)}$. Their zeros coincide under the conditions verified in
Section~6.3, but their magnitudes and their negative intervals do not.

\emph{The Derivative.} In equation \eqref{eq:DStemp}, $\Theta$ is expressed as a total derivative along a specified continuation parameter, $s$—specifically, the control parameter $C$ at a fixed value of $\Sigma$, representing motion along an isogain in the master diagram. This is a directional slope rather than a partial derivative of a state function, as no state manifold with defined held variables has been established. Furthermore, the same value is \emph{not} obtained along different continuations, and Section 6.3 explicitly computes this discrepancy. 

For clarity, we use $\Theta_{s}$ when the path matters and reserve the term ``temperature'' for instances where the sign, rather than the magnitude, is the focus. In summary, for the entire paper, we define $\Theta_{s}^{-1}=\mathrm{d}H/\mathrm{d}U$ as a directional slope evaluated along a particular continuation path. This serves as a thermodynamic-like diagnostic and should not be confused with an absolute inverse temperature, which would require the establishment of a state function, a corresponding conjugate energy variable, and an appropriate stationary ensemble independently.

\emph{The Legendre-like potential.} $\mathcal{F}=U-\Theta_{s}H$ in \eqref{eq:DStemp} is written by analogy with a free energy, but it is not one: $\Theta_{s}$ and $H$ are not conjugate variables of a demonstrated state function, and no variational principle makes $\mathcal{F}$ extremal in the stable state (Section~6.5). We call it a Legendre-like diagnostic and use it only for comparing states at equal $\Theta_{s}$. Two guards follow, stated once here rather than repeated: $\Theta_{s}$ is a
\emph{directional} slope, so every number quoted for it must name its
continuation. In anomalous dispersion, $U$, and therefore $\Theta_{s}$, are
\emph{captured-spectrum} quantities, so every number quoted there must
additionally name $\eta$, $f_{\mathrm{core}}$, $\delta\omega$ and $\omega_{\max}$
(Section~6.5). Where the distinction matters, we write $U_{\mathrm{cap}}$ and
$\Theta_{\mathrm{cap}}$. Elsewhere the shorter symbols are kept, with the
conditioning understood.

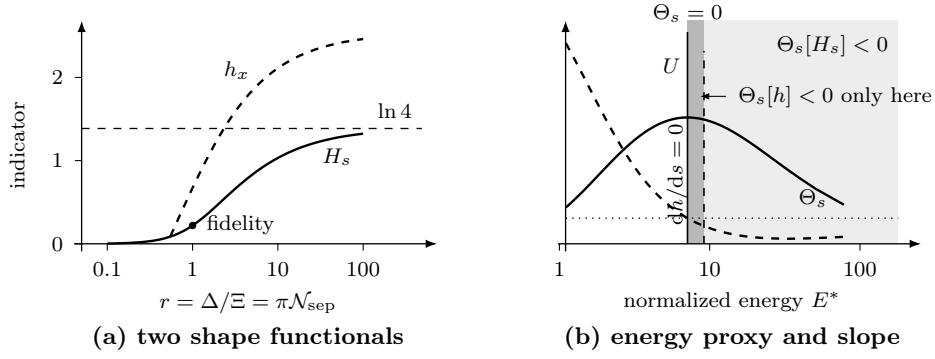
\begin{figure}[t]
\centering
\begin{tikzpicture}[x=1cm,y=1cm,
  ax/.style={draw=black,line width=0.5pt,-{Latex[length=1.5mm]}},
  cv/.style={draw=black,line width=0.9pt}]
\begin{scope}
  \draw[ax] (-0.1,0) -- (4.7,0);
  \draw[ax] (0,-0.1) -- (0,2.95);
  \draw (0.339,0) -- (0.339,-0.09);
  \node[font=\scriptsize,anchor=north] at (0.339,-0.11) {0.1};
  \draw (1.467,0) -- (1.467,-0.09);
  \node[font=\scriptsize,anchor=north] at (1.467,-0.11) {1};
  \draw (2.595,0) -- (2.595,-0.09);
  \node[font=\scriptsize,anchor=north] at (2.595,-0.11) {10};
  \draw (3.723,0) -- (3.723,-0.09);
  \node[font=\scriptsize,anchor=north] at (3.723,-0.11) {100};
  \draw (0,0.000) -- (-0.09,0.000);
  \node[font=\scriptsize,anchor=east] at (-0.11,0.000) {0};
  \draw (0,1.100) -- (-0.09,1.100);
  \node[font=\scriptsize,anchor=east] at (-0.11,1.100) {1};
  \draw (0,2.200) -- (-0.09,2.200);
  \node[font=\scriptsize,anchor=east] at (-0.11,2.200) {2};
  \draw[dashed,line width=0.5pt] (0,1.525) -- (4.5,1.525);
  \node[font=\scriptsize,anchor=south east] at (4.5,1.525) {$\ln 4$};
  \draw[cv] plot coordinates {(0.339,0.004) (0.388,0.004) (0.437,0.005) (0.486,0.007) (0.535,0.008) (0.584,0.010) (0.633,0.012) (0.682,0.015) (0.731,0.018) (0.780,0.021) (0.829,0.026) (0.878,0.032) (0.928,0.038) (0.977,0.046) (1.026,0.055) (1.075,0.066) (1.124,0.080) (1.173,0.095) (1.222,0.112) (1.271,0.133) (1.320,0.156) (1.369,0.181) (1.418,0.210) (1.467,0.242) (1.516,0.277) (1.565,0.314) (1.614,0.353) (1.663,0.394) (1.712,0.437) (1.761,0.482) (1.810,0.526) (1.859,0.572) (1.908,0.617) (1.957,0.662) (2.006,0.706) (2.056,0.750) (2.105,0.792) (2.154,0.833) (2.203,0.873) (2.252,0.911) (2.301,0.948) (2.350,0.983) (2.399,1.017) (2.448,1.049) (2.497,1.079) (2.546,1.108) (2.595,1.135) (2.644,1.161) (2.693,1.186) (2.742,1.209) (2.791,1.230) (2.840,1.251) (2.889,1.270) (2.938,1.288) (2.987,1.304) (3.036,1.320) (3.085,1.335) (3.134,1.348) (3.184,1.361) (3.233,1.373) (3.282,1.384) (3.331,1.395) (3.380,1.405) (3.429,1.414) (3.478,1.422) (3.527,1.430) (3.576,1.437) (3.625,1.444) (3.674,1.450) (3.723,1.456)};
  \draw[cv,dashed] plot coordinates {(1.173,0.098) (1.222,0.207) (1.271,0.315) (1.320,0.423) (1.369,0.529) (1.418,0.635) (1.467,0.739) (1.516,0.841) (1.565,0.942) (1.614,1.040) (1.663,1.135) (1.712,1.228) (1.761,1.318) (1.810,1.404) (1.859,1.487) (1.908,1.566) (1.957,1.642) (2.006,1.714) (2.056,1.783) (2.105,1.847) (2.154,1.909) (2.203,1.967) (2.252,2.022) (2.301,2.073) (2.350,2.122) (2.399,2.167) (2.448,2.210) (2.497,2.250) (2.546,2.287) (2.595,2.323) (2.644,2.355) (2.693,2.386) (2.742,2.415) (2.791,2.442) (2.840,2.466) (2.889,2.490) (2.938,2.511) (2.987,2.532) (3.036,2.550) (3.085,2.568) (3.134,2.584) (3.184,2.599) (3.233,2.613) (3.282,2.626) (3.331,2.638) (3.380,2.650) (3.429,2.660) (3.478,2.670) (3.527,2.678) (3.576,2.687) (3.625,2.694) (3.674,2.701) (3.723,2.708)};
  \fill (1.467,0.242) circle (1.4pt);
  \node[font=\scriptsize,anchor=south west] at (1.52,0.) {fidelity};
  \node[font=\scriptsize,anchor=west] at (3.05,1.15) {$H_{s}$};
  \node[font=\scriptsize,anchor=west] at (1.75,2.30) {$h_{x}$};
  \node[font=\scriptsize,anchor=north] at (2.2,-0.5) {$r=\Delta/\Xi=\pi \Nsep$};
  \node[font=\scriptsize,rotate=90,anchor=south] at (-0.62,1.3) {indicator};
  \node[font=\footnotesize\bfseries,anchor=north] at (2.2,-0.95)
    {(a) two shape functionals};
\end{scope}
\begin{scope}[shift={(6.4,0)}]
  \fill[black!7] (1.612,0) rectangle (4.4,2.95);
  \fill[black!28] (1.612,0) rectangle (1.830,2.95);
  \draw[ax] (-0.1,0) -- (4.7,0);
  \draw[ax] (0,-0.1) -- (0,2.95);
  \draw (-0.082,0) -- (-0.082,-0.09);
  \node[font=\scriptsize,anchor=north] at (-0.082,-0.11) {1};
  \draw (1.905,0) -- (1.905,-0.09);
  \node[font=\scriptsize,anchor=north] at (1.905,-0.11) {10};
  \draw (3.893,0) -- (3.893,-0.09);
  \node[font=\scriptsize,anchor=north] at (3.893,-0.11) {100};
  \draw[cv] plot coordinates {(0.003,0.479) (0.065,0.535) (0.103,0.571) (0.134,0.601) (0.161,0.627) (0.186,0.651) (0.208,0.673) (0.230,0.695) (0.250,0.715) (0.269,0.734) (0.287,0.753) (0.305,0.772) (0.322,0.789) (0.339,0.807) (0.355,0.824) (0.371,0.841) (0.387,0.857) (0.402,0.873) (0.417,0.889) (0.432,0.905) (0.446,0.920) (0.461,0.935) (0.475,0.950) (0.489,0.965) (0.503,0.979) (0.517,0.994) (0.530,1.008) (0.544,1.022) (0.557,1.036) (0.570,1.050) (0.584,1.063) (0.597,1.077) (0.610,1.090) (0.623,1.104) (0.636,1.117) (0.649,1.130) (0.662,1.143) (0.674,1.155) (0.687,1.168) (0.700,1.181) (0.713,1.193) (0.725,1.205) (0.738,1.217) (0.751,1.229) (0.763,1.241) (0.776,1.253) (0.789,1.265) (0.801,1.277) (0.814,1.288) (0.827,1.299) (0.839,1.311) (0.852,1.322) (0.865,1.333) (0.878,1.344) (0.890,1.355) (0.903,1.365) (0.916,1.376) (0.929,1.386) (0.942,1.397) (0.955,1.407) (0.968,1.417) (0.981,1.427) (0.994,1.437) (1.007,1.446) (1.021,1.456) (1.034,1.465) (1.047,1.475) (1.061,1.484) (1.074,1.493) (1.088,1.502) (1.102,1.510) (1.116,1.519) (1.130,1.527) (1.144,1.535) (1.158,1.543) (1.172,1.551) (1.186,1.559) (1.201,1.566) (1.215,1.574) (1.230,1.581) (1.245,1.588) (1.260,1.594) (1.275,1.601) (1.291,1.607) (1.306,1.613) (1.322,1.619) (1.337,1.624) (1.353,1.630) (1.370,1.635) (1.386,1.639) (1.403,1.644) (1.420,1.648) (1.437,1.652) (1.454,1.655) (1.472,1.659) (1.489,1.662) (1.508,1.664) (1.526,1.666) (1.545,1.668) (1.564,1.669) (1.583,1.670) (1.603,1.671) (1.623,1.671) (1.644,1.670) (1.665,1.669) (1.686,1.667) (1.708,1.665) (1.730,1.662) (1.753,1.659) (1.777,1.655) (1.801,1.650) (1.826,1.644) (1.852,1.638) (1.878,1.630) (1.905,1.622) (1.933,1.612) (1.962,1.602) (1.992,1.590) (2.024,1.577) (2.056,1.563) (2.090,1.547) (2.126,1.529) (2.163,1.509) (2.202,1.488) (2.244,1.463) (2.288,1.437) (2.335,1.407) (2.386,1.374) (2.441,1.337) (2.501,1.296) (2.567,1.249) (2.641,1.195) (2.725,1.134) (2.824,1.061) (2.943,0.975) (3.095,0.868) (3.307,0.727) (3.680,0.517)};
  \draw[cv,dashed] plot coordinates {(0.003,2.654) (0.065,2.540) (0.103,2.469) (0.134,2.410) (0.161,2.359) (0.186,2.312) (0.208,2.269) (0.230,2.228) (0.250,2.189) (0.269,2.152) (0.287,2.116) (0.305,2.082) (0.322,2.048) (0.339,2.016) (0.355,1.984) (0.371,1.953) (0.387,1.923) (0.402,1.893) (0.417,1.865) (0.432,1.836) (0.446,1.808) (0.461,1.781) (0.475,1.754) (0.489,1.727) (0.503,1.701) (0.517,1.675) (0.530,1.650) (0.544,1.625) (0.557,1.600) (0.570,1.576) (0.584,1.552) (0.597,1.528) (0.610,1.504) (0.623,1.481) (0.636,1.458) (0.649,1.435) (0.662,1.413) (0.674,1.391) (0.687,1.369) (0.700,1.347) (0.713,1.326) (0.725,1.304) (0.738,1.283) (0.751,1.262) (0.763,1.242) (0.776,1.221) (0.789,1.201) (0.801,1.181) (0.814,1.161) (0.827,1.142) (0.839,1.122) (0.852,1.103) (0.865,1.084) (0.878,1.065) (0.890,1.046) (0.903,1.028) (0.916,1.009) (0.929,0.991) (0.942,0.973) (0.955,0.955) (0.968,0.938) (0.981,0.920) (0.994,0.903) (1.007,0.886) (1.021,0.869) (1.034,0.852) (1.047,0.835) (1.061,0.819) (1.074,0.802) (1.088,0.786) (1.102,0.770) (1.116,0.754) (1.130,0.738) (1.144,0.723) (1.158,0.707) (1.172,0.692) (1.186,0.677) (1.201,0.662) (1.215,0.647) (1.230,0.632) (1.245,0.618) (1.260,0.603) (1.275,0.589) (1.291,0.575) (1.306,0.561) (1.322,0.547) (1.337,0.533) (1.353,0.520) (1.370,0.506) (1.386,0.493) (1.403,0.480) (1.420,0.467) (1.437,0.454) (1.454,0.441) (1.472,0.429) (1.489,0.416) (1.508,0.404) (1.526,0.392) (1.545,0.380) (1.564,0.368) (1.583,0.356) (1.603,0.344) (1.623,0.333) (1.644,0.322) (1.665,0.310) (1.686,0.299) (1.708,0.289) (1.730,0.278) (1.753,0.267) (1.777,0.257) (1.801,0.246) (1.826,0.236) (1.852,0.226) (1.878,0.216) (1.905,0.207) (1.933,0.197) (1.962,0.188) (1.992,0.179) (2.024,0.170) (2.056,0.161) (2.090,0.152) (2.126,0.144) (2.163,0.136) (2.202,0.128) (2.244,0.120) (2.288,0.112) (2.335,0.105) (2.386,0.098) (2.441,0.092) (2.501,0.086) (2.567,0.080) (2.641,0.076) (2.725,0.072) (2.824,0.069) (2.943,0.067) (3.095,0.069) (3.307,0.075) (3.680,0.092)};
  \draw[dotted,line width=0.5pt] (0,0.339) -- (4.4,0.339);
  \draw[line width=0.6pt] (1.612,0) -- (1.612,2.8);
  \node[font=\scriptsize,anchor=south] at (1.612,2.8) {$\Theta_{s}=0$};
  \draw[dashed,line width=0.6pt] (1.830,0) -- (1.830,2.55);
  \node[font=\scriptsize,anchor=south west,rotate=90] at (1.75,0.12)
    {$\mathrm{d}h/\mathrm{d}s=0$};
  \node[font=\scriptsize,anchor=west] at (1.15,2.35) {$U$};
  \node[font=\scriptsize,anchor=west] at (2.95,0.6) {$\Theta_{s}$};
  \node[font=\scriptsize,anchor=north] at (2.2,-0.5) {normalized energy $E^{*}$};
  \node[font=\footnotesize\bfseries,anchor=north] at (2.2,-0.95)
    {(b) energy proxy and slope};
  \node[font=\scriptsize,anchor=north east,align=right] at (4.4,2.90)
    {$\Theta_{s}[H_{s}]<0$};
  \node[font=\scriptsize,anchor=west] at (2.15,1.95) {$\Theta_{s}[h]<0$ only here};
  \draw[line width=0.4pt,-{Latex[length=1.2mm]}] (2.12,1.95) -- (1.78,1.95);
\end{scope}
\end{tikzpicture}
\caption{Indicators of the scale-separation ratio, computed from
\eqref{eq:branches}, \eqref{eq:dimensionless}, \eqref{eq:DSentropy} and
\eqref{eq:DSinternal}; numerical values in Table~\ref{tab:numbers}.
(\textbf{a})~Two functionals of the same ratio $r=\Delta/\Xi=\pi\Nsep$. The
shape indicator $H_{s}$ (solid) increases monotonically and saturates at $\ln4$;
the shape entropy in core units, $h_{x}=\ln(2\arctan r)+H_{s}$ of
\eqref{eq:splitx} (dashed), does not saturate. The two differ by a
shape-dependent term, which is why $H_{s}$ is a \emph{selected} indicator rather
than the entropy of the dimensionless spectrum. Both are level sets of $r$ on
the master diagram, and the fidelity curve is the particular level
$H_{s}\simeq0.220$ (dot). The full differential entropy $h$ at fixed bin adds
$\ln(\Xi/\delta\omega)$ to $h_{x}$ and is not monotone at all: it turns over at
$r\simeq1.45$, Eq.~\eqref{eq:hmax}.
(\textbf{b})~Along an isogain ($\Sigma=0.05$, scalable branch), the
internal-energy proxy \eqref{eq:DSinternal} first grows with $E^{*}$ and then
decays, while $H_{s}$ grows monotonically. The slope $\Theta_{s}$ of
\eqref{eq:DStemp}, evaluated along that isogain, therefore passes through
\emph{zero} at the maximum of $U$---at $r\simeq1.21$, $E^{*}\simeq7.1$, just
beyond the fidelity curve (solid vertical). With $H=H_{s}$ the slope is negative
from there to resonance (light shading); with $H=h$ the zero is unmoved, but
$\mathrm{d}h/\mathrm{d}s$ vanishes at $r\simeq1.45$, $E^{*}\simeq9.2$ (dashed
vertical), so the negative region is the finite window \eqref{eq:negwindow}
(dark shading). Both shadings are properties of these two functionals along this
continuation and are not state-function constructions. Note that it is
$\Theta_{s}$ that vanishes and $1/\Theta_{s}$ that diverges, the mirror image of
the textbook bounded-spectrum case.}
\label{fig:thermo}
\end{figure}

One feature of the shape indicator deserves to be stated on its own, because it is the cleanest statement of the connection to Section~4. \emph{$H_{s}$ depends on the two spectral scales only through their ratio $r$.} It is a monotonically increasing function of $r$ and of nothing else, rising from zero and saturating at
\begin{equation}
  H_{s} \;\longrightarrow\; \ln 4 \simeq 1.386
  \qquad (r\to\infty) ,
  \label{eq:Hsmax}
\end{equation}
which is the ceiling on \emph{this indicator} for a single strongly chirped DS (Figure~\ref{fig:thermo}a)---not a universal entropy ceiling, since
\eqref{eq:splitx} and \eqref{eq:splity} show that the shape entropy proper
carries an extra $r$-dependent term whose limit differs. Three consequences follow at once. The curves $r=\mathrm{const}$ on the master diagram are \emph{iso-$H_{s}$} contours. The fidelity curve $\Xi=\Delta$ of Section~2.3 is the particular contour $H_{s}=0.2201$, so the geometric object that marked the onset of DSR is also an indicator contour. Thus, any statement about $H_{s}$ is, by construction, a statement about the two-scale separation.

That $H_{s}$ saturates is a property of the spectral \emph{shape} and not of the scale separation, and the point deserves emphasis because the two are easily confused. As $r\to\infty$, the truncated Lorentzian converges to the untruncated one, so any scale-invariant functional of the shape must approach a limit. $\ln4$ is that limit for this functional and carries no information about internal degrees of freedom. A functional built to count would behave differently: $S\sim\ln r$ diverges, as the reading of DSR as $\mu_{\mathrm{shape}}\to0^{-}$ at fixed $\Delta$ requires. The two are not in conflict---they measure different things, one the disorder of a normalized distribution over frequency detuning and the other the number of graining cells spanned by the collective scale---and \eqref{eq:Hsmax} should not be read as a bound on the internal degrees of freedom of the pulse.

These are \emph{not} statements about the full differential entropy, and the difference is quantitative. Evaluating $h$ of \eqref{eq:DSentropy} along an isogain of the scalable branch, we find that it rises, turns over, and falls: it reaches a maximum at
\begin{equation}
    r \simeq 1.45 ,
  \label{eq:hmax}
\end{equation}
and decreases monotonically thereafter, diverging logarithmically as
$h\to\ln(4\pi\Xi/\delta\omega)$ in the resonance limit. Two properties of \eqref{eq:hmax} matter, and one non-property should be recorded. It is independent of the choice of $\delta\omega$, since a fixed bin contributes an additive constant. Two operations must be kept apart here, and the difference is not cosmetic. A fixed change of frequency units shifts a differential entropy by a \emph{state-independent} additive constant, and therefore leaves every state derivative---$\Theta_{s}$ included---unchanged. Normalizing the frequency coordinate instead by a \emph{state-dependent} scale, such as $\Xi$ or $\Delta$ at each point of the continuation, changes the reference measure along the path and can alter derivative-based quantities. The first is a choice of units; the second is a choice of what to differentiate. Only the first is made below. It depends only weakly on the isogain ($r=1.449$ at $\Sigma=0.01$, $1.455$ at $\Sigma=0.05$, $1.517$ at $\Sigma=0.4$); and---by the same token---the point at which $h$ passes through zero has \emph{no} invariant meaning, since a change of $\delta\omega$ moves it at will. We accordingly report the turnover and the asymptotic decrease, and attach no significance to any zero crossing. Physically, the two terms of \eqref{eq:splitx} encode competing tendencies: the shape indicator rises because the scale separation
widens, and $\ln(\Xi/\delta\omega)$ falls because the occupied spectrum
contracts onto zero detuning. Only above \eqref{eq:hmax} does the second win.

The phrase ``spectral condensation in a DS raises the entropy'' must
therefore be read as a statement about the shifted shape indicator \(H_s\), not
as a functional-independent statement about entropy. With a fixed-bin
differential entropy, the scale contraction eventually dominates and \(h\)
decreases beyond \eqref{eq:hmax}. The two results are not contradictory: they
answer different questions. A terminology that remains valid in either
convention is that the normalized spectral shape becomes increasingly
core-dominated as the resonance is approached while its occupied frequency
scale contracts. The comparison with Bose--Einstein condensation in
Section~\ref{sec:bec} is formulated at that level.

\begin{table}[t]
\caption{Numerical values used in Sections~6.1--6.4, recomputed for this paper
from \eqref{eq:branches}, \eqref{eq:dimensionless}, \eqref{eq:Estar},
\eqref{eq:DSentropy} and \eqref{eq:DSinternal} on the scalable branch. $r_{U}$
and $r_{h}$ are the maxima of the internal-energy proxy and of the fixed-bin
differential entropy; $E^{*}_{U}$, $E^{*}_{h}$ are the corresponding
normalized energies. The last two columns are the denominators of
\eqref{eq:DStemp} at $r_{U}$, which must be nonzero for the sign reversal to be
convention-independent.}
\label{tab:numbers}
\centering\small
\begin{tabular}{@{}lcccccc@{}}
\toprule
$\Sigma$ & $r_{U}$ & $E^{*}_{U}$ & $r_{h}$ & $E^{*}_{h}$ &
 $\mathrm{d}H_{s}/\mathrm{d}E^{*}|_{r_{U}}$ &
 $\mathrm{d}h/\mathrm{d}E^{*}|_{r_{U}}$ \\
\midrule
0.01 & 1.202 & 7.03 & 1.449 & 9.06 & $3.571\times10^{-2}$ & $9.65\times10^{-3}$ \\
0.05 & 1.208 & 7.12 & 1.455 & 9.16 & $3.543\times10^{-2}$ & $9.52\times10^{-3}$ \\
0.10 & 1.216 & 7.24 & 1.461 & 9.29 & $3.508\times10^{-2}$ & $9.42\times10^{-3}$ \\
0.20 & 1.234 & 7.52 & 1.475 & 9.57 & $3.430\times10^{-2}$ & $9.10\times10^{-3}$ \\
0.40 & 1.284 & 8.28 & 1.517 & 10.32 & $3.244\times10^{-2}$ & $8.32\times10^{-3}$ \\
\bottomrule
\end{tabular}

\medskip
\centering\small
\begin{tabular}{@{}lccccc@{}}
\toprule
$r=\Delta/\Xi$ & 1 & 1.208 & 1.45 & 4 & 16 \\
\midrule
$H_{s}(r)$ & 0.2201 & 0.2813 & 0.3476 & 0.7519 & 1.1313 \\
$h_{x}-H_{s}=\ln(2\arctan r)$ & 0.4516 & 0.5645 & 0.6596 & 0.9752 & 1.1042 \\
\bottomrule
\end{tabular}
\end{table}

\subsection{Energy scaling raises the shape indicator}

The behavior of these potentials along the energy-scaling direction is the
central result. Approaching DSR, the two scales \emph{decouple}: by
Section~5.1, the collective scale $\Lambda=1/\Xi$ diverges while the graining scale $l=\pi/\Delta$ saturates, so $\Nsep$ grows and, by
\eqref{eq:DSentropy}, $H_{s}$ grows with it towards \eqref{eq:Hsmax}
\cite{Kalashnikov2025}. Throughout this subsection, ``entropy'' means $H_{s}$.

That entropy should \emph{rise} as a coherent structure forms is, by itself, not new. It is the signature result of the conservative theories. In wave
condensation, the entropy increase is what compels the field to generate a plane wave or a soliton, because only by minimizing $H_{nl}$ can the fluctuation energy $H_{l}$---and with it the disorder---be maximized
\cite{Picozzi2007,Connaughton2005,Rumpf2001}. In the mean-field NLS ensemble, the entropy is literally the logarithm of the fluctuation kinetic energy left over after the coherent structure has minimized $H$ at fixed $N$, Eq.~\eqref{eq:JTZentropy} \cite{Jordan2000}. Nor is entropy growth with power peculiar to the DS: in the multimode equilibrium theory one has exactly
\begin{equation}
  \left(\frac{\mathrm{d}S^{\mathrm{eq}}}{\mathrm{d}\mathcal{P}}\right)_{u}
  = \frac{M}{\mathcal{P}} > 0
  \qquad (u = U/\mathcal{P}\ \text{fixed}),
  \label{eq:dSdP}
\end{equation}
a relation derived from the equation of state \eqref{eq:eos-conservative} and
corroborated by the cutback measurements of Section~3.2 \cite{Mangini2024}. The statement that must therefore be handled carefully is the comparison between DS spectral condensation and Bose--Einstein condensation. Baudin \emph{et al.} found the \emph{equilibrium} entropy to \emph{decrease} with internal energy, i.e.\ with growing occupation of the fundamental mode, at fixed power \cite{Baudin2020}. However, \eqref{eq:dSdP} shows that along the orthogonal path---increasing power at fixed energy per particle---the entropy rises even in the conservative case. ``Condensation suppresses entropy'' is thus a statement
about one direction in the $(\mathcal{P},u)$ plane, not about condensation as such.

The DS is best understood by asking what $H_{s}$ can and cannot respond to. Because $H_{s}$ is a function of $r$ alone, its variation along any continuation is
\begin{equation}
  \frac{\mathrm{d}H_{s}}{\mathrm{d}E^{*}}
  = H_{s}'(r)\,\frac{\mathrm{d}r}{\mathrm{d}E^{*}} ,
  \qquad H_{s}'(r)>0 ,
  \label{eq:dHsdE}
\end{equation}
so it carries \emph{no} term of the form \eqref{eq:dSdP}---no contribution from the growth of an extensive variable at fixed shape---and responds only to the widening of the scale separation. In the conservative multimode case, the corresponding shape derivative vanishes identically at fixed mode number, and the entropy grows through the extensive term alone. That contrast is the structural point, and it is a statement about which derivative each functional retains, not a decomposition of one entropy into two physically separate pieces. We withdraw the earlier formulation, which presented it as the latter. Its origin is the absence of the clean separation on which both conservative constructions rely---there, coherence and entropy are carried by two distinct components, a condensate and a bath with an immovable ultraviolet cutoff, whereas in a DS the same chirped pulse supplies both the collective scale and the graining scale, so that contracting the one widens the separation from the other. In this respect, the DS is closer to a turbulent system, in which the formation of a large-scale coherent structure feeds small-scale thermalized fluctuations
\cite{Picozzi2014,Nazarenko2011,JordanJosserand2000}.

\subsection{The sign reversal of the energy--entropy slope}

The internal energy does not follow the entropy. Evaluating
\eqref{eq:DSinternal} along an isogain, $U$ first grows with $E^{*}$, passes
through a maximum, and then decays---the decay being forced by the same collapse
$\Xi\to0$ that drives $H_{s}$ up, since $U\to\Xi\Delta/\pi\to0$ in that limit.
The maximum lies at
\begin{equation}
  r \simeq 1.21 ,
  \label{eq:Umax}
\end{equation}
almost independently of $\Sigma$ ($1.202$ at $\Sigma=0.01$, $1.208$ at
$\Sigma=0.05$, $1.234$ at $\Sigma=0.2$, $1.284$ at $\Sigma=0.4$), that is, just
beyond the fidelity curve---so the sign reversal very nearly coincides with the
onset of DSR itself. In normalized energy, the same point is
\begin{equation}
  E^{*} \simeq 7.1\ (\Sigma=0.05) ,
  \qquad 7.0\ (\Sigma=0.01) , \qquad 8.3\ (\Sigma=0.4) ,
  \label{eq:EstarUmax}
\end{equation}
recomputed here from \eqref{eq:branches}, \eqref{eq:dimensionless} and
\eqref{eq:Estar}\footnote{This does \emph{not} agree with the value $E^{*}\approx20$ quoted for the sign reversal in Ref.~\cite{Kalashnikov2025}, although the corresponding ratio
$r\simeq1.2$ does agree. The origin of the remaining discrepancy in $E^{*}$ has not been established. Until the normalization, branch-pairing prescription, continuation parameter and entropy functional used in the two calculations are mapped explicitly onto one another, we report the discrepancy as unresolved.}

Because \eqref{eq:DStemp} is a ratio of two derivatives taken along the same path, $\Theta_{s}$ vanishes wherever $\mathrm{d}U/\mathrm{d}s=0$ and
$\mathrm{d}H/\mathrm{d}s\neq0$. This has a consequence worth isolating, since it is the one result of this section that does not depend on the definitional choices of Section~6.1: \emph{the location of the zero of $\Theta_{s}$, and the fact that $\Theta_{s}$ changes sign there, are the same whether $H$ is taken as the shape indicator $H_{s}$ or as the full differential entropy $h$}. Finiteness and differentiability of the difference are not sufficient for this---the denominator must also be nonzero, and of the same sign, in both cases---so we verified it numerically at the turning point: along the isogain $\Sigma=0.05$,
$\mathrm{d}H_{s}/\mathrm{d}E^{*}=3.5\times10^{-2}$ and
$\mathrm{d}h/\mathrm{d}E^{*}=9.5\times10^{-3}$ at $r=1.208$, both positive, with the same signs and comparable magnitudes throughout
$\Sigma\in[0.01,0.4]$. The same holds for the counting functional $S=\ln r$ introduced in Section~6.1, for which $\mathrm{d}S/\mathrm{d}E^{*}=
1.03\times10^{-1}$ at $r_{U}$ on the isogain $\Sigma=0.05$
($1.04\times10^{-1}$ at $\Sigma=0.01$, $9.1\times10^{-2}$ at $\Sigma=0.4$), so the zero of $\Theta_{s}$ is unmoved for a third functional.

The zero is not, however, unmoved by a change of \emph{path}, and this can be stated in closed form. Because $\Xi^{2}\,(1+5r^{2}/3)=1+C$ identically, the internal-energy proxy reduces to $U=\tfrac12(1+C)\,g(r)$ with $g(r)=(r/\arctan r-1)/(1+5r^{2}/3)$. Since $H_{s}$ depends on $r$ alone, \eqref{eq:DStemp} becomes $\Theta_{s}=[\partial_{C}U\,(\mathrm{d}C/\mathrm{d}r)+\partial_{r}U]/H_{s}'(r)$,
in which the path enters only through $\mathrm{d}C/\mathrm{d}r$ and
$\partial_{C}U=U/(1+C)>0$ always. On any fixed-$C$ continuation, the zero is therefore the universal root $r^{*}=1.44382$ of $g'$, independent of $C$. On the isogain continuation, it is $r\simeq1.21$, and on the branch-pairing curves of Section~6.4, it is $r=1.013$--$1.086$ (Figure~\ref{fig:pairingtheta}). For $C\gtrsim1$ at fixed $C$, there is no interior stationary point at all, $U$ is monotone, and $\Theta_{s}$ never changes sign. The sign reversal is thus a property of a continuation and not of a state, which answers, in the negative, the path-independence question raised in Section~6.1 and recorded among the open items of Section~8.2.

Return to the denominator of \eqref{eq:DStemp}. With $H=H_{s}$ it is
$H_{s}'(r)\,\mathrm{d}r/\mathrm{d}s$, and since $H_{s}'>0$ and $r$ increases
monotonically with $E^{*}$ along the scalable branch, it cannot vanish anywhere
on that branch---which is why, for that functional, the zero of $\Theta_{s}$ is
a simple sign change. The same is not true for $H=h$, whose derivative does
vanish at \eqref{eq:hmax}. We take up that case two paragraphs below. Note also that here it is
$\Theta_{s}$ that vanishes while $1/\Theta_{s}$ diverges---the mirror image of
the textbook bounded-spectrum case, where $S(U)$ has an interior maximum and the
temperature passes through infinity \cite{Ramsey1956}. The DS instead has a
bounded \emph{internal energy}.

What is \emph{not} convention-independent is the extent of the negative region. With $H=H_{s}$, $\mathrm{d}H_{s}/\mathrm{d}s>0$ throughout, so $\Theta_{s}<0$ for all $r>1.21$ and the negative branch continues to the resonance. With $H=h$, the numerator and the denominator vanish at different points:
$\mathrm{d}h/\mathrm{d}s$ changes sign at \eqref{eq:hmax}, so $\Theta_{s}$ diverges there and is positive again beyond, and the negative-slope region is the finite window
\begin{equation}
  1.21 \;\lesssim\; r \;\lesssim\; 1.45
  \qquad (\Sigma=0.05:\ 7.1\lesssim E^{*}\lesssim9.2) ,
  \label{eq:negwindow}
\end{equation}
a window specific to this functional and this continuation.
Within the entropy functionals tested, the qualitative change of slope is
reproducible along this particular isogain continuation. Its location, and
indeed its occurrence, are nevertheless continuation-dependent---Section~6.3
computes the discrepancy explicitly---so it should be regarded as a
path-conditioned diagnostic and not as a universal thermodynamic
transition. Its persistence to arbitrarily
large energy is an artifact of the reduced functional. Both diagnostics
exhibit a qualitative change of trend in the vicinity of the fidelity
region, and that is the statement used in Section~6.4.

We do not identify this sign change with an absolute negative
temperature. Negative optical temperatures are not exotic---they were predicted for highly multimoded nonlinear systems as the states in which power flows towards the \emph{highest} modes \cite{Wu2019}, and have been observed directly, together with the associated thermodynamic processes, in a photonic mesh lattice \cite{Muniz2023}, on the classical statistical-mechanical footing of Ref.~\cite{Ramsey1956}. The sharpest available counterexample is worth placing here rather than leaving to Section~7: in a graded-index multimode fiber, light has been observed to thermalize to genuinely negative-temperature Rayleigh--Jeans equilibrium states \cite{Baudin2023}. Those experiments provide a useful benchmark because their
negative-temperature interpretation is anchored to a bounded Hamiltonian
multimode equilibrium and an identified thermodynamic conjugacy. The DS
construction is different: the stationary pulse belongs to a driven-dissipative
system, the slope depends on the selected continuation, and a conjugate
state-variable structure has not been demonstrated. This does not make the sign
reversal meaningless, nor does it invalidate the earlier use of
``effective negative temperature'' when that qualification is explicit. In the
present review we use the more precise term \emph{negative directional
energy--entropy slope} (or negative-\(\Theta_s\) sector) until a path-independent
state-temperature construction is established. Frameworks that begin from an explicit stochastic dynamics---the steady-state thermodynamics of Langevin systems, for instance, in which a Shannon-entropy relation is derived for transitions between nonequilibrium steady states \cite{HatanoSasa2001}---show
what such a demonstration would require, and the missing ingredient there is precisely the one missing here: an identified stochastic process, rather than a deterministic stationary solution. Two-dimensional hydrodynamic turbulence supplies the other familiar instance, where negative temperature describes a discrete system of statistically independent vortices \cite{Kraichnan1980}---with, as noted in
Section~7, the direction of the energy flux reversed relative to the DS case.

A negative value of $\Theta_{s}[H]$ should therefore be read only as a signed directional diagnostic: along this continuation, $U$ decreases while the selected entropy functional increases. If the entropy-based selection rule proposed below is dynamically valid, this region is a \emph{candidate} for replacement of the single-pulse attractor by a multipulse configuration. Whether that replacement occurs is an independent selection problem, to be settled from transition statistics or from basin and quasipotential information rather than from the sign change. From this point of view, $\Theta_{s}$ cannot be interpreted as a state function, and the negative-slope region is a property of the chosen path rather than of the thermodynamic state. What that candidate configuration is, in normal dispersion, is fixed by the algebra of Section~2.

\subsection{Two branches, energy quantization, and fragmentation}

The quadratic \eqref{eq:branches} has \emph{two} roots at the same $(C,\Sigma)$. This is the structural peculiarity of the normal-dispersion problem. The upper root $P_{0}^{+}$ is energy-scalable and carries DSR. The lower root $P_{0}^{-}$ is not scalable and possesses a conservative-soliton limit. Because the isogains are bent, a set of $n+1$ identical, non-interacting $P_{0}^{-}$ pulses can carry the \emph{same} total energy at the \emph{same} saturated net gain as a single $P_{0}^{+}$ pulse \cite{Kalashnikov2025}. The pairing is one-to-one and discrete in both $C$ and $E^{*}$, and it is worth being explicit that the language which follows is analogical: the branch-dividing curve plays
the role of a ``ground state'' $n=0$, the successive complexes are the
``levels'', and the DSR limit $E^{*}\to\infty$ is the $n\to\infty$ accumulation point. This is branch-pairing algebra---a discreteness of admissible stationary configurations at fixed $(C,\Sigma)$---and not a quantization in either the quantum-mechanical or the thermodynamic sense: no quantum of action is involved, no spectrum of a linear operator, and no equilibrium selection principle. We keep the vocabulary because it is standard in the literature \cite{Renninger2010,Tang2005}. A single scalable pulse can therefore \emph{discharge} its energy into a ladder of unscalable ones without violating either constraint.

This kind of energy quantization has its own established literature, and the adiabatic construction provides an underlying mechanism. The area theorem for dissipative optical solitons already implies that the pulse energy delivered by a normal-dispersion oscillator is quantized rather than continuous \cite{Renninger2010}. Multipulse formation with discrete energy steps is the standard route by which fiber lasers respond to increasing pump \cite{Tang2005,Komarov2005}, and the transitions exhibit hysteresis and multistability between pulse numbers \cite{Komarov2013,Komarov2005}. The statistical reading is older still: the formation and annihilation of individual pulse ``quanta'' in a mode-locked laser was described as proceeding along a \emph{thermodynamic-like pathway}, with the pulse number as the order parameter
\cite{Vodonos2004}. Light-mode condensation in actively mode-locked lasers provides the equilibrium counterpart \cite{Weill2010}. What the two branches add is the identity of the states between which the ladder runs: not merely $n$ and $n+1$ copies of the same solution, but a scalable solution and a complex of unscalable ones drawn from a different algebraic root of the same family.

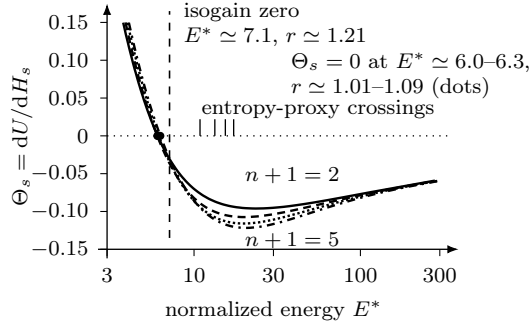
\begin{figure}[t]
\centering
\begin{tikzpicture}[x=1cm,y=1cm,
  ax/.style={draw=black,line width=0.5pt,-{Latex[length=1.5mm]}},
  cv/.style={draw=black,line width=0.9pt}]
  \draw[ax] (-0.12,0) -- (4.65,0);
  \draw[ax] (0,-0.12) -- (0,3.25);
  \foreach \x/\l in {0/3, 1.15/10, 2.2/30, 3.35/100, 4.4/300}{
    \draw (\x,0) -- (\x,-0.09);
    \node[font=\scriptsize,anchor=north] at (\x,-0.11) {\l};}
  \foreach \y/\l in {0/{-0.15}, 0.5/{-0.10}, 1.0/{-0.05}, 1.5/{0}, 2.0/{0.05},
                     2.5/{0.10}, 3.0/{0.15}}{
    \draw (0,\y) -- (-0.09,\y);
    \node[font=\scriptsize,anchor=east] at (-0.11,\y) {$\l$};}
  \draw[line width=0.5pt,dotted] (0,1.5) -- (4.5,1.5);
  \draw[cv] plot coordinates {(0.213,2.997) (0.226,2.942) (0.238,2.888) (0.250,2.834) (0.263,2.782) (0.276,2.729) (0.288,2.678) (0.301,2.627) (0.314,2.576) (0.327,2.526) (0.339,2.477) (0.352,2.429) (0.365,2.381) (0.379,2.334) (0.392,2.287) (0.405,2.241) (0.418,2.196) (0.432,2.151) (0.445,2.107) (0.459,2.063) (0.473,2.021) (0.486,1.978) (0.500,1.937) (0.514,1.896) (0.528,1.855) (0.542,1.816) (0.557,1.777) (0.571,1.738) (0.585,1.700) (0.600,1.663) (0.615,1.626) (0.630,1.590) (0.645,1.555) (0.660,1.520) (0.675,1.486) (0.690,1.452) (0.706,1.419) (0.722,1.387) (0.737,1.355) (0.753,1.324) (0.770,1.293) (0.786,1.263) (0.802,1.234) (0.819,1.205) (0.836,1.177) (0.853,1.150) (0.870,1.123) (0.888,1.097) (0.905,1.071) (0.923,1.046) (0.941,1.021) (0.960,0.997) (0.978,0.974) (0.997,0.951) (1.017,0.929) (1.036,0.907) (1.056,0.886) (1.076,0.866) (1.096,0.846) (1.117,0.827) (1.138,0.808) (1.160,0.790) (1.182,0.773) (1.204,0.756) (1.227,0.740) (1.250,0.724) (1.273,0.709) (1.298,0.695) (1.322,0.681) (1.347,0.667) (1.373,0.655) (1.400,0.643) (1.427,0.631) (1.455,0.620) (1.484,0.610) (1.513,0.600) (1.543,0.591) (1.575,0.583) (1.607,0.575) (1.641,0.568) (1.675,0.562) (1.712,0.556) (1.749,0.552) (1.788,0.547) (1.829,0.544) (1.872,0.542) (1.917,0.540) (1.965,0.539) (2.016,0.540) (2.069,0.541) (2.127,0.544) (2.190,0.548) (2.258,0.554) (2.332,0.561) (2.415,0.571) (2.509,0.584) (2.618,0.600) (2.748,0.621) (2.910,0.649) (3.130,0.689) (3.484,0.755) (4.335,0.908)};
  \draw[cv,densely dashed] plot coordinates {(0.244,3.000) (0.259,2.932) (0.274,2.865) (0.289,2.798) (0.304,2.733) (0.319,2.669) (0.334,2.605) (0.349,2.543) (0.364,2.481) (0.380,2.420) (0.395,2.360) (0.410,2.301) (0.426,2.243) (0.442,2.186) (0.457,2.130) (0.473,2.075) (0.489,2.020) (0.505,1.967) (0.521,1.915) (0.537,1.863) (0.553,1.812) (0.570,1.763) (0.586,1.714) (0.603,1.666) (0.619,1.619) (0.636,1.573) (0.653,1.528) (0.670,1.484) (0.687,1.440) (0.704,1.398) (0.722,1.357) (0.739,1.316) (0.757,1.277) (0.775,1.238) (0.793,1.200) (0.811,1.163) (0.830,1.128) (0.848,1.093) (0.867,1.059) (0.886,1.025) (0.905,0.993) (0.924,0.962) (0.943,0.931) (0.963,0.902) (0.983,0.873) (1.003,0.846) (1.023,0.819) (1.044,0.793) (1.064,0.768) (1.086,0.744) (1.107,0.721) (1.128,0.699) (1.150,0.677) (1.173,0.657) (1.195,0.637) (1.218,0.618) (1.241,0.600) (1.265,0.583) (1.289,0.567) (1.313,0.552) (1.338,0.538) (1.363,0.524) (1.389,0.512) (1.415,0.500) (1.442,0.489) (1.469,0.479) (1.497,0.470) (1.525,0.462) (1.554,0.454) (1.584,0.448) (1.614,0.442) (1.646,0.437) (1.678,0.433) (1.711,0.430) (1.745,0.428) (1.780,0.426) (1.817,0.426) (1.854,0.427) (1.893,0.428) (1.933,0.431) (1.976,0.434) (2.019,0.438) (2.065,0.444) (2.114,0.451) (2.165,0.458) (2.219,0.467) (2.276,0.478) (2.337,0.489) (2.403,0.503) (2.475,0.518) (2.553,0.535) (2.640,0.555) (2.739,0.577) (2.852,0.603) (2.987,0.634) (3.155,0.672) (3.382,0.722) (3.746,0.796) (4.257,0.889)};
  \draw[cv,densely dotted] plot coordinates {(0.271,2.996) (0.287,2.924) (0.302,2.852) (0.318,2.782) (0.334,2.713) (0.350,2.645) (0.366,2.577) (0.383,2.511) (0.399,2.445) (0.415,2.381) (0.432,2.318) (0.448,2.255) (0.465,2.193) (0.482,2.133) (0.498,2.073) (0.515,2.015) (0.532,1.957) (0.549,1.901) (0.566,1.845) (0.584,1.790) (0.601,1.737) (0.619,1.684) (0.636,1.632) (0.654,1.582) (0.672,1.532) (0.690,1.484) (0.708,1.436) (0.726,1.389) (0.744,1.344) (0.763,1.299) (0.781,1.256) (0.800,1.213) (0.819,1.172) (0.838,1.131) (0.857,1.092) (0.876,1.054) (0.896,1.016) (0.915,0.980) (0.935,0.944) (0.955,0.910) (0.975,0.877) (0.995,0.845) (1.016,0.813) (1.037,0.783) (1.057,0.754) (1.078,0.726) (1.100,0.699) (1.121,0.673) (1.143,0.647) (1.165,0.623) (1.187,0.600) (1.209,0.578) (1.232,0.557) (1.255,0.537) (1.278,0.518) (1.301,0.500) (1.325,0.483) (1.349,0.467) (1.374,0.452) (1.398,0.438) (1.423,0.425) (1.449,0.413) (1.475,0.401) (1.501,0.391) (1.528,0.382) (1.555,0.374) (1.582,0.366) (1.610,0.360) (1.639,0.354) (1.668,0.350) (1.698,0.346) (1.728,0.344) (1.759,0.342) (1.791,0.341) (1.823,0.341) (1.856,0.342) (1.891,0.344) (1.926,0.347) (1.962,0.351) (1.999,0.355) (2.037,0.361) (2.077,0.368) (2.118,0.375) (2.161,0.384) (2.205,0.393) (2.251,0.404) (2.300,0.416) (2.351,0.429) (2.404,0.443) (2.461,0.458) (2.521,0.474) (2.586,0.492) (2.655,0.512) (2.731,0.533) (2.814,0.556) (2.906,0.582) (3.011,0.611) (3.133,0.643) (3.279,0.680) (3.465,0.725) (3.726,0.784) (4.193,0.874) (4.381,0.906)};
  \draw[cv,dash dot] plot coordinates {(0.288,2.997) (0.305,2.922) (0.322,2.848) (0.338,2.775) (0.355,2.703) (0.372,2.632) (0.389,2.562) (0.406,2.493) (0.424,2.425) (0.441,2.358) (0.458,2.292) (0.476,2.227) (0.493,2.163) (0.511,2.100) (0.529,2.038) (0.547,1.977) (0.565,1.917) (0.583,1.858) (0.602,1.801) (0.620,1.744) (0.639,1.688) (0.657,1.633) (0.676,1.579) (0.695,1.527) (0.714,1.475) (0.733,1.424) (0.753,1.375) (0.772,1.326) (0.792,1.279) (0.812,1.233) (0.832,1.188) (0.852,1.143) (0.872,1.100) (0.892,1.059) (0.913,1.018) (0.933,0.978) (0.954,0.939) (0.975,0.902) (0.997,0.866) (1.018,0.830) (1.039,0.796) (1.061,0.763) (1.083,0.731) (1.105,0.700) (1.127,0.671) (1.150,0.642) (1.172,0.615) (1.195,0.589) (1.218,0.563) (1.241,0.539) (1.265,0.516) (1.288,0.495) (1.312,0.474) (1.336,0.454) (1.361,0.436) (1.385,0.418) (1.410,0.402) (1.436,0.387) (1.461,0.373) (1.487,0.360) (1.513,0.347) (1.539,0.337) (1.566,0.327) (1.593,0.318) (1.620,0.310) (1.648,0.303) (1.677,0.297) (1.705,0.292) (1.734,0.289) (1.764,0.286) (1.794,0.284) (1.825,0.283) (1.856,0.283) (1.888,0.284) (1.920,0.287) (1.953,0.290) (1.987,0.293) (2.022,0.298) (2.057,0.304) (2.094,0.311) (2.131,0.318) (2.170,0.327) (2.210,0.337) (2.251,0.347) (2.293,0.358) (2.337,0.371) (2.383,0.384) (2.430,0.398) (2.480,0.414) (2.532,0.430) (2.587,0.447) (2.645,0.466) (2.706,0.486) (2.772,0.507) (2.843,0.530) (2.919,0.554) (3.004,0.580) (3.097,0.608) (3.203,0.639) (3.326,0.674) (3.474,0.713) (3.661,0.759) (3.924,0.817) (4.393,0.905)};
  \fill (0.669,1.5) circle (1.5pt);
  \fill (0.664,1.5) circle (1.5pt);
  \fill (0.684,1.5) circle (1.5pt);
  \fill (0.705,1.5) circle (1.5pt);
  \draw[line width=0.6pt,dashed] (0.826,0.15) -- (0.826,3.15);
  \node[font=\scriptsize,anchor=south west,align=left] at (0.88,2.55)
    {isogain zero\\$E^{*}\simeq7.1$, $r\simeq1.21$};
  \node[font=\scriptsize,anchor=west,align=left] at (2.30,2.30)
    {$\Theta_{s}=0$ at $E^{*}\simeq6.0$--$6.3$,\\$r\simeq1.01$--$1.09$ (dots)};
  \foreach \x in {1.229,1.425,1.564,1.674}{
    \draw[line width=0.5pt] (\x,1.5) -- (\x,1.72);}
  \node[font=\scriptsize,anchor=south west] at (1.12,1.6) {entropy-proxy crossings};
  \node[font=\scriptsize,anchor=south] at (2.45,0.74) {$n+1=2$};
  \node[font=\scriptsize,anchor=north] at (2.45,0.36) {$n+1=5$};
  \node[font=\scriptsize,anchor=north] at (2.2,-0.55) {normalized energy $E^{*}$};
  \node[font=\scriptsize,rotate=90,anchor=south] at (-0.85,1.5)
    {$\Theta_{s}=\mathrm{d}U/\mathrm{d}H_{s}$};
\end{tikzpicture}
\caption{The energy--entropy slope $\Theta_{s}$ along the branch-pairing
curves, recomputed here from \eqref{eq:branches}, \eqref{eq:dimensionless},
\eqref{eq:Estar}, \eqref{eq:DSentropy} and \eqref{eq:DSinternal}. Each curve
is the locus on which a single $P_{0}^{+}$ pulse and a complex of $n+1$
identical $P_{0}^{-}$ pulses share the operating point $(C,\Sigma)$ and the
total energy, i.e.\ $E^{*}_{+}(C,\Sigma)=(n+1)E^{*}_{-}(C,\Sigma)$. Solid
$n+1=2$, dashed $3$, dotted $4$, dash-dotted $5$. $\Theta_{s}$ is evaluated
along each such curve and vanishes where $U$ is stationary on it (dots), at
$E^{*}=6.04$, $6.01$, $6.14$ and $6.27$ ($r=1.013$, $1.032$, $1.061$,
$1.086$). The dashed vertical line marks the zero obtained along the isogain
continuation of Section~6.3, $E^{*}\simeq7.1$ at $r\simeq1.21$. The two differ,
which is the content of the path-dependence statement in the text. Ticks above
the axis mark the entropy-proxy crossings obtained above, which all lie at higher energy than the sign reversal. The four curves nearly coincide near
their zeros, so the crossings do not form a well-separated cascade.}
\label{fig:pairingtheta}
\end{figure}

Ref.~\cite{Kalashnikov2025} reports an \(H_s\)-based
single/multipulse crossing at \(E^{*}\approx40\) in Section~III~D
(\(E^{*}\approx30\) in the concluding summary) and a temperature-like sign
change at a lower energy, approximately \(E^{*}\approx20\). The present review
reconstructs the pairing more explicitly. Because \(C\) and \(\Sigma\) specify
the operating point, a \(P_{0}^{+}\) pulse and an \((n+1)P_{0}^{-}\) complex
compared at that point satisfy
\(E^{*}_{+}(C,\Sigma)=(n+1)E^{*}_{-}(C,\Sigma)\). This defines a curve in the
\((C,\Sigma)\) plane for each \(n\). Along those reconstructed curves, the
additive proxy \((n+1)H_s(r_-)\) overtakes \(H_s(r_+)\) at
\(E^{*}=10.85,\,13.33,\,15.43,\) and \(17.29\) for complexes of two to five
pulses, while the directional slope changes sign at
\(E^{*}=6.04,\,6.01,\,6.14,\) and \(6.27\), respectively
(Figure~\ref{fig:pairingtheta}). Thus the qualitative ordering reported in
Ref.~\cite{Kalashnikov2025} is preserved---the slope reversal precedes the
entropy-proxy crossing---although the reconstructed numerical energies are
lower. Because the published article does not document exactly the same
continuation/reconstruction used here, we do not interpret this numerical shift
as a physical contradiction.

Two qualifications are needed. First, a Maxwell construction is more than an intersection of two curves called entropy: it requires a thermodynamic potential, controlled extensive and intensive variables, an extremum principle, and known convexity---none of which is established here (Section~6.5). We therefore call this an \emph{entropy-proxy crossing} and use it as a candidate predictor rather than as a selection rule. Second, the comparison presupposes additivity
\begin{equation}
  H^{(n+1)}(E) \;\stackrel{?}{=}\; (n+1)\,H^{(1)}\!\big(E/(n+1)\big) ,
  \label{eq:additivity}
\end{equation}
\noindent and additivity is exact only if the members of a complex are independent
subsystems. In Ref.~\cite{Kalashnikov2025}, the shared saturable gain is not neglected: the pairing is constructed at \emph{equal} $\Sigma$, so the single pulse and the complex are compared at the same saturated net gain, which is the stationary content of drawing on one reservoir. And the overlap-mediated interaction is empirically small in the regime that supplies the multipulse statistics used here---the
normal-dispersion simulations of Ref.~\cite{Kalashnikov2025} report that the pulses in a set are equal and \emph{well separated} in the vast majority of cases, and it is on those runs that the crossing is evaluated.
Equation~\eqref{eq:additivity} is therefore a working approximation with
evidence behind it in normal dispersion, rather than an assumption made in its absence. Three things nevertheless keep the interaction problem open, and they differ in kind.

\emph{(i) The separability is empirical and parameter dependent.} The
same source records that the probability of unequal or interacting pulses grows with the SAM saturation $\zeta$, and that those regimes are not yet characterized \cite{Kalashnikov2025}. The qualification is therefore about a region of parameter space rather than the construction, and separability must be verified at the operating points where the crossing is evaluated rather than assumed across the diagram.

\emph{(ii) Where pulses do approach, the correction is neither small nor
of fixed sign.} Evaluating the Hamiltonian of the conservative part of a
parameter-free GL equation on a superposition of two pulses gives an effective
interaction potential whose local minima predict bound states at definite
separations. Direct integration confirms them, and the separations---about $3.6$
for the $\pi$-out-of-phase pair in one model, about $4.7$ (in phase) and $6.0$
(out of phase) in the other---are comparable to the internal width of a single
pulse rather than large compared with it \cite{Malomed1997}. Which phase
relation carries the robust bound state \emph{reverses} between the two models,
following the sign of the effective Hamiltonian, so the correction cannot be
argued away as uniformly small. And whether a dense array is more or less
durable than an isolated pulse is itself model-dependent: nonlinear losses
stabilize a densely packed array, whereas under bandwidth-limited gain with
nonlinear dispersion the single pulse outlives one. Two scope limits
belong with this before it is read as a refutation of
\eqref{eq:additivity}: those are simplified GL equations without SAM saturation,
and their pulses are not the strongly chirped, heavily stretched solutions
considered here. What the result describes is the configuration in which
additivity \emph{would} fail---separations of a few internal widths---and not
the well-separated configuration the simulations report. Which of the two the
dynamics selects at a given operating point is precisely what is unsettled.

\emph{(iii) Even perfect separation leaves a choice in the proxy.} For
$n+1$ identical pulses at separations $\tau_{j}$ the field is
$e(\omega)\sum_{j}e^{i\omega\tau_{j}}$, so the spectrum of the complex is the
single-pulse spectrum multiplied by an interference factor with fringe spacing
set by $\tau_{j}$. The additive value $(n+1)H_{s}(r_{-})$ is recovered only
after coarse-graining over those fringes. This is the reference-measure question
of Section~6.1 in another guise---here the resolution at which the complex is
observed---and it is a property of the proxy rather than of the pulses.

An additive proxy prices none of (i)--(iii). The bound-state reading is
better regarded as a
selection principle of a different type---energetic and pairwise, rather than
entropic and collective---and a decisive comparison would have to evaluate both
on the same complex, along the lines set out in Section~8.2.

\emph{A competing prediction, and the channel each mechanism acts on.} This conclusion must be set against a result that appears to contradict it directly. Simulations of a mode-locked laser under resonance conditions report that DSR \emph{suppresses} multipulsing rather than promoting it: as the pump is raised, the rectangular pulse acquires the narrow spectral core described in Section~4.1, its overlap with the finite gain band improves accordingly, it depletes the inversion more efficiently, and the net gain seen by weak radiation elsewhere in the cavity is driven \emph{more} negative, so that no new pulse can be seeded from noise, and the energy of the single pulse grows without bound with
the pump \cite{Komarov2013}. The authors read the word ``resonance'' literally, as the improved resonant coupling of a spectrally narrowed pulse to the gain medium---a reading that the chemical-potential picture of Section~5.1 supports rather than displaces, since $\Xi\to0$ is exactly what narrows the core.

The two statements are compatible because they concern different channels. Suppression is a statement about \emph{nucleation}---whether a fluctuation in the empty part of the cavity can grow---and its criterion is the sign of the net gain outside the pulse, which is the vacuum-stability threshold $\Sigma=0$ of Section~2.3. The two descriptions parameterize gain saturation differently, however, so the direction of motion in $\Sigma$ cannot be read across from one to the other without an explicit mapping: under the terminology used here, $\Sigma\to0^{+}$ \emph{is} approach to the vacuum-stability boundary, and the comparison should not be made until that mapping is supplied. The entropic argument is a statement about \emph{redistribution}---whether energy already carried by one scalable pulse is better held by several unscalable ones---and the paired states at equal total energy \emph{and} equal saturated net gain. It therefore requires no
positive gain in the vacuum at any point, and is untouched by the suppression mechanism. More importantly, the suppression of pulse nucleation from the background does not by itself exclude a different transition, in which an already existing high-energy pulse redistributes into a multipulse attractor. If the entropy-proxy selection rule proposed here is valid, such a redistribution would be a candidate channel under DSR conditions. The two further observations of Ref.~\cite{Komarov2013} are compatible with this interpretation, though they do not establish its entropic origin: the multipulse state is reached from initial conditions rather than by a pump-driven bifurcation, and once several pulses are present they are closely similar in duration, shape, peak power and chirp---which is the empirical content of ``a set of $n+1$ identical, non-interacting $P_{0}^{-}$ pulses'', a phrase to be read as an idealization: pulses sharing one filter and a finite gain-recovery time are approximately identical but not dynamically independent. The shared gain reservoir is not part of that list, since the paired states are constructed at equal $\Sigma$. The
distinction is also experimentally sharp, and worth stating as a test: a
growing continuum background should precede the nucleation transition between pulses; a redistribution transition should not.

This is the nonequilibrium counterpart of the exchange of stability between the two minima of $f(\gamma,y)$ at $\gamma^{*}$ in the noise-driven theory, where the same construction produces coexistence, metastability, and hysteresis between ordered and disordered phases \cite{GordonFischer2002,Gat2004}. A first-order character is not the exclusive property of driven systems: as noted in Section~3.1, classical wave condensation is itself subcritical once the interaction energy discarded by the kinetic closure is restored through a Bogoliubov treatment \cite{Connaughton2005,During2009}.

Conservative soliton theory supplies a precedent for the \emph{process} as well, and it is worth separating from the precedent for the outcome. On an $H$--$Q$ diagram carrying a cusp, a soliton prepared on the unstable branch does not disintegrate: it sheds small-amplitude radiation. It settles onto the stable branch, moving down and to the left on the diagram, and direct simulation confirms the transformation that the concavity criterion predicts \cite{Akhmediev1999HQ}. The DS transition is an event of the same kind---a two-branch family, one branch not sustainable, and a
relaxation carrying the solution from one to the other---with two differences worth keeping in view. There the destination is a single soliton of \emph{lower} energy, the surplus leaving as radiation. Here, it is a complex of several pulses at the \emph{same} total energy and the same saturated net gain. And there, the surviving branch is identified by a theorem. Here an entropy proxy has only been proposed, and the available simulations provide qualitative multipulse statistics rather than a validation of it. 

There is also a conservative precedent for the \emph{outcome}. In multimode fibers with strong random mode coupling, the accessible steady states at intermediate energy are not the globally condensed one but a set of local condensates in higher-order mode groups---the ``glassy'' states of Section~3.2 \cite{Zitelli2024}. Energy distributed over several sub-condensates can be preferred to the same energy placed in one. The DS multipulse complex is the proposed temporal counterpart, with the
selection conjectured to be entropic rather than kinetic---a conjecture that the present calculations motivate but do not test.

The stochastic NGD simulations of Ref.~\cite{Kalashnikov2025} are important because they already display the dynamical phenomenon that the thermodynamic analogy is intended to describe. As energy increases, the final-state statistics move toward multipulsing, while the self-start dynamics show that the single-pulse state becomes progressively less favorable than a multiple-pulse state. This is consistent with a crossover between competing attractors rather than with an abrupt disappearance of the single-DS branch. The entropy-proxy crossing proposed above is therefore best regarded as a candidate marker of that redistribution, not as an independently established cause of it. The key test is conceptual: does the thermodynamic-like indicator locate the change of preferred attractor better than the existence boundary alone?

\subsection{Anomalous dispersion, and the status of the criteria}

Equation \eqref{eq:DSentropy} was derived for a compactly supported
truncated Lorentzian, so it cannot simply be evaluated on the anomalous
spectrum, which has algebraic wings. What those wings do and do not spoil
has to be stated precisely, because the paper has previously stated it too
broadly. For the off-resonant tail $S_{\mathrm{AGD}}\sim|\omega|^{-3}$ of
\eqref{eq:AGDtail}, both the normalization and the differential entropy converge
without any external bound,
\begin{equation}
  \int^{\infty}\!\!\omega^{-3}\,\mathrm{d}\omega<\infty ,
  \qquad
  -\!\int^{\infty}\!\!S\ln S\,\mathrm{d}\omega
  \sim\int^{\infty}\!\!\omega^{-3}\ln\omega\,\mathrm{d}\omega<\infty ,
\end{equation}
whereas the second moment does not:
$\int^{\infty}\omega^{2}S\,\mathrm{d}\omega\sim\int^{\infty}
\mathrm{d}\omega/\omega$ diverges logarithmically. The quantity that requires
the apparatus is therefore not the spectrum but the internal-energy proxy
\eqref{eq:DSinternal}---and, through $\Theta_{s}=\mathrm{d}U/\mathrm{d}H$, every
slope built on it. On the resonance locus itself the situation is worse and the
earlier statement becomes true: the tail softens towards $|\omega|^{-1}$ (this
is the content of \eqref{eq:AGDlogdiv}), and the normalization is then
logarithmically cutoff-sensitive as well. Both facts argue for the capture
protocol below, but they argue for it in different places, and the distinction
should be kept. We therefore derive the anomalous indicators separately.

The spectrum is the stationary-phase envelope of Ref.~\cite{Kalashnikov2026}, which we quote in full rather than through its tail \eqref{eq:AGDtail}: with $\Upsilon(\tilde\omega)=\sqrt{1+4\tilde\chi(\tilde\omega^{2}+D_{a})}>1$,
$D_{a}=-\tilde\Delta_{-}^{2}>0$ and $G$ as in Section~5.2
\begin{equation}
  S_{\mathrm{AGD}}(\tilde\omega) \;\propto\;
  \frac{4\tilde\chi^{2}\,\big[(D_{a}+2\tilde\omega^{2})\Upsilon
        +\tilde\omega^{2}\big]}
       {(\Upsilon-1)\,\Upsilon\,G(\tilde\omega)^{2}},
  \label{eq:AGDspectrum}
\end{equation}
\noindent which reproduces \eqref{eq:AGDtail} as $|\tilde\omega|\to\infty$.
The measurement protocol is then fixed as follows. Because the spectrum is
even in $\omega$, the ordinary one-sided cumulative-distribution quantile is the
wrong object here: it would give $\mathcal{Q}(1/2)=0$ and could not serve as a
positive half-width of the spectral core. What the construction requires is a
\emph{centered cumulative-energy radius},
\begin{equation}
  \mathcal{Q}_{c}(f)
  \;=\;
  \inf\!\left\{\Omega\ge0:\;
  \frac{\displaystyle\int_{-\Omega}^{\Omega}S(\omega)\,\mathrm{d}\omega}
       {\displaystyle\int_{-\infty}^{\infty}S(\omega)\,\mathrm{d}\omega}
  \;\ge\; f\right\} ,
  \qquad 0<f<1 .
  \label{eq:centeredquantile}
\end{equation}
\begin{equation}
  \begin{aligned}
    \omega_{\mathrm{cap}}(E) &= \min\{\mathcal{Q}_{c}(1-\eta),\,
      \omega_{\max}\}, &\omega_{\max}&=\min\{\omega_{\mathrm{f}},
      \omega_{\mathrm{det}}\}, \\
    \omega_{\mathrm{core}}(E) &= \mathcal{Q}_{c}^{(\mathrm{cap})}
      (f_{\mathrm{core}};\omega_{\mathrm{cap}}), &
    \tilde p &= S_{\mathrm{AGD}}\big/\!\!\int_{-\omega_{\mathrm{cap}}}
      ^{\omega_{\mathrm{cap}}}\!\! S_{\mathrm{AGD}}\,\mathrm{d}\omega ,
  \end{aligned}
  \label{eq:AGDprotocol}
\end{equation}
\noindent with $\delta\omega$ the fixed bin. Because $\mathcal{Q}_{c}$ is
referred to the \emph{total} spectral energy, a second radius is needed for
quantities defined inside the capture interval:
\begin{equation}
  \mathcal{Q}_{c}^{(\mathrm{cap})}(f;\omega_{\mathrm{cap}})
  \;=\;
  \inf\!\left\{0\le\Omega\le\omega_{\mathrm{cap}}:\;
  \frac{\displaystyle\int_{-\Omega}^{\Omega}S(\omega)\,\mathrm{d}\omega}
       {\displaystyle\int_{-\omega_{\mathrm{cap}}}^{\omega_{\mathrm{cap}}}
        S(\omega)\,\mathrm{d}\omega}
  \;\ge\; f\right\} ,
  \label{eq:capturedcenteredquantile}
\end{equation}
and we set $\omega_{\mathrm{core}}=\mathcal{Q}_{c}^{(\mathrm{cap})}
(f_{\mathrm{core}};\omega_{\mathrm{cap}})$ in \eqref{eq:AGDprotocol}. Thus
$f_{\mathrm{core}}=1/2$ means that the symmetric interval
$[-\omega_{\mathrm{core}},\omega_{\mathrm{core}}]$ contains one half of the
\emph{captured} spectral energy; it is \emph{not} the median of a one-sided
cumulative distribution, and it is not the half-total-energy radius
$\mathcal{Q}_{c}(1/2)$ either. The distinction is quantitatively immaterial in
the pre-saturation regime to which the results below are restricted, where the
quantile rule rather than $\omega_{\max}$ sets $\omega_{\mathrm{cap}}$ and the
two radii differ by less than $2\%$; it is not immaterial once the extrinsic cap
binds, since the half-captured radius then falls short of the half-total radius
by tens of percent. Every anomalous quantity below inherits this definition through $\omega_{\mathrm{core}}$, $\omega_{\mathrm{cap}}$ and their ratio. The entropy splits as
$h=\ln(\omega_{\mathrm{core}}/\delta\omega)+h_{x}$, exactly as in
\eqref{eq:splitx}. For the internal energy we use only the \emph{moment
definition} of Eq.~\eqref{eq:DSinternal}
\begin{equation}
  U_{\mathrm{cap}}
  \;=\;
  \frac{1}{2}\int_{-\omega_{\mathrm{cap}}}^{\omega_{\mathrm{cap}}}
  \tilde p_{\mathrm{cap}}(\omega)\,\omega^{2}\,\mathrm{d}\omega ,
  \label{eq:AGDinternal}
\end{equation}
and \emph{not} the closed-form expression on the second line of
\eqref{eq:DSinternal}, which is specific to the normal-dispersion truncated
Lorentzian. The slope diagnostic \eqref{eq:DStemp} is then evaluated along the
stated continuation of this captured distribution. The distinction matters
here rather than being pedantic: the $|\omega|^{-3}$ tail of
\eqref{eq:AGDtail} is slow enough that the second moment acquires an explicit
cutoff sensitivity as the resonance is approached, which the Lorentzian closed
form would conceal. The
continuation is $C\to-1/\tilde\chi$ at fixed $(\tilde\chi,\Sigma)$, i.e.\ the
approach to the chirp-control line of Section~5.2.

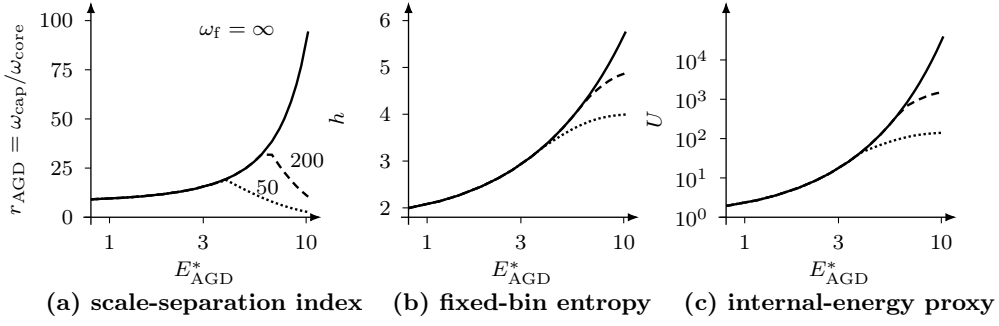
\begin{figure}[t]
\centering
\begin{tikzpicture}[x=1cm,y=1cm,
  ax/.style={draw=black,line width=0.5pt,-{Latex[length=1.5mm]}},
  cv/.style={draw=black,line width=0.9pt}]
\begin{scope}[shift={(0,0)}]
  \draw[ax] (-0.10,0) -- (3.05,0);
  \draw[ax] (0,-0.10) -- (0,2.85);
  \draw (0.246,0) -- (0.246,-0.08);
  \node[font=\scriptsize,anchor=north] at (0.246,-0.10) {1};
  \draw (1.486,0) -- (1.486,-0.08);
  \node[font=\scriptsize,anchor=north] at (1.486,-0.10) {3};
  \draw (2.846,0) -- (2.846,-0.08);
  \node[font=\scriptsize,anchor=north] at (2.846,-0.10) {10};
  \draw (0,0.0) -- (-0.08,0.0);
  \node[font=\scriptsize,anchor=east] at (-0.10,0.0) {0};
  \draw (0,0.65) -- (-0.08,0.65);
  \node[font=\scriptsize,anchor=east] at (-0.10,0.65) {25};
  \draw (0,1.3) -- (-0.08,1.3);
  \node[font=\scriptsize,anchor=east] at (-0.10,1.3) {50};
  \draw (0,1.95) -- (-0.08,1.95);
  \node[font=\scriptsize,anchor=east] at (-0.10,1.95) {75};
  \draw (0,2.6) -- (-0.08,2.6);
  \node[font=\scriptsize,anchor=east] at (-0.10,2.6) {100};
\draw[cv] plot coordinates {(0.000,0.237) (0.414,0.257) (0.661,0.275) (0.959,0.306) (1.184,0.339) (1.372,0.375) (1.620,0.442) (1.791,0.506) (1.901,0.559) (2.043,0.644) (2.157,0.733) (2.256,0.829) (2.394,1.002) (2.493,1.169) (2.559,1.305) (2.646,1.527) (2.718,1.755) (2.782,2.004) (2.873,2.457)};
\draw[cv,densely dashed] plot coordinates {(0.000,0.237) (0.414,0.257) (0.661,0.275) (0.959,0.306) (1.184,0.339) (1.372,0.375) (1.620,0.442) (1.791,0.506) (1.901,0.559) (2.043,0.644) (2.157,0.733) (2.256,0.829) (2.394,0.826) (2.493,0.681) (2.559,0.593) (2.646,0.488) (2.718,0.410) (2.782,0.348) (2.873,0.271)};
\draw[cv,densely dotted] plot coordinates {(0.000,0.237) (0.414,0.257) (0.661,0.275) (0.959,0.306) (1.184,0.339) (1.372,0.375) (1.620,0.442) (1.791,0.492) (1.901,0.433) (2.043,0.361) (2.157,0.307) (2.256,0.262) (2.394,0.206) (2.493,0.170) (2.559,0.148) (2.646,0.122) (2.718,0.103) (2.782,0.087) (2.873,0.068)};
  \node[font=\scriptsize,anchor=west] at (1.30,2.45) {$\omega_{\mathrm{f}}=\infty$};
  \node[font=\scriptsize,anchor=west] at (2.5,0.75) {$200$};
  \node[font=\scriptsize,anchor=west] at (2.05,0.4) {$50$};
  \node[font=\scriptsize,anchor=north] at (1.5,-0.45) {$E^{*}_{\mathrm{AGD}}$};
  \node[font=\footnotesize\bfseries,anchor=north] at (1.5,-0.85) {(a) scale-separation index};
  \node[font=\scriptsize,rotate=90,anchor=south] at (-0.72,1.3) {$r_{\mathrm{AGD}}=\omega_{\mathrm{cap}}/\omega_{\mathrm{core}}$};
\end{scope}
\begin{scope}[shift={(4.2,0)}]
  \draw[ax] (-0.10,0) -- (3.05,0);
  \draw[ax] (0,-0.10) -- (0,2.85);
  \draw (0.246,0) -- (0.246,-0.08);
  \node[font=\scriptsize,anchor=north] at (0.246,-0.10) {1};
  \draw (1.486,0) -- (1.486,-0.08);
  \node[font=\scriptsize,anchor=north] at (1.486,-0.10) {3};
  \draw (2.846,0) -- (2.846,-0.08);
  \node[font=\scriptsize,anchor=north] at (2.846,-0.10) {10};
  \draw (0,0.124) -- (-0.08,0.124);
  \node[font=\scriptsize,anchor=east] at (-0.10,0.124) {2};
  \draw (0,0.743) -- (-0.08,0.743);
  \node[font=\scriptsize,anchor=east] at (-0.10,0.743) {3};
  \draw (0,1.362) -- (-0.08,1.362);
  \node[font=\scriptsize,anchor=east] at (-0.10,1.362) {4};
  \draw (0,1.981) -- (-0.08,1.981);
  \node[font=\scriptsize,anchor=east] at (-0.10,1.981) {5};
  \draw (0,2.6) -- (-0.08,2.6);
  \node[font=\scriptsize,anchor=east] at (-0.10,2.6) {6};
\draw[cv] plot coordinates {(0.000,0.121) (0.414,0.213) (0.661,0.288) (0.959,0.405) (1.184,0.515) (1.372,0.624) (1.620,0.797) (1.791,0.938) (1.901,1.039) (2.043,1.184) (2.157,1.313) (2.256,1.434) (2.394,1.620) (2.493,1.768) (2.559,1.872) (2.646,2.019) (2.718,2.149) (2.782,2.271) (2.873,2.455)};
\draw[cv,densely dashed] plot coordinates {(0.000,0.121) (0.414,0.213) (0.661,0.288) (0.959,0.405) (1.184,0.515) (1.372,0.624) (1.620,0.797) (1.791,0.938) (1.901,1.039) (2.043,1.184) (2.157,1.313) (2.256,1.434) (2.394,1.595) (2.493,1.685) (2.559,1.739) (2.646,1.801) (2.718,1.843) (2.782,1.873) (2.873,1.906)};
\draw[cv,densely dotted] plot coordinates {(0.000,0.121) (0.414,0.213) (0.661,0.288) (0.959,0.405) (1.184,0.515) (1.372,0.624) (1.620,0.797) (1.791,0.934) (1.901,1.005) (2.043,1.094) (2.157,1.159) (2.256,1.209) (2.394,1.267) (2.493,1.299) (2.559,1.316) (2.646,1.333) (2.718,1.344) (2.782,1.351) (2.873,1.359)};

  \node[font=\scriptsize,anchor=north] at (1.5,-0.45) {$E^{*}_{\mathrm{AGD}}$};
  \node[font=\footnotesize\bfseries,anchor=north] at (1.5,-0.85) {(b) fixed-bin entropy};
  \node[font=\scriptsize,rotate=90,anchor=south] at (-0.72,1.3) {$h$};
\end{scope}
\begin{scope}[shift={(8.4,0)}]
  \draw[ax] (-0.10,0) -- (3.05,0);
  \draw[ax] (0,-0.10) -- (0,2.85);
  \draw (0.246,0) -- (0.246,-0.08);
  \node[font=\scriptsize,anchor=north] at (0.246,-0.10) {1};
  \draw (1.486,0) -- (1.486,-0.08);
  \node[font=\scriptsize,anchor=north] at (1.486,-0.10) {3};
  \draw (2.846,0) -- (2.846,-0.08);
  \node[font=\scriptsize,anchor=north] at (2.846,-0.10) {10};
  \draw (0,0.0) -- (-0.08,0.0);
  \node[font=\scriptsize,anchor=east] at (-0.10,0.0) {$10^{0}$};
  \draw (0,0.52) -- (-0.08,0.52);
  \node[font=\scriptsize,anchor=east] at (-0.10,0.52) {$10^{1}$};
  \draw (0,1.04) -- (-0.08,1.04);
  \node[font=\scriptsize,anchor=east] at (-0.10,1.04) {$10^{2}$};
  \draw (0,1.56) -- (-0.08,1.56);
  \node[font=\scriptsize,anchor=east] at (-0.10,1.56) {$10^{3}$};
  \draw (0,2.08) -- (-0.08,2.08);
  \node[font=\scriptsize,anchor=east] at (-0.10,2.08) {$10^{4}$};
\draw[cv] plot coordinates {(0.000,0.149) (0.414,0.226) (0.661,0.290) (0.959,0.389) (1.184,0.485) (1.372,0.581) (1.620,0.735) (1.791,0.864) (1.901,0.959) (2.043,1.095) (2.157,1.219) (2.256,1.337) (2.394,1.521) (2.493,1.670) (2.559,1.777) (2.646,1.929) (2.718,2.065) (2.782,2.193) (2.873,2.391)};
\draw[cv,densely dashed] plot coordinates {(0.000,0.149) (0.414,0.226) (0.661,0.290) (0.959,0.389) (1.184,0.485) (1.372,0.581) (1.620,0.735) (1.791,0.864) (1.901,0.959) (2.043,1.095) (2.157,1.219) (2.256,1.337) (2.394,1.474) (2.493,1.529) (2.559,1.561) (2.646,1.597) (2.718,1.622) (2.782,1.639) (2.873,1.658)};
\draw[cv,densely dotted] plot coordinates {(0.000,0.149) (0.414,0.226) (0.661,0.290) (0.959,0.389) (1.184,0.485) (1.372,0.581) (1.620,0.735) (1.791,0.858) (1.901,0.903) (2.043,0.957) (2.157,0.996) (2.256,1.026) (2.394,1.061) (2.493,1.081) (2.559,1.091) (2.646,1.102) (2.718,1.108) (2.782,1.113) (2.873,1.118)};

  \node[font=\scriptsize,anchor=north] at (1.5,-0.45) {$E^{*}_{\mathrm{AGD}}$};
  \node[font=\footnotesize\bfseries,anchor=north] at (1.5,-0.85) {(c) internal-energy proxy};
  \node[font=\scriptsize,rotate=90,anchor=south] at (-0.72,1.3) {$U$};
\end{scope}
\end{tikzpicture}
\caption{Anomalous-dispersion indicators, computed here from the spectrum \eqref{eq:AGDspectrum} under \eqref{eq:AGDprotocol} with
$\tilde\chi=1.5$, $\Sigma=0.02$, $\eta=0.02$,
$f_{\mathrm{core}}=0.5$, and the continuation $C\to-1/\tilde\chi$ at fixed $(\tilde\chi,\Sigma)$. Solid: no hard cutoff beyond the quantile-capture rule \eqref{eq:AGDprotocol}. Dashed and dotted: the occupied window is capped at $\omega_{\mathrm{f}}=200$ and $50$ respectively.
(\textbf{a})~Without the spectral cutoff, the scale-separation index rises towards the resonance. With extrinsic cutoff, $\omega_{\mathrm{cap}}$
saturates while $\omega_{\mathrm{core}}$ keeps growing and $r_{\mathrm{AGD}}$ \emph{falls}. (\textbf{b})~The fixed-bin differential entropy rises
monotonically and does not turn over, unlike its normal-dispersion counterpart
(Figure~\ref{fig:thermo}b), because here the scale term
$\ln(\omega_{\mathrm{core}}/\delta\omega)$ rises with the shape term instead of
competing with it. (\textbf{c})~The internal-energy proxy likewise rises
monotonically, with no maximum. Hence $\Theta_{s}=\mathrm{d}U/\mathrm{d}h>0$
throughout: the sign reversal of Section~6.3 has no anomalous-dispersion
counterpart on this path.}
\label{fig:agdentropy}
\end{figure}

The outcome turns on a single sign (Figure~\ref{fig:agdentropy}). In normal dispersion, the core collapses,
$\Xi\to0$, so the scale term $\ln(\Xi/\delta\omega)$ \emph{falls} while the shape term rises, and the competition between them produces the turnover \eqref{eq:hmax} and, through the maximum of $U$, the sign reversal of Section~6.3. In anomalous dispersion the occupied spectrum \emph{expands} instead: both
$\omega_{\mathrm{core}}$ and $\omega_{\mathrm{cap}}$ grow towards the resonance,
so the two terms reinforce rather than compete. One point of vocabulary has to
be settled here, because the word ``core'' is doing two jobs. The \emph{two-horn
envelope} is nearly invariant along the near-resonant energy-scaling path---that
is the self-similarity \eqref{eq:selfsimilar}, and it is what Section~5.3 refers
to. The \emph{centered radius} $\omega_{\mathrm{core}}=\mathcal{Q}_{c}^{(\mathrm{cap})}(f_{\mathrm{core};\omega_{\mathrm{cap}})}$
is a different object: as $1+C\tilde\chi\to0$ the tail prefactor of
\eqref{eq:AGDtail} grows, a larger share of the energy moves into the wings, and
the quantile therefore moves outward even while the envelope shape is
unchanged. The two statements are compatible, and they refer to different
continuations---fixed $C$ with varying $E$ in Section~5.3, varying $C$ at fixed
$(\tilde\chi,\Sigma)$ here---but they are not interchangeable, and
$\varrho\sim\omega_{\mathrm{core}}^{-1}$ should not be described as
``core-controlled'' without that qualification. The consequences are that $h$ rises monotonically with no maximum, that $U$ rises monotonically with no maximum,
and therefore that $\Theta_{s}=\mathrm{d}U/\mathrm{d}h>0$ throughout. We find no turnover and no sign reversal anywhere in the admissible window, for $\tilde\chi\in[1.2,3.0]$ and for every $\Sigma$ tested, with or without an extrinsic spectral cutoff. How far that null result generalizes is constrained by the paper's own earlier finding, because Section~6.3 has just shown that $\Theta_{s}$ is \emph{path-dependent}: its zero sits at $r^{*}=1.44382$ on fixed-$C$ continuations, at $r\simeq1.21$ on isogains, at $r\simeq1.01$--$1.09$ on the branch-pairing curves, and nowhere at all for $C\gtrsim1$ at fixed $C$. In normal dispersion, then, whether one finds a sign reversal already depends on which curve one walks along. The anomalous computation walks along exactly one curve---$C\to-1/\tilde\chi$ at fixed $(\tilde\chi,\Sigma)$---with one entropy functional, one quantile protocol and one pair of conventions $(\eta,f_{\mathrm{core}})$. We therefore draw a deliberately restricted conclusion. Under the capture
rule, entropy functional and continuation specified above, the AGD branch
exhibits no analogue of the normal-dispersion entropy turnover or
energy--entropy-slope reversal. A negative result under those conditions is evidence that the particular normal-dispersion entropic criterion of Section~6.3 does not transpose to this continuation along the natural analog of the isogain. It is not evidence that no entropic limitation can exist elsewhere in anomalous parameter space, or for a different physically motivated functional, and we do not claim the stronger statement. Put in the form in which it is meant to be used: the absence of a universal entropy turnover in anomalous dispersion is a \emph{null result}, and within the present reduced description no path-independent thermodynamic selection principle has been established for this branch. That is a stronger and more useful scientific conclusion than any attempt to rescue a universal transition would be, because it is the statement that the proposed quantities are probes whose physical meaning must be tested rather than assumed. What can be said without qualification is narrower: \emph{the particular competition of terms that produces the normal-dispersion turnover---a falling scale term against a rising shape term---has no anomalous counterpart, because there the two terms have the same sign.} That is a structural statement about \eqref{eq:splitx}, and it holds throughout the quantile-capture family investigated here, up to aperture saturation. It is not protocol-free: once $\omega_{\mathrm{cap}}$ locks onto the extrinsic cutoff while $\omega_{\mathrm{core}}$ keeps growing, the shape term turns over while the scale term does not, and the two terms compete again. The claim is therefore about the pre-saturation regime, which is where the capture rule \eqref{eq:AGDprotocol} was designed to sit. 

Three things would strengthen or overturn it: the same computation along a fixed-$C$ continuation and along an anomalous analogue of the branch-pairing curves, if one can be defined; a functional that weights the wings rather than the core, since it is the wings that carry the anomalous divergence \eqref{eq:AGDlogdiv}; and an anomalous branch pair, which the algebra does not at present supply. This is consistent with the anomalous limit being the finite-time dynamical accessibility boundary reported in Ref.~\cite{Kalashnikov2026} and described below, rather than an entropic one, but consistency is all it is: a null result under one continuation, one capture rule and one family of entropy-like functionals does not show that no entropy-related mechanism contributes to the loss of accessibility.

The $r_{\mathrm{AGD}}$-sensitivity is larger than in normal dispersion, and larger in kind. The index behaves as
$r_{\mathrm{AGD}}\sim|1+C\tilde\chi|^{-\nu}$ with
$\nu=0.271$, $0.312$, $0.343$ and $0.357$ for $\eta=0.10$, $0.05$, $0.02$ and
$0.01$ at $f_{\mathrm{core}}=0.5$, and $\nu=0.465$, $0.343$ and $0.211$ for
$f_{\mathrm{core}}=0.3$, $0.5$ and $0.7$ at $\eta=0.02$; asymptotically
$\nu\to1-\eta-f_{\mathrm{core}}$, an empirical envelope rather than a derived law: it reproduces the direction of both trends but overshoots the fitted exponents by $25$--$40\%$ over the range tabulated, so it should be quoted as an ordering rule and not as a formula. The measurement convention therefore enters the \emph{exponent} here, whereas in normal dispersion it enters only as the multiplicative width convention of Section~4.3. Worse, the direction is not safe either: once $\omega_{\mathrm{cap}}$ reaches the extrinsic spectral cutoff, it saturates while $\omega_{\mathrm{core}}$ continues to grow, so $r_{\mathrm{AGD}}$ passes through a maximum and \emph{decreases} (Figure~\ref{fig:agdentropy}a). What survives without qualification is only the ordering, and only while the aperture is not limiting. Any anomalous-dispersion number quoted from these indicators must therefore carry $\eta$,
$f_{\mathrm{core}}$, $\delta\omega$ and $\omega_{\max}$ with it.

The \emph{quantization} half does not transpose either. In anomalous dispersion only the minus branch supports
a strongly chirped solution \cite{Kalashnikov2026}, so there is no scalable and
unscalable pair at the same operating point, and the ladder of
Section~6.4 has no direct counterpart. What is observed instead, within the calculation available here, is a
finite-time dynamical accessibility boundary, identified independently of the
present entropy-like indicators and located by a linearized quantum-noise
calculation rather than by any functional of the spectrum
\cite{Kalashnikov2026}. This does not establish that the anomalous-dispersion
energy limit is generically non-entropic; it shows only that the particular
normal-dispersion proxy developed above does not identify the boundary in this
calculation. Since it is the mechanism quantified by the calculation available here
standing in anomalous dispersion, it is set out in full in Section~6.6 rather
than cited in passing. Whether an entropic criterion can
be formulated for the anomalous-dispersion breakup, and how it would relate to that boundary, is open.

An important caveat accompanies all of these statements. Because the system is
genuinely out of equilibrium---and, unlike the noise-driven model, does not admit
an invariant Gibbs measure once dispersion and reactive nonlinearity are retained
\cite{Gat2004}---the standard equilibrium criteria transfer only partially.
Free-energy minimization does \emph{not} select the stable state: the free energy
$\mathcal{F}$ of \eqref{eq:DStemp} can be evaluated, and the equality of free
energies between single- and multipulse states can be recorded, but its
minimization is not a criterion of dynamic stability, and in the absence of a
Gibbs measure the equality carries none of the meaning a Maxwell construction
would give it \cite{Kalashnikov2025}. Nor does the Lagrangian route restore one. A variational
formulation of \eqref{eq:cqgle} does exist, but only in the extended sense of
Section~2.1, in which the dissipative terms are a source on the right-hand
side rather than part of the functional; its stationarity conditions therefore
reproduce the energy-balance relation and the reduced flow, not a minimum
principle \cite{Ankiewicz2007}. This is the concrete content of the contrast
drawn in Section~2.3: the conservative theory selects between coexisting
branches by a theorem on the concavity of $H(Q)$ \cite{Akhmediev1999HQ}, and
nothing in the dissipative problem plays that role. The entropy and internal-energy crossings are best
understood as \emph{thermodynamic-like} indicators---Maxwell-point and spinodal
analogues---whose predictive content must be corroborated by direct dynamical
simulation, and for which the simulations available so far supply qualitative
support rather than validation of the crossing itself
\cite{Kalashnikov2024,Kalashnikov2025}. Making the status of
these analogies explicit is one of the tasks of this paper.

These predictions are testable with an experimental protocol that need not be
invented, because its instrumentation already exists in the multimode
literature---provided one point of principle is respected. The multimode
measurements report the Boltzmann-type configuration entropy
$\tilde{S}=\sum_{i}\ln|f_{i}|^{2}$ computed from \emph{normalized} occupancies,
which isolates the reshaping of the distribution from the extensive term
\cite{Mangini2024}; the derivation of Section~6.1 uses the Gibbs--Shannon
functional. These are different functionals with different extrema, different
scaling and different sensitivity to near-empty bins, and measuring one does not
test a prediction derived from the other. We therefore state the protocol in
terms of the functional actually used:
\begin{equation}
  H_{\delta\omega} = -\sum_{i}P_{i}\ln P_{i} ,
  \qquad
  P_{i} = \frac{\int_{\mathrm{bin}\,i} p(\omega)\,\mathrm{d}\omega}
               {\int p(\omega)\,\mathrm{d}\omega} ,
  \label{eq:Hbin}
\end{equation}
with a bin width $\delta\omega$ fixed once by the spectrometer and held constant
across the scan, so that \eqref{eq:Hbin} is the discrete counterpart of
\eqref{eq:DSentropy} and inherits its reference measure. Both terms of
\eqref{eq:DSentropy} are then accessible: the shape part from the normalized
$P_{i}$, the scale part from the measured $\Xi$ and $\Delta$. Reporting
$\tilde S$ in addition is useful for comparison with
Ref.~\cite{Mangini2024}, but it should not be presented as a test of
\eqref{eq:Hsmax}.

Transposed to a DS, the mode index is replaced by the spectral bin of
\eqref{eq:Hbin}, the propagation coordinate by the round-trip number, and the
mode-resolved holography by shot-resolved spectroscopy; \eqref{eq:Hbin} can then
be followed as the pump is raised, alongside the pulse-number statistics. Here a
distinction must be drawn that earlier statements of this proposal, our own
included, elided. Dispersive Fourier transformation maps the spectral
\emph{intensity} onto a temporal waveform and delivers single-shot spectra
\cite{GodaJalali2013}; it does not deliver the complex field. It is therefore
sufficient for \eqref{eq:Hbin}, which is built from intensities, and
insufficient for the kernel \eqref{eq:kernel}, which is not---a point that
follows directly from the observation, already made in Section~4.3, that states
with identical intensity profiles can carry different modal weights
\cite{Ponomarenko2004}. Accumulating $J(t_{1},t_{2})$ requires shot-resolved
\emph{field} reconstruction---spectral interferometry, coherent heterodyne or
dual-quadrature detection, or a phase-retrieval technique such as
frequency-resolved optical gating (FROG)
\cite{TrebinoKane1993} or spectral phase interferometry for direct
electric-field reconstruction (SPIDER) \cite{IaconisWalmsley1998}---together with the alignment protocol of Section~4.3.
The two measurements should accordingly be planned as two tiers, as in
Table~\ref{tab:protocol}, and the entropy tier should not be presented as a test
of the mode-count conjecture.

\begin{table}[t]
\caption{Two-tier measurement plan. The first tier is available with existing
single-shot spectroscopy; the second requires phase-sensitive field
reconstruction, and only it can address the conjecture \eqref{eq:calibration}.}
\label{tab:protocol}
\centering\small
\begin{tabularx}{\textwidth}{@{}lX@{}}
\toprule
Target & Minimum measurement \\
\midrule
Fixed-bin spectral entropy $H_{\delta\omega}$, Eq.~\eqref{eq:Hbin}
 & Calibrated shot-resolved spectral intensity (DFT), fixed bins, stated
   noise-floor treatment and $\delta\omega$ sensitivity. \\
Two-time kernel $J(t_{1},t_{2})$ and $\Npr$, Eqs.~\eqref{eq:kernel},
 \eqref{eq:NPR}
 & Shot-resolved complex field or directly measured first-order coherence, with
   removal of timing, carrier-frequency and energy jitter (a shot-dependent
   \emph{global} phase cancels identically in $J$ and need not be removed), with
   raw and conditioned kernels both reported. \\
Multipulse selection (Section~6.4)
 & Pulse-number statistics, branch classification, transition rates and
   residence times at a controlled noise level. \\
Path dependence of $\Theta_{s}$ (Section~6.1)
 & Repeated continuation scans varying one specified physical control at a
   time (pump, filter bandwidth, GDD). \\
   Anomalous-dispersion accessibility boundary (Section~6.6)
 & Survival statistics at controlled detuning from \eqref{eq:AGDlocus},
   both signs, with the exit channel classified---pulse splitting against
   relaxation onto the weakly chirped branch---rather than pooled. \\
\bottomrule
\end{tabularx}
\end{table}

In the multimode case $\tilde{S}$ grows and then saturates once equilibrium is
reached \cite{Mangini2024}; the DS expectation for the shape part is
qualitatively different---continued growth towards the ceiling
\eqref{eq:Hsmax}, against a scale part that turns over at \eqref{eq:hmax}---so
the two contributions should be reported separately rather than summed into a
single number. Three practical warnings carry over. Bins whose measured
occupancy falls below the noise floor must be handled explicitly rather than
silently discarded, as they were in the multimode measurement
\cite{Mangini2024}; the result must be shown to be stable against a change of
$\delta\omega$ over a stated range, since \eqref{eq:Hbin} depends on it; and the
value returned by an equilibrium-inspired functional applied to a
nonequilibrium state should be read as a diagnostic rather than as the entropy
of that state---the bound relation invoked for it in
Ref.~\cite{Mangini2024} holds under conditions that have not been verified
here, and we do not rely on it.

\subsection{Finite-time accessibility of the anomalous branch: a dynamical, one-sided boundary}

Section~6.5 shows that the particular NGD entropy-turnover criterion does not reappear on the AGD continuation studied here. The next question is therefore not ``where is the anomalous thermodynamic threshold?'' but a more basic one: does the analytical strongly chirped AGD solution remain dynamically available throughout its existence window? The noisy calculations of Ref.~\cite{Kalashnikov2026} answer only this robustness question. They do not exclude other physical mechanisms that may ultimately limit energy scaling in a real oscillator.

The two questions are known to come apart in the CQGLE, where analytic pulse solutions can lie outside the finite domains in which they are dynamically stable, and where changes of stability attach to bifurcation and turning points rather than to admissibility \cite{SotoCrespo1997,SotoCrespo2001}. Here the separation matters more than usual, because the strongly chirped branch coexists with the weakly chirped Pereira--Stenflo-type branch of Section~4.2 and with multipulse states.

\emph{The test.} Ref.~\cite{Kalashnikov2026} probes the question numerically under a finite propagation
interval, a specified noise ensemble and an explicit survival criterion, and the protocol is worth stating in full, because every conclusion below is conditional on its cutoffs. The analytical profile $a_{0}(t)$ is reconstructed from the stationary-phase envelope \eqref{eq:AGDspectrum}, perturbed in the rotating frame of \eqref{eq:ansatz}, $a(z,t)=e^{-iqz}[a_{0}(t)+\delta a(z,t)]$, and propagated under the linearization of
\eqref{eq:cqgle},
\begin{equation}
   \partial_{z}\delta a = (iq-\sigma)\,\delta a
  + (\alpha+i\beta)\,\partial_{t}^{2}\delta a
  + (\kappa-i\gamma)\big(2|a_{0}|^{2}\delta a + a_{0}^{2}\delta a^{*}\big)
  + (-\kappa\zeta+i\chi)\big(3|a_{0}|^{4}\delta a
      + 2|a_{0}|^{2}a_{0}^{2}\,\delta a^{*}\big) ,
  \label{eq:bogoliubov}
\end{equation}
which is of Bogoliubov type: the anomalous couplings $a_{0}^{2}\delta a^{*}$ and $|a_{0}|^{2}a_{0}^{2}\delta a^{*}$ are the pair terms of the condensate problem of Section~\ref{sec:dictionary}, here carrying complex rather than real coefficients. The integration is split-step, with the stiff filtering and GDD terms taken in the frequency domain and the $2\times2$ coupling between $\delta a$ and $\delta a^{*}$ in the time domain, so that the stiff part is handled without the step-size restriction that most readily manufactures growth at the highest Fourier modes. That is a mitigation, not a proof of convergence: a quantitative stability map requires a convergence table in step size, window and grid, which has not been published and which we list among the outstanding checks. The initial condition is band-limited Wigner-vacuum noise inside $|\omega|\le\omega_{\mathrm{cap}}$, scaled to the photon number corresponding to one unit of dimensionless energy. A realization
is followed to $z=200$ and counted as surviving if it remains single-pulse, keeps $\|\delta a\|_{2}/\|a_{0}\|_{2}<0.25$, and retains a high spectral-shape correlation with $a_{0}$; the survival fraction $P_{\mathrm{surv}}$ is taken over $N_{\mathrm{shot}}=32$ independent shots per grid point, and a point is called \emph{robust over the simulated interval}---not ``stable'', which we reserve for asymptotic or spectral results---when $P_{\mathrm{surv}}>50\%$
\cite{Kalashnikov2026}.

\begin{figure}[t]
\centering
\includegraphics[width=\textwidth]{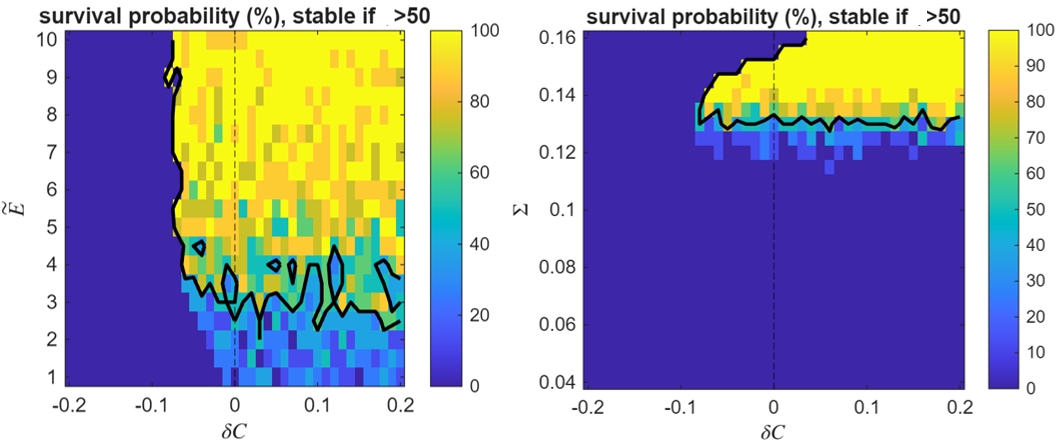}
\caption{Dynamical accessibility of the strongly chirped
anomalous-dispersion branch under quantum-noise perturbation of
\eqref{eq:bogoliubov}. Color is the \emph{finite-time} survival
probability
$P_{\mathrm{surv}}$ over $N_{\mathrm{shot}}=32$ shots, measured at the
fixed propagation distance $z=200$; the black contour is the
\emph{empirical $50\%$ finite-time survival contour}, and the dashed vertical line is the chirp-control resonance
\eqref{eq:AGDlocus}, $\delta C=0$ with $C=-1/\tilde\chi+\delta C$. Common
parameters $\tilde\chi=1.5$, $\kappa/\gamma=0.1$, $\zeta/\gamma=0.39$,
$\alpha=C\beta\kappa/\gamma$. (\textbf{Left})~the $(\tilde E,\delta C)$ plane at
fixed $\Sigma$; (\textbf{Right})~the $(\Sigma,\delta C)$ plane at fixed $\tilde
E=8$. The $\Sigma$ values printed on the axes are those of
Ref.~\cite{Kalashnikov2026} and must be multiplied by four to be read in the
normalization of this paper (footnote to Section~5.2): the left panel is at
$\Sigma=0.56$ here, and the right-hand axis runs over $\Sigma=0.16$--$0.64$
here, against $\Sigma_{\mathrm{th}}=0.667$ at $\delta C=0$. The black contour is the \(P_{\mathrm{surv}}=50\%\) finite-time
survival contour for \(N_{\mathrm{shot}}=32\) and \(z=200\). It should therefore
be read as a practical accessibility/robustness boundary under the specified
noise model, not as an asymptotic orbital-stability boundary. Its pixel-scale
roughness is commensurate with the finite ensemble and grid resolution. Reproduced from
Ref.~\cite{Kalashnikov2026} (Fig.~8), published open access under CC~BY.}
\label{fig:agdstability}
\end{figure}

The sample size here should not be read as the statistical standard for
the entire DS problem. It is the economical per-grid-point ensemble used to map a two-dimensional AGD finite-time survival surface. The earlier NGD study of Ref.~\cite{Kalashnikov2025} used 150 independent stochastic realizations for each reported pulse-number distribution in Fig.~5 and extracted additional
self-start-time statistics in Fig.~6 \cite{Kalashnikov2025}. The two calculations answer different questions: the NGD ensemble probes noise-initiated attractor selection and self-start kinetics, whereas the present AGD ensemble probes local survival of a prepared analytical pulse over a fixed propagation interval. This distinction is essential when the data are reused for Section~4.3: low-dimensional survival or pulse-number probabilities may converge with tens to hundreds of shots long
before a high-rank coherence spectrum does.

One limitation belongs beside the maps rather than in an outlook
section, because it bounds everything read from them. The calculation propagates
a \emph{linearized} stochastic perturbation about an analytical pulse, so it
probes whether small fluctuations grow, decay or remain bounded in the
neighborhood of that solution. It cannot capture the nonlinear processes that
determine global attractor structure---large-amplitude splitting, noise-induced
basin crossing, spontaneous pulse creation, or relaxation into a remote
attractor. The maps can therefore validate local or finite-neighborhood
robustness under the model used; they cannot validate an entropy-based
branch-selection principle of the kind proposed in Section~6.4, and are not
offered as doing so.

\emph{Conceptual reading of the maps.} Figure~\ref{fig:agdstability} supports three conclusions that are more important than the numerical location of any individual contour.

First, \emph{existence is not the same as robustness}. The adiabatic equations allow a wider AGD branch than the noisy propagation actually preserves. This is the AGD counterpart of the NGD self-start result: a mathematically available DS need not be the dynamically favored state.

Second, the robust region is strongly asymmetric about the nominal AGD--DSR locus. Proximity to resonance, or the leading scale-separation factor by itself, therefore cannot be the sole stability variable. This does not contradict the two-scale interpretation of Ref.~\cite{Kalashnikov2026}; it limits its scope. Scale separation may characterize increasing internal susceptibility, but the actual survival boundary also depends on the direction in parameter space and on the full dissipative balance.

Third, within the scanned interval, the AGD calculation does not reveal the same upper-energy loss of single-pulse accessibility that characterizes the NGD multipulse crossover. That difference is physically significant, but it does not prove that AGD has no ultimate energy limit. It means only that the NGD entropy-turnover mechanism should not be promoted to a universal criterion for both dispersion signs. The anomalous branch presently supplies a complementary lesson: the first limitation visible in the calculation is dynamical robustness, not an entropy-proxy crossing.

\begin{figure}[t]
\centering
\includegraphics[width=0.92\textwidth]{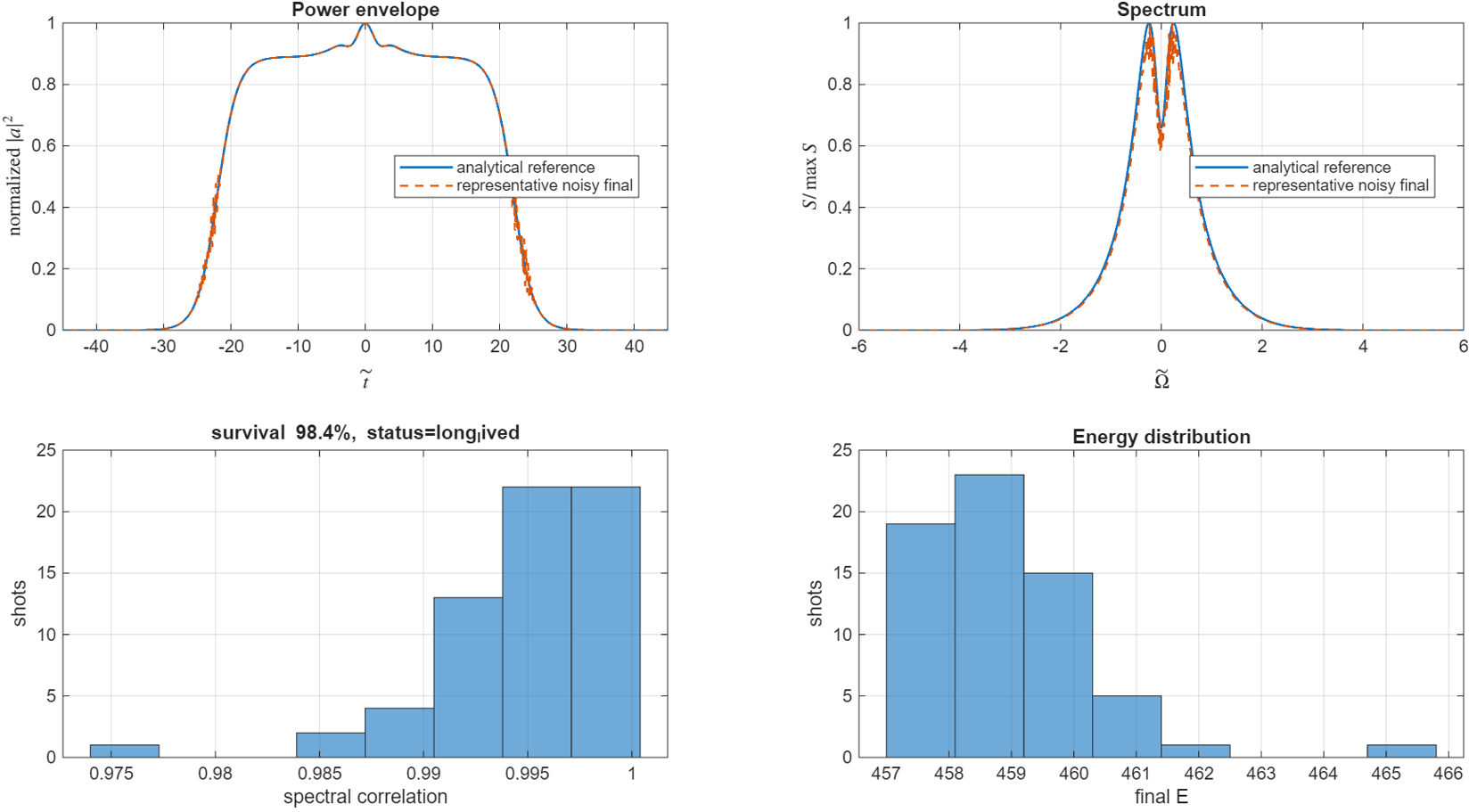}
\caption{A representative point inside the robust region: $\tilde E=8$,
$\delta C=-0.01$, $\tilde\chi=1.5$, and $\Sigma=0.14$ in the normalization of
Ref.~\cite{Kalashnikov2026} ($\Sigma=0.56$ here), over $64$ independent noise
realizations. \emph{Top:} the analytical stationary-phase profile (solid) and a
representative noisy final state (dashed), in power (left) and normalized
spectrum (right). The table-top envelope and the two-horn core are preserved;
the visible residuals sit at the pulse edges and at the central spectral dip.
\emph{Bottom:} distributions of the spectral-shape correlation with the
analytical reference and of the final energy over the retained shots (the
energy axis is in the internal units of the simulation, not in $\tilde E$).
The panel value \(98.4\%\) corresponds to \(63/64\) trajectories
meeting the adopted finite-time survival criterion. The histogram is therefore
used only to document the spread of this finite ensemble; no asymptotic
stability probability or distributional law is inferred from it. The upper row reproduces Fig.~9 of Ref.~\cite{Kalashnikov2026} (CC~BY); the two histogram panels are author-generated data for the present manuscript, from the same ensemble.}
\label{fig:agdrealization}
\end{figure}

\emph{Where the perturbation settles.} Figure~\ref{fig:agdrealization}
shows a representative realization from inside the robust region, and its
interest for the present argument is less the survival statistics---a
spectral-shape correlation clustered above $0.99$, a final-energy spread of about two per cent with a single outlier, and $63$ of $64$ trajectories meeting the finite-time criterion---than \emph{where} the residual noise accumulates. In the realization shown, the largest residual deviations in the time domain sit at the pulse wings and at the steep edges of the table-top, and in the frequency domain at the central dip between the two horns. Both locations were identified in advance, from a quite different direction, in Section~2.2. The edges are where the two stationary points coalesce, and the ordinary stationary-phase approximation loses uniformity, which is why a Chester--Friedman--Ursell uniformization is required there at all. The
central dip is where the $\pm$ saddle contributions interfere, and is
accordingly the most phase-sensitive feature of the entire envelope. The
perturbation is largest exactly where the analytical construction is weakest---in this realization. A single representative shot cannot establish statistical localization of the residual, and the statement is not made at ensemble level here. Converting it into one is cheap and should be done from the stored runs: accumulate $V_{t}(t)=\langle(|a_{j}(t)|^{2}-\langle|a(t)|^{2}\rangle)^{2}\rangle$ and its spectral counterpart $V_{\omega}(\omega)$ over the ensemble, or simply plot the ensemble standard deviation about the mean profile, and test whether the variance is in fact concentrated at the coalescence edges and the interference dip. That is reassuring about the construction, since the table-top and the two-horn core are otherwise reproduced. But it also means that the fine structure of the anomalous spectrum---the very feature that distinguishes the strongly chirped branch from its weakly chirped competitor---is the least robust part of the prediction, and should not be relied upon as an experimental signature without an accompanying noise budget.

\emph{Two things the maps do not settle.} First, they do not identify the
channel. A realization is discarded when it ceases to be single-pulse
\emph{or} when the perturbation norm exceeds the cutoff, and the published summary does not separate the two. The identification of the instability with multipulse fragmentation, rather than with relaxation onto the weakly chirped Pereira--Stenflo branch that occupies the neighboring parameter region (Section~4.2), is therefore an inference and not a measurement. A discriminating observable is available---the weakly chirped branch fills in the central dip and develops oscillatory wings \cite{Kalashnikov2026}, and a fragmenting pulse does neither---but it has not been reported.
A third channel belongs on that list. In GL models a stationary pulse can
lose stability without either fragmenting or changing branch, by a supercritical
Hopf bifurcation into a \emph{breather}: a localized structure of unchanged
pulse number whose peak amplitude executes a limit cycle, with oscillation
amplitude growing as the square root of the distance from threshold
\cite{SakaguchiMalomed2000}. Where this has been characterized the transition is
sharp---stationary pulses and breathers do not coexist, as a supercritical
bifurcation requires---and bound states of two breathers exist, locked with a
$\pi/2$ shift between their internal vibrations. A breathing pulse would be
discarded by the criterion used here whenever its excursions carry the
perturbation norm past the cutoff, yet it is neither fragmentation nor
relaxation onto the weakly chirped branch, and it is separable from both by an
observable the stored realizations already contain: a periodic modulation of the
peak power at fixed pulse number, rather than a monotone departure. Whether the
strongly chirped CQGLE branch admits such a bifurcation at all is open---the
system in which it was characterized is cubic and two-component---but it should
be excluded by inspection of the discarded shots rather than by assumption. Second, and more consequentially for the
program of this paper, the calculation constructs precisely the object
Section~4.3 requires and then discards it. The quantum-noise ensemble is the first of the four candidate ensembles listed in Section~8.2(i). The simulation propagates $32$--$64$ independent realizations of the complex field and reduces each to a survival flag and a scalar correlation. Retaining those fields and accumulating $J(t_{1},t_{2})=\langle a(t_{1})a^{*}(t_{2})\rangle$ over the same shots, after removal of the timing, phase, and energy jitter of Section~4.3, would deliver the coherence kernel \eqref{eq:kernel} and its participation
number \eqref{eq:NPR} at no additional cost in dynamics. Whether $\Npr$ then tracks $r_{\mathrm{AGD}}$ is the calibration \eqref{eq:calibration} on which the statistical reading of the whole framework rests, and this computation is the shortest route to it that we can identify.

\section{Analogies and Their Boundaries: Turbulence and Driven-Open
Condensates}

\subsection{Analogue reasoning as a method}
\label{sec:analogue}

Before the individual correspondences are examined, it is worth saying what kind
of argument they are, because the objection that a driven, lossy, single pulse
``is not a thermodynamic system'' is easy to make and, taken at face value,
proves far too much. An analogy with no stated boundary is decoration; an
analogy with one is a hypothesis---and it is the second kind that is at issue
here. It would equally ignore
the thermodynamic sector of wave turbulence: the wave-kinetic equation admits
Rayleigh--Jeans equilibrium distributions alongside genuinely nonequilibrium,
flux-carrying Kolmogorov--Zakharov solutions, so a driven system may contain a
thermalized spectral sector, even though the Rayleigh--Jeans equilibrium should
not itself be identified with the finite-flux cascade
\cite{Picozzi2007,Picozzi2014}; the thermodynamics of two-dimensional vortex
statistics, whose negative-temperature states Onsager introduced for a
Hamiltonian system with no thermal bath at all \cite{Kraichnan1980,Onsager1949};
the effective temperatures of glasses and active matter, which are known to
depend on the observable and the timescale and are used anyway
\cite{CugliandoloKurchanPeliti1997,Cugliandolo2011}; the stochastic
thermodynamics of small driven systems, built precisely for states that violate
detailed balance \cite{HatanoSasa2001,Seifert2005}; and the thermodynamics of
black holes, where an entropy and a temperature are assigned to an object that
is not a Gibbs state on the strength of a formal correspondence between two sets
of laws. In none of these cases was the response to abandon the vocabulary. It
was to determine which relations survive the loss of equilibrium and which do
not---which is the organizing task of this section and of
Table~\ref{tab:claimstatus}.

The methodological setting is by now a familiar one. Over the past two decades,
\emph{analogue simulation} has become a standard instrument rather than a
rhetorical device: one physical system is used to realize the governing equation
of another, so that consequences derived in one domain can be tested in the
other. Unruh's observation that sound in a transonic flow obeys the wave
equation on a black-hole metric \cite{Unruh1981} grew into a research programme
with its own review literature \cite{BarceloLiberatiVisser2011}, and into
laboratory horizons in flowing condensates \cite{Steinhauer2016} and in optical
fibres \cite{Philbin2008}. Cold atoms, trapped ions and superconducting circuits
are used to realize lattice models whose direct solution is out of reach
\cite{Georgescu2014}. Photonics itself has supplied topological band structures
\cite{Ozawa2019}, photon condensates with a genuine chemical potential
\cite{Klaers2010}, and---most directly relevant here---Kardar--Parisi--Zhang
phase scaling in a one-dimensional polariton condensate
\cite{Fontaine2022}, the same universality class measured earlier in turbulent
liquid crystals \cite{TakeuchiSano2010}. Within nonlinear fibre optics, the
Rayleigh--Jeans equilibrium and the negative optical temperatures of multimode
systems have moved from prediction \cite{Wu2019} to measurement
\cite{Pourbeyram2022,Baudin2020,Muniz2023}, the negative-temperature
Rayleigh--Jeans equilibrium itself having been observed in a conservative
multimode fibre \cite{Baudin2023}, and integrable turbulence and
soliton-gas thermodynamics are now studied on optical fibres as a matter of
routine \cite{ElKamchatnov2005,Suret2024,Fache2025}.

What licenses such transfers is not resemblance but shared structure. An
analogue argument is valid for exactly those consequences that follow from the
part of the description the two systems have in common, and invalid beyond it.
This is why each correspondence below is stated together with the conditions
under which it holds and the point at which it fails, and why
Table~\ref{tab:claimstatus} grades every claim by what would be required to
establish it. The catalog above should therefore be read for what it is.
Black-hole thermodynamics, Onsager vortices, glasses, stochastic
thermodynamics, analogue gravity, photon condensation and KPZ universality
collectively show that concepts developed in one domain can acquire meaningful
analogues in another. They do \emph{not} independently validate the particular
identifications $\mu_{\mathrm{shape}}$, $H_{s}$, $\Theta_{s}$ made here for a
strongly chirped DS. Shared structure motivates a definition; each definition
must then earn its status through its own invariance, conjugacy relation,
ensemble construction, fluctuation relation or predictive success.

Photonics is an unusually good platform for this kind of argument, for four
reasons that bear directly on the present case. At the conservative mean-field level, the
slowly-varying-envelope (paraxial) equations of nonlinear optics and the
Gross--Pitaevskii equation share an NLS-type mathematical structure once the
variables are identified, with the propagation coordinate playing the role of
time. The \emph{driven--dissipative} correspondence used below is less exact: it
becomes term-by-term only after the reservoir elimination, gain expansion,
truncation and noise omission stated explicitly in Section~\ref{sec:dictionary}. The analogy is
therefore structural and conditional rather than an identity of the complete
physical models. The coefficients are tunable over orders of magnitude
and, unlike in most condensed-matter realizations, independently: dispersion,
nonlinearity, filtering and saturable loss can be moved separately, which is
precisely what generates the master diagram of Section~2.3. Single-shot,
time-resolved detection gives access to full distributions rather than ensemble
means \cite{GodaJalali2013}, so the statistical questions raised in Section~4.3
are experimentally answerable in a way they are not for a cold-atom condensate.
And---the point most often overlooked---drive and dissipation here are
\emph{engineered rather than parasitic}. In a laser, gain and loss are the
design variables. That makes the mode-locked oscillator a natural testbed for
driven-open universality \cite{Sieberer2025}, in which the breaking of detailed
balance is the object of study rather than an unwanted correction.

Two consequences follow for how the present work should be read. First, the
analogy is generative, not merely descriptive: the reinterpretation of DSR as
spectral condensation at vanishing $\mu_{\mathrm{shape}}$ predicted three
signatures---saturation of the spectral width, growth of a central Lorentzian
spike, and reversal from shortening to asymptotic stretching---which are
observed \cite{Rudenkov2023}, and it predicts a fragmentation threshold lying
\emph{below} the boundary of existence, which is testable and could fail.
Predictions that could fail are the working criterion for whether an analogy is
doing scientific work. Second, and for the same reason, the framework must identify explicitly
where an analogy ceases to be predictive rather than treating every mismatch as
a failure of the underlying DS theory. Sections~6.5 and 6.6 report that the
entropic mechanism found in normal dispersion does not transpose along the
anomalous continuation examined; Section~\ref{sec:bec} sets out three obstructions to the
condensate correspondence; Section~4 shows that the exact-potential point
\eqref{eq:gibbscubic}--\eqref{eq:gibbsquintic} is a corner of the diagram that is
approached but never occupied. These are not concessions extracted from the
programme. They are what the programme produces when it is run honestly, and
they are the reason the title says \emph{toward}.

The DS framework sits between several better-known descriptions, and it is worth
stating precisely how far each analogy reaches. Two preliminaries sharpen the
exercise. First, one of the correspondences is unusually tight:
Equation~\eqref{eq:cqgle} has the same form, term for term, as the equation used
to describe an incoherently pumped driven-open condensate \emph{only after} a
specific chain of approximations---adiabatic elimination of the excitonic
reservoir, expansion of the saturable gain in the density, truncation at
quartic order, and omission of the noise term---so that the correspondence can
be tabulated rather than asserted, provided those conditions are carried along
with it. Its failures can then be located in particular entries, or in
particular conditions of the reduction, rather than ascribed to a general
mismatch of spirit. Second, a dictionary is of
no use unless its signs are fixed, and the two published versions of the
photonics--BEC dictionary differ by a complex conjugation. We therefore begin
with the dictionary itself, and only then ask what it buys and where it breaks.

\subsection{The dictionary, term by term}
\label{sec:dictionary}

The equation used for an incoherently pumped exciton--polariton condensate,
once the excitonic reservoir has been adiabatically eliminated and the gain
expanded in the density, is
\begin{equation}
  (i-\lambda_{\mathrm{P}})\,\hbar\,\frac{\partial\psi}{\partial T}
  = \Big[-\frac{\hbar^{2}}{2m}\nabla^{2} + V + g|\psi|^{2} + g_{3}|\psi|^{4}
    + \frac{i\hbar}{2}\Big(\frac{R}{1+|\psi|^{2}/n_{s}}-\Gamma\Big)\Big]\psi ,
  \label{eq:ddgpe}
\end{equation}
with $\lambda_{\mathrm{P}}$ the phenomenological energy-relaxation
(``kinetic cooling'') constant and $R/(1+|\psi|^{2}/n_{s})$ the saturable gain
supplied by the reservoir \cite{WoutersCarusotto2007,Wouters2012}.
Equation~\eqref{eq:ddgpe} is referred to in that literature interchangeably as
the \emph{driven-dissipative Gross--Pitaevskii equation} and as the complex
Ginzburg--Landau equation \cite{He2015,Carusotto2013}, and expanding its
saturable gain in powers of the density---precisely the step that produces the
$\kappa(1-\zeta P)P$ term of \eqref{eq:cqgle}---is the standard reduction used,
for instance, in the renormalization-group treatment of one-dimensional
driven-open condensates \cite{He2015}. The relation between the DS problem and
the driven-open condensate problem is therefore closer than a resemblance
between two models: it is a relation between two regimes of one reduced model.
Two qualifications are needed, and the first is algebraic. Equation
\eqref{eq:ddgpe} is not in the form of \eqref{eq:cqgle} until it is divided by
$(i-\lambda_{\mathrm{P}})$, and that division mixes every coefficient:
\begin{equation}
  \hbar\,\frac{\partial\psi}{\partial T}
  = \frac{(-i-\lambda_{\mathrm{P}})}{1+\lambda_{\mathrm{P}}^{2}}
    \big[\hat H + i\hat G\big]\psi ,
  \label{eq:mixing}
\end{equation}
where $\hat H$ collects the Hamiltonian terms of \eqref{eq:ddgpe} and $\hat G$
the gain and loss. Every reactive term therefore acquires a dissipative part and
every gain or loss term a reactive frequency shift, in the fixed proportion
$\lambda_{\mathrm{P}}:1$; kinetic, interaction, potential and gain coefficients
are mixed. A term-by-term reading of the kind tabulated below is recovered only
to leading order in $|\lambda_{\mathrm{P}}|\ll1$---which is the same weak-damping
condition that defines the strongly chirped regime, so the mapping and the
regime of interest are consistent---and after the induced shifts have been
absorbed into the renormalized coefficients. We accordingly describe the
dictionary as an \emph{approximate correspondence valid at small
$\lambda_{\mathrm{P}}$}, not as a term-for-term identity, and Table~1 entries
should be read with the $O(\lambda_{\mathrm{P}})$ corrections implied by
\eqref{eq:mixing}.

The second qualification carried by ``reduced'' is physical and should be stated
with the claim. Equation~\eqref{eq:ddgpe} is itself an effective description,
obtained where the reservoir relaxes fast compared with the condensate
dynamics---a condition that fails in parts of the parameter space, where a
stochastic generalized Gross--Pitaevskii treatment separates KPZ,
soliton-patterned, defect-dominated and reservoir-textured regimes that the
reduced description cannot distinguish \cite{Vercesi2023};
the polariton problem carries a Langevin noise term whose optical counterpart
enters the DS problem only in the stochastic simulations of Section~6.4; the
condensate is two-dimensional and trapped in most experiments, the DS
one-dimensional and free; and the conserved quantities differ, the polariton
density being fixed by a pump--loss balance with an explicit reservoir equation
and the pulse energy by the saturated round-trip gain. What differs at the level
of the solution is the sign of the dispersive coefficient, the presence or
absence of a trap, and, decisively, which branch is of interest: an extended
condensate there, a localized pulse here. Table~\ref{tab:mappingconditions}
collects the conditions under which each entry of the dictionary can be used.

Before the correspondence can be used, its signs must be fixed. Table~1 of
Ref.~\cite{Kalashnikov2024} lists \emph{anomalous} GDD as the counterpart of the
boson kinetic energy and the Kerr nonlinearity as an \emph{attractive}
interaction, whereas Ref.~\cite{Kalashnikov2025} states that the GDD term
corresponds to a boson kinetic energy provided the SPM term describes a
\emph{repulsive} one. Both statements are correct: they identify $\psi$ with
$a^{*}$ and with $a$ respectively, and conjugation reverses the sign of the
kinetic and of the interaction term simultaneously. Only the ratio $\gamma/\beta$
is convention-independent, and it is the ratio that carries the physics:
$\gamma/\beta<0$ (anomalous GDD with self-focusing SPM) is the focusing,
attractive sector that supports bright solitons, while $\gamma/\beta>0$ (normal
GDD) is the defocusing, repulsive sector in which no bright structure exists
without dissipation. Throughout this paper we adopt
\begin{equation}
  \psi \equiv a^{*}, \qquad T \equiv z, \qquad x \equiv t ,
  \qquad\text{whence}\qquad
  \frac{\hbar^{2}}{2m} = -\beta , \quad g = -\gamma , \quad g_{3} = \chi ,
  \label{eq:convention}
\end{equation}
which reproduces the convention of Ref.~\cite{Kalashnikov2024}: anomalous GDD
gives a positive boson mass, self-focusing SPM an attractive two-body
interaction, and---a point used in Section~\ref{sec:thomasfermi}---SPM \emph{saturation}
($\chi>0$ in \eqref{eq:cqgle}) a \emph{repulsive} three-body correction.
Table~\ref{tab:dictionary} sets out the full correspondence, with the
momentum-space reading of each term added in the last column.

Two entries convert the dictionary from a lexicon into a quantitative statement.

\emph{The control parameter is a damping constant.} Since $\alpha$ and $\beta$
both carry the dimension of time squared and $\gamma$ and $\kappa$ that of
inverse power, the control parameter of \eqref{eq:freqODE} factorizes,
\begin{equation}
  C \;=\; \frac{\alpha\gamma}{\beta\kappa}
    \;=\; \underbrace{\frac{\alpha}{\beta}}_{\textstyle \lambda_{\mathrm{P}}}
      \times \frac{\gamma}{\kappa} ,
  \label{eq:Cfactor}
\end{equation}
into a factor with the dimensions and the role of the Pitaevskii damping
constant of the condensate literature
\cite{Pawlowski2017,Wouters2012,Pitaevskii1959,Choi1998}, times the ratio of
reactive to dissipative nonlinearity. The identification requires care with
signs, and we state it with the caveat rather than as a result. In the
convention \eqref{eq:convention} a positive boson mass corresponds to
\emph{anomalous} GDD, $\beta<0$, while spectral filtering requires $\alpha>0$;
hence $\alpha/\beta<0$ there, whereas $\lambda_{\mathrm{P}}$ is introduced in
\eqref{eq:ddgpe} as a positive relaxation constant. The magnitude
$|\alpha/\beta|$ is what plays the role of $\lambda_{\mathrm{P}}$, and the sign
carried by $\alpha/\beta$ records which dispersion regime is in force; a clean
identification, including the sign and the $O(\lambda_{\mathrm{P}})$ shifts of
\eqref{eq:mixing}, requires an explicit linearization and nondimensionalization
of \eqref{eq:ddgpe} about the relevant state, which we have not carried out. We
therefore write $|\lambda_{\mathrm{P}}|=|\alpha/\beta|$ and treat the equality
as provisional. With that reservation, \eqref{eq:Cfactor} reads: the two
conditions defining the strongly chirped regime---$\alpha\ll|\beta|$ and
$\kappa\ll\gamma$---say that the DS is a \emph{weakly damped} condensate held
near the soliton condition $C\simeq1$ by a compensating imbalance between the
two nonlinearities, so that the terminology of
Ref.~\cite{KalashnikovWabnitz2021} is more than metaphorical. The master diagram
of Section~2.3 is \emph{not}, however, ``a plane spanned by a damping constant
and a particle number'': $C$ is a composite ratio of linear and nonlinear,
reactive and dissipative coefficients, of which $|\alpha/\beta|$ is one factor,
and $E^{*}$ is a pulse energy in a system that does not conserve the norm. That
description is withdrawn.

\emph{The cutoff is a radiation resonance with a damped branch.} Linearizing
\eqref{eq:cqgle} about the vacuum gives the dispersion relation of the linear
waves in the complex form
\begin{equation}
  k(\omega) = \beta\omega^{2} - i\big(\sigma+\alpha\omega^{2}\big) ,
  \label{eq:lindisp}
\end{equation}
in the convention $a\propto e^{i\omega t-ikz}$ fixed in Section~2 and used for
the soliton ansatz \eqref{eq:ansatz}, so that $\mathrm{Im}\,k<0$ describes decay
along $z$. (Earlier statements of this relation carry the opposite sign of the
imaginary part, which corresponds to the conjugate convention and is
inconsistent with \eqref{eq:ansatz}.)
Its real part is the branch already used in \eqref{eq:cutoff}: the cutoff is the
detuning at which the DS wavenumber $q$ meets it, so that \eqref{eq:cutoff} is
the optical counterpart of the Landau--Cherenkov construction, in which a
coherent object radiates into a linear branch wherever its own wavenumber
intersects that branch. What the dictionary adds is the imaginary part. A
damping growing as $\omega^{2}$ is exactly what the energy-relaxation term of
\eqref{eq:ddgpe} produces, and the corresponding object in the condensate
literature is the damped Bogoliubov branch
$\omega(k)\simeq c_{s}k-i\lambda_{\mathrm{P}}\hbar k^{2}/2m$ obtained by
linearizing a dissipative GPE about the condensate \cite{Pawlowski2017}. The
identification $|\lambda_{\mathrm{P}}|=|\alpha/\beta|$ of \eqref{eq:Cfactor} is
the statement that the two damping rates coincide in magnitude. In a conservative multimode
system, the ultraviolet cutoff is a property of the waveguide; here it is the
resonance point of a solution with a branch whose damping the same solution
supplies.

Two further things follow from that imaginary part, and the first is
methodological. Equation~\eqref{eq:lindisp} is written for \emph{real}
$\omega$ and \emph{complex} $k$, which is the opposite of the habit inherited
from conservative stability analysis. That choice is not merely a convenience
of the soliton ansatz: it is the prescription the driven-open condensate
literature arrives at independently, and for a reason worth importing. Applied
naively to the real part of the elementary-excitation spectrum of an
incoherently pumped condensate, the Landau criterion predicts a
\emph{vanishing} critical velocity, because that spectrum has no sonic branch
at small wavenumber at all (see below); direct simulation contradicts this and
finds a threshold of order the equilibrium sound speed. The discrepancy is
resolved by observing that in a driven steady state every mode oscillates at
one real frequency fixed by the drive, so the imaginary part of the dispersion
must be reabsorbed into a complex wavenumber $\tilde k(\omega)$ rather than
into a complex frequency; the threshold is then the point at which
$\mathrm{Re}\,\tilde k$ becomes nonzero, and it approaches the equilibrium value
as the damping is reduced \cite{Carusotto2013,WoutersCarusotto2010}. The
convention already adopted in \eqref{eq:lindisp} is therefore the correct one
for a driven-open problem, and we record the precedent rather than leave the
choice looking arbitrary.

The second is quantitative, and it puts a number on a caveat made twice
already. Because $\mathrm{Im}\,k\neq0$, the resonance $q=\mathrm{Re}\,k(\omega)$
is not a point but a line of finite width: the response
$|q-k(\omega)|^{-2}$ peaks at $\omega\simeq\Delta$ with half-width
$\delta\omega=(\sigma+\alpha\Delta^{2})/(2\beta\Delta)$, so that, using
$q=\beta\Delta^{2}$,
\begin{equation}
  \frac{\delta\omega}{\Delta}
  \;=\; \frac{\sigma}{2q} \;+\; \frac{1}{2}\Big|\frac{\alpha}{\beta}\Big|
  \;=\; \frac{\sigma}{2q} + \frac{|\lambda_{\mathrm{P}}|}{2} .
  \label{eq:cutoffwidth}
\end{equation}
The fractional sharpness of the cutoff is thus governed by the same two small
parameters that govern the validity of the dictionary itself---the saturated
net loss measured against the DS wavenumber, and the Pitaevskii damping
constant of \eqref{eq:Cfactor}---so the truncation is sharp exactly where the
strong-chirp reduction is controlled, and blurs as either is relaxed. Two
consequences are carried forward. Equation~\eqref{eq:cutoffwidth} is the
quantitative form of the statement in Sections~2.2 and 4.1 that the
discontinuity at $\pm\Delta$ is a leading-order artifact of stationary phase
rather than a physical edge. And it bounds the precision of every indicator
built on $r=\Delta/\Xi$: the graining scale $l=\pi/\Delta$ is defined only to
fractional accuracy \eqref{eq:cutoffwidth}, which is a floor on the resolution
of any measured or computed $r$, and one that widens on approach to the
vacuum-stability border only through $\sigma/2q$, since $|\lambda_{\mathrm{P}}|$
is fixed along an isogain.

\begin{table}[t]
\caption{The CQGLE \eqref{eq:cqgle} term by term, in the convention
\eqref{eq:convention}. Entries marked $\Leftrightarrow$ are identifications of
terms in the equations; entries marked $\leftrightarrow$ are correspondences of
physical role only. The last column gives the momentum-space reading used in
Sections~3.1 and 7.2.}
\label{tab:dictionary}
\centering
\footnotesize
\begin{tabularx}{\textwidth}{@{}p{2.9cm} p{1.9cm} X p{3.5cm}@{}}
\toprule
\textbf{CQGLE term} & \textbf{Symbol} &
\textbf{Driven-open condensate, Eq.~\eqref{eq:ddgpe}} &
\textbf{Wave-turbulence reading} \\
\midrule
Evolution coordinate & $z$ & $\Leftrightarrow$ time $T$ & slow (kinetic) time \\
Transverse coordinate & $t$ & $\Leftrightarrow$ spatial coordinate $x$ &
  conjugate to $\omega$ \\
Group-delay dispersion & $\beta\,\partial_{t}^{2}$ &
  $\Leftrightarrow$ kinetic energy, $\hbar^{2}/2m=-\beta$: anomalous GDD gives
  $m>0$, normal GDD a \emph{negative effective mass} &
  linear dispersion $\varepsilon=\beta\omega^{2}$, Langmuir-like \\
Self-phase modulation & $\gamma P$ &
  $\Leftrightarrow$ two-body interaction, $g=-\gamma$: self-focusing SPM is an
  \emph{attractive} interaction &
  four-wave vertex; nonlinear frequency shift \\
Quintic phase nonlinearity & $\chi P^{2}$ &
  $\Leftrightarrow$ three-body interaction, $g_{3}=\chi$: SPM saturation
  ($\chi>0$) is the \emph{repulsive} term that arrests collapse of an attractive
  condensate & first correction to the interaction vertex \\
Saturated net loss & $\sigma$ &
  $\Leftrightarrow$ homogeneous dissipation minus pump; $\sigma=0$ is the
  condensation threshold, i.e.\ the vacuum-stability border of the master
  diagram & uniform, $k$-independent sink \\
Spectral filtering & $\alpha\,\partial_{t}^{2}$ &
  $\Leftrightarrow$ energy relaxation (``kinetic cooling''), damping
  $\propto\alpha\omega^{2}$; Pitaevskii constant
  $|\lambda_{\mathrm{P}}|=|\alpha/\beta|$ (provisional; see text) &
  ultraviolet sink; sets the graining scale \\
Saturable nonlinear gain & $\kappa(1-\zeta P)P$ &
  $\Leftrightarrow$ saturable pumping from a finite reservoir expanded to
  quintic order; $\zeta^{-1}\leftrightarrow$ saturation density $n_{s}$ &
  large-scale source; destroys $N$-conservation \\
DS wavenumber & $q=\gamma P_{0}=\beta\Delta^{2}$ &
  $\leftrightarrow$ chemical potential of the condensate, Thomas--Fermi relation
  $\mu_{\mathrm{TF}}=gn$ & soliton wavenumber in the resonance condition \\
Lorentzian width & $-\mu_{\mathrm{shape}}=\Xi^{2}$ &
  $\leftrightarrow$ chemical potential of the quasiparticle gas; $\Xi\to0$ is
  condensation & RJ chemical potential, $n=T/(\varepsilon-\mu)$ \\
Cutoff frequency & $\Delta$ &
  $\leftrightarrow$ ultraviolet cutoff regularizing $g^{(1)}$, here supplied by
  the solution & dissipative wavenumber $k_{d}$ \\
Correlation scales & $l=\pi/\Delta$, $\Lambda=1/\Xi$ &
  $\leftrightarrow$ healing length and condensate size &
  inner and outer scales; $\Nsep=\Lambda/l$ \\
\bottomrule
\end{tabularx}
\end{table}

\begin{table}[t]
\caption{Conditions attached to each cross-field mapping used in this section.
``Defensible core'' states what the correspondence does establish; the last
column states what must be shown before the word \emph{equivalent} is used.
Compare Table~\ref{tab:claimstatus}: every entry here has status \textbf{A} or
\textbf{C}.}
\label{tab:mappingconditions}
\centering\small
\begin{tabularx}{\textwidth}{@{}l X X@{}}
\toprule
Mapping & Defensible core & Required qualification \\
\midrule
CQGLE $\leftrightarrow$ driven-dissipative GPE
 & Both contain dispersion, reactive nonlinearity, saturable gain and loss, and
   share a normal form after gain expansion.
 & Holds after reservoir elimination and quartic truncation; noise, trapping,
   dimensionality and the reservoir equation are not mapped. \\
Spectral filtering $\leftrightarrow$ Pitaevskii damping
 & Both damp high-frequency components as $\omega^{2}$.
 & The identification $\lambda_{\mathrm{P}}=\alpha/\beta$ requires an explicit
   linearization and nondimensionalization, not yet performed about the pulse. \\
DSR $\leftrightarrow$ Thomas--Fermi expansion
 & Growth in extent at nearly fixed peak density, with structure confined to
   the boundary layer.
 & Profile similarity is not a variational Thomas--Fermi limit; healing-length
   separation and local equilibrium are not demonstrated. \\
Chirped DS $\leftrightarrow$ KPZ phase
 & Both concern phase dynamics of a driven nonlinear field described by the same
   reduced equation.
 & KPZ is a stochastic, long-wavelength theory of an extended system; a
   deterministic localized pulse is not automatically a branch of it.
   Universality-class tests on shot-resolved data are required. \\
Narrow spectral core $\leftrightarrow$ BEC
 & Accumulation of weight near the lowest available state.
 & Requires an occupation criterion---a macroscopic eigenvalue of the one-body
   density matrix \cite{PenroseOnsager1956}---not a vanishing fit parameter. \\
Fragmentation $\leftrightarrow$ defect proliferation
 & Both destroy a coherent driven state by nucleating discrete objects.
 & No defect variable, topological charge, or map from pulse number to
   space--time vortices has been constructed. \\
DS phase mode $\leftrightarrow$ Goldstone mode of a driven-open
   condensate
 & Both follow from the same $U(1)$ invariance of a saturably pumped
   equation, and both are neutral.
 & The condensate mode is a \emph{gapless diffusive continuum},
   $\omega\simeq-iDk^{2}$ with $D=c_{s}^{2}/\Gamma$; the DS mode is an
   \emph{isolated zero eigenvalue}. Only the former supports the KPZ
   reduction. \cite{Carusotto2013} \\
\bottomrule
\end{tabularx}
\end{table}

\subsection{Toward turbulence}
\label{sec:turbulence}

Towards turbulence, the correspondence is structural: the RJ-like spectral
profile, the existence of a cutoff supplied by dissipation, the two-scale
correlation hierarchy, and a directed energy flux across scales driven by the
chirp. The direction of the cascade---injection at the spectral centre,
dissipation at the wings---is the reverse of the two-dimensional hydrodynamic
case in which negative temperatures describe statistically independent vortices,
and this reversal is itself informative
\cite{Kalashnikov2025,Nazarenko2011,Kraichnan1980}. The self-organization
scenario of nonlinear wave turbulence, in which solitary structures grow by
absorbing power while radiating small-scale fluctuations until a single large
structure survives in a thermalized sea, provides the closest conservative image
of the DS fragmentation problem run in reverse
\cite{Dyachenko1992,Jordan2000,JordanJosserand2000}.

The filter-free solution of Section~4.1 turns this from an analogy into a
controlled experiment, because it removes the ultraviolet sink and leaves
everything else in place. Ref.~\cite{Kalashnikov2024b} reads its spectrum as carrying two cascades---a direct transport of energy towards $|\omega|\to\Delta$ and an inverse transport of spectral density towards $\omega=0$---and finds numerically that the first has nothing to stop it: a modulated pedestal grows on the spectral wings, with matching perturbation spikes on the pulse edges, and the occupied spectrum expands without bound, the more so when noise is added. The pulse is accordingly only conditionally stable. Restoring $\alpha$ arrests
the outflow, which is the sense in which the filter is the dissipative sink of
the cascade rather than a refinement of the pulse shape. This is the exact
complement of the anomalous-dispersion result of Section~6.5, where the filter supplies the occupied window $\omega_{\mathrm{cap}}$ outright: in normal dispersion it does not set the cutoff, which is kinematic. However, it is still what terminates the cascade and what makes the resulting spectrum thermalized. The two-cascade reading is thus better supported than the Rayleigh--Jeans profile
alone would suggest, and correspondingly the caveat below is sharpened rather than softened: the flux is real, and it is the \emph{stationarity} of the profile, not the flux, that the filter supplies.

The closest dynamical precedent, however, is the \emph{cycle of intermittency}
of optical wave turbulence. Because the point $k=0$ of the nonlinear
Schr\"odinger equation has finite capacity, the inverse particle flux fills it in
finite time and builds a condensate; the condensate is modulationally unstable,
and the instability spawns coherent collapsing filaments that carry the
particles back to the ultraviolet sink, whereupon the cycle repeats
\cite{Dyachenko1992,Newell2001}. Every element of that sequence has a DS
counterpart. The finite capacity of the spectral centre appears here as the
saturation $P_{0}\to\zeta^{-1}$ of Section~5.1; the accumulation at $k\to0$ is
the collapse $\Xi\to0$ of the Lorentzian width; and the role of the collapsing
filaments---coherent structures that transfer the accumulated invariant back to
small scales---is played by the multipulse complexes of Section~6.4. That
coherent, radiating pulses can themselves be the carriers of a turbulent flux,
rather than passive products of it, is an established result of the same school
\cite{Rumpf2009}. Closer still in form, though not in dynamics, is the spectral
incoherent soliton: a structure localized in the \emph{frequency} variable alone,
generated by Langmuir-type wave kinetics in a highly incoherent field
\cite{Picozzi2008SIS}. The DSR ``finger'' of Section~5.1 is its coherent,
dissipatively truncated counterpart, and the comparison is the natural one to
make when asking what part of the DS spectrum is genuinely kinetic in origin.

The boundary of the turbulence analogy should be stated as plainly as its
successes, because it is easy to overstate. The DS spectrum has the RJ
\emph{form}, but the RJ distribution is the \emph{zero-flux} stationary solution
of the wave-kinetic equation, and it is precisely for that reason of limited
relevance to nonequilibrium situations; the flux-carrying stationary solutions of
weak turbulence are the Kolmogorov--Zakharov spectra, which the DS spectrum is
not \cite{Nazarenko2011,Newell2001}. The resolution is that a DS is not a
statistically homogeneous random field at all. Its spectrum follows from a
stationary-phase evaluation of a \emph{coherent, phase-inhomogeneous} pulse
(Section~2.2), and its flux is carried in the time domain by the chirp rather
than in the frequency domain by resonant four-wave interactions. There is
accordingly no $H$-theorem behind the DS RJ profile: the identification of the
normalized spectrum with a distribution over microstates is a proposition
(Section~4.3), corroborated by kinetic theory rather than derived from it, and
the entropy of Section~6 must be checked dynamically for that reason.

\subsection{Toward the noise-driven theory of mode locking}
\label{sec:modelocking}

The closest conceptual precedent is the noise-driven statistical theory of mode locking because it asks the same physical question: how does a fluctuating, driven laser select a pulsed state from the available field configurations? In that theory, a genuine statistical ensemble and a potential can be defined, so pulse formation can be treated as a first-order transition \cite{GordonFischer2002,Gat2004}. The present DS construction is less complete: its entropy and temperature-like quantities are extracted from a stationary chirped pulse and no corresponding Gibbs measure has been established.

The NGD self-start results of Ref.~\cite{Kalashnikov2025} nevertheless make the connection more than formal. As energy increases, the pulse-number statistics shift toward multipulsing and the mean self-start time of the single DS ceases to improve. In the higher-energy range, the two-pulse state can appear faster. This provides a direct dynamical signature of a change in attractor preference. It is also what the thermodynamic-like indicators of Section~6 should seek to describe. Thus, the mode-locking phase-transition theory and the present DS framework are best viewed as complementary levels of description: the former supplies a true
statistical-mechanical construction under special assumptions, while the latter asks whether analogous state-selection signatures can be organized directly on a strongly chirped dissipative branch.

\subsection{Toward Bose--Einstein condensation: three obstructions}
\label{sec:bec}

Towards Bose--Einstein condensation, the correspondence is formal but
incomplete, and Table~\ref{tab:dictionary} allows the incompleteness to be
localized. Three obstructions have to be distinguished, because they have
different status: one is a sign, one is kinematic, and one is thermodynamic.

The first is the sign of the dispersive term. In normal GDD the mapping
\eqref{eq:convention} requires a negative effective mass---equivalently, after
conjugation, a repulsive interaction---so the DS inhabits the \emph{non-soliton}
sector of the corresponding nonlinear Schr\"odinger equation. Deleting the
dissipative terms destroys the structure outright: there is no conservative
limit on the branch of interest, and what survives the transfer is the coherence
structure rather than the ground state. The anomalous-dispersion regime restores
the sign correspondence, which is one motivation for treating both dispersion
signs within a single formalism \cite{Kalashnikov2026}.

The second obstruction appears to be purely kinematic, and if so it is
independent of dissipation altogether. Condensation at fixed power requires the
limit $\mu\to0^{-}$ to be attainable, i.e.\ the integral
$\mathcal{P}=\int g(\varepsilon)\,T/(\varepsilon-\mu)\,\mathrm{d}\varepsilon$ to
converge there, and this is a question about the density of states $g$. In a
graded-index fibre the group degeneracy grows linearly with the group index, so
$g(\varepsilon)\propto\varepsilon$ near the band edge, the integral converges,
and a macroscopic ground-state occupation appears at finite power---which is
what the experiments of Section~3.2 see \cite{Wu2019,Baudin2020,Mangini2022}.
The chirped DS spectrum, by contrast, is effectively one-dimensional with
$\varepsilon\propto\omega^{2}$, so that $g(\varepsilon)\propto\varepsilon^{-1/2}$
and the integral diverges as $|\mu|^{-1/2}$. The limit $\mu\to0^{-}$ therefore
cannot be reached at fixed power at all; it is reached only along a path on
which the power itself diverges. That is exactly DSR: for the truncated
Lorentzian of Section~2, $\mathcal{P}\propto\Xi^{-1}\arctan(\Delta/\Xi)$
diverges as $\Xi\to0$ with $\Delta$ saturating, and $-\mu_{\mathrm{shape}}=\Xi^{2}$ gives
precisely the $|\mu|^{-1/2}$ scaling. Read this way, the unbounded energy of DSR
is not an accident of the dissipative dynamics but the one-dimensional form of
the same statement that forbids condensation of a homogeneous
one-dimensional Bose gas; the dissipative dynamics decides only whether the
divergent path is dynamically accessible. It supplies, in addition, a second and
sharper reason---beyond the sign of the dispersive term---for the failure of the
BEC dictionary in normal dispersion. What the
state looks like \emph{along} that divergent path is the subject of
Section~\ref{sec:thomasfermi}.

The criterion is not new, and recognizing the obstruction as an instance
of it strengthens the statement. In the theory of laser light condensation, a
multimode cavity with noise and \emph{no} nonlinearity condenses when the
density of loss states makes $\int\rho(\varepsilon)\,\mathrm{d}\varepsilon/
(\varepsilon+\varepsilon_{0}-g)$ converge at $\varepsilon=0$; for a spectral
filter behaving as $\varepsilon\propto|\Omega|^{\eta}$ near its minimum this
gives $\rho\propto\varepsilon^{\xi}$ with $\xi=\eta^{-1}-1$, so that
condensation at finite power requires $\xi>0$, i.e.\ a filter sharper than
linear, $\eta<1$ \cite{FischerWeill2012}. The chirped DS carries a parabolic
spectral loss, $\eta=2$, hence $\xi=-\tfrac12$ and
$g(\varepsilon)\propto\varepsilon^{-1/2}$---the exponent obtained above from the
adiabatic spectrum---so the integral diverges and condensation at fixed power is
excluded by exactly the condition that admits it for a sufficiently sharp
filter. The second obstruction is therefore a special case of a criterion
already established for lasers, reached here from the shape of the DS spectrum
rather than from the shape of the filter, and the two routes agree on both the
exponent and the $|\mu|^{-1/2}$ rate of divergence.

The same work is a warning about vocabulary. Laser light condensation
delivers a chemical potential on a loss--gain rather than a photon-energy scale,
negative below threshold and vanishing at a sharp transition from multimode to
single-frequency oscillation, together with spectra that can be made to look
thermal---in a linear, classical, noise-driven system with no thermal
equilibrium, no interaction, and no bosonic statistics
\cite{FischerWeill2012}. A vanishing chemical potential and a sharply narrowing
spectrum are thus not sufficient evidence of condensation in the occupation
sense. This is the strongest available reason not to skip the test stated
next.

The third obstruction is the entropic one, and it is the one on which most
weight has been placed---so it should be stated in the narrower form that
Section~6.1 leaves standing. Spectral condensation raises rather than lowers the
\emph{configuration} entropy, and it does so by widening the scale separation
rather than by populating a bath at fixed cutoff (Sections~4.3 and 6.2): in a
Bose--Einstein condensate, and in RJ condensation at fixed $M$, occupation of
the lowest state suppresses $S$, whereas here $\Xi\to0$ at fixed $\Delta$
decouples the two scales and raises $H_{s}$. The full differential entropy of
the DS spectrum referred to a fixed bin does not behave that way: above
$\Delta/\Xi\simeq1.45$ it falls, as the condensate analogy would lead one to
expect. The genuine distinction is therefore not that DS condensation raises the
entropy \emph{tout court}, but that it possesses two competing entropy
contributions where a condensate at fixed mode number possesses only one.

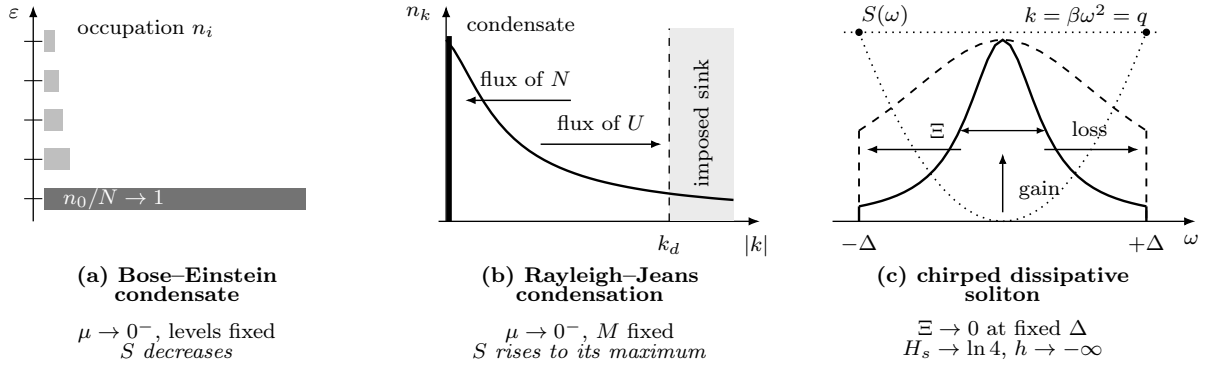
\begin{figure}[t]
\centering
\begin{tikzpicture}[x=1cm,y=1cm,
  ax/.style={draw=black,line width=0.5pt,-{Latex[length=1.5mm]}},
  cv/.style={draw=black,line width=0.9pt},
  fl/.style={draw=black,line width=0.55pt,-{Latex[length=1.4mm]}}]

\begin{scope}
  \draw[ax] (0,0) -- (0,2.85);
  \node[font=\scriptsize,anchor=east] at (-0.05,2.75) {$\varepsilon$};
  \foreach \i in {0,1,2,3,4}{
    \draw[line width=0.4pt] (-0.12,0.30+0.52*\i) -- (0.12,0.30+0.52*\i);}
  \fill[black!55] (0.14,0.16) rectangle (3.60,0.44);
  \foreach \i/\w in {1/0.34, 2/0.25, 3/0.19, 4/0.14}{
    \fill[black!25] (0.14,0.30+0.52*\i-0.14) rectangle (0.14+\w,0.30+0.52*\i+0.14);}
  \node[font=\scriptsize,anchor=west,text=white] at (0.26,0.30) {$n_{0}/N\to1$};
  \node[font=\scriptsize,anchor=west] at (0.45,2.55) {occupation $n_{i}$};
  \node[font=\scriptsize\bfseries,anchor=north,align=center] at (1.9,-0.45)
    {(a) Bose--Einstein\\[-2pt]condensate};
  \node[font=\scriptsize,anchor=north,align=center] at (1.9,-1.18)
    {$\mu\to0^{-}$, levels fixed\\[-2pt]$S$ \emph{decreases}};
\end{scope}

\begin{scope}[shift={(5.46,0)}]
  \draw[ax] (-0.1,0) -- (4.15,0);
  \draw[ax] (0,0) -- (0,2.85);
  \node[font=\scriptsize,anchor=north] at (4.10,-0.05) {$|k|$};
  \node[font=\scriptsize,anchor=east] at (-0.05,2.75) {$n_{k}$};
  \fill[black!8] (2.95,0) rectangle (3.80,2.55);
  \draw[dashed,line width=0.5pt] (2.95,0) -- (2.95,2.55);
  \node[font=\scriptsize,anchor=north] at (2.95,-0.05) {$k_{d}$};
  \node[font=\scriptsize,rotate=90] at (3.40,1.30) {imposed sink};
  \draw[cv] plot coordinates {(0.000,2.400) (0.100,2.272) (0.200,2.103) (0.300,1.920) (0.400,1.741) (0.500,1.577) (0.600,1.431) (0.700,1.302) (0.800,1.190) (0.900,1.092) (1.000,1.007) (1.100,0.932) (1.200,0.867) (1.300,0.809) (1.400,0.758) (1.500,0.712) (1.600,0.671) (1.700,0.634) (1.800,0.600) (1.900,0.570) (2.000,0.542) (2.100,0.517) (2.200,0.494) (2.300,0.473) (2.400,0.453) (2.500,0.435) (2.600,0.418) (2.700,0.402) (2.800,0.388) (2.900,0.374) (3.000,0.361) (3.100,0.349) (3.200,0.338) (3.300,0.328) (3.400,0.318) (3.500,0.308) (3.600,0.299) (3.700,0.291) (3.800,0.283)};
  \draw[line width=2pt] (0.04,0) -- (0.04,2.45);
  \node[font=\scriptsize,anchor=west] at (0.14,2.62) {condensate};
  \draw[fl] (1.65,1.60) -- (0.25,1.60);
  \node[font=\scriptsize,anchor=south] at (1.05,1.66) {flux of $N$};
  \draw[fl] (1.25,1.02) -- (2.85,1.02);
  \node[font=\scriptsize,anchor=south] at (2.05,1.08) {flux of $U$};
  \node[font=\scriptsize\bfseries,anchor=north,align=center] at (1.9,-0.45)
    {(b) Rayleigh--Jeans\\[-2pt]condensation};
  \node[font=\scriptsize,anchor=north,align=center] at (1.9,-1.18)
    {$\mu\to0^{-}$, $M$ fixed\\[-2pt]$S$ \emph{rises to its maximum}};
\end{scope}

\begin{scope}[shift={(10.92,0)}]
  \draw[ax] (-0.35,0) -- (4.45,0);
  \node[font=\scriptsize,anchor=north] at (4.40,-0.05) {$\omega$};
  \draw[dotted,line width=0.6pt] plot coordinates {(0.000,2.500) (0.158,2.101) (0.317,1.736) (0.475,1.406) (0.633,1.111) (0.792,0.851) (0.950,0.625) (1.108,0.434) (1.267,0.278) (1.425,0.156) (1.583,0.069) (1.742,0.017) (1.900,0.000) (2.058,0.017) (2.217,0.069) (2.375,0.156) (2.533,0.278) (2.692,0.434) (2.850,0.625) (3.008,0.851) (3.167,1.111) (3.325,1.406) (3.483,1.736) (3.642,2.101) (3.800,2.500)};
  \draw[dotted,line width=0.6pt] (-0.20,2.50) -- (4.00,2.50);
  \node[font=\scriptsize,anchor=west] at (-0.10,2.72) {$S(\omega)$};
  \node[font=\scriptsize,anchor=east] at (3.95,2.72) {$k=\beta\omega^{2}=q$};
  \fill (0,2.50) circle (1.4pt);
  \fill (3.80,2.50) circle (1.4pt);
  \node[font=\scriptsize,anchor=north] at (0,-0.05) {$-\Delta$};
  \node[font=\scriptsize,anchor=north] at (3.80,-0.05) {$+\Delta$};
  \draw[dashed,line width=0.7pt] plot coordinates {(0.000,1.200) (0.200,1.333) (0.400,1.478) (0.600,1.635) (0.800,1.798) (1.000,1.960) (1.200,2.113) (1.400,2.245) (1.600,2.342) (1.800,2.393) (1.900,2.400) (2.000,2.393) (2.200,2.342) (2.400,2.245) (2.600,2.113) (2.800,1.960) (3.000,1.798) (3.200,1.635) (3.400,1.478) (3.600,1.333) (3.800,1.200)};
  \draw[dashed,line width=0.7pt] (0,1.200) -- (0,0);
  \draw[dashed,line width=0.7pt] (3.80,1.200) -- (3.80,0);
  \draw[cv] plot coordinates {(0.000,0.198) (0.100,0.219) (0.200,0.243) (0.300,0.270) (0.400,0.303) (0.500,0.341) (0.600,0.387) (0.700,0.442) (0.800,0.508) (0.900,0.589) (1.000,0.687) (1.100,0.808) (1.200,0.957) (1.300,1.139) (1.400,1.356) (1.500,1.608) (1.600,1.879) (1.700,2.137) (1.800,2.328) (1.900,2.400) (2.000,2.328) (2.100,2.137) (2.200,1.879) (2.300,1.608) (2.400,1.356) (2.500,1.139) (2.600,0.957) (2.700,0.808) (2.800,0.687) (2.900,0.589) (3.000,0.508) (3.100,0.442) (3.200,0.387) (3.300,0.341) (3.400,0.303) (3.500,0.270) (3.600,0.243) (3.700,0.219) (3.800,0.198)};
  \draw[cv] (0,0.198) -- (0,0);
  \draw[cv] (3.80,0.198) -- (3.80,0);
  \draw[{Latex[length=1.2mm]}-{Latex[length=1.2mm]},line width=0.5pt]
    (1.33,1.20) -- (2.47,1.20);
  \node[font=\scriptsize,anchor=east] at (1.27,1.20) {$\Xi$};
  \draw[fl] (1.90,0.12) -- (1.90,0.90);
  \node[font=\scriptsize,anchor=west] at (1.98,0.42) {gain};
  \draw[fl] (2.45,0.95) -- (3.70,0.95);
  \draw[fl] (1.35,0.95) -- (0.10,0.95);
  \node[font=\scriptsize,anchor=south] at (3.05,1.00) {loss};
  \node[font=\scriptsize\bfseries,anchor=north,align=center] at (1.9,-0.45)
    {(c) chirped dissipative\\[-2pt]soliton};
  \node[font=\scriptsize,anchor=north,align=center] at (1.9,-1.18)
    {$\Xi\to0$ at fixed $\Delta$\\[-2pt]
      $H_{s}\to\ln4$, $h\to-\infty$};
\end{scope}
\end{tikzpicture}
\caption{Three condensations, with different entropy bookkeeping.
(\textbf{a})~Equilibrium Bose--Einstein condensation: as $\mu\to0^{-}$ the
ground level of a \emph{fixed} level structure acquires a macroscopic
occupation and the entropy falls. (\textbf{b})~Rayleigh--Jeans condensation in a
conservative multimode or wave-turbulent system: the inverse flux of the norm
accumulates at $k\to0$ while the direct flux of energy is removed by a sink at
$k_{d}$ imposed from outside---by discretization, by a finite beam area, or by
the waveguide bandwidth (Section~3.1); the mode basis is fixed and the entropy
rises to its equilibrium maximum. (\textbf{c})~Strongly chirped DS: the spectrum
is a Lorentzian of width $\Xi$ truncated at the cutoff $\Delta$, which is fixed
by the resonance of the DS wavenumber $q$ with the linear branch
$k=\beta\omega^{2}$ (dotted, Equation~\eqref{eq:cutoff}); gain enters at
$\omega=0$ and the chirp carries the flux outwards to the lossy wings.
Approaching DSR the spectrum passes from the fidelity condition $\Xi=\Delta$
(dashed) to the ``finger'' $\Xi\ll\Delta$ (solid) while $\Delta$ saturates, so
that the scale-separation index $r=\Delta/\Xi$ diverges. The shape indicator
rises to $H_{s}\to\ln4$, whereas the full fixed-bin differential entropy
contracts, $h\to\ln(4\pi\Xi/\delta\omega)\to-\infty$: the two conventions of
Section~6.1 disagree here, and the disagreement is the content of the panel. The
spectral offset vanishes in all three; only in (c) is the relevant scale
generated by the state, and its identification with a statistical mode count
remains conditional on the ensemble test of Section~4.3.}
\label{fig:threecondensations}
\end{figure}

A prior point of vocabulary belongs here as well. ``Condensation'' in the strict
sense is an occupation statement: the Penrose--Onsager criterion identifies a
condensate through a macroscopically large eigenvalue of the one-particle
density matrix \cite{PenroseOnsager1956}, not through a narrow spectral peak or
a vanishing fit parameter. The DS satisfies the second and has not been tested
against the first---which would require precisely the ensemble kernel
\eqref{eq:kernel} of Section~4.3. We use \emph{spectral condensation} as a
descriptive term for the narrowing at $\mu_{\mathrm{shape}}\to0^{-}$, and mark it
as such.
Dimensionality supplies an additional reason for that caution, and it is
one the driven-open literature has already tested. In a one-dimensional
driven-open geometry---a photonic wire---numerical solution of the same reduced
model shows that first-order coherence decays exponentially in space
\emph{even well above threshold}: at threshold the coherence length rises
sharply and reaches macroscopic values, but it does not diverge, so the state is
a quasicondensate and fails the Penrose--Onsager test outright
\cite{Carusotto2013}. Since the DS problem is one-dimensional, and since the
suppression of intensity fluctuations survives the reduced dimensionality while
the long-range phase order does not, the one quantity that a narrowing spectral
core is least entitled to certify is exactly the one the strict term would
require. This is a third independent reason---beside the linear
laser-condensation counterexample above and the thermalization-rate criterion
below---to keep \emph{spectral condensation} descriptive.
Where a driven-open photonic system falls between the two readings is a
quantitative question, and in a neighbouring system it has been answered. In a
dye-filled microcavity the photon distribution is of Bose--Einstein form only
while the phonon-assisted absorption and emission that thermalize it are fast
compared with the cavity loss; as the loss rate rises, or the cavity cutoff is
detuned far from the molecular line, thermalization fails, the distribution
departs from the equilibrium form, the macroscopic occupation moves into an
\emph{excited} mode, and the behaviour crosses over smoothly into that of a
conventional laser \cite{Kirton2013}. The control parameter is a ratio of a
thermalization rate to a loss rate, and the crossover is smooth rather than a
qualitative divide. A chirped-pulse oscillator sits far on the laser side of
that ratio: its per-round-trip loss is not small, and it has no thermalizing
bath playing the part of the phonons. That is a further reason to keep
\emph{spectral condensation} descriptive; it also says what would have to be
exhibited for the stronger term to be earned---a relaxation channel internal to
the pulse, fast compared with the cavity loss, and an occupation test to go with
it. Figure~\ref{fig:threecondensations} sets the three scenarios side
by side in the form in which the contrast matters.

The asymmetry of evidence should also be stated plainly: the conservative
classical analogue of BEC has been observed and measured \cite{Sun2012},
whereas the DS counterpart is at present supported by closed-form theory,
stochastic simulation, and the three spectral signatures of Section~5
\cite{Rudenkov2023,Kalashnikov2025}.

\subsection{Dissipative soliton resonance as a Thomas--Fermi expansion}
\label{sec:thomasfermi}

The dictionary of Section~\ref{sec:dictionary} does more than locate the failures of the BEC
analogy; it also identifies what the DS \emph{is} along the divergent path
opened by the density-of-states argument, and the answer is a familiar object.
On the vacuum-stability border the adiabatic theory gives $E\to\infty$ with
$P_{0}\to\zeta^{-1}$, $\Delta^{2}\to\gamma/\beta\zeta$ and $\Xi\to0$
(Section~5.1). Under Table~\ref{tab:dictionary} this is a condensate at
\emph{fixed density} whose volume grows with the particle number: the peak power
saturates at the value set by the analogue of the saturation density $n_{s}$,
the effective chemical potential $-\mu_{\mathrm{shape}}=\Xi^{2}$ vanishes, and the added energy is
accommodated entirely by temporal stretching. The flat-topped, linearly chirped
DSR pulse is, in this reading, the optical Thomas--Fermi profile, and the front
picture of Section~5.1---a plateau of fixed height joined to the vacuum by two
edges of energy-independent width---is its natural companion: in a Thomas--Fermi
condensate the interior is set by the local balance $\mu_{\mathrm{TF}}=gn$ and
the essential spatial structure is concentrated in the boundary layer, which mirrors
the division of labour between the spectral core, which supplies $\Xi$ and
collapses, and the pulse edges, which supply $\Delta$ and do not.

Two things follow that are worth stating separately from the analogy itself.
The first is a design statement. If DSR is an expansion at constant density,
then the quantity that fixes the attainable peak power is not the pumping but
the saturation parameter $\zeta$, and energy is harvested by lengthening the
pulse rather than by intensifying it---which is the operational content of the
transition from squeezing to stretching listed among the DSR signatures in
Section~5.1, and the reason the repetition rate, rather than the average
power, is the effective scaling knob for a chirped-pulse oscillator.

The second is that this reading makes the role of the quintic term transparent.
With $g_{3}=\chi$ from \eqref{eq:convention}, SPM saturation is a
\emph{repulsive} three-body correction; in anomalous dispersion, where the
two-body interaction is attractive, it is exactly the ingredient that bounds the
density from above and so permits a plateau to exist at all. The numerical
finding that anomalous-dispersion DSR requires a sufficiently strong quintic
reactive nonlinearity \cite{Chang2009,GreluDSR2010}, and the analytical
condition of Section~5.2 that the resonance locus fall inside the adiabatic
existence window, are then the optical statement of a mechanism standard in
attractive-condensate physics: an attractive condensate has no stable
finite-density branch until a repulsive higher-order term supplies one. In
normal dispersion the two-body term is already repulsive, no such condition
arises, and DSR is intrinsic to the chirped branch---which is precisely the
asymmetry established in Section~5.3.

\subsection{Toward driven-open condensates: coherence, universality, and
symmetry}
\label{sec:drivenopen}

The honest statement about the condensate analogy is that it holds at the level
of \emph{coherence structure under a finite bandwidth}---the ultraviolet cutoff
used to regularize first-order correlation functions in driven-open condensates
plays the same role as the graining scale here, just as $k_{c}$ regularizes the
RJ integrals in classical wave condensation---rather than at the level of
microscopic kinetics
\cite{Picozzi2007,Kalashnikov2026,deLeeuw2014,Carusotto2013}. That places DS
within the broader programme of driven open quantum matter, in which detailed
balance is broken by the simultaneous presence of coherent evolution and
drive/dissipation, and in which universal coherence scaling is the object of
study \cite{Sieberer2025}. Three consequences of the identification made in
Section~\ref{sec:dictionary} are worth drawing out, because they turn that placement into
concrete questions.

\emph{Universality class.} In one dimension the phase of a driven-open
condensate obeys a Kardar--Parisi--Zhang equation, so that its first-order
coherence decays as a stretched exponential with KPZ exponents rather than
algebraically; this has been established theoretically \cite{He2015,Sieberer2025}
and observed in a one-dimensional polariton condensate
\cite{Fontaine2022}; the phase statistics on which such identifications rest
have been mapped numerically \cite{Squizzato2018}, and the field-theoretic
setting is that of driven-dissipative criticality with a nonconserved order
parameter \cite{TauberDiehl2014}. The
model on which those results rest is the quartic-truncated driven-dissipative
GPE---that is, \eqref{eq:cqgle} in anomalous dispersion, without spectral
truncation and on an extended rather than a localized branch \cite{He2015}. The
strongly chirped DS and the KPZ phase of a one-dimensional driven-open
condensate are therefore best described as potentially related asymptotic
regimes of a broader driven nonlinear-field class, rather than as two branches
of one equation. The qualification is not pedantic: the KPZ reduction is a
stochastic, long-wavelength statement about an \emph{extended} phase field,
derived with the noise term retained and the amplitude eliminated, whereas the
DS is a deterministic localized solution of the noise-free equation, and nothing
in the derivation of the former applies to the latter without a fresh
long-wavelength expansion about the pulse. Whether the DS correlation function
of Section~4.1---once promoted to a genuine $g^{(1)}$ by the ensemble of
Section~4.3---carries any trace of the KPZ scaling forms, and whether the
stretched-exponential decay and its exponents survive on a localized branch, is
an open question; the self-similarity \eqref{eq:selfsimilar} is suggestive and
no more, and we return to it in Section~8.

\emph{The mode that carries it, and why the DS has no counterpart of it.}
The KPZ reduction is not an assertion about a generic driven field; it descends
from a specific and well-characterized feature of the linearized problem, and
naming that feature makes the gap to the DS sharper than the qualification just
given. Linearizing \eqref{eq:ddgpe} about a spatially homogeneous condensate at
rest gives an elementary-excitation spectrum
\begin{equation}
  \omega_{\mathrm{B}}(k) = -\frac{i\Gamma}{2}
  \pm \sqrt{\varepsilon_{k}\big(\varepsilon_{k}+2gn\big)-\frac{\Gamma^{2}}{4}} ,
  \qquad
  \varepsilon_{k}=\frac{\hbar^{2}k^{2}}{2m} ,
  \label{eq:diffusivebog}
\end{equation}
in which $\Gamma$ is an effective damping rate set by the distance from the
condensation threshold, $\Gamma=\gamma\,(P/P_{c}-1)/(P/P_{c})$
\cite{WoutersCarusotto2007,Carusotto2013}. Below the crossover wavenumber
$k_{0}\simeq\Gamma/2c_{s}$---the point at which the equilibrium Bogoliubov
branch crosses $\Gamma/2$, with $c_{s}^{2}=gn/m$---the radicand is negative,
$\mathrm{Re}\,\omega_{\mathrm{B}}$ vanishes identically, and the two branches
separate in their imaginary parts. Expanding there gives the two modes that
matter:
\begin{equation}
  \omega_{+}(k)\simeq -\,i\,\frac{c_{s}^{2}}{\Gamma}\,k^{2} ,
  \qquad
  \omega_{-}(k)\simeq -\,i\Big(\Gamma-\frac{c_{s}^{2}}{\Gamma}k^{2}\Big) .
  \label{eq:diffusivelimit}
\end{equation}
The neutral branch $\omega_{+}$ is \emph{diffusive}, not sonic: a phase
perturbation does not propagate as a sound wave but relaxes according to
$\partial_{T}\varphi=D\nabla^{2}\varphi$ with $D=c_{s}^{2}/\Gamma$. The second
branch tends to the finite, purely imaginary value $-i\Gamma$ at $k=0$; its
eigenvector is a spatially uniform density fluctuation that leaves the phase
untouched, and it is the mode that in the laser literature governs
relaxation-oscillation and intensity-fluctuation dynamics.\footnote{We
have verified \eqref{eq:diffusivelimit} symbolically and numerically against
\eqref{eq:diffusivebog}: at $m=g=n=1$, $\Gamma=0.4$ and $k=10^{-3}$ the exact
$\mathrm{Im}\,\omega_{+}=-2.500016\times10^{-6}$ against
$-c_{s}^{2}k^{2}/\Gamma=-2.5\times10^{-6}$, and the sonic estimate
$k_{0}=\Gamma/2c_{s}$ reproduces the exact crossing of
\eqref{eq:diffusivebog} to better than $0.5\%$ for $\Gamma/gn\le0.4$,
degrading to $6\%$ at $\Gamma/gn=1.5$.}

Equation \eqref{eq:diffusivelimit} is the linear precursor of the KPZ
equation quoted above: adding the noise omitted in \eqref{eq:ddgpe} and the
leading nonlinearity $(\nabla\varphi)^{2}$ to the diffusion law turns
$\partial_{T}\varphi=D\nabla^{2}\varphi$ into the KPZ equation, which is why the
universality class is a property of \emph{incoherently pumped} condensates and
not of condensates in general. Stated that way, the disanalogy with the DS
becomes structural rather than a matter of caution. What the KPZ reduction
requires is a \emph{gapless continuum} of phase modes---a branch whose damping
rate vanishes as $k^{2}$ over an extended phase field. A localized DS has no
such continuum. Its neutral modes are the \emph{discrete} zero modes of a
solitary wave: the global phase, guaranteed by the $U(1)$ invariance discussed
below, and the timing, guaranteed by translation in $t$; everything else in the
Bogoliubov problem \eqref{eq:bogoliubov} is either gapped or belongs to the
continuous spectrum of the \emph{background}, whose low-frequency damping rate
is $\sigma$ and not $Dk^{2}$ by \eqref{eq:lindisp}. Two zero modes are not a
hydrodynamic phase field, and no diffusion constant can be read off them. The
correspondence therefore fails at a nameable place: not at the equation, which
is shared, but at the branch, and specifically at the absence on the localized
branch of the gapless diffusive mode from which the whole KPZ construction is
built. That is the sharpest available form of the qualification made in the
preceding paragraph, and it also says what a positive result would have to look
like---a long-wavelength expansion about the pulse that produces a genuinely
gapless mode, presumably associated with a pulse train or a continuum of
interacting pulses rather than with a single one.

\emph{The exit from coherence.} KPZ order in one dimension is destroyed not by a
gradual loss of correlation but by the proliferation of space--time vortices, a
transition into a defect-dominated ``vortex turbulence'' phase \cite{He2017}.
The DS fragmentation of Section~6.4, in which a multipulse state of higher
entropy proxy supersedes the single-pulse state, resembles that scenario in one
respect: in both cases a coherent driven state is destroyed by the appearance of
discrete objects rather than by gradual dephasing, and in both the control
parameter is the distance from the condensation threshold. The resemblance stops
well short of a mapping. No defect variable has been identified on the DS side,
the pulses of a complex carry no topological charge, and the space--time
vortices of Ref.~\cite{He2017} live in an extended phase field with no localized
counterpart; nothing here converts pulse number into defect number. The parallel
is therefore a statement of what a defect-counting description of multipulsing
would have to reproduce, not evidence that one exists.

\emph{Symmetry, and what ``condensation'' can mean.} Equation~\eqref{eq:cqgle}
with saturable gain is invariant under $a\to a\,e^{i\phi}$: the pump is
incoherent, the phase is selected spontaneously, and a neutral phase mode
exists---exactly the situation of an incoherently pumped polariton condensate,
and the reason the condensate vocabulary is applicable at all.
What that neutral mode \emph{is} differs between the two, and the
difference has just been located: in the extended condensate it is the gapless
diffusive branch \eqref{eq:diffusivelimit}, whereas here it is an isolated zero
eigenvalue of \eqref{eq:bogoliubov}. The symmetry is shared; the mode structure
it produces is not. The inference
runs one way only: a neutral global phase follows from $U(1)$ invariance quite
generally, so it is necessary for a condensate and not sufficient, and the
occupation criterion that would be sufficient is the one already stated in
Section~\ref{sec:bec}. The symmetry argument licenses the vocabulary and settles nothing
about the state. Coherently driven
systems---the Lugiato--Lefever description of Kerr microresonators, or a seeded
oscillator, in which an external term $S(t)$ is added to \eqref{eq:cqgle}---have
their phase pinned to the drive: the $U(1)$ symmetry is explicitly broken, no
Goldstone mode exists, and the transition belongs to a different class. The
seeded case, in which a weak coherent injection reshapes the stability islands
of the DS \cite{KalashnikovDDSR2025}, is the natural bridge between the two, and
a reminder that the thermodynamic reading is contingent on the symmetry of the
drive.

Finally, the family of dissipative condensate equations settles, more sharply
than any general disclaimer can, the status of the equilibrium criteria
discussed in Section~6.5. Table~\ref{tab:dgpe} arranges that family by what it
conserves and what it minimizes. The extreme case is the metriplectic
construction: replacing the Poisson bracket by a metriplectic one built on the
projector $\hat{Q}=1-|\psi\rangle\langle\psi|/\|\psi\|^{2}$ yields a dissipative
GPE which conserves the norm \emph{exactly} while the free-energy functional
decreases monotonically, $\mathrm{d}\mathcal{F}/\mathrm{d}t\le0$, so that the
stationary states are the extrema of $\mathcal{F}$ at fixed particle number:
black solitons, which are such extrema, are untouched by the damping, whereas
grey solitons, which are not, decay \cite{Pawlowski2017}. There, free-energy
minimization is a rigorous selection principle. Equation~\eqref{eq:cqgle} has
neither ingredient---the norm is not conserved but fixed a posteriori by the
gain--loss balance, and the dissipative part of the vector field is not generated
by a symmetric bracket acting on an entropy functional---so no monotone
functional is available. This is the structural counterpart of the two
statements made in Section~6.5: that a variational formulation exists but its
minimization does not select the state \cite{Ankiewicz2007}, and that the
concavity theorem of the conservative diagram-of-invariants construction
\cite{Akhmediev1999HQ} has no analogue here. What replaces them is the entropy
and slope indicators of Section~6, together with the dynamical
corroboration that landscape requires.

\begin{table}[t]
\caption{The chirped DS in the family of dissipative condensate models, ordered
by what survives of the equilibrium apparatus. $N$ is the norm (optical power or
particle number), $\mathcal{F}$ the free-energy (Lyapunov) functional. The last
row is Equation~\eqref{eq:cqgle}.}
\label{tab:dgpe}
\centering
\footnotesize
\begin{tabularx}{\textwidth}{@{}p{3.2cm} p{3.5cm} p{1.6cm} p{1.9cm} X@{}}
\toprule
\textbf{Model} & \textbf{Structure} & \textbf{$N$} & \textbf{$\mathcal{F}$} &
\textbf{What selects the stationary state} \\
\midrule
GPE / NLSE
& Hamiltonian
& conserved & conserved
& conservative dynamics; a one-parameter soliton family \\[0.3em]
Pitaevskii-damped GPE
  \cite{Pitaevskii1959,Choi1998}
& $i\partial_{T}\psi=(1-i\lambda_{\mathrm{P}})[\hat{H}_{\mathrm{GP}}-\mu]\psi$
& decays at rate $\lambda_{\mathrm{P}}$ & $\mathrm{d}\mathcal{F}/\mathrm{d}t\le0$
& minimization of $\mathcal{F}$; source of the $\alpha\omega^{2}$ damping
  correspondence, $|\lambda_{\mathrm{P}}|=|\alpha/\beta|$ \\[0.3em]
Metriplectic dissipative GPE
  \cite{Pawlowski2017}
& Poisson bracket replaced by a metriplectic one with projector $\hat{Q}$
& \emph{exactly conserved} & $\mathrm{d}\mathcal{F}/\mathrm{d}t\le0$
& extrema of $\mathcal{F}$ at fixed $N$: the case in which the equilibrium
  criterion survives dissipation \\[0.3em]
Stochastic (projected) GPE
  \cite{Proukakis2008}
& damping and noise related by a fluctuation--dissipation relation
& fluctuating &---
& detailed balance; thermal equilibrium at $(T,\mu)$ \\[0.3em]
\textbf{Driven-dissipative GPE $\longleftrightarrow$ CQGLE-type reduced
  description}
  \cite{WoutersCarusotto2007,He2015}
& saturable gain, loss, energy relaxation; no potential
& fixed by gain $=$ loss & none
& dynamical attractor selection. The mapping is a reduction, not an
  identity: it depends on the reservoir, saturation, trapping, dimensionality
  and noise assumptions listed in Table~\ref{tab:mappingconditions}, and the
  cited polariton models are generalized Gross--Pitaevskii descriptions rather
  than the same equation. The entropy- and slope-based crossings are candidate
  diagnostics constructed in this review; direct dynamical branch-selection
  validation remains open, and the references in this row do not supply
  it \\
\bottomrule
\end{tabularx}
\end{table}

\section{Conclusions, Open Questions, and Outlook}

\subsection{What the framework establishes}

The central conclusion of this work is not that a DS should be regarded as an equilibrium thermodynamic system. It is that several pieces
of DS physics can be organized consistently once three levels are kept
separate: the \emph{existence} of a stationary solution, its \emph{dynamical robustness}, and its \emph{noise-driven accessibility}. The adiabatic theory addresses the first; the AGD survival maps address the second; the NGD pulse-number and self-start statistics address the third. Thermodynamic-like quantities are useful insofar as they help relate these levels rather than replace their distinct dynamical meanings.

Within that interpretation, the main results are:
\begin{enumerate}[leftmargin=*,itemsep=0.3em]
\item Strong chirp produces a genuine two-scale internal structure. In NGD the short scale is tied to the intrinsic spectral edge, whereas in AGD
the absence of a hard cutoff makes the corresponding scale operational
and dissipation-dependent. This difference must be retained in any
unified discussion of the two dispersion regimes.
\item The ratio of the two scales is a useful structural coordinate of the
DS. Interpreting it as an effective number of internal degrees of
freedom is an additional hypothesis, not a prerequisite for the
adiabatic theory or for the observed dynamical trends.
\item The entropy-, internal-energy- and temperature-like quantities derived from the spectrum are branch and path diagnostics. In NGD, they identify crossovers that occur before the ideal DSR asymptotic is reached and can be compared with the observed growth of multipulse accessibility. They should not be described as equilibrium state functions or as a proven cause of pulse breakup.
\item The NGD stochastic results provide the clearest dynamical support for the framework: with increasing energy, probability moves toward
multipulse states and the self-start kinetics show that a two-pulse
attractor can become easier to reach than the single-pulse one. The
physically relevant limit can therefore precede the loss of stationary
single-pulse existence.
\item AGD gives a complementary result. The analytical branch can exist
beyond the region in which it is robust to noise, but the studied
continuation shows no counterpart of the NGD entropy turnover. The AGD
results therefore argue against a universal thermodynamic cutoff and in
favor of a regime-dependent competition between existence, robustness,
and state selection.
\item The analogies with wave turbulence, mode-locking thermodynamics, BEC, and driven-open condensates remain useful when they are used to identify common organizing principles---spectral redistribution, scale separation, metastability and competing macrostates---rather than to claim literal thermodynamic equivalence.
\end{enumerate}

\subsection{Open questions}

Four conceptual questions remain.

\emph{(i) What is the physical status of the ``thermodynamic'' variables?}
For a deterministic DS, entropy and temperature-like quantities are constructed
from a chosen spectral description and, in the present formulation, depend on
the continuation path. Their strongest possible role is therefore predictive:
if they consistently anticipate a change in attractor preference, they provide
a useful nonequilibrium thermodynamic language even without becoming equilibrium
state functions.

\emph{(ii) Does the two-scale structure acquire a statistical meaning?}
The separation of correlation scales is established by the adiabatic solution.
What is not established is that the ratio of those scales counts statistically
independent internal degrees of freedom. A future ensemble-coherence
measurement can decide that question. Failure of such a calibration would
weaken the quasiparticle language, but it would not invalidate the two-scale DS
structure itself.

\emph{(iii) How fundamental is the NGD--AGD asymmetry?}
NGD combines scale separation with a stochastic crossover toward multipulsing
and with entropy-like turnovers along the continuations studied here. AGD, by
contrast, presently shows a robust-region boundary without an analogous
thermodynamic turnover. Whether this difference reflects a genuinely different
energy-limiting mechanism or only the particular continuation and stability test
used so far remains open.

\emph{(iv) What actually selects the breakup channel in a laser?}
Multipulsing is only one possible termination of energy scaling. The decisive
experiment or simulation is therefore one that follows the laser from noise or
from a controlled perturbation and identifies which attractor is reached, how
its formation time changes, and whether the thermodynamic-like indicators track
that change. The NGD self-start results already provide a partial example of
such a test; an analogous attractor-selection study in AGD would be especially
informative.

\subsection{Outlook}

Three directions seem most promising to us, and they are quite different in character.

The first direction is experimental and dynamical. The central comparison is not between two entropy conventions but between structural indicators and actual state selection. Shot-resolved spectra can follow the evolution of the spectral shape, while simultaneous pulse-number statistics and self-start or transition times show which attractor the laser selects. Repeating the NGD logic of Ref.~\cite{Kalashnikov2025} across the predicted crossover, and extending it to AGD, would provide a direct test of whether the thermodynamic-like variables carry predictive content. Phase-sensitive coherence measurements would be a valuable second step if one wishes to test the stronger claim that the two-scale ratio corresponds to an effective statistical degree count.

The second direction concerns design. Read as a Thomas--Fermi expansion (Section~\ref{sec:thomasfermi}), DSR is an expansion at approximately constant peak density: the gain-saturation parameter limits the peak power, while additional energy is accommodated primarily by pulse lengthening. For the class of gain-clamped oscillators considered here, reducing the repetition rate can therefore provide an effective route to higher single-pulse energy at constrained average power, although this is not a universal principle of dissipative lasers. The fidelity curve $\Xi=\Delta$ marks both entry into the scale-separated subregion in which DSR is approached and a locus of favorable external compressibility. The latter statement has the local adiabatic meaning established in Section~2.2: the spectral-phase curvature \eqref{eq:specphase2} is stationary at zero detuning there; it is not a global optimization theorem for the compressed pulse duration. Finally, if the entropy proxy survives the tests proposed in Section~8.2, its crossing in Section~6.4 may delimit the useful operating region before the stationary branch loses existence. Whether these criteria transfer quantitatively to all-normal-dispersion fiber lasers, whose parameters occupy different regions of the master diagram, requires a systematic parametric study.

The third direction is conceptual and extends beyond laser physics. The most interesting feature of the strongly chirped DS is that its internal scale separation changes with the state while the system remains driven and dissipative. This makes the DS a useful model for asking a broader nonequilibrium question: when can structural measures derived from a coherent nonlinear object act like thermodynamic coordinates for the competition between attractors? The answer need not be a literal temperature or entropy in the equilibrium sense. A successful framework would organize measurable changes in accessibility, metastability, and fragmentation across different driven nonlinear systems.

\section*{Acknowledgements}
 This work was supported by Norges Forskningsr{\aa}d (\#303347 (UNLOCK), \#326503 (MIR)), and ATLA Lasers AS. The authors thank Dr. Alexander Rudenkov and Prof. Evgeni Sorokin for the fruitful discussions.

\end{document}